\documentclass[aps,twocolumn,prx,reprint,a4paper,showpacs,amsmath,amssymb,superscriptaddress,floatfix,footinbib]{revtex4-2}

\usepackage{hyperref}
\hypersetup{colorlinks,linkcolor=blue,urlcolor=blue,citecolor=blue}

\usepackage{bbm}

\usepackage[normalem]{ulem}
\usepackage{tabularx}
\usepackage{graphics,graphicx}
\usepackage{epsfig}
\usepackage{xcolor}
\usepackage{float}
\usepackage{adjustbox}

\definecolor{pythonBlue}{rgb}{0.12156862745098039, 0.4666666666666667, 0.7058823529411765}
\definecolor{pythonOrange}{rgb}{1.0, 0.4980392156862745, 0.054901960784313725}
\definecolor{pythonGreen}{rgb}{0.17254901960784313, 0.6274509803921569, 0.17254901960784313}

\usepackage{dsfont}

\usepackage{amsmath,amssymb,mathrsfs}
\usepackage{physics}

\definecolor{pink2}{RGB}{254, 1, 154}

\newcommand{\ci}{\mathrm{i}}
\newcommand{\me}{\mathrm{e}}
\newcommand{\pmt}{t^{>}}

\begin{document}
\title{Localization, fractality, and ergodicity in a monitored qubit}

\author{Paul P{\"o}pperl}
\affiliation{\mbox{Institute for Quantum 
Materials and Technologies, Karlsruhe 
Institute of Technology, 76021 Karlsruhe, 
Germany}}
\affiliation{\mbox{Institut f\"ur Theorie 
der Kondensierten Materie, Karlsruhe 
Institute of Technology, 76128 Karlsruhe, 
Germany}}
\author{Igor V. Gornyi}
\affiliation{\mbox{Institute for Quantum 
Materials and Technologies, Karlsruhe 
Institute of Technology, 76021 Karlsruhe, 
Germany}}
\affiliation{\mbox{Institut f\"ur Theorie 
der Kondensierten Materie, Karlsruhe 
Institute of Technology, 76128 Karlsruhe, 
Germany}}
\author{David B. Saakian}
\affiliation{\mbox{A.I. Alikhanyan National Science Laboratory (Yerevan Physics Institute) Foundation,
Yerevan 375036, Armenia}}
\author{Oleg M. Yevtushenko}
\affiliation{\mbox{Institut f\"ur Theorie 
der Kondensierten Materie, Karlsruhe 
Institute of Technology, 76128 Karlsruhe, 
Germany}}

\begin{abstract}

We study the statistical properties of a single two-level system (qubit) subject to repetitive ancilla-based measurements. This setup is a fundamental minimal model for exploring the intricate interplay between the unitary dynamics of the system and the nonunitary stochasticity introduced by quantum measurements, which is central to the phenomenon of measurement-induced phase transitions. We demonstrate that this ``toy model’’ harbors remarkably rich dynamics, manifesting in the distribution function of the qubit’s quantum states in the long-time limit. We uncover a compelling analogy with the phenomenon of Anderson localization, albeit governed by distinct underlying mechanisms. Specifically, the state distribution function of the monitored qubit, parameterized by a single angle on the Bloch sphere, exhibits diverse types of behavior familiar from the theory of Anderson transitions, spanning from complete localization to almost uniform delocalization, with fractality occurring between the two limits. By combining analytical solutions for various special cases with two complementary numerical approaches, we achieve a comprehensive understanding of the structure delineating the ``phase diagram'' of the model. We categorize and quantify the emergent regimes and identify two distinct phases of the monitored qubit: ergodic and nonergodic. Furthermore, we identify a genuinely localized phase within the nonergodic phase, where the state distribution functions consist of delta peaks, as opposed to the delocalized phase characterized by extended distributions. Identification of these phases and demonstration of transitions between them in a monitored qubit are our main findings.

\end{abstract}

\maketitle

\section{Introduction}

Statistical properties of random systems constitute one of the most outstanding topics of theoretical physics, which keeps attracting a lot of attention, in spite of a long history. Such an interest results, in particular, from the beauty and fundamental importance of the physics of phase transitions. One seminal example is the theory of the Anderson transition between localized and delocalized phases \cite{Anderson58, evers_2008}. Its archetypal setup involves noninteracting particles in a random potential.
When strong disorder is introduced in the system the wave functions of particles become spatially localized. Conversely, for weak disorder, the wave functions may spread uniformly throughout the system, exhibiting extended behavior. 
The transition between the two phases is characterized by critical fluctuations of wave functions, leading to the formation of a multifractal spectrum.

Recently, a similar type of phase transition governed by randomness has been discovered in hybrid random quantum circuits, subject to both unitary evolution and local measurements (monitoring) \cite{Li2018a, Skinner2019a, Chan2019a, Szyniszewski2019a, Fisher2022}. 
In monitored systems, randomness is introduced by the quantum probabilistic nature of measurement outcomes rather than by the disordered potential. The stochastic nonunitary evolution of the system emerges from the inherently projective nature of measurements.
The influence of measurements on the dynamics of quantum systems has attracted considerable attention, largely because of recent advances in the field of quantum information processing. Regardless of the particular quantum hardware, the nuisance of environmental noise is a formidable challenge \cite{Aharonov2000a, Preskill2018a, Bharti2022}. In this context, it is crucial that measurements can act both as a tool to monitor the properties of a quantum system and as a source of controllable disturbances.  

It is the interplay of the unitary evolution of the system with the nonunitary one governed by measurements that gives rise to measurement-induced entanglement phase transitions, a phenomenon originally predicted and studied in the context of quantum circuits \cite{Li2018a, Skinner2019a, Chan2019a, Szyniszewski2019a, Li2019a,  Bao2020a, Choi2020a, Gullans2020a, Gullans2020b, Jian2020a, Zabalo2020a, Iaconis2020a, Turkeshi2020a, Zhang2020c, Nahum2021a, Ippoliti2021a, Ippoliti2021b, Lavasani2021a, Lavasani2021b, Sang2021a, Fisher2022, Block2022a, Sharma2022, Jian2023, Kelly2023}. These transitions and related phenomena were subsequently investigated in a variety of models, including free fermionic systems \cite{Cao2019a, Alberton2021a, Chen2020a, Tang2021a, Coppola2022, Ladewig2022, Carollo2022, Buchhold2022, Yang2022, Szyniszewski2022, Buchhold2021a, Buchhold2021a, VanRegemortel2021a, Youenn2023, Loio2023, Turkeshi2022b, Kells2023, Fava2023, Swann2023, Merritt2023, Poboiko2023, poboiko2023a, chahine2023, jin2023}, Ising spin systems \cite{Lang2020a, Rossini2020a, Biella2021a, Turkeshi2021, Tirrito2022, Yang2023, Weinstein2023, Murciano2023, Sierant2022a, Turkeshi2022a}, various models with interactions  \cite{Tang2020a, Goto2020a, Fuji2020a, Jian2021a, Altland2022, Doggen2022a, Doggen2023}, and disordered systems exhibiting Anderson or many-body localization \cite{Szyniszewski2022, Lunt2020a, Paul2023, Yamamoto2023}. Recent experimental works have reported on measurement-induced phase transitions in systems with trapped ions \cite{Noel2022a} and in superconducting qubit arrays \cite{Koh2022, Hoke2023}.

A direct link between measurement-induced entanglement transitions and the phenomenon of Anderson localization was recently established in Refs.~\cite{Fava2023, Swann2023, Poboiko2023, poboiko2023a, chahine2023} by deriving nonlinear sigma-models for monitored free fermions. These field theories share significant similarities~\cite{Poboiko2023} with those used in the context of Anderson localization. Particularly noteworthy is the striking resemblance between the predicted entanglement transition for free fermions in spatial dimensions larger than one ($D>1$) and the Anderson transition observed in disordered systems of dimension $D+1$ \cite{poboiko2023a, chahine2023}. In the field of measurement-induced dynamics, the entanglement transition can be considered as a ``metal-insulator transition'' for quantum information. In this analogy \cite{poboiko2023a}, mutual information serves as a counterpart to dimensionless conductance, shedding light on the evolving properties of the quantum system under the influence of measurements. 

Importantly, both the Anderson transition and the measurement-induced entanglement transition are phenomena that take place in infinite systems (thermodynamic limit). The main question we are addressing here pertains to whether a macroscopic spatial size or a macroscopically large number of degrees of freedom is a necessary condition for the observation of complex behaviors driven by \textit{randomness}, thereby leading to transitions between distinct phases. More specifically, 
our primary focus lies in ascertaining whether a transition akin to the Anderson localization-delocalization transition can be driven by repeated measurements in a \textit{microscopic} quantum system (cf. Ref.~\cite{snizhko_2020}).

In the present paper~\footnote{In parallel to this work, a monograph containing a chapter about the project was written and submitted as a doctoral thesis by P.P.~\cite{thesis_paul}}, we come across such complexity even in a single monitored two-level system---a qubit (e.g., a Loss-DiVincenzo spin qubit \cite{Loss1998}), i.e., in the smallest possible quantum system with nontrivial dynamics. The thermodynamic limit that is necessary for true phase transitions can be reached here in the limit of infinite observation time and, correspondingly, an infinite number of measurements. 
Our setup comprises two two-level systems interacting with each other, one of them representing the qubit, while the other serves as the detector, see Sec.~\ref{sect:model}. 
This is arguably the simplest model implementing ``ancilla-based'' generalized measurements.
Experimentally, monitoring and steering of qubit quantum trajectories has become feasible~\cite{qubit_quantum_traj_exp,quantum_jumps,quantum_trajectory_tracking_nn,topological_transition_weak_measurements_experiment,topological_transitions_optical_experiment}.

The concept of variable-strength (generalized) measurements, involving the coupling of a system to a two-state detector (ancilla) followed by a projective measurement of the detector, boasts a rich heritage across a variety of topics, such as dephasing, weak values, counting statistics, quantum control and state engineering, as well as already mentioned measurement-induced entanglement transitions \cite{Gurvitz1997, Korotkov1999, Korotkov2001, Goan2001, Shpitalnik2008, Romito2008, Taranko2012, li_2014, Barbarino2019, Bao2020a, Esin2020, snizhko_2020, Kumar2020, Roy2020, Turkeshi2021, Herasymenko2021, Doggen2022a, dubey_2023, Tirrito2022, Friedman2022, Rosenow2022, Silveri2023, Doggen2023, MedinaGuerra2023, Morales2023}. Historically, a setup with a two-site system monitored by a detector first appeared, perhaps, in Refs.~\cite{Gurvitz1997} and \cite{Korotkov1999}, where a double quantum dot was electrostatically coupled to a point contact in a conducting channel. It is also worth noting that recent experimental investigations into measurement-induced entanglement transitions \cite{Noel2022a, Koh2022} also employ ancillas for measurements on qubit systems, in particular, in superconducting quantum processors with mid-circuit readout \cite{Koh2022}. 

We demonstrate that this simple measurement model  
possesses extremely rich dynamics induced by the interplay of unitary evolution and stroboscopic measurements. This shows up, in particular, in the statistical properties of the distributions of quantum states of the monitored qubit, which correspond to different quantum trajectories. We find that, in the long-time limit, the probability of finding a qubit in a given state right after the measurement can be described by the time-averaged distribution function, $W(\theta)$, of a \textit{single} angle variable parameterizing the state, see definitions in Sec.~\ref{sec:OneSite}. This is because the measurement protocol attracts the quantum trajectories to a specific one-dimensional manifold on the Bloch sphere of the qubit. One relevant observable, where the state statistics manifests itself, is the expectation value of the occupation of the monitored level (site) of the qubit. Upon varying the strength of coupling between the system and the detector, as well as the measurement period, $ W(\theta) $ exhibits a remarkably diverse and nontrivial behavior. 
It is worth mentioning that recent work \cite{snizhko_2020} predicted a cascade of dynamical transitions in a similar model approaching the quantum Zeno limit of frequent measurements (continuous monitoring). Here, we are interested in the stationary phase diagram of the model in the full parameter space, which includes stroboscopic measurements with a finite period.  

\begin{figure}[tb!]
    \centering
    \includegraphics{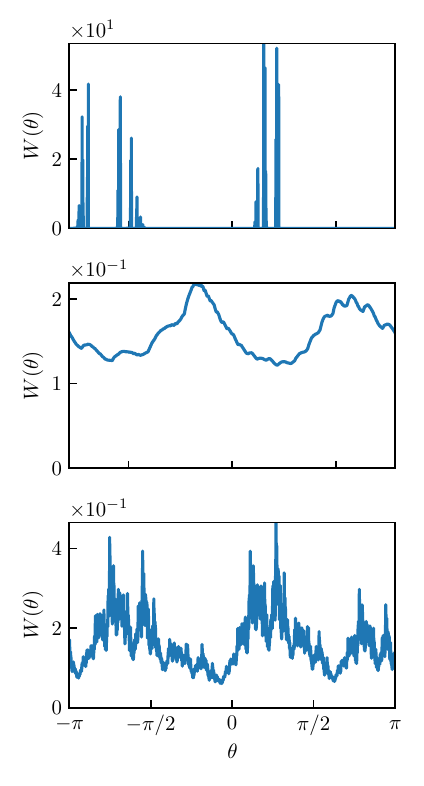}
        \caption{Distributions \(W(\theta)\) that look localized (\textit{upper panel}), extended and smooth (\textit{middle panel}), or fractal (\textit{lower panel}).}
    \label{fig:distributions_introduction}
\end{figure}

One could argue that such complexity is kind of similar to that arising during the evolution of chaotic
systems \cite{lichtenberg_1992}. However, there is a crucial difference between the chaotic and monitored systems:
The complexity in the Hamiltonian or dissipative chaotic systems results from the intrinsic nonlinearity of the evolution
or nonintegrability induced by the system boundaries;
an additional source of randomness is not needed there. On the contrary, the complexity of the monitored system is
unavoidably related to the quantum-mechanical uncertainty in the measurement outcomes (Born's rule) and disappears if one 
applies a deterministic post-selection procedure. Technically, the nonlinearity in our problem emerges at the level of a 
functional master equation governing the distribution functions for our stochastic maps.

The importance of measurement-induced randomness points towards properties of disordered systems and the Anderson localization transition rather than to chaotic systems in the present context.
Having in mind the above-mentioned link between the measurement-induced transitions and Anderson localization in large systems, it is very natural to anticipate that, in the limit of the infinite observation time, statistical properties of $ W(\theta) $ should be reminiscent of those of the wave functions in disordered systems. 
Indeed, we observe that $ W(\theta) $ may look localized (sharply peaked, see a representative example in the upper panel of 
Fig.~\ref{fig:distributions_introduction} and the first two panels of Fig.~\ref{fig:different_distributions}), 
delocalized (see a representative example in the middle panel 
of Fig.~\ref{fig:distributions_introduction} and last three panels of
Fig.~\ref{fig:different_distributions}) and even fractal (self-similar, see a representative example in 
the lower panel of Fig.~\ref{fig:distributions_introduction} and Fig.~\ref{fig:fractal}).

However, this visual similarity between the wave-function statistics and the statistics of the states of the monitored qubit is not conclusive. The classification of various types of behavior of the monitored qubit and identification of distinct phases is only possible by using a combination of 
several mutually complementary
quantitative indicators. Inspired by the striking parallels to the statistics of wave functions and the local density of states in the theory of Anderson localization, we employ here the standard indicators of the Anderson transition: 
(A)~the participation ratio and its scaling, Sec.~\ref{sec:PR};
(B)~the support of $ W (\theta) $, Sec.~\ref{sec:support};
(C)~the typical value of $ W (\theta) $, Sec.~\ref{sec:order};
(D)~the Hausdorff dimension of the curve $ W (\theta) $, Sec.~\ref{sec:Hausdorff}.
In addition, we characterize the stochastic evolution resulting in the steady-state distribution $ W (\theta) $ by the 
(E) ergodicity marker of the corresponding Markov process, Sec.~\ref{sec:Ergo}, which is common in the studies of chaotic dynamical systems.

By combining the analytical solutions for several special cases, which are related to various commensurability conditions, with numerical approaches, we understand the overall structure of the phase diagram of the monitored qubit, see Fig.~\ref{fig:phase_diags}. The extensive numerical analysis of the indicators has allowed us to categorize and quantify the rich variety of regimes exemplified in Figs.~\ref{fig:distributions_introduction} and \ref{fig:different_distributions}. 
Using indicator (E), we have identified two distinct phases of the monitored qubit: \textit{ergodic} and \textit{nonergodic}. Furthermore, we have found a genuinely \textit{localized} phase within the nonergodic phase, where the state distribution functions consist of delta peaks, as opposed to the \textit{delocalized} phase characterized by $W(\theta)$ with nonzero support. Identification of distinct phases and demonstration of transitions between them in a monitored qubit are our main findings.

Our analysis of the qubit toy model signifies
that measurement-induced transitions are 
characterized by an order parameter that is not a conventional scalar quantity but rather a distribution function. This distinguishing feature again resonates with the peculiarities of the Anderson transition, where the order parameter is represented by a distribution function of the local density of states~\cite{evers_2008}. Although the ``single-spin'' system we study cannot demonstrate the genuine entanglement transition, the discovered complexity of the model manifested in the probability distribution $W(\theta)$ suggests that in macroscopic monitored systems additional ``hidden'' transitions are feasible, which could possibly be observed in various distribution functions rather than average quantities.  

The paper is organized as follows. In Sec.~\ref{sec:basic_definitions_concepts}, we introduce the model and measurement protocol. The evolution of the qubit state on the Bloch sphere is discussed in terms of measurement operators. We further define the angle distribution function (ADF) $W(\theta)$ and explain its manifestation in various averages (stationary solution of a Master equation and time average of a single typical quantum trajectory). In Sec.~\ref{sec:single_step_solution}, the measurement operators of our model are derived. Based on these operators, we explain why the time evolution can be described asymptotically in terms of a single angle variable. We obtain the ADF analytically for several special parameter choices in Sec.~\ref{sub_sec:special_cases}. The quantifiers of localization, fractality, and ergodicity are introduced in Sec.~\ref{sec:generic_distribution}, where we illustrate and explain their properties using generic ADFs obtained from numerical simulation. Section~\ref{sec:localization_ergodicity_fractality} is dedicated to the systematic numerical calculation of these indicators in the parameter plane of the model. Diagrams of the indicators in the parameter plane are shown and their structure is explained in terms of the analytically understandable special cases. The emergence of different phases in the parameter plane is demonstrated. 
We discuss the relation of our results to previous works on similar topics, implications of our findings to experiment, as well as further directions, in Sec.~\ref{sect:Disc} and conclude in Sec.~\ref{sect:Concl}. Technical details are relegated to Appendices.

\section{Basic definitions and concepts}
\label{sec:basic_definitions_concepts}

\subsection{Model and measurement protocol}
\label{sect:model}

\begin{figure}[tb!]
    \centering
    \includegraphics[width=0.5\linewidth]{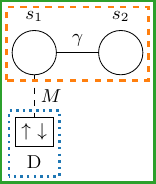}
    \caption{Scheme of the model: particles can occupy two sites, $ s_{1,2} $, which are connected
             by tunneling with the strength $ \gamma $. The site $ s_1 $ interacts with a two-level
             system (detector D), with the coupling strength being $ M $. In the text we refer to the the chain of sites \(s_{1, 2}\) (orange, dashed box) as ``the system'', the two level system D (blue, dotted box) is called ``the detector'', and the entirety of system and detector (green, solid box) is ``the setup''.}
             \label{Fig_Model}
\end{figure}

The model we study is illustrated in Fig.~\ref{Fig_Model}. The entire setup (green, solid box) consists of a system (orange, dashed box) and a detector (blue, dotted box). The two-level system is represented by two tunnel-coupled sites $s_1$ and $s_2$, which can be occupied by a single spinless electron (one can equivalently consider any other realization of a qubit as the system). 
One of the two sites (labeled by 1) is coupled to the two-state detector that can be a single spin-1/2 (or, equivalently, another pair of sites occupied by a single electron). The particular form of the coupling between the system and the detector is chosen to conserve the number of electrons in the system; in our case, the detector monitors the occupation of site $s_1$.

The specific model to be analyzed is described by the following Hamiltonian:
\begin{align}\label{2-Sites-Ham}
      \hat{H}& = \hat{H}_s+\hat{H}_\text{sd}, \\
      \hat{H}_s&=\gamma \hat{a}^\dagger_1 \hat{a}_2 + \text{H.c.},
      \quad
      \hat{H}_\text{sd}= M (\hat{a}^\dagger \hat{a})_1 \hat{\tau}_x \, .
\end{align}
Here, $\hat{H}_s$ is the system Hamiltonian,  
$\hat{H}_\text{sd}$ couples the system and the detector, $\hat{a}_{1,2} $ ($ \hat{a}^\dagger_{1,2} $) are the fermionic annihilation (creation) operators on sites $s_1$ and $s_2$, the Pauli matrix $ \hat{\tau}_x $ acts in the detector space, and $\gamma$ and $M$
are the tunneling and interaction constants, respectively. The occupation of site $s_1$ facilitates transitions between the two levels of the detector.
If the detector is realized, e.g., with two auxiliary sites (cf. Ref.~\cite{Doggen2022a}), the chosen form of $\hat{H}_\text{sd}$ would correspond to the hopping between the auxiliary sites modulated by the density on $s_1$ of the system.
We assume that the system sites are unbiased, i.e., they have the same on-site energies. The value of the on-site energy determines the origin and we have set
it to zero. Besides, we disregard an unimportant phase of $\gamma$, which can be gauged out. 
The dimension of the Hilbert space of the setup (two-level system plus two-state detector) is four. 
The orthonormal basis vectors of the system can be chosen as follows:
\[
  | 1 \rangle \equiv | 1, 0 \rangle , \quad | 0 \rangle \equiv | 0, 1 \rangle ,
\]
where the state $ | 1, 0 \rangle $ ($ | 0, 1 \rangle $) shows whether the first (second) site is occupied.

Within our protocol, the detector is initiated in a given state $ | - \rangle $ at time $ t_0 = 0 $,  
such that the initial state of the entire setup (measured system plus detector) reads as
\[
  \mbox{Initial setup's state:} \quad \Psi_0=
  \psi_0 \otimes | - \rangle ;
\]
where $ \psi_0 $ is the initial state of the system.
Measurements are performed stroboscopically at time instants $ t_j = j T, \ j = 1, 2, 3, \ldots $ ($T$ is the measurement period), when the detector is projected on one of its two states in the basis of its initial state. The probabilities of the detector readouts are given by the standard Born's rule. The projection to state $ | - \rangle $ (with
probability $ P_-$) will be called a ``no-click'' event; the projection to the flipped state $ | + \rangle $ (with probability $ P_+$) will be referred to as a ``click'' event. The system's state $\psi(t_j)$ after the projection of the detector depends on the outcome $\pm$ (``post-measurement state''):
\begin{align}
  \mbox{Setup state at}\ t_j\!: \quad\! \Psi(t_j)=\!
  \begin{cases}
       \psi_+(t_j) \otimes | + \rangle \ \  \text{click}; \\
  \psi_-(t_j) \otimes | - \rangle \ \  \text{no click}.
  \end{cases}
 \end{align}
After each projection, the detector is {\it reinitialized} in the \(\ket{-}\) state at post-measurement times $ \pmt_j \equiv t_j + 0 $:
\begin{align}
  \mbox{Setup state at}\ \pmt_j\!: \quad\! \Psi(\pmt_j)=\!
         \psi(t_j) \otimes | - \rangle \   
   \end{align}
This corresponds to ``resetting'' in the problem, which removes memory effects~\cite{Doggen2023} that would appear if the joint system-detector evolution between times $\pmt_j$ and $t_{j+1}$ started with the fiducial state of the detector after projection at $t_j$.     
   
The dynamics of the entire setup between 
$ t_0 $ and $ t_1 $, as well as between any two successive measurements, can be described by the unitary evolution operator, 
$$ \hat{U}_T = \exp(- \ci \hat{H} T ).$$ 
This unitary evolution entangles the system and the detector, so that, generically, the state of the setup is not a product state for all times except for $t_j$ (and $\pmt_j$). However, after each measurement on the detector, the total wave function collapses into a separable product of the system and detector states. Importantly, the unitary evolution of the setup is governed by the full Hamiltonian \eqref{2-Sites-Ham} at all times between the successive measurements, combining both the system's own tunneling dynamics and the one induced by the system-detector coupling. This should be contrasted with other models known in the literature (see, e.g., Ref.~\cite{snizhko_2020}), where the periods of ``measurement dynamics'' (governed only by $\hat{H}_\text{sd}$, with $\hat{H}_s$ switched off) follow the periods of free (decoupled from the detector, $\hat{H}_\text{sd}$ switched off) evolution of the system. In such models, the measurement Kraus operators are independent of the system Hamiltonian (would not involve $\gamma$ for the system considered). Our model is thus more realistic for studying the interplay and competition of the measurement-induced dynamics with the system dynamics. It is this type of competition that results in the measurement-induced entanglement transitions, see the discussion of a related toy model in Ref.~\cite{Buchhold2021a}.

\subsection{Electron states and post-measurement mapping}

\begin{figure}[tb!]
    \centering
    \adjustbox{trim=1.8cm 1.8cm 1.8cm 1.8cm}{\resizebox{\linewidth}{!}{\includegraphics{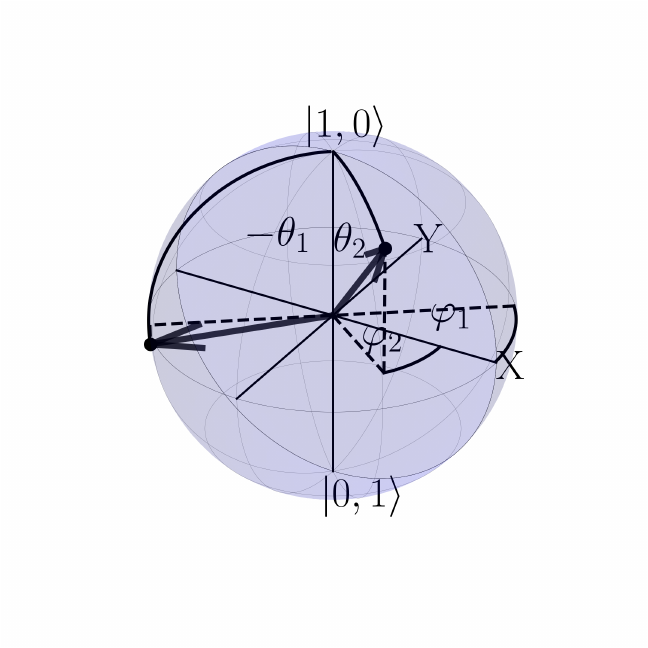}}}
    \caption{States on the Bloch sphere: \({\theta = 0}\) is the north pole, corresponding to the state \(\{{\alpha = 1,\,\beta = 0}\}\) [see Eq.~\eqref{eq:PolarAngles}]; \({\theta = -\pi}\) is the south pole, corresponding to the state \(\{{\alpha = 0,\,\beta = 1}\}\). The dots on the surface marked by arrows indicate states \(\{{\alpha=\cos(3\pi / 26),\,\beta= \me^{-\ci \pi/5}\sin(3\pi / 26)}\}\) and \(\{{\alpha=\cos(-7\pi / 26),\, \beta=\me^{\ci \pi/5}\sin(-7\pi / 26)}\}\).
       }
    \label{fig:bloch_sphere_schematic}
\end{figure}

In what follows, we consider the system states at times $t_j$, $\psi(t_j)$, which are normalized two-component spinors that can be written as 
\begin{equation}
\label{eq:InitStat}
    \psi = \alpha | 1 \rangle + \beta | 0 \rangle \equiv \Big(\begin{array}{ll}
         \alpha  \\
         \beta 
    \end{array}\Big), \quad | \alpha |^2 + | \beta |^2 = 1 \, .
\end{equation}
After omitting the overall phase of $ \psi $, the corresponding amplitudes can be parameterized by two angle variables, $ \theta \in
[-\pi, \pi) $ and $ \varphi \in [ - \pi / 2, \pi / 2 ]$:
\begin{equation}
  \label{eq:PolarAngles}
  \alpha = \cos(\theta / 2) , \quad \beta = e^{\ci \varphi} \sin(\theta /2) \, .
\end{equation}
Now, we can map the Hilbert space of the system onto the space of unit vectors,
\begin{align}
\mathbf{b}\equiv \left(\begin{array}{ll}
         \sin\theta \, \cos\varphi  \\
         \sin\theta \, \sin\varphi  \\
         \quad \cos\theta
    \end{array}\right), 
    \label{eq:Bloch-vector}
\end{align}
starting at the origin and pointing to 
points on the Bloch sphere in three dimensions, see Fig.~\ref{fig:bloch_sphere_schematic} for examples of states on the Bloch sphere. Note that the resulting parameterization of the Bloch sphere differs from the one based on conventional Euler angles, where the azimuthal angle varies from $0$ to $\pi$.  

The coefficients (or angles) parameterizing the state $\psi$ take some values at the initial time, ${\alpha(t_0)=\alpha_0}$, ${\beta(t_0)=\beta_0}$
[or, equivalently, $ \theta(t_0) = \theta_0, \, \varphi(t_0) = \varphi_0$] and change during the quantum evolution. Since between the measurement events, the system is entangled with the detectors, these coefficients are only defined at post-measurement times $ \pmt_j $. Such 
evolution at discretized time instants is described by the mapping that depends on random measurement outcomes:
\begin{eqnarray}
    \{ \alpha_{j-1}, \beta_{j-1} \}&\to&
      \begin{cases}
           \{ \alpha^{(-)}_j, \, \beta_j^{(-)} \}, &\text{no click},  \\
           \{ \alpha^{(+)}_j, \,  \beta_j^{(+)} \}, &\text{click},
      \end{cases}
      \label{eq:Mapping} 
      \end{eqnarray}
      where
      \begin{eqnarray}
    \alpha_{j-1} &\equiv& \alpha(\pmt_{j-1}), \quad 
    \beta_{j-1} \equiv \beta(\pmt_{j-1}) . 
    \label{eq:alphabeta}
\end{eqnarray}
The probabilities of the measurement readout at $ \pmt_j $ generically depend on 
the state of the system at time $\pmt_{j-1}$, i.e., on $\alpha_{j-1}$ and $\beta_{j-1}$. 
To reflect this property, we introduce notations 
\begin{equation} P_j^{(\mu)} \equiv P_\mu ( \alpha_{j-1}, 
\beta_{j-1}), 
\label{eq:Ppm}
\end{equation}
with $ \mu = \pm $ for click- and no-click outcomes.
The mapping (\ref{eq:Mapping}) can be formulated in the matrix form:
\begin{equation}
\left(\begin{array}{ll}
         \alpha_{j}^{(\mu)}  \\
         \beta_{j}^{(\mu)}
    \end{array}\right)
     = \frac{1}{\sqrt{P_j^{(\mu)}} }\, 
       \hat{M}_\mu \left(\begin{array}{ll}
         \alpha_{j-1}  \\
         \beta_{j-1}
    \end{array}\right).
\label{eq:MappingMatr}       
\end{equation}
Singling out the normalization factor with $P_j^{(\mu)}$ in this matrix equation renders the matrices $ \hat{M}_\pm $ (Kraus operators) independent of the system's state.
Note that $ \alpha $ and $ \beta $ on the right-hand side of Eq.~\eqref{eq:MappingMatr} are not marked by the upper index $ (\mu) $
because the outcome at $ \pmt_{j-1} $ (whether it is ``click'' or ``no-click'') is
not important for obtaining the state at $ \pmt_{j} $.
This is a consequence of the ancilla's resetting after its projection, which also implies that the probabilities $P_j^{(\mu)}$ do not depend explicitly on the previous measurement outcome (hence, only one outcome label $\mu$). 

\begin{figure}[tb!]
    \centering
    \includegraphics{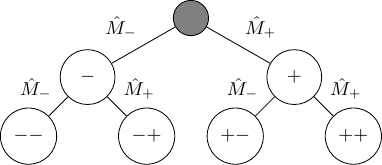}
    \caption{Exponential branching of states generated from the initial state (gray) by different combinations of post-measurement matrices \(\hat{M}_\pm\) (the $j=2$ trajectories are shown). After the \(j\)-th measurement, there are \(2^j\) different endpoints (in principle, some endpoints can describe the same state). Each generated state (including all endpoints) is labeled by a bitstring showing the sequence of measurement outcomes along the quantum trajectory leading to the state. }
    \label{fig:measurement_tree}
\end{figure}

If we start with the state determined by $ \alpha_0 $ and $ \beta_0 $ and
explore all quantum trajectories of length $j$, we come across the tree-like graph 
with the branching number 2 and $ 2^j $ endpoints~\footnote{In principle, there might be some fine-tuned cases where some endpoints for different branches correspond to the same state}, see Fig.~\ref{fig:measurement_tree}. 
Each endpoint of the graph represents a quantum state of the system obtained after acting on 
$ \{ \alpha_0, \beta_0 \}^\text{T} $
by a random product 
\begin{equation} \hat{\mathbb{M}}(\mathcal{T})\equiv \hat{M}_{\mu_j} \ldots \hat{M}_{\mu_2} \hat{M}_{\mu_1}
\label{eq:Mproduct}
\end{equation}
of $2\times 2$ matrices $\hat{M}_{+}$ and $\hat{M}_{-}$. It
corresponds to a given sequence of click and no-click outcomes -- a ``quantum trajectory'' parameterized by the $k$th ``bit-string'' $\mathcal{T}_k\equiv \{\mu_1,\mu_2,\ldots,\mu_j\}$, where $k=1,...,2^j$ labels one of the endpoints of the tree, see Fig.~\ref{fig:measurement_tree}.
Each quantum trajectory $\mathcal{T}_k$ is weighted
by the total Born's probability, 
\begin{equation} 
{\cal P}_{\rm Born}(\mathcal{T}_k) \equiv \prod_{i =1}^j 
P_i^{(\mu_i)}, \quad \mu_i\in \mathcal{T}_k.
\label{eq:BornPj}
\end{equation}
Following Eq.~(\ref{eq:MappingMatr}), the normalization factor of the resulting state is given by the square root of the same total probability (\ref{eq:BornPj}). 

It is worth noting that random products of matrices \cite{Furstenberg1963} appear in various contexts ranging from Anderson localization in one-dimensional arrays of impurities  \cite{Berezinskii1979, Perel1984, Dmitriev1989, Pastur-Book, Comtet2013} to biological evolution and computer science (see, e.g., Refs.~\cite{Hufton2016, saakian_2017, saakian_2018, infinite_series_of_singularities, david_random_matrices_2} and references therein). However, in most of the applications, the focus is the Lyapunov exponent characterizing such random products. In particular, the maximum Lyapunov exponent determines the localization length for low-dimensional disordered systems described by the transfer-matrix techniques \cite{BeenakkerRMP97}. In our case, the renormalization factor introduced by the total Born's probability does not allow the state vector to change its length, so that we are not interested in the Lyapunov exponent of the random matrix $\hat{\mathbb{M}}$. Instead, we are focusing on the statistics of the resulting system's states as parameterized by the angles $\theta_j$ and $\varphi_j$. Furthermore, the probabilities of applying the two matrices to the state-vector are generically state-dependent in our case, in contrast to most works on random products of matrices. Below, we will analyze the statistics of these Bloch-vector angles in the limit of infinite time, $ j \to \infty $.

\subsection{Angle distribution function and long-time limit}

\begin{figure}[tb!]
    \centering
    \adjustbox{trim=1.8cm 1.8cm 1.8cm 1.8cm}{\resizebox{0.97\linewidth}{!}{\includegraphics{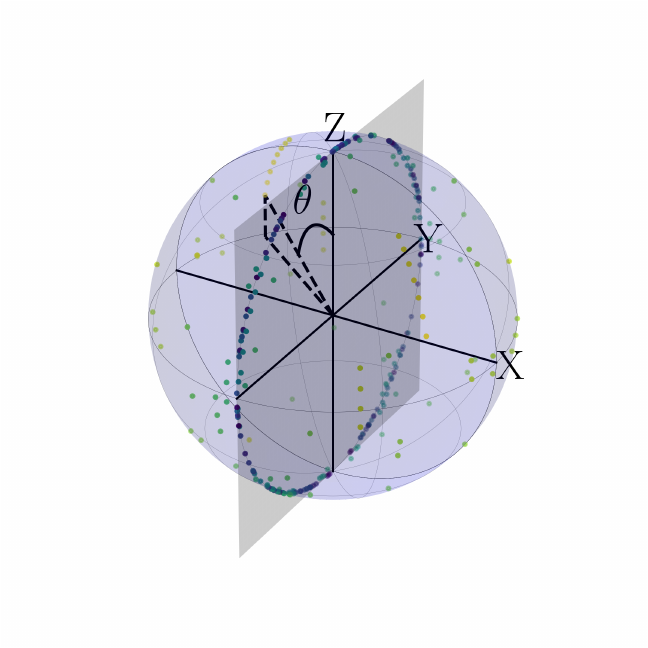}}}
    \vspace*{0.5cm}
    \caption{Example of a trajectory (colorful markers) on the Bloch-sphere (light blue) converging to the GC. 
             The parameters are \(M = 2.92\) and \(T = 1\) and the trajectory is initialized at \((\theta, \varphi) = (1.3, 2.5)\). 
             Scattered dots correspond to the positions of the quantum trajectory immediately after measurements. The time instances are color-coded, with light (dark) markers corresponding to early (late) times. The (YZ)-plane (GC-plane) is indicated in gray.
            }
    \label{fig:bloch_sphere_trajectory}
\end{figure}

Owing to the probabilistic nature of measurement outcomes, the model requires a statistical description. 
We characterize the system by the statistics of its pure states at 
times $ \pmt_j $, i.e., by the statistics of both Bloch-sphere angles. In most of the setups studied in the present paper,
the statistics of quantum trajectories in the long-time limit turn out to be fully described by the statistics of angle $ \theta $. In particular, in the case of finite hopping, $ \gamma \ne 0 $, we encounter an attraction of almost all quantum trajectories 
to the circle formed at the intersection of the Bloch sphere with the $YZ$-plane,
see Sec.~\ref{sec:two_sites_half_filling} below. 
This plane is parameterized by $ \varphi = \pi/2 $, so that $\alpha = \cos(\theta / 2)$ and $\beta = \ci\,\sin(\theta /2)$ on this manifold, see Fig.~\ref{fig:bloch_sphere_trajectory}. In what follows, we refer to it as  
the {\it Grand Circle} (GC).
Importantly, the GC is an invariant manifold of evolution: starting from the state belonging to the GC, the system never escapes from it.

This important property of generic evolution in our setup allows us to use the distribution function of the 
single angle $\theta$ to describe the statistics of the system's states in 
the long-time limit.  In view of this special role of the GC, we will, for simplicity, focus on quantum trajectories starting on the GC, without going into detail of the transient behavior of quantum trajectories approaching the GC. We will discuss those special fine-tuned cases, for which the evolution of the system does not have the GC as an attractor, separately.       

The matrix mapping \eqref{eq:MappingMatr} for the GC can be equivalently rewritten in terms of the discrete mapping of angle $\theta_{j-1}\to \theta_j$ in the course of quantum evolution:
\begin{eqnarray}
    \theta_j = 
      \begin{cases}
          \Theta_-(\theta_{j-1}) &\text{with probability}\ P_-(\theta_{j-1}),   \\
           \Theta_+(\theta_{j-1}) &\text{with probability}\ P_+(\theta_{j-1}).
      \end{cases}
      \label{eq:Theta-Mapping} 
      \end{eqnarray}
Here, the functions $\Theta_\pm(\theta)$ 
are derived from the form of the matrices $\hat{M}_\pm$. The 
probabilities of click and no-click outcomes, $\mathcal{P}_\pm$, are functions of $\theta$ corresponding to Eq.~(\ref{eq:Ppm}) via Eq.~\eqref{eq:PolarAngles} with $\varphi=\pi/2$. The explicit forms of these functions are presented in Appendix~\ref{App:GC-evolution}.
It is also useful to introduce the functions for the inverse (``retrospective'') mapping $\theta_j \to \theta_{j-1}$, describing the angles from which a given angle $\theta$ is obtained by application of $\hat{M}_-$ or $\hat{M}_+$ 
(when these are invertible matrices):
\begin{eqnarray}
    \theta_{j-1} &=& 
      \begin{cases}
          \mathcal{F}_-(\theta_j) &\text{before no-click outcome},   \\
          \mathcal{F}_+(\theta_j) &\text{before click outcome}.
      \end{cases}
       \label{eq:F-Mapping} \end{eqnarray}
The following relations clearly hold for invertible mappings:       
\begin{eqnarray}
\theta&=&\mathcal{F}_\mu[\Theta_\mu(\theta)]
=\Theta_\mu[\mathcal{F}_\mu(\theta)].     
      \end{eqnarray}

To characterize all possible endpoints of a quantum trajectory after \(j\) steps, we consider the probability distribution of these states. Specifically, we take the angle $ \theta_k $ for each endpoint (labeled by $k=1,\ldots, 2^j$) of the above-described tree-like evolution graph, thus accounting for all possible quantum trajectories $\mathcal{T}_k$ of length $ j $, and sum the corresponding $\delta $-functions with their Born weights \eqref{eq:BornPj}:
\begin{align}
    W_j^{(\varepsilon)} (\theta | \theta_0) &= \sum_{k=1}^{2^j} {\cal P}_{\rm Born}(\mathcal{T}_k)\, 
        \delta^{(\varepsilon)}(\theta - \theta_k ) \, ,
    \label{eq:W-theor}    \\
    W_j (\theta | \theta_0) &\equiv \lim_{\varepsilon\rightarrow  0} W^{(\varepsilon)}_j (\theta | \theta_0). \label{eq:w_non_reg}
\end{align}
Here, $ \theta_0 $ in the argument marks the starting point of the graph, i.e., the $\theta$-angle of the initial
state on the GC. Both $\theta_k$ (for each $k$) and ${\cal P}_{\rm Born} $ generically depend on $\theta_0$ (exceptions will be discussed separately).
The superscript \((\varepsilon)\) on both sides of Eq.~\eqref{eq:W-theor} denotes a regularization of the \(\delta\)-function parametrized by \(\varepsilon\). This procedure is important, because the distribution of pure states for any number of measurement steps $j$ and also for \(j \rightarrow \infty\) is a set of delta-peaks. Taking the limit \(\varepsilon \rightarrow 0\) at the end of the calculation allows one to define a continuous distribution.
Evidently, $ W_j (\theta | \theta_0) $ is normalized:
\begin{equation}
    \int_{-\pi}^{\pi} {\rm d} \theta \, W_j (\theta | \theta_0) = 1 .
\end{equation}

The structure of the angle distribution is similar to the pattern of the local density of states in disordered systems (see also Sec.~\ref{sec:order} below for more details). The local density of states captures the weight of states of a given energy at a given position, while the angle distribution captures the weight of quantum trajectories at a given angle, thus quantifying the probability to find the system in the state parameterized by this angle. In a finite closed disordered system, the local density of states as a function of energy is represented as a series of exact delta-peaks located at the eigenenergies of the system, similarly to Eq.~(\ref{eq:W-theor}). The weights of these peaks are determined by the amplitudes of the corresponding eigenfunctions at the point where the density of states is calculated [the counterpart of ${\cal P}_{\rm Born}(\mathcal{T}_k)$ in Eq.~(\ref{eq:W-theor})]. With increasing size of the disordered system, the density of energy levels increases (similarly, the density of the endpoints $\theta_k$ of a quantum trajectory generically increases with increasing the number of steps $j$). However, the distance between levels with high weight may remain finite as a result of the spatial localization of eigenstates.    
The inhomogeneity of a disordered system leads to a nontrivial structure of the density of states, while the complex structure of the angle distribution is determined by the \(\hat{M}_\pm\) maps and the corresponding Born probabilities.

The distribution function of the local density of states plays a role of the functional order parameter in the theory of Anderson transitions~\cite{evers_2008}. The transition to the metallic state in the thermodynamic limit is characterized by changing the character of the locally defined spectrum from quasi-discrete to continuous. Importantly, in order to obtain the continuous spectrum from an infinite set of weighted \(\delta\)-functions, one introduces infinitesimal broadening of the states, and sends the width of \(\delta\)-functions to zero at the end of the calculation, after taking the thermodynamic limit. We adopt a similar procedure in our case; this once again underscores the relation of the present problem to Anderson localization.

We do not know a priori whether or not the distribution of states keeps a memory of the initial state in the limit $ j \to \infty $. As we are going to show, both situations (dependence on and independence of the initial angle) are possible, depending on the parameters of the setup. We introduce a time-independent steady-state distribution in the
long-time limit (when such a limiting function exists) as 
\begin{equation}
  \widetilde{W}(\theta | \theta_0) \equiv  \lim_{\varepsilon\to 0} W^{(\varepsilon)} (\theta | \theta_0)
\end{equation}
where
\begin{equation}
   W^{(\varepsilon)} (\theta | \theta_0) = \lim_{j \to \infty} W_j^{(\varepsilon)} (\theta | \theta_0) \, .
   \label{eq:W-steady}
\end{equation}
It is worth emphasizing that the ``thermodynamic limit'' $j\to \infty$ is taken in Eq.~(\ref{eq:W-steady}) first. 

Below, we will encounter situations (e.g., period-2 trajectories, Sec.~\ref{sec:period_2_traj}) when the limit $j\to \infty$ does not exist, implying the absence of a unique steady state. In such a case, it is convenient to introduce the ``time-averaged'' distribution of states via
\begin{align}
    W(\theta|\theta_0)\equiv \lim_{\varepsilon\to 0}\lim_{J\to \infty}
   \frac{1}{J} \sum_{j=1}^J W_j^{(\varepsilon)}(\theta|\theta_0).
   \label{eq:Wbar}
\end{align}
This distribution is uniquely defined by the system parameters and the initial angle \(\theta_0\). The above equation is, therefore, a suitable definition for a general \textit{angle distribution function} (ADF) that describes the asymptotic state distribution of the system for the parameter tuple \((M / \gamma, T \gamma, \theta_0)\).
If the limit \(j \rightarrow \infty\) in Eq.~\eqref{eq:W-steady} exists, the stationary state distribution \(\widetilde{W}(\theta | \theta_0)\) is equivalent to the ADF.
In what follows, we will use the definition~\eqref{eq:Wbar} when discussing the steady-state distributions.  

For any given quantum trajectory on the GC, the probability of a single-step transition 
\(\theta_j \rightarrow \theta_{j+1}\) is determined by Born's rule and is equal to the corresponding probability $P_\mu(\theta_j)$.
If the quantum trajectory visits the vicinity of every point \(\theta\) with \({W}(\theta | \theta_0) > 0\) many times, all possible transitions between the discretized angle intervals are probed. The probability of any transition is then repeatedly sampled according to the Born rule, which means that the time-averaged distribution of states from a single quantum trajectory should converge to the ADF. If this is the case, the ADF has a simple interpretation in terms of quantum trajectories: If a single trajectory is observed for a sufficiently long time, the fraction of time it spends in a certain interval of the GC is determined by the integral of \(W(\theta|\theta_0)\) over that interval. The long-time behavior of almost any quantum trajectory is in this case completely described by the model parameters, being independent of outcome sequences. This time-averaging [cf. Eq.~(\ref{eq:Wbar})] is, in particular, naturally implemented in the numerical simulations based on the Monte-Carlo procedure.

Following this reasoning, we can investigate a distribution for an individual quantum trajectory $\mathcal{T}$ of length $m$---without performing any explicit average over outcomes (or trajectories):
\begin{align}
    W_{\mathcal{T}}^{(\varepsilon)}(\theta | \theta_0) \equiv \frac{1}{m \varepsilon }\int_{\theta - \varepsilon / 2}^{\theta + \varepsilon / 2}\dd{\theta'} \sum_{j=0}^{m} \delta\bigl(\theta_{\mathcal{T}}(\pmt_j) - \theta'\bigr), \label{eq:time_average_gc_distribution}
\end{align}
where \(\theta_{\mathcal{T}}(\pmt_j)\) is the $\theta$ angle at time $\pmt_j$ for the quantum trajectory and \(\theta_{\mathcal{T}}(t_0) = \theta_0\). Equation~\eqref{eq:time_average_gc_distribution} allows one to approximate the distribution~\eqref{eq:Wbar}, if any typical path for sufficiently large \(m\) reproduces the distribution~\eqref{eq:Wbar}, as described above. This approach serves as a basis for the Monte-Carlo numerical simulations. 

In any numerical analysis, representation of the ADF necessitates discretization of the angles. As a result, $\delta$-functions in Eq.~(\ref{eq:W-theor}) are necessarily regularized.
We will discuss the procedures of numerical evaluation and characterization of $W (\theta | \theta_0)$ in Sec.~\ref{sec:generic_distribution}.
Instead of taking the limit \(\varepsilon \rightarrow 0\) after introducing peak broadening of order \(\varepsilon\), we will use the equivalent regularization
\begin{align}
    W_j(\theta | \theta_0) \equiv \lim_{\varepsilon \rightarrow 0} \frac{1}{\varepsilon} \int_{\theta -  \varepsilon / 2}^{\theta + \varepsilon / 2} \dd{\theta} \lim_{\varepsilon' \rightarrow 0}W_j^{(\varepsilon')}(\theta | \theta_0).
\end{align}
where \(\lim_{\varepsilon' \rightarrow 0} W^{(\varepsilon')}\) is the set of unregularized \(\delta\)-peaks from Eq.~\eqref{eq:W-theor}. At finite \(\varepsilon\) this defines a discretized distribution function, as for example in Eq.~\eqref{eq:time_average_gc_distribution}.

\subsection{Master equation}

The time-dependent distribution of states Eq.~\eqref{eq:w_non_reg}, can be obtained from an iterative integral master equation (ME) with the initial condition \(W_0 = \delta(\theta - \theta_0)\):
\begin{align}
    &W_j (\theta)=W_j^{(+)} (\theta)+W_j^{(-)}(\theta),
    \notag
    \\
    &W_j^{(\mu)}(\theta)
    =\int_{-\pi}^{\pi}\!\!\!\dd\theta^\prime\, 
    W_{j-1}(\theta^\prime)\, P_\mu(\theta^\prime)\, \delta\bigl(\theta-\Theta_\mu(\theta^\prime)\bigr).
    \label{eq:Wj-int}
\end{align}
The distributions $W_j^{(\pm)}$ here describe the endpoints of the evolution graph obtained after the click (no-click) readout at the last step. For brevity, we skip the initial-angle argument of $W$ and encode $\theta_0$ in the initial condition $W_0(\theta)=\delta(\theta-\theta_0)$; note that the ME has the same form for arbitrary $W_0$.
In fact, in most of the cases we consider, the limiting distribution will not depend on the initial angle.

The recurrence (\ref{eq:Wj-int}) yields the following steady-state integral equation: 
\begin{align}
   &W(\theta)= \int_{-\pi}^{\pi}\dd\theta^\prime\, 
    W(\theta^\prime)\notag
    \\
    &\times \Big[P_+(\theta^\prime)
    \delta\bigl(\theta-\Theta_+(\theta^\prime)\bigr) 
    +P_-(\theta^\prime)
    \delta\bigl(\theta-\Theta_-(\theta^\prime)\bigr) \Big].
    \label{eq:W-steady-int}
\end{align}
This equation is fulfilled, in particular, by the ADF~\eqref{eq:Wbar}, as evident from substituting Eq.~\eqref{eq:Wj-int} into Eq.~\eqref{eq:Wbar}.

In general, the stationary ME can also have solutions that do not correspond to an ADF, if the steady state is degenerate.
As we will see below, the existence of a unique steady state depends on the setup's parameters. However, in most situations, the steady state is nondegenerate, allowing us to infer the ADF from it. In the following, we focus on such cases when the steady state of the ME is described by the ADF. In this case, the \(\theta_0\) argument in the ADF can be dropped and the ADF can be investigated with the help of the ME. We will point out special cases when such a treatment is not justified.

Performing the integration in Eq.~\eqref{eq:W-steady-int} by using (for invertible maps) the identity
$$\delta\bigl(\theta-\Theta_\mu(\theta^\prime)\bigr)=
\left|\dv{\mathcal{F}_\mu(\theta)}{\theta}\right|
\,\delta\bigl(\theta^\prime-\mathcal{F}_\mu(\theta)\bigr),$$
we arrive at the functional equation for the steady-state distribution: 
\begin{align}
W(\theta)&=W^{(+)}(\theta)+W^{(-)}(\theta),
\notag \\
W^{(\mu)}(\theta)
&=W[\mathcal{F}_\mu(\theta)]\,
P_\mu[\mathcal{F}_\mu(\theta)]\,
\left|\dv{\mathcal{F}_\mu(\theta)}{\theta}\right| .
\label{eq:W-funct}
\end{align}
This type of functional equations~\cite{numerical_solution_of_circle_equation} was addressed in the literature devoted to random products of matrices (cf.~Refs.~\cite{saakian_2017,saakian_2018,infinite_series_of_singularities}), where the probabilities of applying the matrices are typically state-independent. In the present problem, the dependence of $P_\mu$ on $\theta$ is dictated by the quantum-mechanical Born's rule.

\subsection{Characteristic features of the ADF, quantitative indicators, and classification of phases and regimes}

The solution to the functional equation (\ref{eq:W-funct}) cannot be obtained analytically in a closed form, except for some fine-tuned special cases. In particular, the reduction of the master equation to the Fokker-Planck form (cf. Refs.~\cite{snizhko_2020,dubey_2023}) is possible in the limiting case of frequent measurements, when the system's own evolution governed by $\hat{H}_s$ is slow compared to the measurement rate, so that the change in $\theta$ after the no-click measurement is small. Another possibility is related to various kinds of commensurability in the setup's parameters, which leads to simplifications in the functions $P_\mu(\theta)$ and $\mathcal{F}_\mu(\theta)$. In general, however, the solution can be obtained only numerically. 
Our strategy below is to identify the relevant special cases, where the analytical treatment is possible, and guess the overall ``phase diagram'' describing different types of ADF's behavior based on the exactly (or nearly exactly) solvable cases. The rest of the parameter space will be analyzed numerically.  

Experience from related works (e.g.,  Refs.~\cite{numerical_solution_of_circle_equation}, \cite{infinite_series_of_singularities}, \cite{snizhko_2020}) suggests that the steady-state angle distribution $W(\theta)$ and, thus, the ADF can show a rich phenomenology of characteristic features, being either smooth or singular. Specifically, it may exhibit, for instance (cf. Fig.~\ref{fig:distributions_introduction}), 
\begin{itemize}
    \item[(i)] isolated narrow peaks with vanishing background between them (akin to ``localized phase'' in the terminology of Anderson localization);
\item[(ii)] a smooth background covering all the angles, with fluctuations on top (akin to ``ergodic metallic phase'');
\item[(iii)] 
coexisting regions of nonzero values separated by segments where the function vanishes (akin to ``granular metal'');
\item[(iv)] 
a fractal-like pattern of 
singularities distributed over zero or nonzero background. 
\end{itemize} 
Indeed, in our analysis, we encounter 
all these types of behavior, see Secs.~\ref{sub_sec:special_cases} and \ref{sec:localization_ergodicity_fractality}.

In what follows, we will establish a classification of the regimes existing in our setup and demonstrate the existence of finite areas (distinct phases) in the overall two-dimensional phase diagram. For this purpose, we will fix the value of the hopping matrix element $\gamma$ in $\hat{H}_s$ and vary the parameters $\gamma T$
and $M/\gamma$. For each point in this two-dimensional parameter space, based on the steady-state angle distribution, we will calculate the following quantitative indicators:
\begin{enumerate}
\item Participation ratio $\mathcal{R}$ and its scaling exponent upon coarse-graining, Sec.~\ref{sec:PR};
\item Support of $W(\theta)$: the fraction of angles yielding a given fraction of the total probability, Sec.~\ref{sec:support};
\item Position of the maximum in the histogram of heights for the discretized angle distribution, Sec.~\ref{sec:order};
\item Hausdorff dimension of the curve $W(\theta)$ obtained by covering it with square boxes, Sec.~\ref{sec:Hausdorff};     
\item Ergodicity marker of the Markov process derived from Eq.~\eqref{eq:W-funct} for a given discretization scale, Sec.~\ref{sec:Ergo}.
\end{enumerate}
These indicators distinguish between localized and extended distributions and describe the degree of nonergodicity and fractality in a given setup. Although none of them can serve simultaneously as both the necessary and the sufficient condition of localization or delocalization, the combination thereof is foreseen to give a consistent description yielding a well-defined phase diagram of the system's state in the long-time limit.

\section{Single-step solution}
\label{sec:single_step_solution}

In this Section, we analyze the quantum dynamics between the initial time $ t_0 = 0 $ and the first post-measurement time $ \pmt_1 $, or, equivalently, between two successive post-measurement times $ \pmt_j $
 and $ \pmt_{j+1} $.
In terms of the matrix formulation of the problem, we elaborate on a single-step solution for the mapping given by Eqs.~(\ref{eq:Mapping},\ref{eq:MappingMatr}), and (\ref{eq:Theta-Mapping}).
We start with a warm-up example of the system with decoupled sites, $ \gamma = 0 $, and study the state of the first site. We use this simple example to set the stage, illustrating the basic steps of the approach. Much more important is the general case with finite tunneling, $ \gamma \ne 0 $. 
We will explore it for fixed half-filling where a single electron tunnels between sites 
\(s_1\) and \(s_2\).

\subsection{Warm-up exercise: $ \gamma = 0 $}
\label{sec:OneSite}

The model with $ \gamma = 0 $ was discussed extensively in Ref.~\cite{Doggen2022a} (see also Refs.~\cite{Roy2020} and \cite{MedinaGuerra2023} for a different form of $\hat{H}_{\rm sd}$); here, we repeat the derivation, as it turns out to be instructive for compreheding the general case of $\gamma\neq 0$.
If there is no tunnelling between the sites, we can consider only the first site--the one that is coupled
to the detector. With a single electron residing in the two-site system, this site can be either empty, the state $ | 0 \rangle $, or occupied, the state $ | 1 \rangle $. 
The initial state of the system, given by Eq.~(\ref{eq:InitStat}), can be prepared by hybridizing the two sites at times $t<0$ with the subsequent switching off the hopping between the sites at $t=0$. The repeated measurements will then disclose whether the first site is occupied or not. 
Using the parameterization by angles $\theta$ and $\varphi$, Eq.~(\ref{eq:PolarAngles}),
the initial expectation value of occupancy of the first site is equal to $ \cos^2(\theta_0/2) $.

Using the basis
\begin{equation}
    \begin{split}
      \hat{b}_1^{(1)} &= | 1 \rangle \otimes | + \rangle , \quad
      \hat{b}_2^{(1)} = | 1 \rangle \otimes | - \rangle , \\
      \hat{b}_3^{(1)} &= | 0 \rangle \otimes | + \rangle , \quad
      \hat{b}_4^{(1)} = | 0 \rangle \otimes | - \rangle ,
  \end{split} \label{OneSite-Basis}
\end{equation}
we construct the matrix form of the Hamiltonian and the evolution operator
\begin{align}\label{OneSite-Ham}
  \hat{H} &=
  \left(
    \begin{array}{cccc}
      0 & M & 0 & 0 \\
      M & 0 & 0 & 0 \\
      0 & 0 & 0 & 0 \\
      0 & 0 & 0 & 0
    \end{array}
  \right), \\
  \hat{U}_{T} &=
  \left(
    \begin{array}{cccc}
      \cos(M T)      & - \ci \sin(M T) & 0 & 0 \\
       - \ci \sin(M T) & \cos(M T)     & 0 & 0 \\
                 0          &        0             & 1 & 0 \\
                 0          &        0             & 0 & 1
    \end{array}
  \right) ;
\end{align}
here $ T \equiv t_{k+1} - t_k $.
The setup state between the initial time and the first measurement reads
\begin{eqnarray}
    \Psi( 0 \le t < T )\!&=&\!  \alpha_0 | 1 \rangle \otimes 
                          \Bigl[ \cos(MT) | - \rangle - \ci \sin(MT) | + \rangle \Bigr] \cr
                        &+&\!        
                          \beta_0 | 0 \rangle \otimes | - \rangle \ .
\end{eqnarray}
Right after the first measurement  and reinitialization of the detector, the setup
state becomes
\begin{eqnarray}
\!\!\! \Psi^{(-)}_1\!&=&\!\!\frac{1}{\sqrt{P^{(-)}_1}}
    \Bigl[
    \alpha_0 \cos(MT) | 1 \rangle 
    + \beta_0 | 0 \rangle \Bigr] \otimes | - \rangle , 
\label{eq:Psi-min-1site} 
                    \\
\!\!\! \Psi^{(+)}_1\!&=&\!\!\frac{1}{\sqrt{P^{(+)}_1}}
\Bigl[ - \ci\, \alpha_0 \sin(MT) | 1 \rangle \Bigr] \otimes | - \rangle ,
\end{eqnarray}
where \(\Psi^{(\mu)}_1 \equiv \Psi^{(\mu)}(\pmt_1)\) 
denote the setup state after click, $ \mu = + $, and no-click, $ \mu = - $, outcomes,
and $ P^{(\pm)}_1 $ are probabilities of the click and no-click outcomes of the first measurement,
\begin{equation}
    P^{(+)}_j = | \alpha_{j-1} |^2 \sin^2(MT), \quad 
    P^{(-)}_j = 1 - P^{(+)}_j \, .
\end{equation}
This procedure is repeated for later post-measurement times and yields the mapping
\begin{align}
\label{OneSite_mapping-}
      \{\alpha_j^{(-)},\ \beta_j^{(-)}\}&=\frac{1}{\sqrt{P^{(-)}_j}} \{\alpha_{j-1} \cos(MT), \beta_{j - 1}\}, \\
      \{\alpha_j^{(+)},\ \beta_j^{(+)}\} &=\frac{1}{\sqrt{P^{(+)}_j}} \{-\ci\, \alpha_{j - 1} \sin(MT), 0\}.
      \label{OneSite_mapping+}
\end{align}
In terms of matrices $\hat{M}_{\mu}$, the mapping is given by
\begin{align}
    \hat{M}_{-}& = \left(
          \begin{array}{cc}
             \cos(MT) & 0  \\
          0 & 1
          \end{array}
        \right),
        \\
        \hat{M}_{+}& = \left(
          \begin{array}{cc}
             -\ci\, \sin(MT) & 0  \\
          0 & 0
          \end{array}
        \right). 
\end{align}
Note that $\hat{M}_{+}$ is a projecting matrix.

Equations~(\ref{OneSite_mapping-}) and (\ref{OneSite_mapping+}) show that the stroboscopic values of the angle 
$ \theta $ defined at the post-measurement times, $ \theta_j $, change during the quantum evolution 
while those of $ \varphi $ are equal to its initial value, $ \varphi_j = \varphi_0 $. 
Besides, $ \varphi_0 $ has no effect on the evolution of $ \theta $. In particular,
\begin{eqnarray}
    \Theta_- (\theta) & = & 2 \arctan \bigl[ \cos(MT) \tan(\theta/2) \bigr], \\
    \Theta_+ (\theta) & = & 0,
\end{eqnarray}
regardless of the value of $ \varphi_0 $.
If we are interested in an observable that is not sensitive to $ \varphi_0 $, e.g., the occupation of the first site, we can choose any value of $ \varphi_0 $
and explore the evolution of the trajectory parameterized by a single angle $ \theta $.

\subsection{Main model: $ \gamma \ne 0 $}
\label{sec:two_sites_half_filling}

In the previous Section, we have set $ \gamma = 0 $ such that the expectation value of the occupation of the measured site is changed only by the measurement backaction. 
Let us now take into account the quantum dynamics of the system (governed by $\hat{H}_s$) between two successive measurements, which is due to finite tunneling between the measured and non-measured sites. 
Finite tunneling introduces a new energy scale 
$\gamma $, or 
\begin{equation}
    Y =  \sqrt{M^2 + 4 \gamma^2 }.
    \label{eq:Y-def}
\end{equation} 
In this case, none of the angles parameterizing the system state remains constant in the post-measurement mapping. The states could 
be anywhere on the Bloch sphere and should be parameterized by both Bloch sphere angles, \(\theta\) and \( 
\varphi\). However, as we show below, almost all trajectories are attracted to the GC in the
long-time limit and, in this limit, the quantum trajectories can be parameterized only
by the angle $ \theta $:
\begin{equation}
    \psi_{\rm GC}
= \{\cos(\theta / 2),\, \ci \sin(\theta / 2)\}^{\rm T} .
\label{eq:GCstate}
\end{equation}

The first steps of the description of two tunnel-coupled sites interacting with the detector are
very similar to those explained in Sec.~\ref{sec:OneSite}. 
We use the basis
\begin{equation}
    \begin{split}
  \hat{b}_1^{(2)} &= | 1, 0 \rangle \otimes | + \rangle , \quad
  \hat{b}_2^{(2)} = | 1, 0 \rangle \otimes | - \rangle , \\
  \hat{b}_3^{(2)} &= | 0, 1 \rangle \otimes | + \rangle , \quad
  \hat{b}_4^{(2)} = | 0, 1 \rangle \otimes | - \rangle ,
  \end{split} 
  \label{2Sites_Basis}
\end{equation}
and construct the matrix Hamiltonian 
\begin{equation}\label{2Sites_Ham}
  \hat{H} =
  \left(
    \begin{array}{cccc}
      0 & M & \gamma & 0 \\
      M & 0 & 0 & \gamma \\
      \gamma & 0 & 0 & 0 \\
      0 & \gamma & 0 & 0
    \end{array}
  \right).
\end{equation}
Next, we choose the initial state in a full analogy with the $\gamma=0$ case,
\begin{eqnarray}\label{InitState_2-sites}
  \Psi_0 = \Bigl( \alpha_0 | 1, 0 \rangle + \beta_0 | 0, 1 \rangle \Bigr) \otimes | - \rangle , \\
  \alpha_0 = \cos(\theta_0/2), \quad \beta_0 = e^{\ci \varphi_0} \sin(\theta_0/2) ,
\end{eqnarray}
and solve the evolution equation between measurements.
The phases of amplitudes $ \alpha $ and $ \beta $ in $\psi$ now enter
expressions for $ \Psi^{(\mu)} (\pmt_j) $ in a nontrivial way. Therefore, the system state should generically be parameterized by two angles:
\begin{equation}\label{2Site_ElState}
    \psi(\pmt_j) = \cos\left(\theta_j/2\right) | 1\rangle + 
 \me^{\ci \varphi_j}  \sin\left(\theta_j/2\right) | 0 \rangle \, .
\end{equation}
However, we show at the end of this section that $\varphi_j\to \pi/2$ for $j\gg 1$, yielding the GC state, Eq.~(\ref{eq:GCstate}), in the long-time limit for almost all quantum trajectories.


The discrete evolution of the system state is described by the matrix mapping in the form of Eq.~(\ref{eq:MappingMatr}). 
Expressions for the probabilities $ P_j^{(\mu)} $ are rather cumbersome and, since they are of secondary importance for the current explanations, we present them 
in Appendix~\ref{app:2Site_Mapping}. The ``no-click'' and ``click'' matrices 
have the following form (see Appendix~\ref{app:2Site_Mapping} for algebraic details): 
\begin{align}\label{eq:Mgen-}
      \!\!\hat{M}_- &\!=\!
        \left(
          \begin{array}{cc}
            c_M c_Y - \frac{M}{Y} s_M s_Y & - \ci\, \frac{\sqrt{Y^2-M^2}}{Y}\, c_M s_Y\\[0.2cm]
           - \ci\, \frac{\sqrt{Y^2-M^2}}{Y}\, c_M s_Y & c_M c_Y + \frac{M}{Y} s_M s_Y
          \end{array}
        \right),
\\
\label{eq:Mgen+}
      \!\!\ci\, \hat{M}_+ &\!=\! 
        \left(
          \begin{array}{cc}
             s_M c_Y + \frac{M}{Y} c_M s_Y & -\ci\, \frac{\sqrt{Y^2-M^2}}{Y} \, s_M s_Y  \\[0.2cm]
           -\ci\, \frac{\sqrt{Y^2-M^2}}{Y}\, s_M s_Y & s_M c_Y - \frac{M}{Y} c_M s_Y
          \end{array}
        \right),
\end{align}
where we have introduced short-hand notations:
\begin{align}
    c_M&\equiv \cos(MT/2), \quad c_Y\equiv \cos(YT/2), \\
    s_M&\equiv \sin(MT/2), \quad s_Y\equiv \sin(YT/2).
    \label{eq:SHN}
\end{align}
Determinants of these matrices read
\begin{align}
  {\rm det} \hat{M}_- &= c_M^2 - \frac{M^2}{Y^2} s_Y^2 , \\  {\rm det} \hat{M}_+ &= - s_M^2 + \frac{M^2}{Y^2} s_Y^2 \, .
\end{align}

Other properties of matrices $ \hat{M}_\mu $ are described in Appendix~\ref{app:2Site_Mapping_Matrs}.
Let us briefly recapitulate here the most important ones which will be used for the analysis of the
system dynamics. The matrices $ \hat{M}_\pm $ are symmetric, $ \hat{M}_\pm = \hat{M}_\pm^{\rm T} $,
not Hermitian, but generically invertible. The latter property allows one to find the previous 
state by backward-time evolution of the current state with the account of the measurement outcomes.
The exception includes those system parameters, at which $ {\rm det} \hat{M}_\pm = 0 $, see 
Fig.~\ref{Fig:ProjectingMatr}, such that at least one of the matrices is a projector up to the normalization 
of its nonzero eigenvalue, and the evolution cannot be inverted.
Depending on the parameters $ M/\gamma $ and $ T\gamma $, eigenvalues of $ \hat{M}_- $ and 
$ \ci \hat{M}_- $ can be either complex-valued (and then
complex-conjugated to each other), or purely real and generically not equal to each other. 
In the former case, both eigenvectors point to the equator of the Bloch sphere, $ \theta = \pm \pi/2 $,
while, in the latter one, they point to the GC.

\begin{widetext}
\begin{minipage}{\linewidth}
 \begin{figure}[H]
\centering
\resizebox{0.8\columnwidth}{!}{\includegraphics{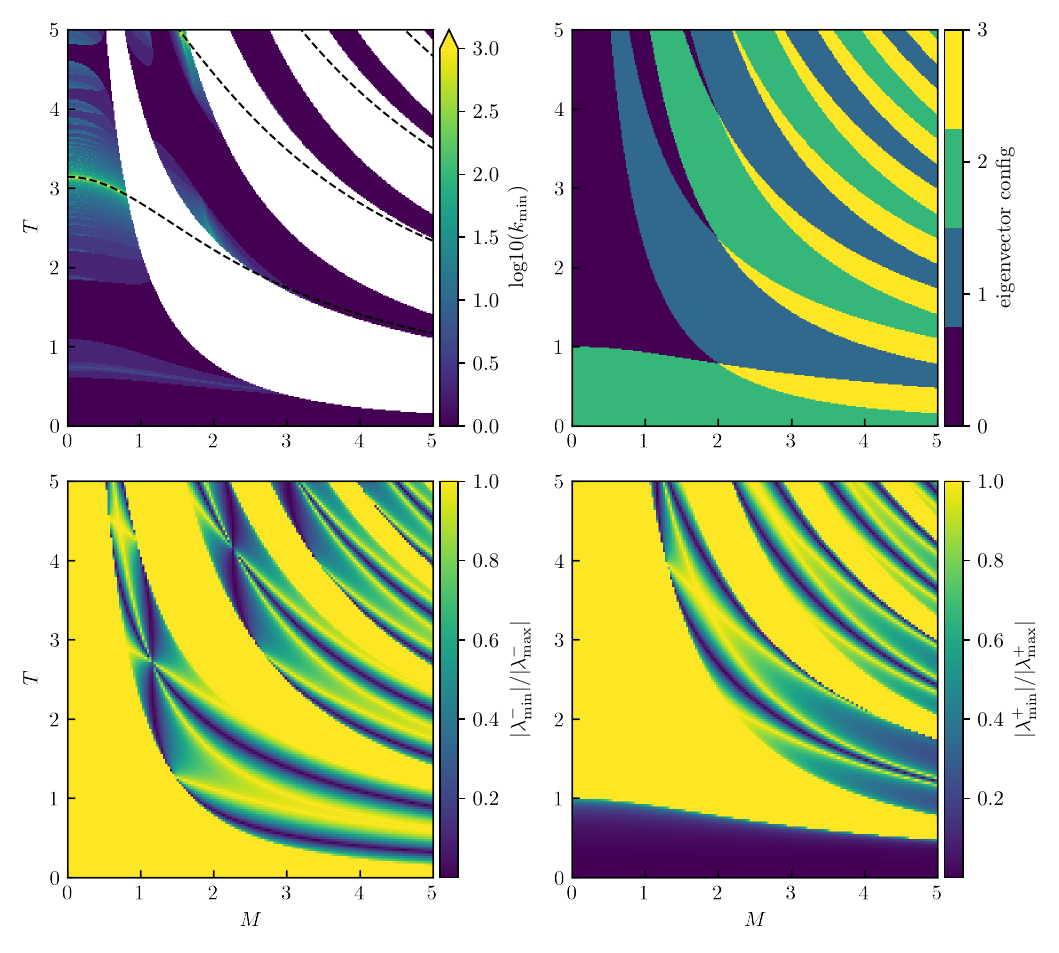}}
\caption{\textit{Upper left panel:} Minimum values of the power $ k $, at which both eigenvectors of \(\hat{M}_+^k \hat{M}_-\) point to GC up to errors of the order of the numerical precision. We have analyzed \(k \in [0, 10^3]\). 
White regions correspond to \(k = 0\). Black dashed lines correspond to the special (frozen) case \(YT =2l \pi\) with $ l $ being integer. \textit{Upper right panel}: Configurations of eigenvectors of $\hat{M}_\pm$. (0): all eigenvectors off GC; (1): only eigenvectors of \(\hat{M}_-\) on GC; (2): only eigenvectors of \(\hat{M}_+\) on GC; (3): eigenvectors of both matrices on GC. \textit{Lower panels}: Ratios of eigenvalues of the matrices.}
\label{fig:gc_convergence_numerics}
\end{figure}
\end{minipage}
\end{widetext}

As we emphasized above, the GC plays a special role in our consideration. 
Multiplying the matrices  
$ \hat{M}_\pm $ with the column vectors $ \{ \cos(\theta / 2), \ci\, \sin(\theta / 2) \}^{\rm T} $ 
and restoring the normalization, one proves that the GC is an invariant manifold of the post-measurement mapping.
Furthermore, a generic trajectory is attracted to the GC, see an example in Fig.~\ref{fig:bloch_sphere_trajectory}.  Arguments supporting the
attraction to the GC are given in Appendix~\ref{App_2Site_GC-Attraction}. The main point of this
consideration is that there always exist products of matrices, $ \hat{\mathbb{M}}(k,l) = \big( \ci
\hat{M}_+ \big)^k \big(\hat{M}_-\big)^l $, whose eigenvectors point to the GC and, hence, these products tend to project the electron state to the GC in the limit of an infinite number of measurements 
Figure~\ref{fig:gc_convergence_numerics} shows minimum powers $ k $ at which the product \(M_+^k M_-\) starts projecting the state to the GC. 
More precisely, the invertibility of the maps together with the GC forming an invariant set for the inverse maps means that a trajectory can never truly arrive at the GC. However, the closer they get to the GC, the harder it is to escape from its vicinity, because of the smoothness of the map. This allows us to restrict ourselves mainly to the initial conditions located on 
the GC and to parameterize the quantum trajectories in terms of a single angle $ \theta $. 
It is worth mentioning that attraction to the GC is a consequence of the equivalence of the two on-site energies in the system. 
If the on-site energies are different and the symmetry between the sites is broken, attraction to the GC disappears.

\section{Analysis of solvable cases}
\label{sub_sec:special_cases}

The definition of the ADF, Eq.~\eqref{eq:Wbar} implies averaging over all quantum trajectories, which is generically very nontrivial at $ \gamma \ne 0 $ and often requires extensive numerical simulations. We have managed to develop the (mostly) analytical description for finite tunneling only in several solvable cases described in this Section. We focus
mainly on the GC and point out extensions of our results beyond the GC wherever necessary.

The special cases that can be addressed analytically at $ \gamma \ne 0 $ are related to (i) commensurability effects, (ii) the existence of periodic orbits on the GC, or (iii) the projecting nature of matrices $ \hat{M}_\pm $. We will show how localization in 
$ W(\theta) $ emerges in these cases. The structure of the entire phase
diagram of the system is essentially determined by these special cases.

\subsection{Localization at $ \gamma = 0 $}
\label{sec:Loc-gamma0}

Before delving into the analysis of $\gamma\neq 0$ cases, let us return to the simple model of Sec.~\ref{sec:OneSite} and find the ADF, $ W(\theta) $ for  $ \gamma = 0 $. The initial state is assumed to be $ \theta(t=0) = \theta_0 $ and $ \varphi_0 = 0 $ (the latter equality makes intermediate equations 
shorter). Dependence of the final state on $ \varphi_0 $ will be trivially restored at the end. Equation~(\ref{OneSite_mapping+}) suggests that, after the very first click event, the system state is projected to $ | 1 \rangle $: the measurement reveals that site $s_1$ is occupied with probability one and it remains occupied for all later
post-measurement times (since there no hopping to site $s_2$). In this case, the system is in the state
$ | 1 \rangle $ at $ j \to \infty $ and the ADF becomes
$ \delta(\theta)$. If the quantum trajectory contains only no-click outcomes,
the final state at $ j \to \infty $ is $ | 0 \rangle $, see Eq.~(\ref{OneSite_mapping-}),
which yields another contribution to the ADF: $ \delta(\theta - \pi) $. 

Let us calculate the probability of this specific (``null-measurement'') outcome, 
\begin{equation} {\cal P}^{\rm null}_L = 
\prod_{j=1}^{L} P_j^{(-)}, 
\label{eq:Pnull-def}
\end{equation}
where $ L $ is the length of the no-click-trajectory.
Recalling that we have chosen $ \Im[\alpha_j] = \Im[\beta_j] = 0 $, the following equality holds true:
\begin{eqnarray}
\label{eq:OneSite_Norm}   \left[\alpha^{(-)}_j\right]^2\!+\!\left[\beta^{(-)}_j\right]^2\!=\! 
   \frac{\left[\alpha^{(-)}_{j-1}\right]^2\!\cos^2(MT) +\! \left[\beta_{j-1}^{(-)}\right]^2}{P^{(-)}_j}.
        \end{eqnarray}
The normalization of the post-measurement system state requires the left-hand side of Eq.~(\ref{eq:OneSite_Norm})---and, hence, the right-hand side---to be unity. 
Combining Eqs.~(\ref{OneSite_mapping-}) and (\ref{eq:OneSite_Norm}), we find:
\begin{eqnarray}
    {\cal P}^{\rm null}_L & = & \cos^2(\theta_0/2) \cos^{2L}(MT) + \sin^2(\theta_0/2), 
    \label{eq:Pnull-L}\\[0.2cm]
                          & \Rightarrow & \
    {\cal P}^{\rm null}_{L \to \infty} = \sin^2(\theta_0/2).
    \label{eq:Pnull-inf}
    \nonumber
\end{eqnarray}
This shows that the ADF is given by
\begin{equation}
\label{OneSite_Distr}
     W(\theta | \theta_0 )
     = \cos^2(\theta_0/2) \delta(\theta) + \sin^2(\theta_0/2) \delta(\theta - \pi) \, .
\end{equation}
The weights of delta functions coincide with the initial probabilities for the first site to be 
occupied, $ \cos^2(\theta_0/2) $, and empty, $ \sin^2(\theta_0/2) $. 
Thus, repeated generalized measurements of one site yield, in the limit of an infinite number of measurement steps, the same distribution of final states as a single projective measurement. 

An exception is a trivial case where the probability of the click event vanishes and there is only one quantum trajectory consisting of no-click events: 
$$ M T = \pi l, $$ 
with $ l $ being integer.
In this special case, $ \theta_j = \theta_0 $ for all $ j $ and $ W(\theta | \theta_0) = 
\delta(\theta - \theta_0) $ coincides with the initial ADF.
Another special (``commensurate'') case is realized for 
$$MT=\pi l/2$$ 
with integer $l$. According to Eq.~(\ref{eq:Psi-min-1site}), a single no-click outcome in this case also immediately projects the system state (now, to $ | 0 \rangle $), thus, both matrices $M_\pm$ are projectors then. This implies that the generalized measurement becomes a strong (projective) measurement (cf. Ref.~\cite{Doggen2022a}).

If instead of a single initial state with fixed $ \theta_0 $ one prepares a set of states
characterized by an initial distribution of $ \theta_0 $, $ W(\theta | \theta_0)$ in  Eq.~\eqref{OneSite_Distr}
should be averaged over $ \theta_0 $:
\begin{eqnarray}
\label{OneSite_DistrAver}
    W(\theta) & = & \left\langle \cos^2\left(\theta_0 / 2\right) \right\rangle_{\theta_0} \delta(\theta)  + \\ 
    & & +
\left\langle \sin^2\left(\theta_0 / 2\right) \right\rangle_{\theta_0} 
\delta \left( \theta - \pi \right).
\nonumber
\end{eqnarray}
The distribution functions (\ref{OneSite_Distr},\ref{OneSite_DistrAver}) fully describe the statistics
of the occupation of the measured site:
since $ \varphi $ does not influence evolution of $ \theta $, the full ADF is
factorized into a product $W(\theta)\times\delta(\varphi - \varphi_0 )$,
and similarly for $ W $ being averaged over the initial angles.

To conclude this section, let us reiterate that the qubit is always found within one of two states, $ \{ \theta = 0,\ \varphi = \varphi_0 \} $
and $ \{ \theta = \pi,\ \varphi = \varphi_0 \} $, at $ \gamma = 0 $ in the limit $ j \to \infty $. The initial condition for 
$ \theta_0 $ can only change the relative weights of these two states but is unable to modify the structure of the ADF 
containing only two $ \delta $-peaks. 
Such a distribution falls into our operational definition of the localized phase. The localization of the angle $\theta$ can also be considered as a measurement-induced steering~\cite{Roy2020, Rosenow2022} of an arbitrary state of the qubit to the states with $ \theta = 0,\pi$. This example also allowed us to pinpoint two special cases: $MT=\pi l$ and $MT=\pi l/2$, corresponding to ``frozen dynamics'' and strong-measurement dynamics, respectively. In what follows, commensurability conditions of this sort will enable us to identify special cases of dynamics that will determine the phase diagram of the model at $\gamma\neq 0$.

\subsection{Effects of (in)commensurability and commensurability transitions on GC and beyond}
We now return to the general case of finite hopping \(\gamma \neq 0\).
\subsubsection{Frozen case}
\label{sec:frozen}

Let us start with a simple case, where 
$$ Y T = 2 \pi q, $$ 
with $ q $ being integer. Using equations from Appendix~\ref{app:2Site_Mapping}, it is straightforward to show that the outcome probabilities are state-independent and the electron post-measurement states differ from previous ones only by a total phase:
\begin{align}\label{2Site_Frozen}
  P^{(\pm)} & = \frac{1}{2} \big[1 \mp \cos(MT)\big],
  \\
  \psi^{(+)}(\pmt_j) & = \left( \ci \, {\rm sign}\, s_M \right)^j
       ( \alpha_0 | 1 \rangle + \beta_0 | 0 \rangle ) \, , \\
  \psi^{(-)}(\pmt_j) & = \left( \ci \, {\rm sign}\, c_M \right)^j
       ( \alpha_0 | 1 \rangle + \beta_0 | 0 \rangle ).
       \end{align}
Thus, $ \theta $ is not changed by the post-measurement mapping and its ADF coincides
with the initial distribution at $ t = 0 $: 
\begin{equation}
    W_j(\theta | \theta_0) = \delta (\theta - \theta_0) \, .
\end{equation}
Moreover, similar to the $\gamma=0$ case, we can trivially restore the entire distribution
function of states even beyond the GC: both angles $\theta$ and $\varphi$ are frozen at their initial values.
Hence, the initially localized distribution remains unchanged at any $ j=1,2,\ldots $.
One can refer to this commensurate case as the {\it ``frozen'' case}.

\subsubsection{Shift case}
\label{sec:shift}

The outcome probabilities also do not depend on the system state if 
$$ M T = q \pi $$ 
with $ q $ being integer: 
\begin{eqnarray}
  P^{(\pm)}  =  \left( \frac{M}{Y} \right)^2 s_Y^2, \quad
  P^{(\mp)}  =  1 - P^{(\pm)} .
  \label{eq:Pqpm}
\end{eqnarray}
The choice of the upper index of $ P^{(\pm)} $ depends on the parity of $ q $. To be more specific, choosing odd $ q = 2 l + 1 $, we arrive at the following expressions for the matrices $ \hat{M}_\pm $:
\begin{eqnarray}
       \hat{M}_- & = & (-1)^{l+1}\frac{M}{Y} s_Y \, \hat{\sigma}_3 ,
       \\
       \hat{M}_+ & = & 
\ci \, (-1)^{l+1} \left( c_Y\, \hat{\sigma}_0 - \ci \frac{2\gamma}{Y} 
s_Y\, \hat{\sigma}_1 \right),
  \label{2Site_Commensurate}
\end{eqnarray}
where $ \hat{\sigma}_{0,1,3} $ denote Pauli matrices acting
in the system state.
These matrices satisfy the relations:
\begin{eqnarray}
\!\!\!\!\! \hat{M}_\mu^\dagger \hat{M}_\mu & = & | {\rm det}(\hat{M}_\mu) |^2 \hat{\sigma}_0, \quad
\hat{M}_+^{-p} \propto \hat{M}_- \hat{M}_+^p \hat{M}_- , 
\end{eqnarray}
where \(p\) is integer. Equation~\eqref{2Site_Commensurate} suggests that any product of matrices 
$ \hat{M}_\pm $ can be reduced to one out of four possible forms:
\begin{equation}\label{Matr_MT-pi}
  \hat{M}_+^p, \ \ \hat{M}_+^p \hat{\sigma}_3, \ \ \hat{\sigma}_3 \hat{M}_+^p , \ \ \hat{\sigma}_3 \hat{M}_+^p \hat{\sigma}_3 \, .
\end{equation}

\begin{figure}[tb!]
    \centering
    \includegraphics{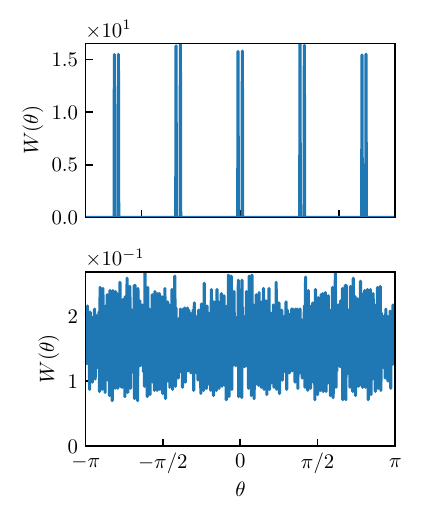}
    \vspace{-0.5cm}
        \caption{Time histogram of a post-measurement trajectory in the shift case \(MT = \pi\) over \(10^5\) time instances. By tuning the shift angle to a multiple of \(\pi\), a localized distribution emerges (\textit{upper panel}). If an incommensurate shift angle is chosen, all points on the GC are eventually covered (\textit{lower panel}).}
    \label{fig:distributions_shift}
\end{figure}

This is another rare case where there is no attraction to the GC. Nevertheless, the evolution of a GC state, Eq.~(\ref{eq:GCstate}),
is remarkable.
Matrix $ \hat{\sigma}_3 $ trivially inverts the sign of the angle $\theta$. Matrices $ \hat{M}^p_+ $ shift this angle:
\begin{equation}
     \hat{M}^p_+ \psi_{\rm GC} \propto 
            \{\cos(\theta/2 + p \phi), \, \ci \sin(\theta/2 + p \phi)\}^{\rm T},
\end{equation}
where $ \phi $ is the phase of the eigenvalue 
$ \upsilon_+ $ (the upper index of $ \upsilon_+ $ is not important), see Appendix \ref{app:2Site_Mapping_Matrs}.
If the phase $ \phi $ is commensurate, $ \phi = p_1 \pi / p_2 $, the matrices $ \hat{M}_+ $ yield $ p_2 $-periodic
(or $ 2 p_2 $-periodic) trajectories along the GC. Hence, the initially localized distribution of $ \theta $ remains
localized at infinite time. Examples of these scenarios are presented in Fig.~\ref{fig:distributions_shift}.

The set of the phases $ \phi $ that correspond to localization is somewhat similar to the set of parameters that yield
the structure of the well-known Hofstadter butterfly \cite{hofstadter_1976}, 
where the density of states is also determined by certain commensurability. Specifically, the structure of the energy levels in this model is governed by the ratio of magnetic 
flux through a lattice cell and the flux quantum. 
For rational values of this parameter entering the Harper equation \cite{harper_1955}, the density of states is represented by a finite set of peaks, similar to the structure of $W(\theta)$ for commensurate values of the shift angle (upper panel in Fig.~\ref{fig:distributions_shift}).
If the phase $ \phi $ is not commensurate, the shifts $ p \phi $ fill the entire GC at infinite time. This breaks 
any initial localization of the $ \theta $ distribution 
(see an example in lower panel of Fig.~\ref{fig:distributions_shift}).
Whether or not the GC becomes filled homogeneously 
or the ADF becomes fractal similar to the spectrum of the Harper equation in the incommensurate regime
depends on the combinatorics of various quantum trajectories and can be checked numerically. 
The case of 
$ M T = 2 l \pi $ is analogous to the above case of $ M T = ( 2 l + 1 ) \pi $ up to reshuffling 
$ \hat{M}_+ \leftrightarrow \hat{M}_-$.

We note in passing that the post-measurement states in quantum trajectories at $ M T = q \pi $ are located on two 1D circles even beyond the GC.
For example, in the example of odd $ q $, the trajectories are governed by the intersection
of the Bloch sphere with two planes being parallel to the GC plane and located symmetrically with respect to the GC. The matrix $ \hat{M}_- $ reflects the state with respect to the GC plane, while $ \hat{M}_+ $ rotates the
state around the x-axis. 
The expression for a single step of the rotation is rather cumbersome and we omit it for 
the sake of brevity. Let us emphasize that such rotations can also be commensurate or incommensurate.

\subsection{Period-2 trajectory}
\label{sec:period_2_traj}
Above, we have considered two cases where the localized nature of $ W(\theta) $ at $ j \to \infty $
is related to localized initial conditions -- quantum dynamics and measurements are unable to
destroy localization due to commensurability. 
Let us now analyze other solvable examples where
localization of the ADF emerges at long times regardless of the nature of
the initial conditions.

In this Section, we focus on the setup with 
$$ Y T = ( 2 l + 1 ) \pi, $$ 
where the post-measurement map again simplifies:
\begin{align}
\hat{M}_- & =
 \frac{(-1)^{l+1}}{Y}\, \hat{m}_- ,\quad
\hat{M}_+  =
\frac{\ci (-1)^l}{Y} \hat{m}_+ , 
        \\
\hat{m}_- &\equiv
 2\ci\gamma c_M \sigma_1 + M s_M \sigma_3 \, , \\
\hat{m}_+ &\equiv
 2\ci\gamma s_M \sigma_1 - M c_M \sigma_3 \, 
\end{align}
For the GC, the probabilities of outcomes take the form:
\begin{align}
P_j^{(\pm)} & = \frac{1}{2} \pm 2 \gamma
\frac{M \sin(2\theta_{j-1}) \sin (M T) + \gamma  \cos (M T)}{Y^2} \, .
\end{align}
Noting that 
$$ \hat{m}_\pm \{ 1/\sqrt{2}, \pm \ci /\sqrt{2} \}^{\rm T} \propto \{ 1/\sqrt{2}, 
\mp \ci /\sqrt{2} \}^{\rm T}, $$ 
we conclude that the states 
$ \{ 1/\sqrt{2}, \pm \ci /\sqrt{2} \}^{\rm T} $ form a period-2 trajectory of the matrices
$ \hat{m}_\pm $.  

Moreover, this is a limit cycle that generically attracts quantum trajectories. In fact, it is a kind of weak attraction: a typical trajectory will always return to any previously visited point on the GC; it is just that the most time is spent around the attractive points. The origin of attraction can be explained as follows:
since $ \hat{m}_\mu^2 \propto \hat{\sigma}_0 $, any product of the matrices $ \hat{M}_\mu $ reduces to one of four possible forms:
\begin{align}
\label{2Site_Period2-M_chain-1}
  \left( \hat{M}_+ \hat{M}_{-} \right)^k &\propto \left( \hat{m}_+ \hat{m}_{-} \right)^k \, , \\
      \hat{M}_{-} \left( \hat{M}_+ \hat{M}_{-} \right)^k &\propto \hat{m}_{-} \left( \hat{m}_+ \hat{m}_{-} \right)^k \, , \\
\label{2Site_Period2-M_chain-2}
  \left( \hat{M}_- \hat{M}_{+} \right)^k &\propto \left( \hat{m}_- \hat{m}_{+} \right)^k \, , \\
      \hat{M}_{+} \left( \hat{M}_- \hat{M}_{+} \right)^k &\propto \hat{m}_{+} \left( \hat{m}_- \hat{m}_{+} \right)^k \, .
\end{align}
The eigenvalues of the products $\hat{m}_\mu \hat{m}_{-\mu} $ are real numbers $ 2\eta M \gamma - Y^2 \sin(M T)/ 2 $. 
Importantly, absolute values of the eigenvalues 
are different, with the maximum being $ 2 M \gamma + Y^2 | \sin(M T)| / 2 $. 
The eigenvectors of these products are $ \{ 1 / \sqrt{2}, \pm \ci / \sqrt{2}\}^{\rm T} $ and point to the intersections (two 
opposite points) of the equator with the GC. In the limit of large time, typical quantum trajectories are characterized by $ k^2 $ that is of the order of a large number of measurements for a long-time quantum trajectory, $k^2\sim j\gg 1$. 
This follows from the analogy with classical random walks in 1D, where, following the diffusion law, the mean-square displacement (the counterpart of $ k^2 $) is proportional to time (the counterpart 
of the number of measurements), see Appendix~\ref{App_2Site-RandomWalks} for details.

If the typical value of $ k $ is large, one can keep only the ``main'' eigenvector in the expansion of the matrix products:
\begin{align} 
  &k\gg 1:\ \ \left( \hat{M}_\mu \hat{M}_{-\mu} \right)^k \propto ( M \gamma + Y^2 | \sin(M T) | / 2 )^k 
  \notag 
  \\
  &\times
\Bigl[
\mbox{ either }  \{ 1 / \sqrt{2}, \ci / \sqrt{2}\}^{\rm T}
\mbox{ or } \{ 1 / \sqrt{2}, - \ci / 
\sqrt{2}\}^{\rm T}\Bigr].
\end{align}
Thus, the matrices $ \left( \hat{M}_\mu \hat{M}_{-\mu} \right)^k $ project any state onto one of the two states $ \{ 1 / \sqrt{2}, \pm \ci / 
\sqrt{2}\}^{\rm T} $. 
Finally, we note that
\begin{equation}
  \hat{m}_{\mu} \{ 1 / \sqrt{2}, \pm \ci / \sqrt{2}\}^{\rm T} \propto \{ 1 / \sqrt{2}, \mp \ci / \sqrt{2}\}^{\rm T},
\end{equation}
which means that, in the long-time limit, all combinations in Eqs.~(\ref{2Site_Period2-M_chain-1}-\ref{2Site_Period2-M_chain-2}) with $ k \gg 1 $ project any state onto one of two states:
$ \{ 1 / \sqrt{2}, \pm \ci / \sqrt{2}\}^{\rm T} $.

Thus, we conclude that, if $ Y T = (2 l + 1) \pi $, the typical quantum trajectories generate two universal (independent of the initial state) peaks in the ADF $W(\theta)$ at $ \theta = \pm \pi / 2 $. This can be regarded as steering an arbitrary quantum state to these specific states~\cite{Roy2020, Rosenow2022}.
Numerical simulations confirm the above semi-phenomenological explanation, see Fig.~\ref{fig:special_distributions}.

\begin{figure}[ht]
\begin{center}
       \includegraphics{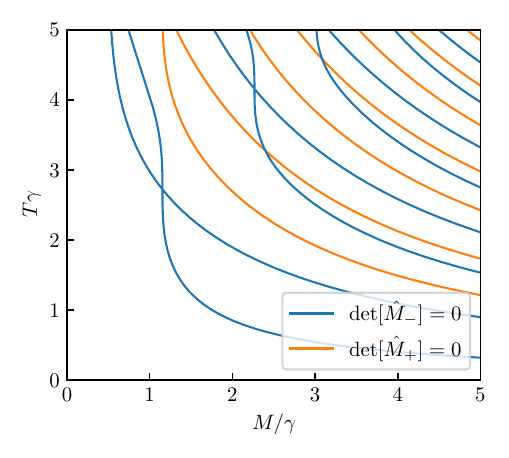}
\end{center}
   \caption{
        \label{Fig:ProjectingMatr}
        Lines on the plane of dimensionless parameters $ \{ M / \gamma, T \gamma \} $, where $ {\rm det} \hat{M}_\pm = 0 $. 
           }
\end{figure}

\subsection{Projecting matrices}
\label{sec:projective_case}

There is a special case of real eigenvalues of the matrices $ \hat{M}_\pm $, where one of them is zero
and, therefore, the corresponding matrix is a projector.
This holds true if the determinant of one of the matrices vanishes:
\begin{align}
\label{2Site_Proj1}
\left| c_M \right| &= ( M / Y )  \left| s_Y \right|\quad \text{for} \quad \mu = -, \\
\label{2Site_Proj2}  
\left| s_M\right| &= ( M / Y ) \left| s_Y \right| \quad \text{for} \quad \mu = +.
  \end{align}
Lines, where ${\rm det} \hat{M}_\pm = 0 $ are shown in Fig.~\ref{Fig:ProjectingMatr}.

\begin{figure}[tb!]
    \centering
    \includegraphics{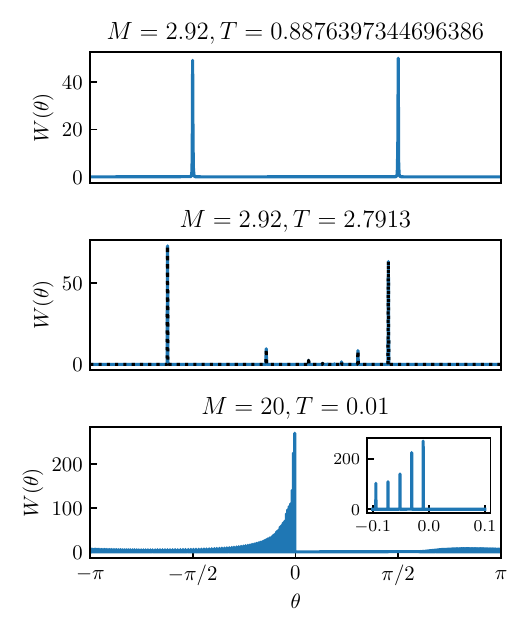}
    \vspace{-0.75cm}
    \caption{Examples of GC-angle distributions $W(\theta)$ for the period-2 and projective cases (see Secs.~\ref{sec:period_2_traj} and~\ref{sec:projective_case}, respectively). Both panels show the numerical results (blue) obtained for \(\gamma=1\) with the different \(M\) and \(T\)-values shown in the plot titles. \textit{Upper panel}: Period-2-case, with \(TY = \pi\). In agreement with our prediction (Sec.~\ref{sec:period_2_traj}), the distribution consists of two peaks of (approximately) equal height, located at angles \(\theta = \pm \pi / 2\). 
\textit{Middle panel}: Projective case, where one of the matrices \(M^{\pm}\) has a vanishing eigenvalue, projecting to the ``main eigenvalue'' (the leftmost peak). 
The blue curve was obtained numerically by performing a time average over a single random Monte-Carlo post-measurement trajectory. 
The dotted curve is the analytical prediction, Eq.~\eqref{eq:proj_solution}, truncated at 20 terms. \textit{Lower panel:} Almost projective case in the limit \(M / \gamma \rightarrow \infty\) and \(T \gamma \rightarrow 0\), where the matrix  $\hat{M}_+$ projects to $\theta=0$, while matrix $\hat{M}_-$ introduces a small shift. The inset shows the same distribution in the interval \(\theta \in [-0.1, 0.1]\).}
    \label{fig:special_distributions}
\end{figure}

Let, for instance, $ \hat{M}_- $ be the projector. After the very first no-click event along
the quantum trajectory, the system is projected to the eigenstate $ \Upsilon_{p}^{(-)} $ of $ \hat{M}_- $ corresponding
to the nonzero eigenvalue and ``eigenangel'' $\theta_-$. The subsequent click event occurs with the probability 
$ P_+ = 1 - (M/Y)^2 \sin^2(Y T) $ and drives the system out of $ \Upsilon_{p}^{(-)} $. 
The asymptotic angle distribution will consist of the highest peak governed by $ \Upsilon_{\cal P} $,
its smaller satellite resulting from $ \hat{M}_+ \Upsilon_{p}^{(-)} $, even smaller satellite of the
second generation $ \hat{M}_+^2 \Upsilon_{p}^{(-)} $, etc. Application of $\hat{M}_-$ to any of the satellites moves the state back to $ \Upsilon_{p}^{(-)} $ and the process starts over. The resulting distribution is the limiting case of the pattern predicted in Ref.~\cite{infinite_series_of_singularities}.

Such a structure is insensitive to the
initial distribution and, if there is a gap between the main peak and the satellite, as well as between
the satellites, $ W(\theta) $ appears localized at $ j \to \infty $. Numerical simulations confirm this scenario, see the middle panel of Fig.~\ref{fig:special_distributions}. 
If the determinant of one of the matrices is close to zero but not equal to zero, the structure
of localized peaks becomes more and more smeared with increasing deviation from conditions 
(\ref{2Site_Proj1}-\ref{2Site_Proj2}).

The ME (\ref{eq:W-funct}) for $\hat{M}_-$ projecting to $ \Upsilon_{p}^{(-)} $ is formally solved by the following ADF:
\begin{align}
    W(\theta) &= \mathcal{N}\Bigl\{ \delta\!\left(\theta -\theta_{-}\right)\notag
    \\
    &+\sum_{n=1}^{\infty} \delta\!\left(\theta - F^{(n)}_{+}(\theta_{-})\right) {\prod_{j=1}^n} P^{(+)}\!\left(F^{(j - 1)}_{+}(\theta_{-})\right)\Bigr\}, \label{eq:proj_solution}\\
    \mathcal{N}^{-1} &=1+\sum_{n=1}^{\infty} \prod_{j=1}^n P^{(+)}\!\left(F^{(j - 1)}_{+}(\theta_{-})\right),\end{align}
    where 
    \begin{align}
    F^{(j)}_{\mu}(\theta) &:= \begin{cases}
        \theta, & j = 0,\\
        \Theta_\mu\!\left(F^{(j - 1)}_{\mu}(\theta)\right), & j > 0.
    \end{cases}
\end{align}
Here, the normalization factor \(\mathcal{N}\) contains information about all of the satellite peaks, which are projected back onto the mean peak by an application of matrix $\hat{M}_-$ and develop again by sequences of $\hat{M}_+$. In the stationary state, the weight of the main peak is determined by the projections from all satellites, weighted by the respective probabilities. The weight of the satellites is, in turn, balanced against the main peak. It can be seen that this is achieved by first generating all satellites and then renormalizing this pattern. A good approximation to Eq.~\eqref{eq:proj_solution} can be obtained by truncating the sum, since higher-order terms are exponentially suppressed, see Fig.~\ref{fig:special_distributions}.

An interesting case of projective matrices is realized in the frequent-measurements limit \(M/\gamma \to \infty\) (so that $ Y \simeq M$) and \(T\gamma \to 0\), with $M^2 T/\gamma$ kept constant. In this limit, the matrix $\hat{M}_+$ becomes projecting, whereas $\hat{M}_-$ induces very small shifts: $|\Theta_-(\theta)-\theta|\ll 1$. The ME (\ref{eq:W-funct}) then reduces to a Fokker-Planck equation with resetting, see Refs.~\cite{snizhko_2020,dubey_2023} for details. The limiting ADF contains an extended region in a finite domain of angles; beyond this domain $W(\theta)=0$ (see the lower panel of Fig.~\ref{fig:special_distributions}). 
Importantly, however, the smooth curve in this distribution appears only in the limit $T\gamma \to 0$; for any finite $T$, the extended region comprises well-defined isolated peaks (see inset in Fig.~\ref{fig:special_distributions}).

In the special case where 
\begin{eqnarray}
    \label{eq:DoubleProj1}
    M T & = & \frac{\pi}{2} + \pi l, \ l = 0, \pm 1, \pm 2, \ldots \\
    \label{eq:DoubleProj2}    
    \left| \frac{\pi}{2} + \pi l \right| | s_Y | & = & \frac{Y T}{\sqrt{2}} \, ; \\
    & \Rightarrow & {\rm det} \hat{M}_+ = {\rm det} \hat{M}_- = 0 ;
    \nonumber
\end{eqnarray}
both matrices
are projectors and $ W(\theta) $ consists of two peaks generated by these projectors:
\begin{equation}
\label{eq:DP-W}
    W(\theta) = \sum_{\mu = \pm } A_\mu \delta( \theta - \theta_\mu ).
\end{equation}
Here $ \theta_\pm $ are the angles that correspond to eigenstates with 
nonzero eigenvalues of the projectors. Using equations from Appendix~\ref{app:2Site_Mapping_Matrs},
one can show that the probabilities of applying the matrix $\hat{M}_-$ at angle $\theta_+$ (and vice versa) 
are determined by choice of the solution of Eqs.~(\ref{eq:DoubleProj1}) and (\ref{eq:DoubleProj2})
\begin{equation}
\label{eq:CrossProb}
   P_{\pm} ( \theta_\mp ) = 1 - 2 c_Y^2 \, ,
\end{equation}
and are generically nonzero. This means that the ``double-projecting limit'' is not equivalent to the 
strong-measurement one (cf.~the case of ${MT=\pi l/2}$ for ${\gamma=0}$ in Sec.~\ref{sec:Loc-gamma0}).

The weights $ A_\pm $ depend on the dwelling time of a given peak, i.e., on 
the typical length of sequences $ \hat{M}_- \hat{M}_- \ldots \hat{M}_- $ and $ \hat{M}_+ \hat{M}_+ 
\ldots \hat{M}_+ $. Since the escape probabilities are the same for both peaks, Eq.~(\ref{eq:CrossProb}),
we  conclude that
\[
  A_\pm = 1/2.
\]
Equation~(\ref{eq:DP-W}) gives an example of localization. Since the peak positions do not depend on the initial state, we come across the measurement-induced steering \cite{Roy2020}.

\section{Numerical simulations: Characterization of the angle distribution}
\label{sec:generic_distribution}

In order to explore the phase diagram beyond the analytically solvable special cases, we have performed numerical studies of the model.
A combination of two complementary methods---(i) solution of a discretized ME and (ii) Monte-Carlo simulations (see Appendixes~\ref{sec:methods} and \ref{sec:stationary_vs_time_average})---has allowed us to systematically explore the ADF in the \((M, T)\) parameter plane (fixing \(\gamma = 1\)).
The approach based on the iterative solution of the discretized ME is our main working tool, see Appendix~\ref{sec:master_equation}. The convergence of the numerically obtained distributions to the ADF as defined and motivated in Sec.~\ref{sec:basic_definitions_concepts} is extensively discussed in Appendix~\ref{sec:stationary_vs_time_average}. In essence, excellent agreement between results from the two completely different approaches proves their validity.
We consider the domain \(M, T \in (0, 5]\), which turns out to include a broad range of regimes. Our understanding of generic distributions is based on the consideration of four cases (frozen, shift, period-2, and projective) in Sec.~\ref{sub_sec:special_cases}, which shape the rich ``phase diagram'' of the model.

To give an impression of various angle distributions for generic parameters, we fix (arbitrarily) \(M = 2.92\) and present six different distributions, corresponding to different values of \(T\) in Fig.~\ref{fig:different_distributions}. Comparing several distributions, there is an immediate observation: For $T = 2.5$ and $2.7$, 
\(W(\theta)\) has heavy peaks around a few points and is close to zero at most angles. At \(T = 3.0\), there are some high peaks, but the distribution has small finite values at all angles. For $T = 3.0554, 3.1$, and \(3.722\), the distribution is close to uniform (note the scale of the \(y\)-axis) but features an intricate structure on smaller scales. Based on the immediate visual difference between the distributions, it is tempting to refer to them as \textit{localized} and \textit{delocalized}. We now proceed with a more detailed description of the indicators that allow one to characterize the localized and delocalized angle distributions.

\begin{widetext}
    \begin{minipage}{\linewidth}
        \begin{figure}[H]
            \centering
            \resizebox{0.95\linewidth}{!}{\includegraphics{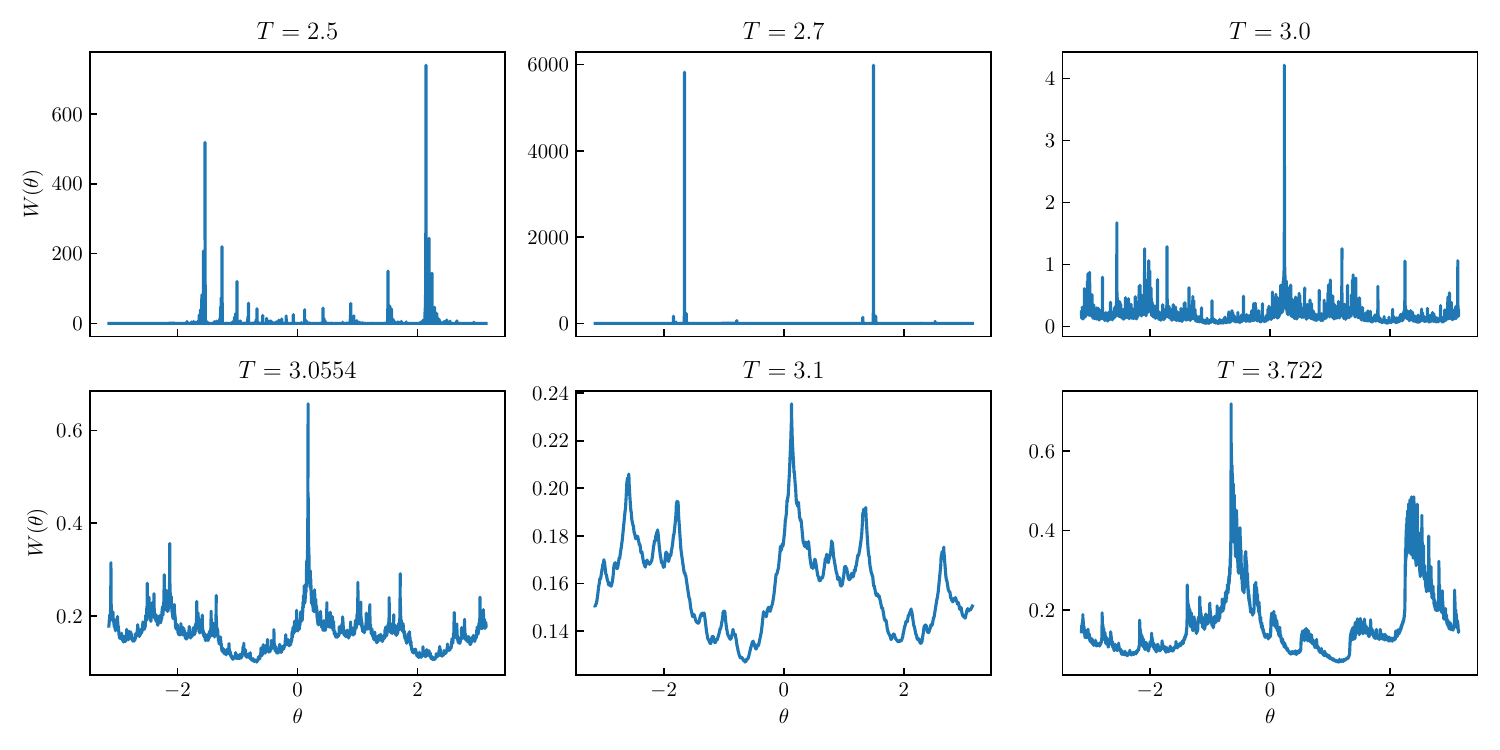}}
            \vspace*{-0.5cm}
            \caption{Different GC distributions obtained for \(M = 2.92\) by solving the discretized Master equation (\(N = 10^5\) grid cells) numerically for different values of \(T\), starting from a homogeneous initial condition.}
            \label{fig:different_distributions}
        \end{figure}
    \end{minipage}
\end{widetext}

\subsection{Participation ratio and its scaling}    
\label{sec:PR}
    
An ADF could be referred to as localized if there is a large chance to find the angle within a small interval on the GC [heavy peaks in \(W(\theta)\)]. If no such subset exists, the distribution is delocalized. To quantify this, we calculate a \textit{participation ratio} (PR) for discretized angles
    \begin{align}
        \mathcal{R}_N(W) &\equiv \left[\sum_{i = 1}^{N} {\rm Pr}_i^2\right]^{-1}\!\! = \frac{N^2}{(2\pi)^2} \left\{ \sum_{i=1}^N [W^{(\Delta \theta)}(\theta_i)]^2\right\}^{-1}, \label{eq:participation_ratio}
    \end{align}
 where 
 \begin{equation}\text{Pr}_i \equiv \int_{c_i} \dd{\theta} W(\theta)
 \label{eq.Pri-def}
 \end{equation}
are the probabilities 
obtained from integrating the ADF over \(N\) 
discretization cells $c_i$ of equal size $\Delta\theta=2\pi/N$ (see Appendix~\ref{sec:methods} for details).    
A perfectly localized distribution \(W_{\rm loc}(\theta) = \delta(\theta - \theta_\text{loc})\) gives \(\mathcal{R}_N(W_{\rm loc}) = 1\), while a uniform distribution \(W_\text{uf}(\theta) = (2 \pi)^{-1}\) gives \(\mathcal{R}_N(W_\text{uf}) = N\). Accordingly, large (small) values correspond to more delocalized (localized) distributions.

In addition to the PR, we also calculate its scaling with the number of coarse-graining cells, \(N_{g}\). Namely, we take the ADF at the highest available 
resolution 
(characterized by $ N $), superimpose a broader grid (characterized by $ N_{g} $), and 
sum up the terms $\mathrm{Pr}_i^2$ in each broader cell---as opposed to numerically solving the ME 
at every considered discretization level with \(N\) cells separately.
The scaling can be described approximately by a power-law
\begin{equation}
        \mathcal{R}_N \propto N_{g}^{\zeta} \quad \zeta \in [0, 1],
        \label{eq:PR_scaling_exp}
\end{equation}
with the PR exponent \(\zeta\).

In the preceding Sections, we referred to (de)localization as a property of wave functions in space in disordered systems. In a one-dimensional Anderson-localized system, the probability amplitude \(|\psi_i|^2\) corresponding to an arbitrary eigenstate of the disordered Hamiltonian falls off exponentially with the distance to the center site. In this case, a participation ratio is calculated as 
    \begin{align}
        \tilde{\mathcal{R}}_L(\psi) \equiv \left[\sum_{i=1}^L |\psi_i|^4\right]^{-1}
    \end{align}
where \(L\) is the system size. In a localized system, the PR becomes independent of the system size if \(L\) exceeds the localization length $\xi$, whereas delocalization is defined by $$\tilde{\mathcal{R}}_L \rightarrow L$$
in the thermodynamic limit.
Our model has a fixed size (the angle $\theta$ is compact), so that the role of the system size $L/\xi\to \infty$ is played by the number of the GC discretization grid cells \(N\). In the present problem, the counterpart of the modulus-squared amplitude of the wave function is the stationary probability Pr$_i$ on the GC. In a sense, increasing the number of ``bins'' in our model is similar to decreasing $\xi$ (by increasing the strength of disorder) in the Anderson-localization problem.
    
The scaling of the PR with \(N\) does not generally have an equivalent meaning as the scaling of the PR with \(L\) in the context of Anderson localization. For example, for a box distribution
    \begin{align}
        W(\theta) &= \begin{cases}
            1 / |I_1|, & \theta \in I_1,\\
            0, & \text{otherwise},
        \end{cases}
    \end{align}
which can be arbitrarily narrow, \(|I_1| \ll 1\),
we get \(\mathcal{R}_N(W) \propto N\), as the fraction of the GC covered by the distribution is independent of \(N\). The situation is different, if the distribution is given by a sum of delta-peaks: in this case, the discretized distribution \(W^{(\Delta \theta)}(\theta)\) becomes narrower as \(\Delta \theta \rightarrow 0\), and the PR is constant as a function of \(N\). In this sense, a distribution would only be localized, if its support on the GC decreased with increasing resolution. There can be situations where distributions are localized in the sense of a small PR value (narrow peaks), but delocalized in the sense that the support is independent of the discretization (if \(\Delta \theta\) is sufficiently small to resolve the distribution). Such distributions can be regarded as ``metallic grains'' (coexisting delocalized and localized regions), which are not generally expected in the problem of Anderson localization in homogeneously disordered systems. At the same time, we know of two special cases (period-2 trajectories and the projective case), where the distribution \(W(\theta)\) is localized in the strict sense: the observed peak width scales to zero with increasing the resolution of discretization. Without additional analytical arguments, based only on the analysis of PR, we are limited by the minimum resolution \(\Delta \theta\) when distinguishing the true localization from the localization in the sense of metallic grains.

\subsection{Support of the angle distribution}
\label{sec:support}
    
To further characterize the ``localized-looking'' ADF, we introduce another observable \(\mathcal{S}^\mathcal{C}_N\), which captures the minimum support \(N_\mathcal{C} / N\) needed to cover a fraction \(\mathcal{C} \leq 1\) of the total probability:
    \begin{align}
        \mathcal{S}^\mathcal{C}_N(W) &\equiv \frac{N_\mathcal{C}}{N} \label{eq:support_def},\\
       N_\mathcal{C}:\quad  \sum_{i=0}^{N_c}\,[\text{sorted}(\textbf{Pr})]_i &\geq \mathcal{C},
    \end{align}
where \(\text{sorted}(\textbf{Pr})\) are descendingly sorted probabilities $\{\text{Pr}_i\}$ and \(N_\mathcal{C}\) is the smallest integer, such that the inequality is fulfilled. If the support is one, the distribution can be regarded as extended; otherwise, it can be localized or ``granular''. It is worth emphasizing, however, that this indicator explicitly depends on the cut-off scale $\mathcal{C}$, which should be compared with the scale introduced by the discretization resolution in numerical simulations. 

\subsection{The typical height of the distribution}
\label{sec:order}
    
In the theory of Anderson transitions~\cite{evers_2008}, the transition between insulating and metallic phases manifests itself in the local density of states~\cite{evers_2008,ldos_anderson_transition,anderson_transition_numerics,anderson_transition_introduction},
    \begin{align}
        \rho_i(\epsilon) &= \sum_n \delta(\epsilon - \epsilon_n) |\bra{i}\ket{\epsilon_n}|^2. \label{eq:LDOS}
    \end{align}
On the localized side, there is just a small number of wave functions contributing to the sum~\eqref{eq:LDOS} at any given site \(i\) (because most wave functions are localized away from \(i\) and their contribution to the local density of states at a given site is exponentially suppressed). As a result, the typical value of the local density of states (essentially the value at the maximum of the distribution) vanishes~\cite{evers_2008,ldos_anderson_transition,anderson_transition_introduction}. On the metallic side close to the transition, many states contribute at any site, because the eigenfunctions \(\bra{i}\ket{\epsilon_n}\) are extended. The typical value becomes finite and the distribution is spread around the typical value~\cite{anderson_transition_numerics,anderson_transition_introduction}. The order parameter of the Anderson transition is then a functional one: the distribution function of the local density of states. Characterizing this distribution function by the corresponding typical value, one can invent a simple-minded scalar ``order parameter'' that could be of practical use for a variety of questions in the context of Anderson transitions.

Inspired by this experience, we introduce a third observable to distinguish localized ADFs from delocalized. Making use of the similarity between the patterns of the local density of states and the angle distribution in our case, we introduce the height distribution $H(h_j)$ of
    \begin{align}
        h_i&= \text{min}(W^{(\Delta \theta)}) + (2i + 1) \Delta h,
        \label{eq:histogram_of_heights}\\
        \Delta h &\equiv \frac{\text{max}[W^{(\Delta\theta)}] - \text{min}[W^{(\Delta\theta)}]}{2 N_h},
    \end{align}
where \(i \in [0, N_h - 1]\) and \(N_h \ll N\) is the number of distinct heights, see Fig.~\ref{fig:histogram_of_heights}.    
This is a discretized version of the distribution of the ``local density of states'' (the local density of post-measurement trajectories on the GC): We count the number of bins, where the probability distribution at a given resolution lies within a given window of values (a ``histogram of heights''). 

   \begin{widetext}
        \begin{minipage}{\linewidth}
    \begin{figure}[H]
            \centering
            \includegraphics{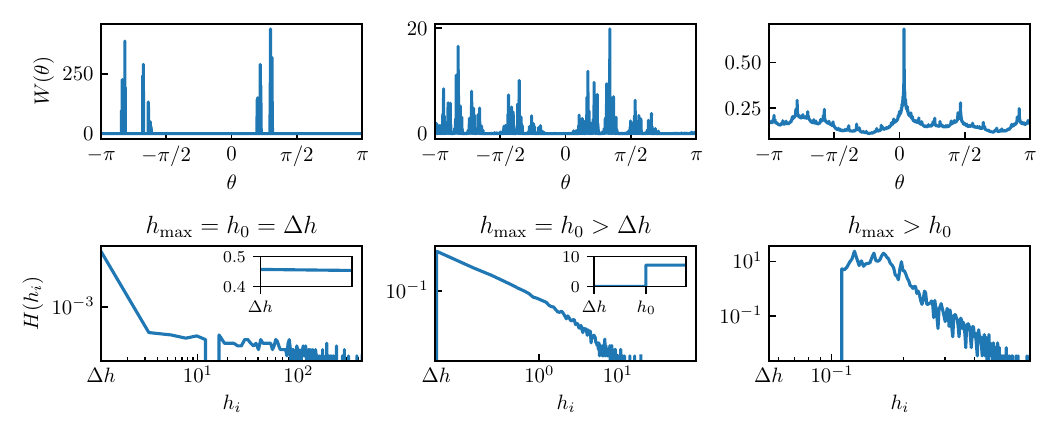}
            \vspace*{-1cm}
            \caption{Examples of histograms of heights falling into the three different categories introduced in Sec.~\ref{sec:order}. \textit{Upper panels:} Distributions obtained from the master equation with \(N = 10^5\), at \((M = 2.263, T = 3.498)\), \((M = 0.990, T = 1.811)\), \((M = 4.052, T = 3.768)\) from left to right. \textit{Lower panels:} The corresponding histograms of heights. From left to right categories 1 (localized, maximum at the ``numerical zero,'' \(h = \Delta h\), see inset), 2 (maximum at the left boundary $h_0$ of the histogram but at \(h> \Delta h\), see inset), and 3 (delocalized, maximum not at the left boundary). Here, \(h_0= \min(W^{(\Delta \theta)}) + \Delta h\) is the height corresponding to the leftmost bin in the histogram of heights with \(H(h_i) > 0\).}
            \label{fig:histogram_of_heights}
        \end{figure}
    \end{minipage}
    \end{widetext}

We analyze the typical value of this distribution by considering the position of its maximum \(h_{\rm max}\), such that \(H(h_{\rm max}) = \max(H)\). Based on numerical results, we distinguish three categories:
    \begin{itemize}
        \item[1:]\quad \(h_{\rm max} = h_0 =     \Delta h\),
        \item[2:]\quad \(h_{\rm max} = h_{0} > \Delta h\),
        \item[3:]\quad \(h_{\rm max} > h_{0}\).
    \end{itemize}
Here \(h_0= \min(W^{(\Delta \theta)}) + \Delta h\) is the height corresponding to the leftmost bin in the histogram of heights with \(H(h_i) > 0\), see Eq.~\eqref{eq:histogram_of_heights}.
The first category implies a vanishing typical value in analogy to the insulating phase of a disordered system. 
The third category means a finite typical value in analogy to the metallic phase of a disordered system. The second, intermediate case corresponds to a nonvanishing typical value, however, at the left boundary of the distribution. This category describes, for instance, isolated peaks on top of a nonzero background in the ADF.

\subsection{Fractal dimension}    
\label{sec:Hausdorff}

\begin{figure}[tb!]
        \centering
        \resizebox{\linewidth}{!}{\includegraphics{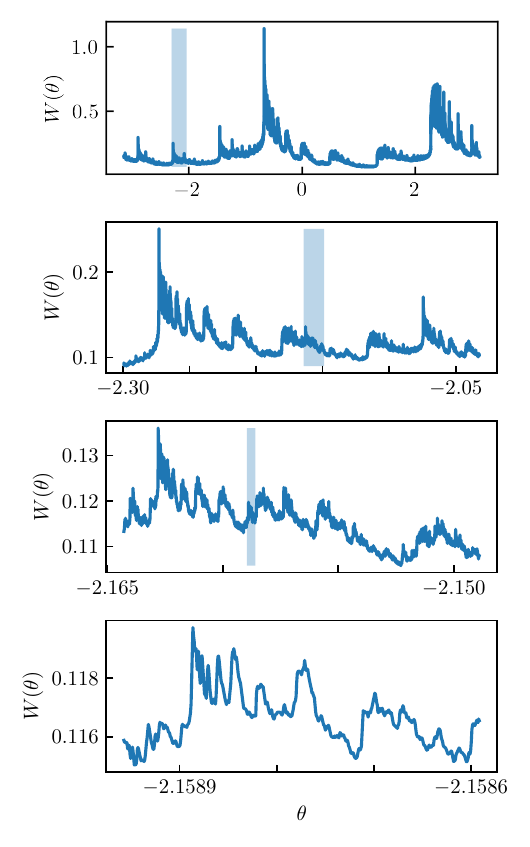}}
\caption{Example of a fractal ADF. Angle distribution for \(M = 2.92\), \(T\approx 3.729\) is calculated from the ME with \(N=10^7\) grid cells, starting from a uniform initial condition. The upper panel shows \(W(\theta)\) on the entire GC, \(\theta \in [-\pi, \pi)\). The lower panels show progressively smaller segments of the GC, with the respective interval indicated by blue shading in the preceding panel.}
        \label{fig:fractal}
    \end{figure}

Having introduced observables to quantify ``localization'', we now take a closer look at the apparent substructure in some of the distributions. As an example, we consider \(M = 2.92\), \(T \approx 3.729\) in Fig.~\ref{fig:fractal} at high grid resolution \(N = 10^7\). The upper-left panel shows the entire distribution \(W(\theta)\) with \(\theta \in [-\pi, \pi)\). The other panels show sections of the distribution taken from progressively smaller intervals on the GC. The blue shaded areas indicate the intervals that are displayed in the next (zoom-in) panels. Remarkably, these four sections look similar to each other, suggesting that the distribution ``repeats itself'' on different scales, with the interval considered in the lower panel corresponding to \(3 \cdot 10^{-4} / (2 \pi) \approx 5 \cdot 10^{-5}\) fractions of the GC. Numerically, we cannot further resolve this pattern without going to larger \(N\). A heuristic argument suggests that this self-similarity can exist on any scale, rendering the distribution \textit{fractal}. 

We quantify fractality of the distribution by calculating its Hausdorff fractal dimension \(d\): Overlaying \(W(\theta)\) with a uniform grid of \(m^{-1} \times m^{-1}\) cells, we count the number of cells \(C(m)\) required to fully cover the curve~\footnote{We cover the curve resulting from connecting the data points, not the points themselves. A single spike of height \(h\) thus contributes \(h / m\) boxes, not one box.}. The relation
    \begin{align}
        C(m) \propto \left(\frac{1}{m}\right)^d, \quad m \rightarrow 0,
        \label{eq:definition_fractal_dim}
    \end{align}
defines the \textit{box counting dimension} \(d\)~\cite{box_counting}. If the structure can be fully resolved at finite \(m\), we get \(d=1\). The dimension \(1 < d < 2\) corresponds to a fractal structure. Numerically calculating the fractal dimension, we cannot increase \(m\) above the number of grid cells \(N\) without trivializing the box-counting dimension. Thus, any curve with \(d > 1\) should turn out to scale trivially when the box size is reduced beyond our numerical resolution. 

The emergence of fractality of the distributions
can be exemplified by a heuristic consideration of
the vicinity of the projective cases 
(cf. Ref.~\cite{infinite_series_of_singularities}).
Let \(\Theta_{+}(\theta)\) map a large fraction of the GC to a narrow interval around its main eigenangle, resulting in a slightly broadened peak. The matrix \(\hat{M}_-\) translates this peak to another angle interval, slightly ``distorting'' the peak shape (because \(\mathcal{F}_-^\prime\) and \(P^{-}\) are not constants). If we start from a peaked distribution around the main eigenangle, the translating map generates a set of decaying ``peak clones'' on the GC. Many of these peak clones are reflected back onto the main peak by the almost projecting map. Self-consistency requires those modulations to be translated to the secondary peaks as well [cf. Eq.~\eqref{eq:proj_solution}]. Recursively applying this argument suggests that the stationary limit is given by a fractal.     

Generally, this mechanism is not limited to the projective cases. As soon as there is some back-and-forth copying between two points, fractality can emerge. Quantifying when exactly this breaks down towards the uniform cases requires a more careful consideration, which we do not provide here.
A related analysis was performed in Ref.~\cite{infinite_series_of_singularities} for correlated random products of two matrices, where the appearance of singular peaks at ``strong'' eigenangles, as well as their ``cloning'', was found at the condition analogous to $P_{\pm}(\theta_\pm)|\mathcal{F}^\prime_\pm(\theta_\pm)|=1$. 

The emergence of fractality is another parallel to the theory of Anderson transitions: At an Anderson transition, the wave function of the system becomes multifractal~\cite{evers_2008}, which means that its self-similarity can be characterized by a whole set of nontrivial fractal dimensions by attributing a fractal dimension \(f(\alpha)\) to a subset of points of the wave function that is characterized by scaling as~\(L^{-\alpha}\) with the system size~\cite{evers_2008,singularity_spectrum_fractals,fractal_dimension_singularity_powerlaw,geometric_multifractality}. The function \(f(\alpha)\) is called the singularity spectrum and can be extracted from the scaling of moments of the wave function (like the inverse participation ratio) with the system size~\cite{singularity_spectrum_fractals,evers_2008}. The box-counting dimension of the entire wave function is closely related to the singularity spectrum but contains less information~\cite{geometric_multifractality}.

\subsection{Ergodicity indicator for the Markov process}    
\label{sec:Ergo}
    
Finally, we address \textit{ergodicity} of the Markov chain defined by the matrix \(\hat{\mathcal{M}}_N\), 
\begin{equation}
    [\hat{\mathcal{M}}_N]_{ik} \equiv \frac{1}{\Delta \theta}\sum_{\mu \in \{+, -\}}  P_\mu(\theta_k) |f_\mu(c_i)\cap c_k| .
\label{eq:discretization} 
\end{equation}
Here, \(f_\mu(c_i)\) is the image of \(c_i\) under the action of map $\mathcal{F}_\mu(\theta\in c_i)$, Eq.~(\ref{eq:F-Mapping}), which is continuous on the periodic interval \([-\pi, \pi)\) and invertible for all sets of parameters except for the projective limit; $ |\ldots| $ denotes the length of the corresponding angle interval. The sparse matrix \(\hat{\mathcal{M}}_N\) relates the cell probabilities $\text{Pr}_i$, 
\begin{equation}
    \text{Pr}_i = \sum_{k=1}^{N}[\hat{\mathcal{M}}_N]_{ik} \text{Pr}_k,
    \label{eq:Pri-M}
\end{equation}
in the  discretized ME, see Appendix~\ref{sec:methods}
for details.

There exist several different notions of ergodicity for Markov processes in the literature~\cite{different_notions_of_markov_ergodicity,geometric_ergodicity}. In the following, we call the system ergodic, if the Markov process is irreducible~\footnote{The apparently most common notion of ergodicity of a Markov chain requires aperiodicity as well~\cite{different_notions_of_markov_ergodicity}. This would exclude for example our period-2-trajectory case. Irreducibility and aperiodicity together imply a unique stationary distribution of the Markov process. However, by including a time average in the definition of the GC distribution, we can define a universal distribution for a set of parameters without this requirement.}. 
Irreducibility means that any state \(i\) (bin \(c_i\)) can be reached from every state 
\(k\)~\cite{different_notions_of_markov_ergodicity}.
If this is the case, the ADF is probed by any typical post-measurement trajectory implying ergodicity of the dynamical system~\cite{ergodic_theory_dyn_sys}.  In our notation, for any \(i_0, k_0\) there exists a natural number \(s\) such that
    \begin{align}
        \sum_{l=1}^N\left[(\hat{\mathcal{M}}_N)^s \right]_{i_0l} \delta_{lk_0} > 0.
    \end{align}
For the coarse-grained ADF, ergodicity implies:
    \begin{enumerate}
        \item Support of the ADF on the entire GC.
        \item A unique stationary state of the ME.
        \item Equivalence of the ADF, stationary state~\footnote{Note that power-iteration does not necessarily work.}, and time average of a typical quantum trajectory.
        \item Independence of the ADF of the initial angle.
    \end{enumerate}
    
To numerically check whether the process is irreducible for given parameters, we can consider its Markov matrix \(\hat{\mathcal{M}}_N\) as the transition matrix of a directed graph
\begin{eqnarray}
G_N & := & (V_N, E_N), \\
V_N & := & \{1, \ldots, N\}, \notag \\
E_N & := & \{ (k, i),\, \mbox{ such that } [\hat{\mathcal{M}}_N]_{ik} > 0; \ k,i \in V_N \},
\nonumber        
\label{eq:graph_from_transition_matrix}
\end{eqnarray}
where the $ N $ nodes from the set \(V_N \) represent the grid cells \(c_i\) and the edges from the set $ E_N $ 
represent transitions \(k \rightarrow i\)  between the cells with finite (nonzero) probability in the matrix \(\hat{\mathcal{M}}_N\).
Ergodicity of the Markov process with transition matrix \(\hat{\mathcal{M}}_N\) is equivalent to \(G_N\) having a single strongly connected component (SCC) containing all \(N\) nodes. The SCC is a maximal set of nodes \(v \subseteq V_N\), such that for every pair of nodes \(s, s' \in v\) there is a path \(s\rightarrow s'\) (and also \(s' \rightarrow s \)) that only traverses edges between nodes in \(v\). Here, ``maximal'' means that no further nodes can be added to the set without breaking the latter property. 

The SCCs of a graph \(G_N\) can be calculated efficiently, within \(\mathcal{O}(|V_N| + |E_N|)\) operations~\cite{scc_alg_1,scc_alg_2}
\footnote{We use the Python library \textit{NetworkX}~\cite{networkx} to find SCCs}. 
Our ergodicity indicator has two values: $1$ (ergodic, \(G_N\) has a single SCC) and $0$ (nonergodic, \(G_N\) has more than one SCC). It is however not obvious whether or not this finite-\(N\) indicator can be used to classify the continuous process.
Indeed, ergodicity can be an artifact of discretization. As an extreme example, suppose we discretized the entire GC into a single cell \(c_1 = [-\pi, \pi)\). This cell forms a single SCC, thus corresponding to an ergodic process. This dismisses any sub-intervals of the GC that may prove unreachable at higher discretization, which would render the process nonergodic. 

Importantly, nonergodicity in \(\mathcal{M}_N\) does have implications for the continuous process. To see this, a simple example of a discrete, nonergodic process is helpful: let us assume that the set of nodes \(V_N\) of the graph \(G_N\) splits into two SCCs \(v_1\) and \(v_2\). These SCCs can be either (i)~disconnected, or (ii)~connected only by edges in one direction, e.g. \(v_1 \rightarrow v_2\).
Edges in both directions \(v_1 \rightarrow v_2\) \textit{and} \(v_2 \rightarrow v_1\) are not possible because this would mean that \(G_N\) consists only of a single SCC.

    In both situations (i) and (ii), the GC subset represented by the nodes from \(v_2\),
        \(I_2 \equiv \bigcup_{i \in v_2}c_{i}\),
    forms an invariant subset. This subset is also invariant for the continuous post-measurement evolution on the GC regardless of the value of $ N $ for which it was found, see Appendix~\ref{sec:non_erg_elaboration} for details. 
    The above argument can easily be generalized to situations where \(G_N\) splits into more than two SCCs, the key insight being that, by construction, there always exists at least one SCC that corresponds to an invariant subset of the dynamical process.
    
    We stress that the finite value of $N$ in this argument \textit{does not} imply any discretization of the dynamical process. Indeed, the mapping used in the construction of a finite-$N$ graph $G_N$ is the mapping of continuous intervals of the angles (rather than a mapping of discrete interval labels), which preserves the continuous nature of the stochastic process. Discretization with finite $N$ refers here only to the resolution of the instrument (a ``microscope'') employed to explore the continuous process by constructing a corresponding graph.
    
With the above prerequisites, we define nonergodicity of the continuous process in the following way:
The process is nonergodic, if an invariant subspace of the dynamical process can be found that does not include the entire GC. Otherwise, the process is ergodic (for any discretization, paths between any two bins exist). 
Importantly, once our ``microscope'' detects such an invariant subset in the continuous process at some $N$, there is no need to further enhance its resolution: nonergodicity of the continuous process is already demonstrated.
Finally, it is worth mentioning that nonergodicity does not imply degeneracy of the stationary state. In the above example with two SCCs, both cases (i) and (ii) are nonergodic, but only case (i) corresponds to a degenerate stationary state.

\subsection{Summary of regimes and respective indicators}

We have introduced four indicators which are  suitable for identifying the behaviour of the ADF:

1. {\it Localization and delocalization} are reflected by the participation ratio and its scaling, the 
support measure, and the typical value of the ``histogram of heights''. These indicators have been used 
based on a loose analogy of the shape of the asymptotic angle distributions (possibly at different 
discretizations) with localized and delocalized wave functions in a disordered system. In the truly
localized regime, we may expect $ {\cal R} \to 1 $, $ \zeta \to 0 $ regardless of the smallness of the
bin size.

2. {\it Fractality and self-similarity} of the ADF curves are described by their box-counting Hausdorff dimension. 

3. {\it Ergodicity and nonergodicity} of a discrete Markov process corresponding to the transition matrix 
\(\hat{\mathcal{M}}_N\) is described by the connectivity of the graph $ G_N $.

\section{Numerical results: (De)localization, (Non)ergodicity, \newline 
and fractality}
\label{sec:localization_ergodicity_fractality}

In this Section, we present the results of our numerical study of the structure of the ``phase diagram'' in the $M$-$T$ plane for $\gamma=1$.
We obtain the stationary angle distribution $W(\theta)$ from the ME for a high-resolution grid in the domain ${0<\{M,T\}<5}$, see Appendix~\ref{sec:master_equation} for details. We classify the distribution into (non)ergodic, (de)localized, 
and fractal types by means of the characterization schemes described in Sec.~\ref{sec:generic_distribution}.    

\subsection{Cross-section of the phase diagram}

We first investigate a single cross-section of the ``phase diagram'' through the parameter space at fixed \(M=2.92\), considering 640 equally spaced values of \(T\) in the interval \(T\in[10^{-3}, 5]\), see Figs.~\ref{fig:pr_cross_section1} and~\ref{fig:pr_cross_section2}.  The distributions were obtained using the ME method with \(N=10^5\) grid cells, starting from uniform distributions and iterating for up to \(10^4\) steps. Different special conditions are indicated by vertical lines in Figs.~\ref{fig:pr_cross_section1} and~\ref{fig:pr_cross_section2}, see captions of these Figures. Additionally, distributions for projective \(T\)-values were calculated from Eq.~\eqref{eq:proj_solution}.

\subsubsection{Delocalization and localization regions}

\begin{figure}[ht]
    \includegraphics{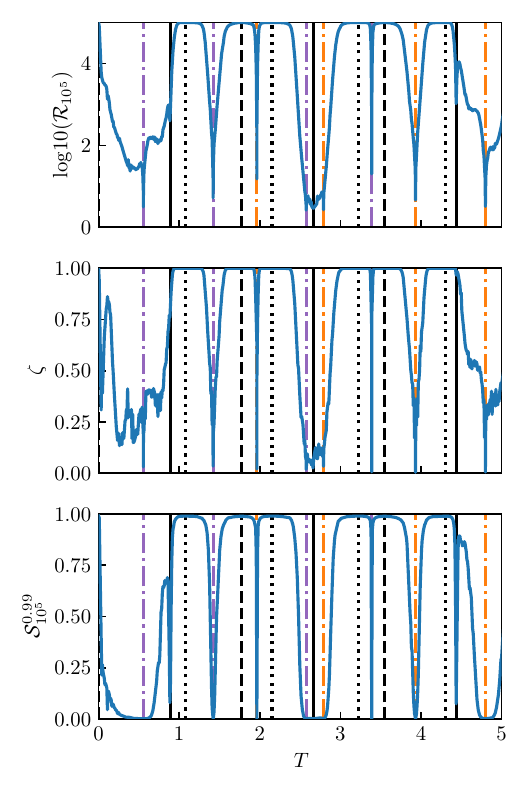} \vspace*{-0.5cm}
    \caption{Indicators of (de)localization in a cross-section of the $M$-$T$ ``phase diagram'' with \(T \in [10^{-3}, 5]\) at \(M=2.92\). 
    All curves were obtained numerically from a discretized ME with \(N = 10^5\) grid cells and a maximum of \(10^4\) iteration steps, starting from a uniform initial distribution. \textit{Upper panel:} Participation ratio \(\mathcal{R}_{N}\) [defined in Eq.~\eqref{eq:participation_ratio}]. Large (small) values correspond to delocalization (localization). \textit{Middle panel:} Scaling exponent $\zeta$ of the PR    
    [Eq.~(\ref{eq:PR_scaling_exp}].  
    \textit{Lower panel:} Support \(\mathcal{S}_{N}^{C}\) with $ C = 0.99 $
    [defined in Eq.~\eqref{eq:support_def}], which makes a direct link between (de)localization and (non)ergodicity.  
    Solid vertical lines: period-2 cases \(YT = (2k + 1) \pi\) with \(k \in \mathbb{N}_0\). Dashed lines: frozen cases \(TY = 2k \pi\). Dotted lines: shift cases \(MT = k \pi\). Dash-dotted lines: projective limit defined by conditions~\eqref{2Site_Proj1} (purple) or \eqref{2Site_Proj2} (orange).
        }
    \label{fig:pr_cross_section1}
\end{figure}

The upper and middle panels of Fig.~\ref{fig:pr_cross_section1} show, respectively, the PR, $\mathcal{R}_N$, and the PR scaling exponent, 
$\zeta$, calculated for $N=10^5$. First, we note that the PR exponent essentially follows the behavior of the PR and, therefore, 
we can focus on the dependence $ {\cal R}_N(T)$. All regions with the values of $ {\cal R}_N $ that are attributed to localization or delocalization, 
can be understood based on the properties of the special cases.

In the limit $T\to 0$, as well as at the values of $T$ that correspond exactly to the frozen (dashed lines) and shift (dotted lines) cases, the PR indicates delocalization: $\mathcal{R}_N\approx N$. However, this is an artefact of the uniform initial condition. 
At \(T\to 0\), the detector state is Zeno-frozen (cf. Ref.~\cite{snizhko_2020}), since there is no (joint) unitary time evolution between the measurements. Therefore, the detector state never changes, and the measurement outcome is always no-click. At \(T = 0\), the matrix \(\hat{M}_{-}\) acts trivially on the system state, such that it remains frozen as well (though not in an eigenstate of the projective density measurement).

Similarly, exactly along the frozen cases, both matrices $\hat{M}_\pm$ act trivially and the system state never changes, while the analysis of the ME indicates a delocalized distribution. In these cases, any state is an eigenstate of the Markov matrix, and these degenerate stationary states do not correspond to the actual (frozen) ADF, which is given by the initial state.

In the shift case, the ME also always has a uniformly delocalized stationary state, as every bin has exactly the same incoming contributions. The actual ADF depends on the shift angle \(\phi\) and the initial condition: If the shift angle is commensurate with \(\pi\), \(n \phi = 2 \pi\) for some \(n \in \mathbb{N}\), the ADF only has support on a finite set of points that depends on the initial condition. If the angle is not commensurate with \(2\pi\), any point on the invariant manifold can be approached arbitrarily closely and the ADF is delocalized. In any (shift) case, generic post-measurement quantum trajectories do not converge to the GC.

Thus, the indicators based on the solution of the discretized ME with a uniform 
initial condition may fail in the special cases. In particular, they may overlook the localized points: ``fake delocalization'' at special points can emerge as a result of averaging over localized distributions. 
In this regard, the structure of the matrix $\hat{\mathcal{M}_N}$ can provide additional valuable insights into the expected behavior 
of the ADF, see Appendix~\ref{sec:methods}.

The situation is different already in the immediate vicinity of the frozen and shift cases: Freezing of state and Bloch-angle is clearly broken away from 
the commensurability points.
The deviation lifts the exact degeneracy of eigenvalues in the frozen case, and the ME approach can be used---agreeing well with the MC time average (Appendix~\ref{sec:stationary_vs_time_average}). Note that the maximum (for given $N$) value of the PR is achieved for finite segments of $T$-values around special commensurability points, suggesting the existence of delocalized phases. Specifically, the PR indicates regions of true delocalization of width \(\Delta T \approx 1/2\) with almost saturated PR values \({\mathcal{R}_{N} \approx N = 10^5}\), where the distributions are close to uniform.

The localized behavior, \({\mathcal{R}_{N}\ll N}\), is correlated with period-2 trajectories (solid lines) and the projective limit (dash-dotted lines). 
The behavior of the indicators shows, however, that localization at the period-2 trajectory is destroyed by slight deviations from these commensurability points. Perturbatively, we might expect that a product \(\hat{M}_\pm\hat{M}_\mp\) still has eigenstates close to the period-2 peaks. However, the attraction to 
the period-2 peaks relies on long chains of products \(\hat{M}_+\hat{M}_-\), which emerge from the contraction argument \(\hat{M}_\pm^2 = \mathds{1}\), 
see Sec.~\ref{sec:period_2_traj} and Appendix~\ref{App_2Site-RandomWalks}. If this contraction is not fulfilled exactly, the deviations are accumulated
in long chains. This may lead to the broadening of the peaks in the ADF.

In the projective limit, the solution of the ME is given by a discrete set of delta-peaks, whose strength decays exponentially with the number of necessary transitions from the main projective peak, resulting in apparent localization. Indicators in Fig.~\ref{fig:pr_cross_section1} 
demonstrate that the stability of this projective localization with respect to variation in \(T\) depends on 
the value of \(T\), in particular, through the ``interaction'' with neighboring commensurability conditions. Specifically, 
for projective lines at $ T \approx 0.6,\,2.5,\, 2.8 $, and $4.8$,
we come across relatively wide regions of small PR values. Regions around $ T \approx 1.5$ and $3.9$ lie within narrow ``valleys'', while those at $ T \approx 1.4$ and $3.3$ are enclosed by the shift and frozen lines and only show a sharp dip. 
At the same time, the combined effect of projective and period-2 cases favors localization in a relatively broad range of $T$, see the region around $T=2.7$.
Interestingly, all broad regions of apparent localization observed in Fig.~\ref{fig:pr_cross_section1}
correspond to cases with a strong hierarchy in both post-measurement matrices.

The support measure 
\(\mathcal{S}^{C}_{N}\) with $ C  = 0.99 $ shows behavior similar to that of the PR and the PR exponent. 
Since nonergodic regions turn out to correlate with low values of this measure this establishes a 
link between (de)locatization and ergodic properties of the steady-state distributions.    
Finally, yet another indicator of (de)localization---the typical value of the ``histogram of heights'' shown in 
Fig.~\ref{fig:pr_cross_section2}---also exemplifies the correlations between the indicators oriented on the
(de)localization and (non)ergodicity.

In summary, such defined ``localization'' manifests in analogs of several common localization measures, all of them yielding similar regions in the cross-section, where the distribution appears localized. 
Whether these regions correspond to genuine localization (with the width of the ADF peaks tending to zero in the continuous limit, as it does exactly at the projecting limit or for period-2 trajectories) or to granularity (with the peak width saturating with increasing the resolution of discretization) requires additional analysis, which is beyond the scope of this paper. The consideration of (non)ergodicity and fractality below sheds more light on this question.

\subsubsection{Ergodicity and fractality}

\begin{figure}[ht!]
    \includegraphics{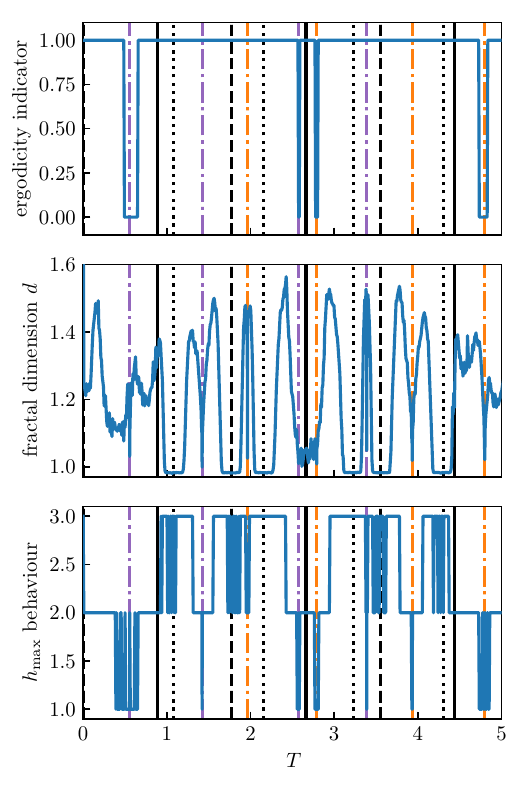} \vspace*{-0.9cm}
    \caption{(Non)ergodicity and fractality vs. localization in the cross-section of the $M$-$T$ ``phase diagram'' with \(T \in [10^{-3}, 5]\).
    All curves were obtained from discretized ME,
    parameters are the same as in Fig.~\ref{fig:pr_cross_section1}.
    \textit{Upper panel:} Ergodicity indicator of the Markov process corresponding to \(\hat{\mathcal{M}}_{N}\): 0 (1) indicates nonergodicity (ergodicity) in terms of connection of graph \(G_N\),   Eq.~\eqref{eq:graph_from_transition_matrix}.
    \textit{Middle panel:} Fractal dimension \(d\) from box counting [Eq.~\eqref{eq:definition_fractal_dim}]; \({d=1}\) corresponds to a one-dimensional curve, \({1<d<2}\) gives a nontrivial Hausdorff dimension of the curve. \textit{Lower panel:} Category of the position of the maximum of the height histogram [Eq.~\eqref{eq:histogram_of_heights}]; 1 and 3 indicate localization and delocalization, respectively; 2 indicates peaks on top of the extended background. These categories establish a connection between localization and nonergodicity as well as delocalization and ergodicity, see discussion in the text.
    The vertical lines denote the same cases as in Fig.~\ref{fig:pr_cross_section1}.
       }
    \label{fig:pr_cross_section2}
\end{figure}

We move on to consider the ergodicity indicator 
shown in the upper panel of
Fig.~\ref{fig:pr_cross_section2}. The plot shows whether or not the Markov process corresponding to \(\hat{\mathcal{M}}_{N}\) with $ N = 10^5 $ is ergodic [whether or not \(G_{N}\) defined in Eq.~\eqref{eq:graph_from_transition_matrix} has a single SCC]. 
Importantly, nonergodicity does not necessarily imply localization.
An example is given by a granulated case where delocalized grains are embedded in a localized background, which is characterized by the ADF not having full support. Such a distribution could look similar to the curve in the lower panel of Fig.~\ref{fig:special_distributions}, where a gap of very low density opens between intervals where the distribution is finite.
As argued above, nonergodicity of the discrete process does imply nonergodicity of the continuous process---there are invariant disconnected subspaces on the GC. However, an analogous statement does not directly hold true for ergodic regions: ergocity of a discrete process does not necessarily mean ergodicity in the continuous limit. 
Nevertheless, the extension of ergodic regions at finite discretization observed in Fig.~\ref{fig:pr_cross_section2} suggests that genuine ergodic phases should exist in our parameter space.
Indeed, these regions are remarkably stable with respect to the number of discretization cells, implying their stability and finiteness in the continuous limit.
This is quite natural, in fact, once the existence of \textit{finite} nonergodic regions has been established for continuous processes: all the rest should be then regarded as ergodic.

Most of the considered values of \(T\) in the cross-section correspond to ergodic distributions.
However, we find nonergodic intervals of $T$ in the vicinity of those projective (dash-dotted) lines,
where both post-measurement matrices have strong eigenvalue hierarchies, 
i.e., the vicinity of the double projective case. To understand this, consider two small but finite intervals around the two ``strong'' eigenangles: \(I_\pm := [\theta_{\pm} - \delta_{\pm},\, \theta_{\pm} + \delta_{\pm}]\). 
A strong eigenvalue hierarchy means that a large fraction of the GC is mapped into the vicinity of the eigenangle by the corresponding map (see Fig.~\ref{fig:me_matrix_structure}, upper-right panel). In particular,
    \begin{align}
        \Xi_{\mu}(I_{\mu}) \subset I_{\mu}, \label{eq:non_ergodic_cond_1}
    \end{align}
where $\Xi_{\mu}(I_{\mu})$ is the image of the interval $I_{\mu}$ under the action of $\Theta_\mu(\theta\in I_\mu)$.  
If these ``attractive regions'' of the intervals \(I_\mu\) are sufficiently large to include the respective other eigenangle, the condition
    \begin{align}
        \Xi_{\mu}(I_{-\mu}) \subset I_{\mu} \label{eq:non_ergodic_cond_2}
    \end{align}
is fulfilled. In this case, no 
measurement operator can facilitate escape from the intervals \(I_{\mu}\), which thus form an invariant subset for the post-measurement state~\footnote{The existence of such a region is constructively proven with analogous reasoning, if we find more than one SCC in \(G_{N}\) at any \(N\).}.

The fractal (box-counting) dimension displayed in the middle panel of Fig.~\ref{fig:pr_cross_section2}
is trivial, \(d \approx 1\), around the frozen and shift cases. This is expected, since we learned from the PR values that these cases correspond to almost uniform distributions. For all projective cases, the fractal dimension shows dips to \(d\approx 1\). Based on the structure of the ADF that shows a series of exponentially shrinking peaks 
in this limit [see middle panel of Fig.~\ref{fig:special_distributions} and Eq.~\eqref{eq:proj_solution}], we indeed expect trivial scaling \(C(m)\propto 1 / m\) in the exact projective limit. Some of the observed dips are extremely narrow, which suggests an increased susceptibility of the vicinity of the projective case towards forming a fractal pattern, in agreement with the argument in Sec.~\ref{sec:Hausdorff}. Some of the period-2 cases also correspond to dips in the fractal dimension.
This agrees with our expectation that, exactly in this case, there are only two peaks in the ADF and, therefore,  
$ d = 1 $.
All other cases have a nontrivial fractal dimension---this appears to be the generic case in our system.

Although (non)ergodicity and (de)localization are not necessarily correlated types of behavior, a connection is established by the behavior of the support and the maximum in the histogram of heights.
Interestingly, regions of nonergodicity are correlated with the first category of the height indicator (analogous to the manifestation of the insulating phase in the local density of states). Around all of the nonergodic projective cases, the maximum ventures into the first category. Additionally, around the projective case at \(T \approx 4\), there is a small first-category dip, which is not present in the ergodicity cross-section. Further, the behavior of \(h_{\rm max}\) fluctuates, which can be due to numerical limitations~\footnote{There are two discretization steps, one to find the stationary solution of the discretized Master-equation and another one to find the histogram of heights from which the maximum is extracted.}. The surrounding parameter regions of category-one behavior are ``transitional'' category-two regions (this does not have an analog in the theory of Anderson transitions). Finally, ergodic regions correspond to category-three behavior of heights (delocalization in the Anderson picture) with occasional fluctuations into the second category around special lines.

\subsection{Phase diagram of the monitored qubit}    
\label{sec:PhaseDiagram}

Having investigated a generic cross-section in the \(M\)-\(T\)-parameter space, we proceed with studying (de)localization, fractality, and (non)ergodicity in the whole parameter plane. 
Figure~\ref{fig:phase_diags} shows the same quantities as in Figs.~\ref{fig:pr_cross_section1} and~\ref{fig:pr_cross_section2}, calculated for a \(160 \times 160\) grid in the domain \(M, T \in [10^{-2}, 5]\) at $ N = 10^4 $. The analysis of the entire diagrams is pretty much similar to that of the cross-sections, which we have elaborated in great detail.

The upper-left panel shows the PR values \(\mathcal{R}_{N}\). Again, the localized and delocalized behavior can be distinguished based on the special cases, in analogy to what was discussed before. Delocalized phases are found around frozen and shift cases (dashed and dotted lines): as discussed above, the ME with a uniform initial condition yields delocalization along these special lines due to averaging over localized trajectories. In the vicinity of these lines, genuine delocalization quickly sets in for all quantum trajectories. Almost uniform delocalization around frozen and projective cases is remarkably stable with respect to changes in parameters, manifesting in broad yellow ``bands'' around commensurate lines.

\begin{widetext}
  \begin{minipage}{0.85\linewidth}
   \begin{figure}[H]
    \centering
     \includegraphics{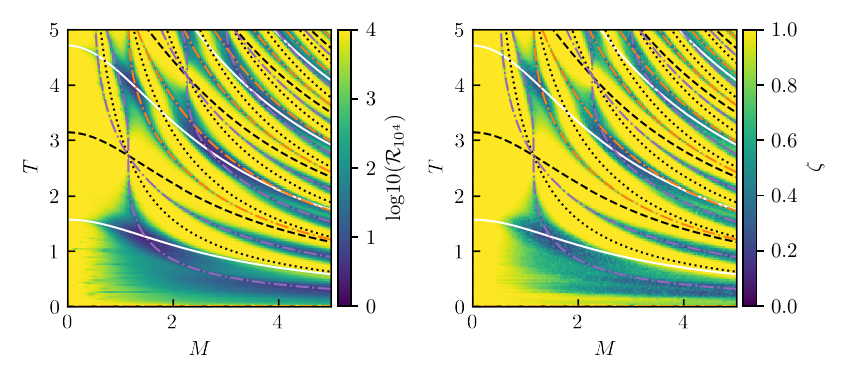}
      \vspace{-0.3cm}
       \includegraphics{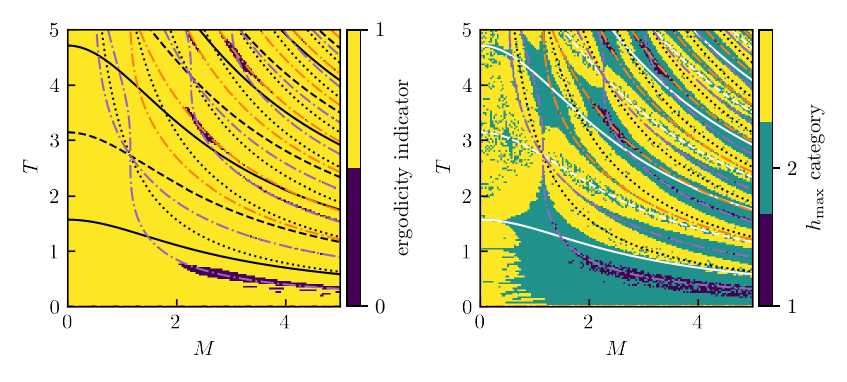}
         \includegraphics{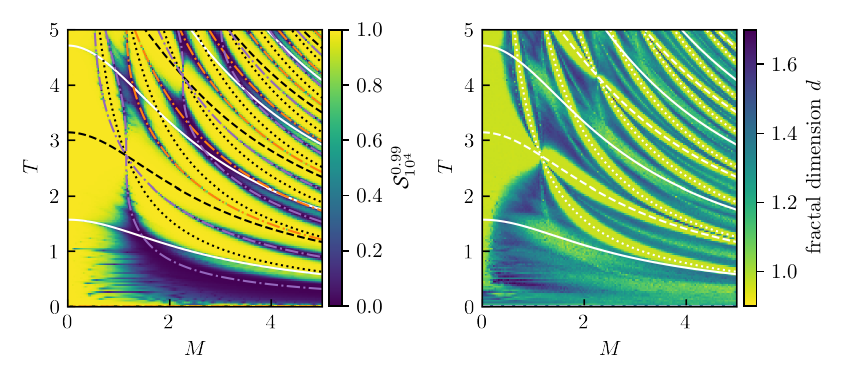}
\caption{Diagrams illustrating localization, fractality, and ergodicity of the distributions obtained by using the ME method. All panels were generated with \(N = 10^4\) grid cells and a maximum of \(10^4\) iteration steps starting from a uniform initial distribution. \textit{Upper-left panel:} participation ratio, Eq.~\eqref{eq:participation_ratio}; large (small) values correspond to delocalization (localization). \textit{Upper-right panel:} Scaling exponent $\zeta$ of the PR [Eq.~\eqref{eq:PR_scaling_exp}]; 0 (1) indicate localization (delocalization). \textit{Middle left panel:} Ergodicity marker of the Markov process corresponding to the transition matrix \(\hat{\mathcal{M}}_{N}\), indicating whether the graph defined in Eq.~\eqref{eq:graph_from_transition_matrix} has a single strongly connected component; yellow (purple) regions are ergodic (nonergodic).
\textit{Middle right panel:} Category of the position of the maximum of the height histogram,  Eq.~\eqref{eq:histogram_of_heights}; 1 and 3 indicate localization and delocalization, respectively; 2 indicates peaks on top of the extended background.
\textit{Lower-left panel:} Support  \(\mathcal{S}_{N}^{C}\), Eq.~\eqref{eq:support_def}, with $ C = 0.99 $. \textit{Lower-right panel:} Fractal dimension \(d\) of the distributions, 
Eq.~\eqref{eq:definition_fractal_dim}; \({d=1}\) corresponds to a one-dimensional curve, \({1<d<2}\) describes a fractal pattern.
Lines show special cases described in Sec.~\ref{sub_sec:special_cases}. Solid lines mark period-2 cases: \(YT = (2k + 1) \pi\) with \(k \in \mathbb{N}_0\). Dashed lines denote frozen cases: \(YT = 2k \pi\). Dotted lines denote shift cases: \(MT = k \pi\). Dash-dotted lines correspond to the projective limit for \(\hat{M}_-\) [purple, Eq.~\eqref{2Site_Proj1}] and for \(\hat{M}_+\) [orange, Eq.~\eqref{2Site_Proj2}].
            }
            \label{fig:phase_diags}
        \end{figure}
    \end{minipage}
\end{widetext}

Localization is observed around the projective limit (orange and purple lines) and period-2-trajectories (solid lines).
The period-2 trajectory crosses the delocalized bands through narrow ``bridges'' of localization (for example at \(T = 0.3, M \approx 4.7\)). The projective lines are surrounded by a narrow or broad region of localization, depending on the parameters. The case where both GC maps $\hat{M}_\pm$ are almost projective is special in that such regions always correspond to apparent localization in finite bands. Support measure and PR scaling phase diagrams are visually very similar to the PR diagram. 

Comparing ergodicity of the discrete process (middle-left panel of Fig.~\ref{fig:phase_diags}) with the \(h_{\rm max}\) 
indicator (middle-right panel of Fig.~\ref{fig:phase_diags}) we observe that the locations of the nonergodic regions are completely correlated 
with the locations of the most localized regions. There is an astonishing agreement between category three of the height distribution and uniform delocalization, as well as between category one and nonergodicity. Categories one and three are separated by the ``transient'' category two. 

The fractal dimension as a function of \(M\) and \(T\) (lower-right panel of Fig.~\ref{fig:phase_diags}) confirms our conclusions 
from the analysis of the cross-section. Extended regions of almost-uniform distributions around frozen and shift cases correspond 
to a trivial dimension $ d = 1 $. Away from these regions, the fractal dimension is nontrivial (except for the vicinity of the projective lines, which are not drawn to avoid covering fine lines of \(d \approx 1\)).

As discussed above, the regions where both matrices \(\hat{M}_\mu\) have a strong eigenvalue hierarchy (double-projecting limit) are also special for ergodicity. As expected, in the vicinity of double-projective points (intersections of orange and blue curves in Fig.~\ref{Fig:ProjectingMatr})  we find finite regions of nonergodicity embedded into the mostly ergodic phase diagrams. To estimate the expected extension of these regions, we consider the following conditions for a nonergodic region:
    \begin{enumerate}
        \item Both matrices \(\hat{M}_\mu\) have eigenvectors on the GC, establishing the existence of a region according to relation~\eqref{eq:non_ergodic_cond_1};
        \item The eigenangles $\theta_{-\mu}$  lie within the attractive region of the ``partner'' maps $\hat{M}_\mu$, fulfilling the relation~\eqref{eq:non_ergodic_cond_2}. This corresponds to~\footnote{Technically, none of the discussed properties of the maps prevents the second derivative from changing sign between those points, such that attraction at the ``wrong'' eigenangle does not imply attraction everywhere between the eigenangles. We ignore this possibility in our estimate.}
            \begin{align}
            \label{eq:Pheno_NonErgodic}
                \left|\Theta_{\mu}'(\theta_{-\mu})\right|
                < 1.
            \end{align}
    \end{enumerate}
    
The nonergodicity criterion covers all regions which we find to be nonergodic at finite discretization (and which belong to category one of the height indicator), see the left panel of Fig.~\ref{fig:ergodicity_calc}. Furthermore, it predicts that the regions of nonergodicity actually extend further in the parameter space (which would only be visible at higher discretization in our numerical procedures).

Under certain conditions, a direct connection between nonergodicity and localization can be demonstrated. Given a nonergodic process with images
\(\Xi_\pm(I) \subseteq I\) for the invariant subset \(I\), the inequality
\begin{align}
    \sum_{\mu \in \{-, +\}} \left| \pdv{\Theta_\mu(\theta)}{\theta}\right| \leq c < 1, \quad \theta \in I \label{eq:nonergodicity_localization}
\end{align}
guarantees a localized distribution, starting from an initial state that is nonzero within the invariant subset, e.g.,
\begin{align}
    W_0(\theta) \equiv \begin{cases}
        1 / |I|, & \theta\in I,\\
        0, & \text{else}. 
    \end{cases}
\end{align}
To see this, we consider the support \(S_1\) of the distribution after applying the protocol once:
\begin{align}
    |S_1| = \left| \Xi_+(I) \cup \Xi_-(I) \right| \leq \sum_\mu \int_I \dd{\theta} \left| \pdv{\Theta_\mu(\theta)}{\theta}\right| < c |I|.
\end{align}
This argument can be iterated indefinitely with the support shrinking exponentially with the number of iterations. Thus, the limiting distribution function consists of a set of \(\delta\)-peaks on the invariant subset and is, therefore, localized.

The above argument provides a sufficient condition for localization of the ADF within the chosen invariant subset $I$. Generically, there can be
several disjoint invariant sets, $I_j$, in the GC and a similar analysis 
should be performed for each set.
If the localization condition (\ref{eq:nonergodicity_localization}) 
holds true only for some $I_j$,
a ``granular'' structure, where the ADF is localized 
in some invariant subsets and delocalized over other angles,
is not forbidden.
If the initial angle $\theta_0$ belongs to a ``localized subset'' of this 
structure, the state of the system is localized in the long-time limit.

More interesting is the truly localized regime where
Eq. (\ref{eq:nonergodicity_localization}) is satisfied 
in all $I_j$.
At this point, we note that, besides for the double-projective points, the condition~\eqref{eq:nonergodicity_localization} can never be satisfied for the entire GC, because the maps \(\Theta_{\pm}\) are bijections from the GC to the GC: The trivial map \(\Theta(\theta) = \theta\) has slope one everywhere, and any map that has a slope smaller than one necessarily has the slope greater than one somewhere, such that the image of the GC is equal to the GC. However, 
this does not contradict the existence of true localization, since regions with \(\sum_\mu | \Theta_\mu'(\theta) | \geq 1\) can be located in the complement \(\bar{U} \equiv [-\pi, \pi) \backslash U\) of the union $U$ of all $I_j$. Crucially, $ \bar{U} $ is by construction unstable because of ``leakage'' into invariant subsets
\footnote{
See the condensation graph in Fig.~\ref{fig:nonergodic_graph_example_large} of Appendix~\ref{sec:non_erg_elaboration} as an example of such leakage to the blue supernode. Without leakage those complementary regions would themselves form an invariant subset.
} 
and, thus, cannot be part of the support of the converged distribution.
We conclude that the corresponding steady-state 
distributions are fully localized, since localization (in the sense of absence of states) within the complement $\bar{U}$ is established in the limit \(N\rightarrow \infty\).

We numerically confirmed that the condition~\eqref{eq:nonergodicity_localization} holds for most of the regions of nonergodic parameters, implying that the associated ADFs feature \(\delta\)-peaks on the constructed invariant subsets.  
To evaluate this condition numerically, we calculate the invariant subspaces for the nonergodic parameters identified in Fig.~\ref{fig:phase_diags} as the leaf nodes of the corresponding graphs (see also Sec.~\ref{sec:basic_definitions_concepts} and Appendix~\ref{sec:non_erg_elaboration}). Parameters for which the condition is fulfilled for \textit{all} invariant subspaces are indicated in the right panel of Fig.~\ref{fig:ergodicity_calc} as dark regions, indicating the localized phases within the nonergodic phase
\footnote{
We note that, without further conditions, it is possible to miss invariant subspaces with a given finite resolution \(N\). For these missed intervals, the localization condition may not be fulfilled, and the ``granular'' structure may appear
in the limit \(N \rightarrow \infty\). At the same time, since Eq.~\eqref{eq:nonergodicity_localization} gives only a sufficient condition for localization, it is possible that the localized phase extends beyond the dark regions in the right panel of Fig.~\ref{fig:ergodicity_calc}
.}.

\begin{widetext}
    \begin{minipage}{\linewidth}
       \begin{figure}[H]
            \centering
            \includegraphics{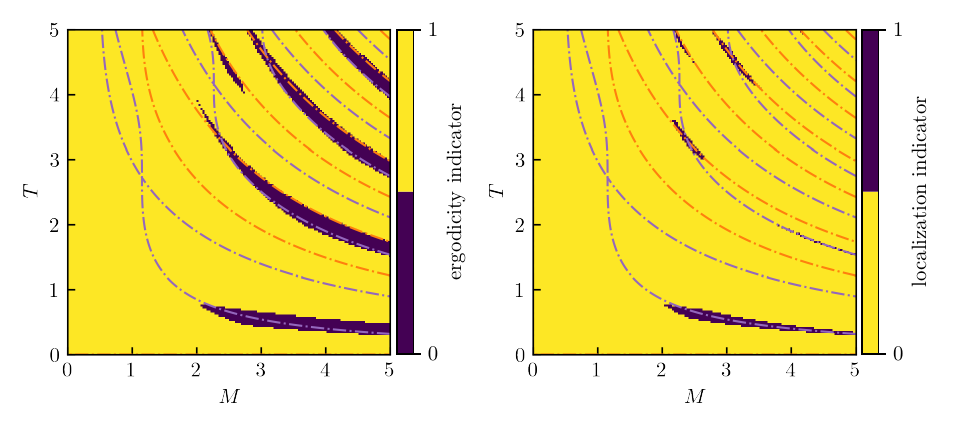}
            \caption{Evaluation of nonergodicity and localization conditions. \textit{Left panel:} Ergodic and nonergodic regions obtained by using the phenomenological nonergodicity condition in the continuous Markov process, Eq.~(\ref{eq:Pheno_NonErgodic}). Purple and orange lines correspond to projecting \(\hat{M}_-\) and \(\hat{M}_+\) matrix, respectively. The \(\hat{M}_+\) matrix has a strong eigenvalue hierarchy in the lower purple region as well, see Fig.~\ref{fig:gc_convergence_numerics}. The color code is the same as in the middle left panel of Fig.~\ref{fig:phase_diags}. \textit{Right panel:} Localization condition~\eqref{eq:nonergodicity_localization} evaluated for all the invariant subspaces obtained as the leaf nodes in nonergodic graphs \(G_N\), see Sec.~\ref{sec:generic_distribution} and Appendix~\ref{sec:non_erg_elaboration}. The condition for localization is fulfilled in the dark purple regions.}
            \label{fig:ergodicity_calc}
        \end{figure}
    \end{minipage}
\end{widetext}

\subsection{Summary of the numerical analysis}

Combining analytical solutions for the special cases with extensive numerical analysis, we have managed to identify various distinct regimes and phases in the monitored qubit's statistics and describe its phase diagram. The analytical approach has been used to build the skeleton of the diagrams, the numerical study has allowed us to understand the properties of the generic cases.

Analysis of the PR has allowed us to distinguish localized and delocalized regimes. The behavior of the system with respect to (de)localization can be qualitatively understood in terms of the special cases (Sec.~\ref{sub_sec:special_cases}). From the number of SCCs in the graph of the discretized Markov process, we have found ergodic and nonergodic phases. We have established the existence of finite nonergodic phases in the continuous process; hence, ergodic phases also survive in the continuous case. Thus, we predict a transition between the ergodic and nonergodic phases. The behavior of \(h_{\rm max}\) establishes a connection between (de)localization and (non)ergodicity ~\footnote{With our semi-analytical explanation of different regimes in the parameter space, we do not describe every detail of the phase diagrams. One particularly prominent feature is visible in the lower-left corner of the \(h_{\rm max}\), PR, and fractal dimension diagrams: Regions of delocalization are broken by almost horizontal lines of localization (category two for \(h_{\rm max}\), \(d \approx 1\) for the fractal dimension). This frequent change in the behavior of the distributions may be attributed to crossings of other types of commensurability, which we did not address here.}. Further, we have identified the regions of a truly localized phase inside the nonergodic phase. Whether the nonergodicity-ergodicity and localization-delocalization transitions are equivalent in the monitored qubit is a challenging question, which we relegate for future work. Finally, we have found that fractal distributions are rather generic for solutions of a continuous functional equation governing the ADF.

The phase diagrams that are presented in this section, which reveal fractality and transitions in the statistics of a single monitored qubit, are our {\it main results}.

\section{Discussion}
\label{sect:Disc}

\subsection{Relation to previous work: Continuous measurements limit}
\label{sec:VIIA}

In the context of weak measurements, the concept of continuous measurements~\cite{Korotkov1999, Korotkov2001, Goan2001} of a qubit has been extensively discussed in the literature, both from the theoretical (see, e.g., Refs.~\cite{li_2014, snizhko_2020, Kumar2020, dubey_2023} and references therein) and experimental ~\cite{qubit_quantum_traj_exp,quantum_jumps,quantum_trajectory_tracking_nn,continuous_error_correction_experiment,continuous_control_experiment} perspective. In our notations, the corresponding limit is $ M / \gamma \to \infty $ 
and  $ T \gamma \to 0 $, with the product $M^2 T/\gamma$ being fixed. In this limit, after averaging over quantum trajectories, the evolution of a qubit is described by the famous Lindblad equation. 

A system-detector setup similar to ours was theoretically considered in the continuous limit in Ref.~\cite{snizhko_2020} (see also Refs.~\cite{Kumar2020, dubey_2023} 
and the related earlier work \cite{li_2014}). Focusing on the onset of the quantum Zeno effect, the authors of Ref.~\cite{snizhko_2020} have demonstrated that this onset is characterized by a series of dynamical transitions, which also manifested in the steady-state 
probability distribution of states (analogous to our ADF). It has been discovered that one of these transitions is 
seen as an opening of a region of forbidden states in the ADF. 

Let us interpret this finding in terms of our classification of regimes 
and phases. The relevant example of the ADF is shown in the lower panel of Fig.~\ref{fig:special_distributions}. 
It implies establishing nonergodicity: the stationary ADF splits into two disconnected sectors. Specifically, a 
delocalized segment of angles is embedded in the background of zero height. This is not genuine localization in 
our language, as the delocalized region does not shrink to zero in the continuous limit. Following the list of 
expected types of behavior given in Sect.~\ref{sec:basic_definitions_concepts}, one can refer to such a regime as a ``granulated 
metal''.

We note in passing that there is
another transition predicted in Ref.~\cite{snizhko_2020} which reveals itself as the emergence of a singularity in the ADF.
The vicinity of the projective cases is associated with the emergence of power-laws in the ADF (cf.~Ref.~\cite{infinite_series_of_singularities, david_random_matrices_2}), which further develop into the fractal pattern by the mechanism of ``cloning.'' This type of possible transitions was not addressed in our analysis of phase diagram, as we focused on the indicators directly related to localization and ergodicity.

The consideration in Ref.~\cite{snizhko_2020} has focused on the far lower-right corner of the parameter plane, which is far beyond the range presented in our phase diagrams. 
Our study of the whole parameter plane reveals extended areas that correspond
to ergodic and nonergodic phases of the dynamical process. Physically, the 
strength of the system-detector coupling and a finite time interval of their joint evolution conspire nontrivially 
to give rise to regimes that are much different from the quantum Zeno effect.

For instance, in the double-projective case, the system state can be deduced from the outcome of a single detector measurement, without knowing the state before the measurement. However, the probabilities of the detector readouts are independent of the previous measurements---in contrast to the Zeno regime. Instead, the system mimics a ``quantum coin'' with just two different states which are realized at random. In the frozen cases, the system state is completely unaffected by the detector readouts. Clearly, deviating from the continuous limit (stroboscopic measurements) adds complexity to the behavior and physical interpretation of the stationary probability distribution. This point is further underlined by astonishing parallels between the ADF and wave functions in the theory of Anderson transitions.

On a more technical level, allowing for arbitrary \(M\)
and \(T\) can lead to a nonlinear implicit master equation describing the dynamical process on the GC. At any point on the GC, both measurement operators can induce large jumps of the state, which give rise to extremely complex typical quantum trajectories. This new (compared to the continuous limit) ingredient is reflected, for example, in fractal peak structures that are associated with back-and-forth cloning of peaks in the ADF, or in the relevance of commensurability conditions. The entire parameter space is structured by special cases, where the associated Markov process reduces to a one-dimensional random walk (period-2, shifting, or projective cases), complete standstill (freezing), or displays mechanisms of projection and translation associated with the continuous limit~\cite{snizhko_2020}---but different in that both measurement operators still correspond to finite-angle steps.

Despite these differences, the above-mentioned works, as well as the present paper,
demonstrate the importance of the study of the distribution functions for identifying hidden dynamical transitions that are hardly observed in the averaged properties of a monitored qubit. This key message is anticipated to be relevant also for measurement-induced dynamics in large systems with a macroscopic number of degrees of freedom.

\subsection{Implications of our findings}
\label{sec:Applications}

Our results have several immediate implications. We have already pointed out that the attraction of the quantum trajectories to the period-2 
limiting cycle or to two main angles in the double-projective case realizes passive steering of the monitored qubit. The attraction of
the quantum trajectories to the GC, where all states have the form $ \{ \cos(\theta/2), \ci \sin(\theta/2)\}^{\rm T} $, results in an effective 
fine-tuning of the tunneling amplitude to a purely imaginary quantity, $ \gamma \to \ci | \gamma | $.

The intriguing question is whether regimes and phases visible in the diagrams of Fig.~\ref{fig:phase_diags} are experimentally detectable.
The important argument in favor of observability is the broad range of system parameters where the regimes and phases show themselves. However, the phases can not be detected by monitoring conventional observables: Consider an observable that depends
on the angle $ \theta $ and introduce its mean value in terms of the ADF
\[
  \langle A \rangle_\theta = \int_{- \pi}^{ \pi} \dd{\theta} A(\theta) W(\theta)  \, .
\]

Two common examples are the expectation value of the site occupation numbers, $ N_1 = \cos^2(\theta/2) $ and $ N_2 = \sin^2(\theta/2) $, and the formally introduced entropy of entanglement between the sites: $ S_{2} = -{\rm Tr} 
[\hat{\rho}_2 \log(\hat{\rho}_2)]$, where $\hat{\rho}_2 \equiv {\rm Tr}_1 \hat{\rho} $, and $ \hat{\rho} $ is the density matrix of the qubit. Note that $ S_{2} $ does not describe any entanglement of particles, since there is only one electron in the system; nevertheless, this quantity carries some useful information about the monitored qubit. Using the results
of Appendix~\ref{AppendixA}, it is straightforward to show that $ S_2(\theta) $ at post-measurement times can be
expressed via $ N_{1,2} (\theta) $ as 
$ S_2 = - [ N_1 \log(N_1) + N_2 \log(N_2) ] $ 
and is a smooth function of $ \theta $.

Note that $ \langle N_{1,2} \rangle_\theta $ and $ \langle S_2 \rangle_\theta $ do not necessarily distinguish between the localized and strongly delocalized regimes. Indeed, the same expectation value  $N_1=1/2$ can be obtained for both the uniform ADF and the localized ADF with two 
equal peaks at $\theta=\pm \pi/2$. For the same reason---the smooth dependence of these observables on $\theta$---their averaged values are also unable to reflect the ergodicity-to-nonergodicity transition. 

Instead, as emphasized at the end of Sec.~\ref{sec:VIIA}, the transition can be observed in ``tomographic'' measurements of individual quantum trajectories \cite{qubit_quantum_traj_exp,quantum_jumps,quantum_trajectory_tracking_nn,snizhko_2020}, which yields the full state distribution function. 
It would also be interesting to relate the transition to the statistics of sequences of measurement outcomes (cf. Refs.~\cite{li_2014,dubey_2023}). A search for convenient observables that are sensitive to the transitions in the statistical properties of the monitored microscopic qubit remains an open question.

\subsection{Outlook}
\label{sec:Outlook}

We have demonstrated that the study of the smallest nontrivial models of ancilla-based measurements uncovers a lot of
interesting properties of dynamical processes, physical implications of indirect measurements, and even the possibility 
of phase transitions in the dynamical behavior. Let us list some questions, which we leave for future studies, including 
possible extensions of our model and methods.

Introducing variations to specifics of our model, for example, imbalanced energy levels of the qubit, or a different kind 
of coupling to the ancilla, may generate a family of interesting quantum systems. It is a priori not clear whether or not all
models can be described by a single angular variable.
Another natural modification of the protocol would be to randomize the times between consecutive measurements, such that the protocol would be characterized by a single period $T$ only on average.
One can also omit the reinitialization of the detector, allowing the system to keep a memory of the measurement outcome (which corresponds to considering ``correlated random products'' of post-measurement matrices~\cite{infinite_series_of_singularities} in our approach).
Robustness of the various regimes and phases as well as of the predicted phase transition to these
modifications
is worth investigating.

Regardless of the model details, it is interesting to understand better the mechanism of fractality, if it appears.
To this end, one can use the locator expansion, e.g., close to the double-projective case of our setup, see modern implementations of the method in Refs.~\cite{yevtushenko_virial_2003,yevtushenko_2007,kronmuller_2010}.
Exploring the possibility of multifractality in the category-two regions of the height indicator and analyzing fractality in terms of the singularity spectrum (see also the comment at the end of Sec.~\ref{sec:Hausdorff}) may establish an even closer connection of transitions in the monitored qubits to the theory of Anderson localization. Regarding the numerical analysis of multifractality, a challenge of calculating scaling exponents for higher moments is the necessity to obtain the distribution at higher resolution \(N\) than used for the participation ratio. Besides fractal exponents of the limiting \(N\rightarrow \infty\) distribution, it could also be interesting to investigate the transient ``dynamical'' scaling of fractal exponents with the number of time steps of the protocol.

The similarities between Anderson (de)localized wave functions and the state distribution of the monitored qubit
gives a hint to their common mathematical origin. A search for such an origin may start from a field-theoretical connection between Anderson and measurement-induced transitions which has been recently reported 
in Refs.~\cite{Poboiko2023,poboiko2023a}.
Understanding (de)localization in the continuous case based on the Master equation could rigorously establish the 
localization-delocalization transition in the continuous model.

Especially interesting further studies are related to increasing the number of particles in the system which can be achieved
by considering either a two-site chain with spinful fermions or bosons, or by slightly enlarging the number of sites that could then accommodate more spinless fermions. 
This would allow one to explore the interrelations between the ergodicity/nonergodicity (or possible localization)
transitions in the monitored qubit with the entanglement between particles, thus going towards the field of measurement-induced entanglement transitions. 
Finally, it would be interesting to understand the influence of interparticle interactions in larger setups on
the phases and regimes discovered in the present work.

\section{Summary and Conclusions}
\label{sect:Concl}

We have studied statistical properties of a single qubit under the influence of stroboscopic ancilla-based measurements. 
The detector (ancilla) is represented by another two-level system. The qubit and the detector are coupled and evolve
unitarily between the measurements. The detector is projectively measured at the end of each interval of the unitary
evolution and, after this, is re-initialized in a given state.
This protocol defines the two measurement (Kraus) operators (represented by $2\times 2$ matrices in our model), which 
map a qubit state onto one of two possible new post-measurement states, depending on the measurement outcome.
Any quantum trajectory of the discrete-time evolution of the qubit's post-measurement state is described by a random 
product of these measurement matrices, with the probabilities of applying them dictated by Born's rule. 

The qubit state can be parameterized by two angles on the Bloch sphere. We showed that, in the long time limit, the 
quantum trajectories in our model are generically attracted to 
a one-dimensional circle on the Bloch sphere (Grand Circle -- GC), which is described by a single angle, $ \theta $.
We have characterized statistics of the qubit states
in terms of the ADF $W(\theta)$ -- the angle distribution function on the GC, Eq.~\eqref{eq:Wbar}. 
We have investigated the ADF by combining the analytical approach, suitable for some special cases, with two complementary 
numerical methods (solution of a discretized master equation and Monte-Carlo simulations of individual quantum trajectories).
We have revealed the richness of the dynamics of the monitored qubit and demonstrated remarkable similarities to
Anderson localization in disordered systems. Namely, as the system parameters are varied, the ADF of a monitored qubit 
exhibits a number of regimes reminiscent of those in the theory of Anderson transitions, appearing either localized or 
delocalized or even fractal. 

Motivated by the analogy with Anderson localization, we have used several standard indicators from the localization theory and
supplemented them with those from the theory of random evolution. The indicators encompass the participation ratio [Eq.~\eqref{eq:participation_ratio}] and its 
scaling exponent [Eq.~\eqref{eq:PR_scaling_exp}], the support measure [Eq.~\eqref{eq:support_def}], 
the typical value of the 
state distribution (Fig.~\ref{fig:histogram_of_heights}), the box-counting dimension of the curve representing the ADF [Eq.~\eqref{eq:definition_fractal_dim}], and the ergodicity indicator that characterizes the structure of Markov matrices [Eq.~\eqref{eq:graph_from_transition_matrix}] for the qubit evolution.
Analysis of their properties has allowed us to quantitatively describe 
$W(\theta)$, despite fundamental distinctions between the mechanisms underlying localization in a monitored qubit and in disordered systems. 

Our combined approach has resulted in a solid classification of various emergent regimes and phases, which are reflected in 
the phase diagrams, Fig.~\ref{fig:phase_diags}. The analytical consideration of special cases (in particular, related to commensurability of the stroboscopic measurements) has been used to build the skeleton of the diagrams, while the numerical study has allowed us to understand the properties of the generic cases.
Concretely, we have established the existence of ergodic and nonergodic phases and identified regimes of localization and delocalization behavior of the ADF. In a large portion of the parameter space, $W(\theta)$ exhibits fractality, which ranges from strong to weak, cf. discussion in Refs.~\cite{kravtsov_2010,kravtsov_2011,kravtsov_2012} and references therein. We have predicted a transition between
the ergodic and nonergodic phases and demonstrated genuine localization inside the nonergodic phase. This is our main result that may pave the way to the theory of various transitions
in monitored qubits.
Our work highlights the importance of studying the distribution functions of quantum states, as they can reveal concealed transitions that remain unnoticed when focusing on the averaged properties of systems subject to quantum measurements.

\acknowledgements

We thank Jonas Karcher, Alexander Mirlin, and Kyrylo Snizhko for useful suggestions and comments. In addition, I.V.G. is grateful to his co-authors of Refs.~\cite{Roy2020, Herasymenko2021, Doggen2022a, Doggen2023, MedinaGuerra2023} and \cite{Morales2023} for earlier stimulating discussions on the ancilla-based measurement protocols. 
P.P. and I.V.G. acknowledge support by the Deutsche Forschungsgemeinschaft (DFG) via the grant GO 1405/6-1. D.B.S. was supported by the SCS of Armenia via grants No. 21T-1C037 and 20TTAT-QTa003.

\appendix

\widetext

\section{Main equations describing the case $ \gamma \ne 0 $}
\label{AppendixA}

In this Appendix, we describe the derivation of the one-step mapping for the case $\gamma\neq 0$, see Sec.~\ref{sec:two_sites_half_filling}. 

\subsection{Post-measurement mapping for two tunneling coupled sites}
\label{app:2Site_Mapping}

To find the post-measurement mapping we first solve the evolution equation in the time interval $ 0 < t < T $:
\begin{align}
 \me^{-\ci \hat{H} t} | 1, 0 \rangle \otimes | - \rangle & = 
 a^{(+)}_{10} | 1, 0 \rangle \otimes | + \rangle + a^{(-)}_{10} | 1, 0 \rangle \otimes | - \rangle +
     a^{(+)}_{01} | 0, 1 \rangle \otimes | + \rangle + a^{(-)}_{01} | 0, 1 \rangle \otimes | - \rangle \, ,
     \\
     \me^{-\ci \hat{H} t} | 0, 1 \rangle \otimes | - \rangle & = 
 b^{(+)}_{10} | 1, 0 \rangle \otimes | + \rangle + b^{(-)}_{10} | 1, 0 \rangle \otimes | - \rangle +
   b^{(+)}_{01} | 0, 1 \rangle \otimes | + \rangle + b^{(-)}_{01} | 0, 1 \rangle \otimes | - \rangle \, . 
     \end{align}
     This yields the setup state
\begin{align}
  \Psi(0 < t < T) & =
     \left[
       \left( \alpha_0 a^{(+)}_{10}(t) + \beta_0 b^{(+)}_{10}(t) \right) \ket{1, 0} + \left( \alpha_0 a^{(+)}_{01}(t) + \beta_0 b^{(+)}_{01}(t) \right) \ket{0, 1}
     \right] \otimes \ket{+} \nonumber \\
    & +
     \left[
       \left( \alpha_0 a^{(-)}_{10}(t) + \beta_0 b^{(-)}_{10}(t) \right) | 1, 0 \rangle + \left( \alpha_0 a^{(-)}_{01}(t) + \beta_0 b^{(-)}_{01}(t) \right) | 0, 1 \rangle
     \right] \otimes | - \rangle ,
   \label{app:2Sites_FreeEvol}
\end{align}
where 
\begin{equation}
\begin{split}
  a^{(+)}_{10}(t) = - \ci \left( \sin\frac{Mt}{2} \cos\frac{Yt}{2} + \frac{M}{Y} \cos\frac{Mt}{2}\sin\frac{Yt}{2} \right) , \quad a^{(+)}_{01}(t)& = - \frac{2\gamma}{Y} \sin\frac{Mt}{2} \sin\frac{Yt}{2}, \\
  a^{(-)}_{10}(t) = \cos\frac{Mt}{2} \cos\frac{Yt}{2} - \frac{M}{Y} \sin\frac{Mt}{2} \sin\frac{Yt}{2} , \quad 
  a^{(-)}_{01}(t)& =  - \ci \frac{2\gamma}{Y} \cos\frac{Mt}{2} \sin\frac{Yt}{2},
  \end{split} 
  \label{app:2Sites_Evolution-1}
\end{equation}
and
\begin{eqnarray}
\label{app:2Sites_Evolution-2}
         b^{(\mu)}_{10}(t) &=& a^{(\mu)}_{01}(t), \quad
         b^{(\mu)}_{01}(t) = a^{(\mu)}_{10}(t) - \frac{M}{\gamma} a^{(-\mu)}_{01}(t). 
\end{eqnarray}
The electron state after the first measurement depends on the measurement outcome:
\begin{equation}
  \Psi_{1}^{(\pm)} =
    \frac{1}{\sqrt{P_1^{(\pm)}}}                   
                  \left(
       \left[ \alpha_0 a^{(\pm)}_{10}(T) + \beta_0 b^{(\pm)}_{10}(T) \right] \ket{1, 0} +
       \left[ \alpha_0 a^{(\pm)}_{01}(T) + \beta_0 b^{(\pm)}_{01}(T) \right] \ket{0, 1} 
                  \right) . 
\label{app:2Site_PM} 
\end{equation}
Straightforward calculations yield the following expressions for the outcome
probabilities of the two-site model with finite tunneling:
\begin{eqnarray}\label{app:2Site_P-Click}
  {P}_j^{(+)} 
             & = & \left|\alpha_{j-1} a^{(+)}_{10}(T) + \beta_{j-1} b^{(+)}_{10}(T)\right|^2 +
                   \left|\alpha_{j-1} a^{(+)}_{01}(T) + \beta_{j-1} b^{(+)}_{01}(T)\right|^2 
                   \\
                   &=& \frac{1}{2} -   
         \frac{4 \gamma^2+ M^2 \cos(Y T)}
                   {2 Y^2} \cos (M T) 
                   \notag
\\  &+&   \frac{M}{Y^2}
               \left\{
                  Y (|\alpha_{j-1}|^2 - |\beta_{j-1}|^2) \cos\frac{YT}{2}
                  - 4 \gamma \sin\frac{YT}{2} \Im[\alpha_{j-1} \beta_{j-1}^*]
               \right\} \sin\frac{YT}{2} \sin (M T) ,
   \\
    \label{app:2Site_P-NoClick}
  {P}_j^{(-)} 
             & = & \left|\alpha_{j-1} a^{(-)}_{10}(T) + \beta_{j-1} b^{(-)}_{10}(T)\right|^2 +
                   \left|\alpha_{j-1} a^{(-)}_{01}(T) + \beta_{j-1} b^{(-)}_{01}(T)\right|^2 = 
                   1 - {P}_j^{(+)} .
\end{eqnarray}
Generalizing Eqs.~(\ref{app:2Sites_FreeEvol}-\ref{app:2Site_PM}) to later times, we obtain
the mapping of the coefficients $ \alpha_j $ and $ \beta_j $ at the post-measurement time
$ \pmt_j $ to the successive one $ \pmt_{j+1} $:
\begin{align}
\label{app:2Sites_Mapping}
  \{ \alpha_{j+1}^{(\mu)}, \beta_{j+1}^{(\mu)} \}^{\rm T} &= 
     \bigl[ P_j^{(\mu)}(\alpha_j, \beta_j) \bigr]^{-1/2} 
        \hat{M}_\mu \, \{ \alpha_j, \beta_j \}^{\rm T} ,\qquad 
      \hat{M}_\mu \equiv
        \left(
          \begin{array}{cc}
           a_{10}^{(\mu)} & a_{01}^{(\mu)} \\
           a_{01}^{(\mu)} & b_{01}^{(\mu)}
          \end{array}
        \right) \Bigl|_{t=T}.
\end{align}

\subsection{Properties of matrices $\hat{M}_\pm$}
\label{app:2Site_Mapping_Matrs}

The explicit forms of the mapping matrices $ \hat{M}_\pm $ are given by
\begin{equation}\label{2Site_M_-}
       \hat{M}_- =
        \left(
          \begin{array}{cc}
           \cos\frac{MT}{2} \cos\frac{YT}{2} - \frac{M}{Y} \sin\frac{MT}{2} \sin\frac{YT}{2} & - \ci\, \frac{\sqrt{Y^2-M^2}}{Y}\, \cos\frac{MT}{2} \sin\frac{YT}{2}\\[0.2cm]
           - \ci\, \frac{\sqrt{Y^2-M^2}}{Y}\, \cos\frac{MT}{2} \sin\frac{YT}{2} & \cos\frac{MT}{2} \cos\frac{YT}{2} + \frac{M}{Y} \sin\frac{MT}{2} \sin\frac{YT}{2}
          \end{array}
        \right),
\end{equation}
\begin{equation}\label{2Site_Mgen_+}
       \hat{M}_+ = -\ci
        \left(
          \begin{array}{cc}
             \sin\frac{MT}{2} \cos\frac{YT}{2} + \frac{M}{Y} \cos\frac{MT}{2}\sin\frac{YT}{2} & -\ci\, \frac{\sqrt{Y^2-M^2}}{Y} \, \sin\frac{MT}{2} \sin\frac{YT}{2} \\[0.2cm]
           -\ci\, \frac{\sqrt{Y^2-M^2}}{Y}\, \sin\frac{MT}{2}\sin\frac{YT}{2} & \sin\frac{MT}{2} \cos\frac{YT}{2} - \frac{M}{Y} \cos\frac{MT}{2} \sin\frac{YT}{2}
          \end{array}
        \right).
\end{equation}
Eigenvalues 
\begin{align}
    \label{EigenVal-gen}
  \upsilon_\mu^{(\eta)}  =  \frac{1}{2} \left( {\rm Tr}[\hat{M}_\mu] -
\eta\, \sqrt{ {\rm Tr}^2[\hat{M}_\mu] - 4\, {\rm det}[\hat{M}_\mu] }                                      \right)
\end{align}
and (not normalized) eigenvectors $\Upsilon_-^{(\eta)}$
of $ \hat{M}_\pm $ read as follows:
\begin{eqnarray}
  \upsilon_-^{(\eta)}
& = & \cos\frac{MT}{2} \cos\frac{YT}{2} - 
\eta \sin\frac{YT}{2} \sqrt{ \frac{M^2}{Y^2} - 
\cos^2\frac{MT}{2}} , 
\\
  \upsilon_+^{(\eta)}
& = &- \ci \left[\sin\frac{MT}{2}  \cos\frac{YT}{2} + \eta \sin\frac{YT}{2} \sqrt{ \frac{M^2}{Y^2} - \sin^2\frac{MT}{2} } \right] ,
\\
\Upsilon_-^{(\eta)} & = & \left\{
\frac{ (M/Y) \sin(M T/2) + \eta \sqrt{ (M/Y)^2 - \cos^2(M T/2)} }{(2 \gamma / Y ) \cos(M T/2)}, \ci
\right\}^{\rm T} , 
\\
\Upsilon_+^{(\eta)} & = & \left\{
\frac{ (M/Y) \cos(M T/2) + \eta \sqrt{ (M/Y)^2 - \sin^2(M T/2)} }{(2 \gamma / Y ) \sin(M T/2)}, -\ci
\right\}^{\rm T} ,
\end{eqnarray}
with $ \eta = \pm $ distinguishing the two eigenvalues (and eigenvectors) of a given matrix. 
Since
\begin{equation}
\label{Mmatr-Conj}
  \left( \hat{M}_\mu \right)^* = - \mu \hat{\sigma}_3 \hat{M}_\mu \hat{\sigma}_3 ,
\end{equation}
we conclude that eigenvalues (eigenvectors) exist ``in pairs''
\begin{equation}
  \left\{ \upsilon_\mu ; \Upsilon_\mu \right\} \quad \mbox{and} \quad \left\{ -\mu \upsilon_\mu^* ; \hat{\sigma}_3 \Upsilon_\mu^* \right\} .
\end{equation}
This suggest that either
\begin{equation}
\label{ComplSpectr}
  \mu \left( {\rm Tr}^2[\hat{M}_\mu] - 4\, {\rm det}[\hat{M}_\mu] \right) > 0 \ \Rightarrow \
  \upsilon_\mu^{(\eta)} = -\mu \left( \upsilon_\mu^{(-\eta)} \right)^* \quad 
  \mbox{(complex eigenvalues with the same modulus)} 
\end{equation}
or
\begin{equation}
\label{RealImSpectr}
  \mu \left( {\rm Tr}^2[\hat{M}_\mu] - 4\, {\rm det}[\hat{M}_\mu] \right) < 0 \ \Rightarrow \
  \upsilon_\mu^{(\eta)} = -\mu \left( \upsilon_\mu^{(\eta)} \right)^*
  \quad \mbox{(purely real/imaginary eigenvalues for $ \mu = - / + $)} .
\end{equation}
In the latter case, $ \Upsilon_\mu^{(\eta)} = \hat{\sigma}_3 \left( \Upsilon_\mu^{(\eta)} \right)^* $, which means 
that $ \Upsilon_\mu^{(\eta)} $ has a purely real first element and a purely imaginary second one.
Properties (\ref{Mmatr-Conj}-\ref{RealImSpectr}) hold true for an arbitrary product of the
matrices $ \prod_j \hat{M}_{\mu_j} $.

Generically, i.e., excluding the degenerate case, vectors
$ \Upsilon_-^{(\pm)} $ (and $ \Upsilon_+^{(\pm)} $) are orthogonal:
\begin{equation}
 \label{2Site_OrthVect}
 \left( \Upsilon_\mu^{(\eta)} \right)^{\rm T} \cdot \Upsilon_\mu^{(-\eta)} = 0 ,
\end{equation}
since the matrices $ \hat{M}_\pm $ are symmetric. Using the orthogonality condition (\ref{2Site_OrthVect}), one
can expand matrices $ \hat{M}_\pm $ in terms of eigenvalues and eigenvectors:
\begin{eqnarray}
  \label{2Site_MatrEigenExp}
  \hat{M}_\mu & = & \sum_{\eta = \pm} v_\mu^{(\eta)} \hat{\cal P}_\mu^{(\eta)} \, , \quad
  \hat{M}_\mu^k = \sum_{\eta = \pm} \left( v_\mu^{(\eta)} \right)^k \hat{\cal P}_\mu^{(\eta)} \, , \quad
  \hat{M}_\mu^k \hat{M}_{-\mu}^l = \sum_{\eta, \eta' = \pm} \left( v_\mu^{(\eta)} \right)^k \left( v_{-\mu}^{(\eta')} \right)^l
\hat{\cal P}_\mu^{(\eta)} \hat{\cal P}_{-\mu}^{(\eta')},
\end{eqnarray}
where, $ \hat{\cal P}_\mu^{(\eta)} $ are projectors on the eigenvectors of the matrix $ \hat{M}_\mu $:
\begin{eqnarray}
    \hat{\cal P}_\mu^{(\eta)} & \equiv & \frac{ \Upsilon_\mu^{(\eta)} \otimes \left( \Upsilon_\mu^{(\eta)} \right)^{\rm T} }
{\left( \Upsilon_\mu^{(\eta)} \right)^{\rm T} \cdot \Upsilon_\mu^{(\eta)} }. 
  \nonumber
\end{eqnarray}

\section{Attraction of a generic quantum trajectory to the Grand Circle}
\label{App_2Site_GC-Attraction}

\begin{figure}[H]
             \centering  \includegraphics{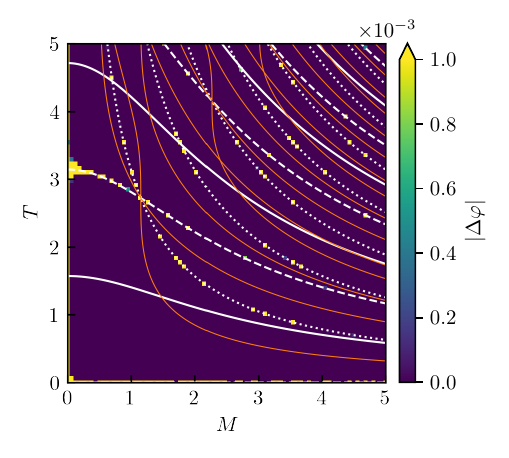}
        \caption{Attraction of quantum trajectories to the GC: Bloch-angle difference \(|\Delta \varphi| := \min(|\varphi - \pi / 2|, |\varphi + \pi / 2|)\) from the GC after \(10^5\) steps of time evolution with the MC method, starting from a random state on the Bloch sphere, indicating, where states converge to the GC.}
    \label{fig:Monte-phase}
    \end{figure}

In this Appendix, we consider the possible origins of the attraction of a generic quantum trajectory to the GC (cf. Figs.~\ref{fig:bloch_sphere_trajectory} and \ref{fig:Monte-phase}). The case, where all eigenvectors of the matrices $ \hat{M}_\pm $ belong to the GC, is trivial, as can seen from the expansion of matrices in terms of these eigenvectors. Let us discuss two other cases in detail.

\subsection{Two eigenvectors pointing to the GC and the other two pointing to the equator}

Let the eigenvectors of the matrix $ \hat{M}_\mu \equiv \hat{M}_{EQ} $ belong to the equator and eigenvectors of $ \hat{M}_{-\mu} \equiv \hat{M}_{GC} $ point to the GC.
The matrix $ \hat{M}_{EQ} $ itself is unable to provide the attraction to the GC. 
This is clear from the properties of a sequence of these matrices, as follows from Eq.~(\ref{2Site_MatrEigenExp}):
\begin{equation}
    \left( \hat{M}_{EQ} \right)^k = \sum_\eta \left( v_{EQ}^{(\eta)} \right)^k \hat{\cal P}_{EQ}^{(\eta)} \, .
\end{equation}
The $k$th power of the ``equator matrix,'' $ \left( \hat{M}_{EQ} \right)^k $, is a matrix with the eigenvectors located on
the equator and generically complex eigenvalues [except for commensurate cases where $ {\rm arg}\left( v_\mu^{(\eta)} \right) =
l \pi / k $]. There is no reason to assume that such a matrix is able to project an arbitrary state onto the GC. 
Numerical results confirm the absence of attraction in postselected realizations, where there is no matrix $ \hat{M}_{GC} $.

Next, consider the product
\begin{equation}
  \hat{M}_{EQ}^k \hat{M}_{GC}^q =
    \left( v_{GC}^{(+)} \right)^q \left( \hat{M}_{EQ}^k \bar{\Upsilon}_{GC}^{(+)} \right) \otimes
           \left( \bar{\Upsilon}_{GC}^{(+)} \right)^{\rm T} +
    \left( v_{GC}^{(-)} \right)^q \left( \hat{M}_{EQ}^k \bar{\Upsilon}_{GC}^{(-)} \right) \otimes
           \left( \bar{\Upsilon}_{GC}^{(-)} \right)^{\rm T} ,
\end{equation}
where the bar means that the eigenvectors are normalized. Since the GC is the invariant manifold, the vectors
$ \Upsilon_{GC}^{(\eta,k)} \equiv \hat{M}_{EQ}^k \bar{\Upsilon}_{GC}^{(\eta)} $ also belong to the GC, such that
$ \hat{M}_{EQ}^k \hat{M}_{GC}^q $ with $  q > 1 $ generically projects a state $ | \psi \rangle $ onto the
vector $ \Upsilon_{GC}^{(\eta,k)} $ that corresponds to the larger eigenvalue $ v_{GC}^{(\eta)} $. 
Convergence
(and the speed) of the attraction is determined by the ratio 
$ {\rm max} \left| v_{GC}^{(\eta)} \right| / {\rm min}
\left| v_{GC}^{(\eta)} \right| $ and by the typical value of the exponent $ q $.
The former approaches unity if the
eigenvectors $ \left( \bar{\Upsilon}_{GC}^{(\eta)} \right)^{\rm T} $ are close to intersections of the GC with
the equator. The typical value of $q$
can be estimated as $q_{\rm typ} \sim  \langle 1/P_{EQ} \rangle $ (the typical length of the continuous
chain of GC matrices).
The position of the attractor depends on the typical value of the exponent $ k $, which can be estimated as
$ \langle 1/P_{GC} \rangle $ (the typical length of the continuous chain of equatorial matrices).

\subsection{All eigenvectors pointing to the equator}

If the endpoints of eigenvectors of both matrices $ \hat{M}_\pm $ are located on the equator, the attraction to GC can also
be provided only by some products
\begin{equation}
  \hat{\mathbb{M}}(k,l) = \left( \ci \hat{M}_+ \right)^k \hat{M}_-^l ;
\end{equation}
here, the factor ``$\ci$'' has been added only to simplify equations. Numerical simulations show that, for any values of $ M $ and $ T $
(excluding the trivial commensurate case $ Y T = 2 n \pi $ ), at which $ \hat{M}_\pm $ are equatorial matrices, one can find
a pair of exponents $ \{ k, l \} $ that make the inequality
\begin{eqnarray}
   {\rm tr}^2 [ \hat{\mathbb{M}}(k,l) ] - 4\, {\rm det} [ \hat{\mathbb{M}}(k,l) ] < 0 \quad \Rightarrow \quad
  - 2 <
  \sum_{\eta, \eta' = \pm} \left( v_\mu^{(\eta)} \Bigl/ \left| v_\mu^{(\eta)} \right| \right)^k
 \left( v_{-\mu}^{(\eta')} \Bigl/ \left| v_{-\mu}^{(\eta')} \right| \right)^l
{\rm tr} \left[ \hat{\cal P}_\mu^{(\eta)} \hat{\cal P}_{-\mu}^{(\eta')} \right]
      < 2
\end{eqnarray}
hold true. This means, that the eigenvalues of $ \hat{\mathbb{M}}(k,l) $ are real and $ \hat{\mathbb{M}}(k,m) $ is the GC matrix,
which is able to project an arbitrary state onto the GC.

\section{Evolution on the Grand Circle}
\label{App:GC-evolution}

In this Appendix, we consider the evolution of the state parameterized by angle $\theta$ on the GC of the Bloch sphere under 
the application of matrices $\hat{M}_+$ and $\hat{M}_-$ to the state vector $(\cos\theta,\, \ci\sin\theta)$. 
The click and no-click matrices transform angle $\theta$ according to [cf. Eq.~\eqref{eq:Theta-Mapping}]
\begin{align}
\tan\Theta_+(\theta)&= \frac{(Y c_Y s_M-M c_M s_Y) \tan\theta -\sqrt{Y^2-M^2}\, s_Y s_M}
{Yc_Y s_M+M c_M s_Y + \sqrt{Y^2-M^2}\, s_Y s_M \tan\theta},
\label{eq:App1}
\\
\tan\Theta_-(\theta)&= \frac{(Y c_Y c_M + M s_M s_Y) \tan\theta -\sqrt{Y^2-M^2}\, s_Y c_M}
{Y c_Y c_M - M s_M s_Y + \sqrt{Y^2-M^2}\, s_Y c_M \tan\theta},
\label{eq:App2}
\end{align}
where $c_Y,\ c_M,\ s_Y,\ s_M$ are defined in Eqs.~(\ref{eq:SHN}). 
The inverse functions $\mathcal{F}_\pm(\theta)$, defined in Eq.~\eqref{eq:F-Mapping}, satisfy the following relations:
\begin{align}
\tan\mathcal{F}_+(\theta)&=\frac{(Y c_Y s_M+M c_M s_Y) \tan\theta + \sqrt{Y^2-M^2}\, s_Y s_M}
{Y c_Y s_M-M c_M s_Y - \sqrt{Y^2-M^2}\, s_Y s_M \tan\theta},
\label{eq:App3}
\\
\tan\mathcal{F}_-(\theta)&=\frac{(Y c_Y c_M-M s_M s_Y) \tan\theta + \sqrt{Y^2-M^2}\, s_Y c_M}
{Y c_Y c_M+M s_M s_Y - \sqrt{Y^2-M^2}\, s_Y c_M \tan\theta}.
\label{eq:App4}
\end{align}
To obtain explicit mappings \(\theta \mapsto \theta\) one has to perform a case analysis to take into account the quadrant of the state vector in the cartesian plane, for example by employing the two-argument function \(\textit{atan2}\)~\cite{organick1966fortran}.
The click and no-click probabilities read as: 
\begin{align}
P_\pm(\theta)=\frac{1}{2} &\mp \frac{Y^2-2 M^2 s_Y^2}{2 Y^2} \left(c_M^2-s_M^2\right) 
\pm \frac{2 M c_Y s_M c_M}{Y^2}
\left[ Y c_Y \cos(2\theta) + \sqrt{Y^2 - M^2} s_Y \sin(2\theta)\right].
\label{eq5}
\end{align}
When the eigenvectors of $\hat{M}_\pm$ belong to the GC, their
``eigenangles'' $\theta_\mu^{(\eta)}$ ($\eta=\pm 1$) are expressed as:
\begin{align}
    \tan\theta_+^{(\eta)}&=-\frac{\sqrt{Y^2 - M^2} s_M}{M c_M + \eta   \sqrt{M^2 - Y^2 s_M^2}},
    \label{eq:App6}
    \\
   \tan \theta_-^{(\eta)}&=\frac{\sqrt{Y^2 - M^2} c_M}{M s_M + \eta   \sqrt{M^2 - Y^2 c_M^2}}.
    \label{eq7}
\end{align}

\section{Period-2 trajectory on GC: Analogy with random walks in 1D}
\label{App_2Site-RandomWalks}

In this Appendix, we analyze the quantum trajectories for the period-2 case (see Sec.~\ref{sec:period_2_traj}) in light of the analogy with a random walk.
If the commensurability condition \(YT = \pi(2l + 1)\) with \(l \in \mathbb{N}\) is fulfilled, $\hat{M}_\mu^2=\mathds{1}$. 
Then, all possible products of \(\hat{M}_+\) and \(\hat{M}_-\)---corresponding to all possible quantum trajectories---simplify to be proportional to
	\begin{align}
		\begin{cases}(\hat{M}_+ \hat{M}_-)^k \text{ or } (\hat{M}_- \hat{M}_+)^k, & {N_t\text{ even}}\\
		\hat{M}_+(\hat{M}_- \hat{M}_+)^{k} \text{ or } \hat{M}_-(\hat{M}_+ \hat{M}_-)^{k}, & {N_t\text{ odd}}
		\end{cases} \label{eq:period-2-states}
	\end{align}
after \(N_t\) time evolution steps. Here \(k\) is an integer with \(k \in [0, \lfloor N_t / 2 \rfloor]\). 

From Eq.~\eqref{eq:period-2-states}, we already know all states with a finite probability after \(N_t\) steps of time evolution. All that is left to do is to determine the probability of ending up in a specific state, taking into account all possible quantum trajectories that lead to this point. For this purpose it is instructive to draw the possible matrix products as a tree:
\begin{figure}[H]
    \centering
    \includegraphics{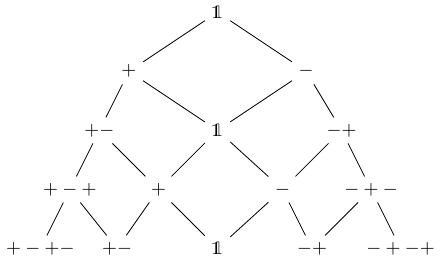}
\end{figure}
Due to the reduction rule \(\hat{M}^2_\mu = \mathds{1}\) the exponentially growing number of states in the generic case (cf. Fig.~\ref{fig:measurement_tree}) reduces to linear growth for the period-2 case.

Generally, for angle dependent transition probabilities \(P_{\pm}(\theta)\), it is still difficult to determine the weight of a given node in above tree as there are still exponentially many possible outcome sequences that traverse the tree. Therefore, as a first step, consider a simplified situation with \(P_{+}(\theta) = P = 1 - P_{-}(\theta)\).
We limit our consideration to even \(N_t\); the generalization to odd \(N_t\) is achieved by performing one more step of time evolution. If \(P = 1 - P = 1 / 2\) counting the number of paths leading to one node in the tree suffices to determine the weight of the corresponding state, because each path of length \(N_t\) has a probability of \(2^{-N_t}\). The counting generates the binomial coefficients:
\begin{figure}[H]
\centering
\includegraphics{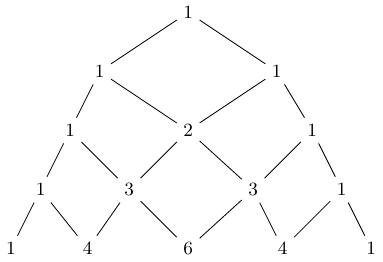}
\end{figure}
We thus find a distribution of the following form (note that this is not an ADF):
\begin{align}
    W_{N_t}(\theta| \theta_0) &=  \frac{B[N_t, N_t / 2]}{2^{N_t}}\, \delta(\theta - \theta_0)\! +\! \sum_{n = 1}^{N_t / 2} 
    \frac{B[N_t, (N_t + 2n) / 2]}{2^{N_t}} 
    \left\{\delta\left(\theta - \text{angle}\left[(\hat{M}_+\hat{M}_-)^n \psi_0\right]\right)  + (+ \leftrightarrow -)\right\},\label{eq:the_distribution_period_2}
\end{align}
where \(B[n,m]\) denotes the binomial distribution, $\psi_0$ is the initial state of the system (characterized by angle $\theta_0$) represented as column vector [cf. Eq.~\eqref{eq:MappingMatr}], and we introduced the short-hand notation 
$\text{angle}[\psi]$ for the angle $\theta$ of state $\psi$ on the GC.

It was shown in Sec.~\ref{sub_sec:special_cases} that the matrix product \((\hat{M}_{-\mu}\hat{M}_{-\mu})^n\) attracts the state to either \(\theta = \pi / 2\) or to \(\theta = -\pi / 2\) (depending on \(\mu\)) for large \(n\). This means that we can find a number \(n_1\), such that \(\text{angle}[(\hat{M}_{\mu}\hat{M}_{-\mu})^l\psi_0] \in [\pi / 2 - \varepsilon, \pi / 2 + \varepsilon] \cap [-\pi / 2 - \varepsilon, -\pi / 2 + \varepsilon] \equiv I_\varepsilon\) for any small \(\varepsilon\) and \(l \geq n_1\). We compare the probability within these \(\varepsilon\)-regions to the probability anywhere else on the GC:
    \begin{align}
        \frac{\int_{[-\pi, \pi) \backslash I_\varepsilon}\dd{\theta} W_{N_t}(\theta|\theta_0)}{\int_{I_\varepsilon}\dd{\theta} W_{N_t}(\theta|\theta_0)} &= \frac{B[N_t, N_t / 2] + 2\sum_{n = 1}^{n_1} B[N_t, (N_t + 2n) / 2]}{2 \sum_{n = n_1}^{N_t} B[N_t, (N_t + 2n) / 2]} \notag \\
        &\leq \frac{(1 + 2 n_1)B[N_t,N_t / 2]}{2^{N_t} - (1 + 2 n_1)B[N_t,N_t / 2]}= \frac{ \frac{\Gamma(N_t / 2 + 1 / 2)}{\sqrt{\pi} \Gamma(N_t / 2 + 1)} (1 + 2 n_1)}{ 1 - \frac{\Gamma(N_t / 2 + 1 / 2)}{\sqrt{\pi} \Gamma(N_t / 2 + 1)} (2 n_1 + 1)}\overset{N_t \rightarrow \infty}{\rightarrow} 0.
    \end{align}
This proves that the distribution converges to arbitrarily narrow peaks around the angles \(\theta = \pm \pi /2\) for \(N_t \rightarrow \infty\). Since the tree structure is completely symmetric between the two attractive points for \(P_{+} = P_-\), the time-averaged ADF reads after regularizing the \(\delta\)-peaks of individual trajectories
\begin{align}
    W(\theta) &= \frac{1}{2} \left( \delta(\theta - \pi / 2) + \delta(\theta + \pi / 2) \right). \label{eq:tADF_period2}
\end{align}

In order to generalize this result to imbalanced and nonconstant \(P_{\pm}\), it is useful to think of the tree structure as a one-dimensional random walk. The states of the random walk represent the different states that can be reached by forming combinations of \(\hat{M}_+\) and \(\hat{M}_-\) matrices: 
\begin{figure}[H]
    \centering
    \includegraphics{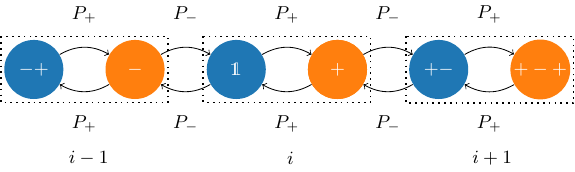}
\end{figure}
Clearly, for constant \(P_{+} \neq P_-\) the random walk on this chain still follows a diffusion law without a preferred direction. The distribution of states at any given time can still be found exactly by diagonalizing the transition matrix
\begin{align}
			[T_{i, j}] &= \begin{pmatrix}
				0 & \delta_{i, j} P + (1-P) \delta_{i, j - 1}\\
				\delta_{i, j} P + (1-P) \delta_{i, j + 1} & 0
			\end{pmatrix}
\end{align}
where indices label the unit cells (dotted boxes) and the matrix elements are written in the \(A-B\) space of orange and blue nodes. Without calculating the state distribution explicitly, we can see
from the symmetry of the random walk that the time-averaged ADF in this case still converges to the form~\eqref{eq:tADF_period2}.

In the general case of angle-dependent probabilities \(P_\pm(\theta)\), the calculation of the distribution of states more tedious. However, it stands to reason that the probability for the random walker to be found within the central region of the chain (around the node labeled \(\mathds{1}\)) still vanishes as \(N_t \rightarrow \infty\). If the random walker is sufficiently far from the central region, all states are close to the two special points \(\pm \pi / 2\). In this situation, both matrices mediate transitions to the vicinity of the respectively opposite special point, such that the time evolution is again described by a simple random walk in state space, corresponding to the limiting time-averaged ADF~\eqref{eq:tADF_period2}. We thus conclude that this distribution describes the period-2 case, as long as there is no other special mechanism confining the state to the central region.

\section{Numerical methods}
\label{sec:methods}

In this Appendix, we describe the two numerical approaches that we have used 
to obtain the ADF on the GC: solution of a discretized ME and Monte-Carlo simulations.

\subsection{Discretized Master equation}
\label{sec:master_equation}

Let us discretize the ME, aiming to solve it numerically (cf.  Refs.~\cite{numerical_solution_of_circle_equation,david_random_matrices_2}, where numerical approaches to solving similar implicit functional equations on a discretized circle were discussed). 
To discretize the ME, we partition the GC into \(N\) equally-sized subintervals (grid cells) \(c_i\):
\begin{align}
    c_i &:= \left[\theta_i -  \Delta \theta / 2,\, \theta_i + \Delta \theta / 2\right], \quad i \in [0, N - 1],\qquad 
    \theta_i = - \pi + (2i + 1) \Delta \theta / 2, \qquad \Delta \theta = \frac{2\pi}{N}.
\end{align}
With this, the discrete Markov process is derived as
\begin{align}
    \widetilde{\text{Pr}}_i &:= 
    \int_{\theta_i-\Delta\theta/2}^{\theta_i+\Delta\theta/2} \dd{\theta} W(\theta)  \approx 
    \sum_{k=1}^{N}[\hat{\mathcal{M}}_N]_{ik} \widetilde{\text{Pr}}_k, \quad
    [\hat{\mathcal{M}}_N]_{ik} \equiv \frac{1}{\Delta \theta}\sum_{\mu \in \{+, -\}}  P_\mu(\theta_k) |f_\mu(c_i)\cap c_k| .\label{eq:derivation_discrete_me}
\end{align}
Here \(\widetilde{\text{Pr}}_i\) is the ``true'' probability corresponding to integrated weight of the stationary solution \(W(\theta)\) to the continuous ME~\eqref{eq:W-steady} in the bin \(c_i\). To obtain a set of linear equations for the \(\{\widetilde{\text{Pr}}_i\}\) from the bin integrals of $W(\theta)$, we used Eqs.~\eqref{eq:W-funct} and assumed that 
\(P_\mu(\theta)\) is constant on the scale of \(\Delta \theta\). This corresponds to an expansion to the first order in \(\Delta \theta\).

Using the Markov matrix \([\hat{\mathcal{M}}_N]\) derived from the ME, we define another set of probabilities \(\{\text{Pr}_i\}\) as the stationary solution of the discrete Markov process
\begin{align}
    \mathbf{Pr} := \hat{\mathcal{M}}_N \mathbf{Pr}.\label{eq:discrete_me}
\end{align}
Finding the stationary state \(\mathbf{Pr}\) of the discrete ME~\eqref{eq:discrete_me} is equivalent to solving a set of \(N\) linear algebraic equations for the ``vector'' of probabilities \(\mathbf{Pr}=\{\mathrm{Pr}_1, \ldots, \mathrm{Pr}_N\}\) from Eq.~(\ref{eq:Pri-M}), with the matrix $\hat{\mathcal{M}}_N$ defined in Eq.~\eqref{eq:discretization}: 
In our numerical calculations, the defining equation for $\mathbf{Pr}$ is Eq.~(\ref{eq:Pri-M}); with this vector, we obtain the approximate version of $W(\theta)$ that is shown in our plots via the relation
 \begin{align}
     W(\theta) \approx \sum_{i=1}^N \theta[-\theta + (\theta_i + \Delta \theta / 2)] \theta[\theta - (\theta_i - \Delta \theta / 2)] \frac{\text{Pr}_i}{\Delta \theta}.
 \end{align}
In the limit \(\Delta \theta \rightarrow 0\), the approximation in the derivation of the discrete ME~\eqref{eq:derivation_discrete_me} becomes exact, such that \(\mathbf{Pr} \overset{\Delta \theta \rightarrow 0}{\rightarrow} \widetilde{\mathbf{Pr}}.\) In this limit, the infinite set of probabilities \(\widetilde{\mathbf{Pr}}\) recovers the true distribution \(W(\theta)\) via the relation \(\widetilde{\text{Pr}}_i / \Delta \theta \overset{\Delta \theta \rightarrow 0}{\rightarrow} W(\theta_i)\). Therefore also
 \begin{align}
     \text{Pr}_i / \Delta \theta \overset{\Delta \theta \rightarrow 0}{\longrightarrow} W(\theta_i).
 \end{align}

The number of nonzero matrix elements in row \(i\) of the Markov matrix $\hat{\mathcal{M}}_N$  [or, equivalently, the number of cells
that overlap with the image \(f_s(c_i)\) of $c_i$] can be estimated as follows:
\begin{align}
|f_\mu(c_i)| / \Delta \theta &= |\mathcal{F}_\mu(\theta_i + \Delta \theta / 2) - \mathcal{F}_\mu(\theta_i - \Delta \theta / 2)| / \Delta \theta 
\approx |\mathcal{F}_\mu'(\theta_i)|.
\end{align}
As the maps \(f_\mu\) are invertible away from the projective limit, this derivative is typically of order one. 

\begin{figure}[H]
       \begin{center}
      \includegraphics{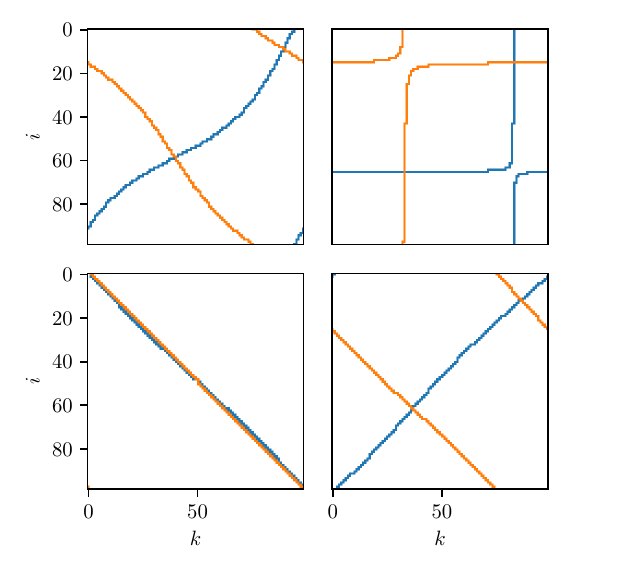}
       \end{center}
        \caption{Structure of the matrix \(\hat{\mathcal{M}}_{10^2}\) [defined in Eq.~\eqref{eq:discretization}] obtained from discretizing the master equation for parameters \(M = 2.92\), \(T=3.0554\) (\textit{upper-left panel}), \(M = 2.410\), \(T = 3.2554\) (\textit{upper-right panel}), \(M = 1.979\), \(T=2.2663\) (\textit{lower-left panel}), and \(M = 2.258\), \(T = 1.4369\) (\textit{lower-right panel}). 
        The axes \(i\) and \(k\) show the row- and column indices of the matrix. Nonzero matrix elements $[\hat{\mathcal{M}}_{10^2}]_{ik}$ form colorful ``bands'' (appearing as thin curves on this scale) corresponding to the matrices $ \hat{M}_+ $ (orange bands) and $ \hat{M}_- $ (blue bands).
                }
    \label{fig:me_matrix_structure}
    \end{figure}

The matrices $\hat{\mathcal{M}}$ defined in Eq.~(\ref{eq:discretization}) are sparse.
Figure~\ref{fig:me_matrix_structure} shows examples of matrices $\hat{\mathcal{M}}_N$, for \(N= 10^2\) and four different parameter sets. Finite matrix elements
\([\hat{\mathcal{M}}_{N}]_{ik}\) are shown in orange and blue. 
Each matrix features two continuous ``bands'' of nonzero entries, corresponding to \(f_\pm\).
As the maps \(f_\mu\) are invertible, both bands are monotonous and cover all rows and columns, being periodic on a torus. These bands have small widths, rendering
\(\hat{\mathcal{M}}_N\) extremely sparse.

The upper-left panel corresponds to generic parameters. In the upper-right panel, each of the mapping matrices \(\hat{M}_{\pm}\) has a strong hierarchy between its eigenvalues (``almost projective''), see Sec.~\ref{sec:projective_case}, such that a large interval of the GC is mapped to close vicinity of the dominant eigenvalue---note the steep slopes of the ``bands'' near \(j \approx 250,\, 750\). The parameters in the lower-left panel are close to the frozen case, \(YT \approx 2.03\), see Sec.~\ref{sec:frozen}. In the frozen case, we have \(\hat{M}_{\pm} \propto \mathds{1}\), such that also \(\hat{\mathcal{M}}_N \propto \mathds{1}\), since no transition away from a given bin takes place. The lower-right panel is close to the shift case, \(MT \approx 1.03\), see Sec.~\ref{sec:shift}. In the shift case, one of the matrices reflects \(\theta \rightarrow -\theta\) on the GC. The other matrix causes a constant shift of \(\theta\), that depends on the parameters. 

It should be noted that the generic case in the upper-left panel looks somewhat similar to the almost-shift case in the lower-right panel. This supports the idea that the generic case can be understood based on the different special cases we analyzed: there is always a ``similar'' special case. In particular, if the bands in the upper-left panel were deformed further, they might resemble, for example, the projective case (upper-right panel) instead. This approximate similarity will allow us to comprehend the overall ``phase diagram'' based on the (analytical) knowledge about special cases, represented in the $M$-$T$ plane by a set of curves, see Sec.~\ref{sec:localization_ergodicity_fractality}.

A stationary solution of the discretized ME corresponds to an eigenvector of \(\hat{\mathcal{M}}_N\) with eigenvalue \(1\). As can be seen from Eq.~\eqref{eq:discretization}, all entries of 
\(\hat{\mathcal{M}}_N\) are positive. Furthermore, probability conservation is manifestly built into the ME,
\begin{align}
    \sum_{i=1}^N [\hat{\mathcal{M}}_N]_{ik} &= 1.
\end{align}
For this reason, \(\hat{\mathcal{M}}_N\) is a positive Markov matrix and, as such, necessarily has an eigenvector corresponding to the eigenvalue 1~\cite{markov_matrices}: At least one stationary state of the ME always exists, regardless parameters and discretization. Such a state can be efficiently found for sparse matrices
\(\hat{\mathcal{M}}_N\) by repeatedly applying the matrix to an arbitrary initial state \(\mathbf{Pr}_0\) until convergence is reached (``power iteration'')---provided that the eigenvalue \(\lambda = 1\) is not degenerate and there is no other eigenvalue of modulus one~\cite{power_iteration_1, power_iteration_2}.

Assuming a hierarchy between the eigenvalues of \(\hat{\mathcal{M}}_N\), convergence to the stationary state happens at an exponential rate, which is proportional
to the difference \(\Delta \lambda=|\lambda_1| - |\lambda_2|\) between the dominant and the second-dominant eigenvalues (largest and second-largest eigenvalues by modulus),
\begin{align}
    ||\mathbf{Pr} - (\hat{\mathcal{M}}_N)^{n} \mathbf{Pr}_0|| \propto \exp\left(-a_0 \Delta \lambda\, n\right). 
    \label{eq:convergence_me}
\end{align}
where $n$ is the iteration number and \(a_0\) is a positive number.

There are two cases where the ME method can run into problems. (i) If the eigenvalue \(\lambda = 1\) is degenerate, the converged state depends on the initial state.
In this case, we cannot make an immediate connection to the asymptotic behavior of quantum trajectories.
(ii) If there is only one eigenvector corresponding to \(\lambda = 1\), but other eigenvectors have \(|\lambda'| = 1\), the power iteration can converge to a state that does not correspond to a stationary solution of the ME.

In practice, we are able to find nondegenerate stationary states by power iteration efficiently for most parameter sets.
There are two exceptions, corresponding to lines of special parameters:
\begin{itemize}
    \item Frozen trajectories are maximally degenerate, as every basis state (a probability vector comprising ${N-1}$ zeros and one unity) solves the discretized ME.
    \item For certain ``periodic'' trajectories an iterative solution does not necessarily converge at all, as it can get stuck switching periodically between
    different angle bins (say, if there is a period-2 trajectory between two bins, there is a vector $\mathbf{Pr}$, such that $\hat{\mathcal{M}}_N \mathbf{Pr} \neq \mathbf{Pr}$ but \((\hat{\mathcal{M}}_N)^2 \mathbf{Pr} = \mathbf{Pr}\), corresponding to \(\lambda = -1\)).
    The nondegenerate solution to the ME in this case corresponds to balanced occupation of the two peaks.
\end{itemize}

The structure of the matrix $\hat{\mathcal{M}}_N$ allows us to understand important features of the resulting ADFs regarding localization or delocalization, see Sec.~\ref{sec:localization_ergodicity_fractality}. For example, in the lower-left panel of Fig.~\ref{fig:me_matrix_structure} we show the structure of the Markov matrix close to the frozen case: The degenerate diagonal ``band'' of matrix elements splits into two separate bands which are slightly shifted from the main diagonal and bent towards opposite sites (crossing close to the middle and the ends). On the discretized level, it can be understood that a slight shift of the bands away from the main diagonal should lead to delocalization, since it essentially introduces transitions between neighboring grid cells. If the cells are smaller than the distance of the band to the diagonal at some point, then there is a next-nearest-neighbor transition, but the other band can facilitate a back-transition to the cell in between, and so on. At the same time, perturbing the frozen case, the outcome probabilities \(P_\pm\) are almost independent of the state, such that the stationary state is almost translationally invariant. From the trajectory point of view, we can think of the state performing short-distance hops on the GC in a random direction, eventually covering many points on the GC in a diffusive fashion. Similarly, in the shift case, existing commensurability is also broken by a perturbation, leading to delocalization around the corresponding curve.
    
As a consequence of the above mechanism, the PR would reveal a sharp jump moving onto or off a frozen line, if we had chosen a localized initial state. Dynamically, this jump would appear as a crossover, since the distance from the commensurate line controls the ``diffusion coefficient'' in a post-measurement trajectory. Close to the frozen case, the time average of a post-measurement trajectory converges slowly. Similarly, the difference between the two dominant eigenvalues of the Markov matrix is controlled by the distance to the frozen line, such that the convergence rate of power iteration goes to zero as the commensurability is approached.

\subsection{Monte-Carlo simulation}
\label{Sec:MonteCarlo}

The idea of the Monte-Carlo (MC) approach here is to simulate (at most) a few post-measurement trajectories for a given initial state 
\(\theta_0\), by randomly drawing measurement outcomes according to the Born rule, and performing time evolution according to the maps 
\(\hat{M}_\pm\) corresponding to the chosen outcomes. Specifically, for every measurement time \(t_j\), we draw a random number \(q(t_j) \in [0, 1]\) (uniformly distributed). Considering the probability \(P_j^{(+)}\) to apply the \((+)\) map at that time (depending on the current position on the GC), we evolve the state with \(\hat{M}_+\), if \(q(t_j) \leq P_j^{(+)}\), and with \(\hat{M}_-\) otherwise. Importantly, one instance of the simulation follows one individual quantum trajectory (corresponding to a pure state). The resulting post-measurement trajectories are then time averaged; a histogram of visited angles is obtained. 

This procedure provides information about the coarse-grained distribution \(W^{(\varepsilon)}(\theta|\theta_0)\), if the outcome average (containing exponentially many terms in the number of time steps) is well described in terms of \textit{typical} post measurement trajectories, see Eq.~(\ref{eq:time_average_gc_distribution}) of the main text. 
As this numerical procedure is based on randomly sampling possible outcomes according to their quantum mechanical probabilities, we refer to this method as MC simulation.
The stability of the obtained histogram with respect to the number of initial time evolution steps, as well as the number of successive time points after the initial phase, indicates that the chosen \(m\) is sufficient for convergence.

We use the MC method to
\begin{enumerate}
    \item verify that quantum trajectories generically converge to the GC (see Figs.~\ref{fig:bloch_sphere_trajectory} and \ref{fig:Monte-phase});
    \item compare the time average over a single trajectory to the ME result (see Fig.~\ref{fig:phase_diag_gc_conv} in Appendix~\ref{sec:stationary_vs_time_average}).
\end{enumerate}
While the MC method is useful owing to its simplicity, it converges rather slowly compared to the Markov method discussed below (with decreasing \(\varepsilon\)).

In Appendix~\ref{sec:stationary_vs_time_average}, we numerically confirm the convergence of generic post-measurement trajectories to the GC and find good agreement between the time-averaged distribution obtained from the MC simulations and the power iterated solution of the ME. The stationary state of the discretized ME is nondegenerate and its eigenvalue \(\lambda = 1\) is always well separated from the other eigenstates of the Markov matrix~\eqref{eq:discretization}, except for the vicinity of frozen and shift cases. This means that the GC distribution is generically independent of the initial state and can be obtained from the time average or stationary state equivalently.
The general independence of the GC distribution from the initial state allows us to drop the initial state argument \(\theta_0\) from \(W(\theta| \theta_0)\) and investigate the distributions \(W(\theta)\) within the chosen parameter range. All following distributions are obtained with the verified ME approach, allowing us to go to higher \(N\) and parameter grid resolutions, see Fig.~\ref{fig:different_distributions}.

    \begin{minipage}{\linewidth}
        \begin{figure}[H]
            \centering
             \includegraphics{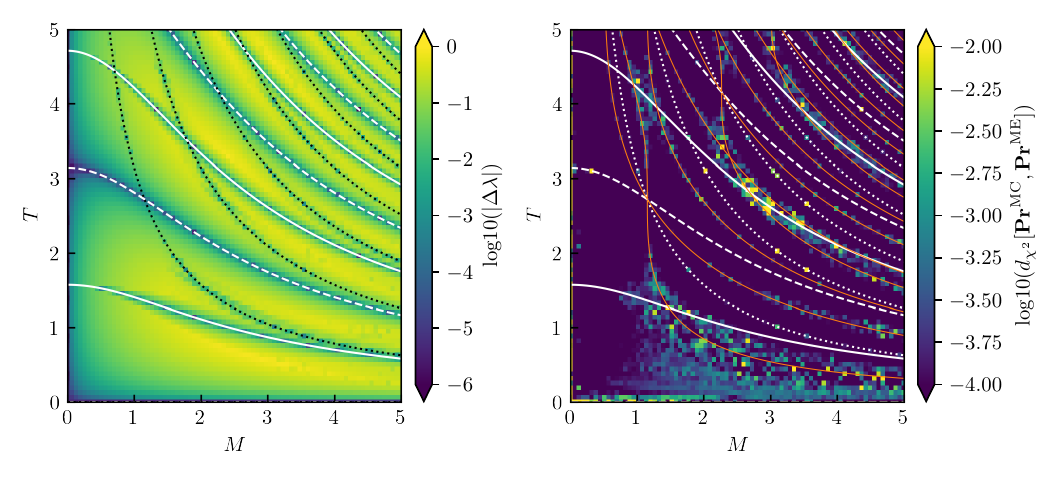}
            \caption{Characterization and comparison of discretized ME and MC methods for obtaining the GC distributions. \textit{Left panel}: Difference of moduli between the first and second dominant eigenvalues of the ME matrix \([\hat{\mathcal{M}}_{10^3}]\). The dominant eigenvalue is always equal to one. \textit{Right panel}: The  \(\chi^2\)-distance \eqref{eq:chi2_dist} between GC distributions with \(N = 10^3\) grid cells, obtained from the MC and ME methods. For the MC simulation, we start for each parameter tuple from a random state on the Bloch sphere and perform time evolution for \(10^7\) steps. We obtain the distribution from the last \(10^7 - 10^5\) states. The ME is solved starting from a uniform distribution on the GC with \(N=10^5\) bins, iterating for up to \(10^4\) steps. The converged distribution is then coarse-grained on \(N = 10^3\) grid cells. Solid lines: period-2 cases \(YT = (2k + 1) \pi\) with \(k \in \mathbb{N}_0\). Dashed lines: frozen cases \(TY = 2k \pi\). Dotted lines: shift cases \(MT = k \pi\). Orange lines: projective limit, where an eigenvalue of one of the post-measurement matrices \(\hat{M}_{\pm}\) becomes zero. }
        \label{fig:phase_diag_gc_conv}
        \end{figure}
    \end{minipage}
\vspace{0.5cm}

\section{Comparison of stationary solution and time average}
\label{sec:stationary_vs_time_average}

We start by verifying the applicability of the MC and ME approaches for calculating the GC distribution. 
In the left panel of Fig.~\ref{fig:phase_diag_gc_conv}, we show the difference 
\(|\Delta \lambda|\) between the two dominant eigenvalues of \(\hat{\mathcal{M}}_{10^3}\). While the largest eigenvalue always fulfills \(\lambda = 1\), the difference between this eigenvalue and the second-largest in modulus determines the validity of the power iteration procedure as described in Sec.~\ref{sec:master_equation}. We note that \(|\Delta \lambda|\) approaches zero towards the curves corresponding to the frozen and shift cases, as well as the period-2 trajectories. This confirms the exceptions discussed in Sec.~\ref{sec:master_equation}---maximum degeneracy for the frozen- and shift cases, and the existence of a solution with (\(\lambda = -1\)) in the period-2 case. Deviating from these lines of special parameters, \(|\Delta \lambda|\) increases rapidly (note the logarithmic scale). This means that the discrete ME, away from the special lines, has a unique solution, which can be found efficiently using power iteration.

Next, we calculate distributions from both ME (power iteration) and MC simulation (time average over a single post-measurement trajectory) on \(N=10^3\) grid cells. For the MC simulation, starting from a random state on the Bloch sphere for every parameter set, we perform time evolution for \(10^7\) steps, using the last \(9.9 \cdot 10^6\) steps to obtain the GC distribution (having checked that generic trajectories are converged to the GC after \(10^5\) steps, see Fig.~\ref{fig:gc_convergence_numerics}).

Assuming that a single MC trajectory captures the probability distribution \(W^{(\Delta \theta)}(\theta)\), we can estimate the number of time steps necessary for convergence. An approximate requirement for ``qualitative'' convergence of the MC distribution is then to have a large number of sampling points per bin, \( n_{\rm GC} / N \gg 1\) where \(n_{\rm GC}\) is the number of time steps on the GC from which the distribution is obtained. To get sufficient accuracy for every component of \(\mathbf{Pr}\), we need to require
\begin{align} n_{\rm GC}\!\! \min_{i \in \{1, \ldots, N\}}(\text{Pr}_i) = n_{\rm GC} \Delta \theta \min_{\theta}\left[W^{(\Delta \theta)}(\theta)\right]\gg 1.
\end{align}
We thus estimate that we get 1\%-accurate components of the probability vector if \( \Delta \theta W^{(\Delta \theta)} \gtrsim 10^2 / n_{\rm GC} \approx 10^{-5}\), which should give a good qualitative picture of the distributions.

\begin{figure}[htbp!]
    \centering
    \includegraphics{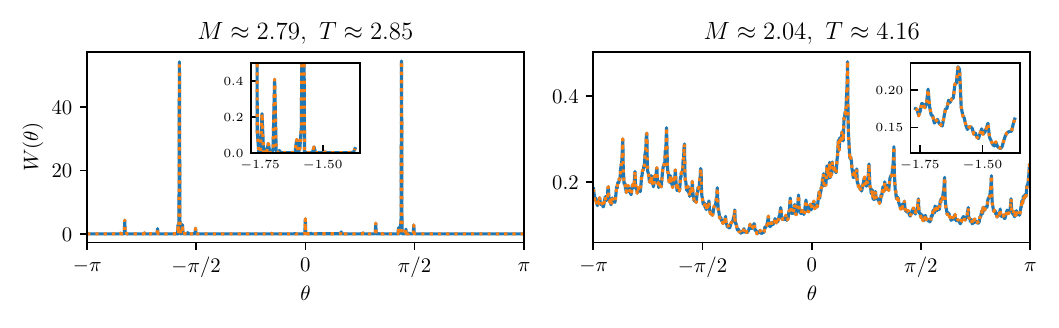}
    \caption{Examples for ADFs~\eqref{eq:Wbar} obtained as the stationary state of the discrete master equation~\eqref{eq:discrete_me} (solid blue line) and the time average of a single Monte-Carlo post-measurement trajectory~\eqref{eq:time_average_gc_distribution} (dotted orange line).}
    \label{fig:comparison_mc_me_dists}
\end{figure}

We compare the distributions obtained from the MC simulation (time average of random quantum trajectories) to the stationary states of the discretized ME in the right panel of Fig.~\ref{fig:phase_diag_gc_conv}. We obtain the ME results using \(10^5\) grid cells and up to \(10^4\) iteration steps. Using a uniform distribution as an initial state for power iteration, the iterated solution is guaranteed to contain all stationary states, in case the stationary state is degenerate. If this is the case, the iterated state cannot be the same as the time-averaged state from a single post-measurement trajectory. The iterated distribution is coarse-grained to \(N=10^3\) cells. The distributions are compared by calculating the \(\chi^2\)-distance between the probability vectors from ME and MC calculations,
\begin{align}
    d_{\chi^2}[\mathbf{Pr}^{\rm MC},\, \mathbf{Pr}^{\rm ME}]\equiv\frac{1}{2}\sum_{i=1}^{N}\frac{\left(\text{Pr}_i^{\rm MC} - \text{Pr}_i^{\rm ME}\right)^2}{\text{Pr}_i^{\rm MC} + \text{Pr}_i^{\rm ME}} \in [0, 1].
    \label{eq:chi2_dist}
\end{align} 

We observe good agreement, \({d_{\chi^2} < 10^{-2} \ll 1}\), for most values of parameters. Some exceptions are related to the frozen and shift cases (note bright markers exactly on dashed and dotted lines). If the freezing condition is perfectly fulfilled, both ME and MC methods preserve the initial state, which is localized at one point of the GC for MC evolution, and a uniform distribution on the GC for the ME calculation. In the numerical comparison, we ignore the fact that the MC initial condition does not necessarily lie on the GC and just compare the distributions over \(\theta\), keeping in mind that the frozen cases should be excluded from further analysis. This gives a large deviation \(d_{\chi^2} \overset{N \rightarrow \infty}{\rightarrow} 1\) between the distributions. In the shift cases, there is again no convergence of the MC trajectory to the GC, while the ME calculation takes place entirely on the GC. Therefore, these cases must also be excluded from the comparison between MC and ME distributions.

Other points where the difference between MC and ME distributions is large are related to a strong eigenvalue hierarchy for \(\hat{M}_{+}\) or \(\hat{M}_-\). Vanishing \(\lambda^\mu_{{\rm min}}\) for \(\mu \in \{+, -\}\) corresponds to the orange lines in the plot, and for small \(M\) \(\lambda^+_{{\rm min}} / \lambda^+_{{\rm max}}\) becomes small (see the dark blue region in Fig.~\ref{fig:phase_diag_gc_conv}, right panel). In the projective limit, the GC distribution can be calculated to high accuracy from a small number of terms in Eq.~\eqref{eq:proj_solution} and is independent of the initial angle. This distribution is also the unique stationary state of the ME. However, writing the ME according to Eq.~\eqref{eq:discretization} would no longer be  valid, as the projective GC map is not invertible. This can also lead to inaccuracies close to the projective case. 

Other inaccuracies with \(d_{\chi^2} \sim 10^{-2}\) do not correspond to systematic problems with the methods but only to insufficient convergence (number of time points for the MC method, number of grid cells for the ME method). To demonstrate this, we show in the left panel of Fig.~\ref{fig:comparison_mc_me_dists} a comparison between MC and ME distributions at one of the yellow parameter points with \(d_{\chi^2} \sim 10^{-2}\), but increase the number of time steps in the MC simulation to \(10^8\), and the grid size for the ME to \(10^6\), coarse graining both results to \(\Delta \theta = 2 \pi / 10^{3}\) to approximate the corresponding \(W^{(\Delta \theta)}\). As the inset shows, the distributions agree perfectly. Note that \(d_{\chi^2} \sim 10^{-2}\) is still good agreement between the distributions, sufficient for the following numerical investigations. In the discussed example distribution it corresponds to a slightly different weight distribution in the heavy right peak between the two methods.

We have thus verified the convergence of our MC distribution and confirmed the validity of obtaining the GC distribution from the ME approach. A comparison between MC and ME calculations shows agreement between the distributions in all but the frozen and shift cases, demonstrating that generic trajectories initiated at random states on the Bloch sphere eventually evolve according to a unique distribution on the GC. The ME-matrix \(\hat{\mathcal{M}}_{10^3}\) has for all but the frozen, shift, and period-2 cases a single eigenvalue of modulus one. The nondegeneracy of the stationary probability vector confirms the uniqueness of the GC distribution at the chosen grid size. In the special frozen and shift cases, the stationary distribution is not unique and depends on the initial state. In the period-2 case, we can analytically calculate the stationary distribution, see Sec.~\ref{sec:period_2_traj}. In this case, the distribution \(W(\theta|\theta_0)\) is unique and independent of the initial state, but it cannot be reliably found using power iteration. It should be instead obtained either from the MC simulation or by explicitly finding the eigenvector of \(\hat{\mathcal{M}}_{N}\) corresponding to \(\lambda = 1\).

\section{Nonergodicity in the discrete and continous process}
\label{sec:non_erg_elaboration}

In this Appendix, we explain in detail how finding nonergodicity in the process described by a finite matrix \(\hat{\mathcal{M}}_N\) at any number of discretization cells \(N\) has immediate implications for the continuous process. To see this, a simple example of a  nonergodic process is helpful: Let us say that the set of nodes \(V_N\) of the graph \(G_N\) splits into two strongly connected components (SCCs) \(v_1\) and \(v_2\)
    \begin{align}
        v_1, v_2 &\subset V_N \quad v_1 \cap v_2 = \emptyset \quad v_1 \cup v_2 = V_N
    \end{align}
Consider the directed edges that lead either from a node in \(v_1\) to a node in \(v_2\) or vice versa. In general, two situations are possible:
    \begin{itemize}
        \item[(i)] No such edges exist.
        \item[(ii)] Edges exist only in one direction, say \(v_1 \rightarrow v_2\) without loss of generality.
    \end{itemize}
Edges in both directions \(v_1 \rightarrow v_2\) \textit{and} \(v_2 \rightarrow v_1\) are not possible, because this would mean that \(G_N\) consists only of a single SCC.
In both cases (i) and (ii), the subset of the nodes in \(v_2\)
    \begin{align}
        I_2 \equiv \bigcup_{i \in v_2}c_{i}  \quad c_i = [\theta_i - \Delta \theta / 2, \theta_i + \Delta \theta / 2]
    \end{align}
    forms an invariant subset for the continuous post-measurement evolution on the GC: A trajectory with \(\theta_0 \in I_2\) can never transition into the set \(v_1\),
    \begin{align}
            \Xi_\mu(I_2) \subseteq I_2 \quad \Rightarrow \quad \theta_j \notin \bigcup_{i \in v_1} c_i\quad  \forall j \in \mathbb{N}
    \end{align}
     as none of the maps \(\hat{M}_\pm\) will allow for such a transition by construction of \(G_N\) via \(\hat{\mathcal{M}}_N\).

   \begin{figure}[tbhp!]
        \centering
        \includegraphics[width=0.4\linewidth]{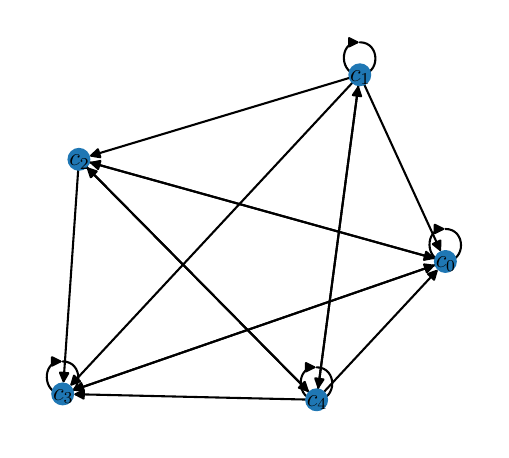}
        \caption{Example of a graph \(G_5\) corresponding to an ergodic Markov process described by \(\hat{\mathcal{M}}_5\). It can be verified that every node can be reached from every node, such that all nodes belong to the same SCC.}
        \label{fig:ergodic_graph_example}
    \end{figure}
    
    \begin{figure}[tbhp!]
        \centering
        \includegraphics[width=0.9\linewidth]{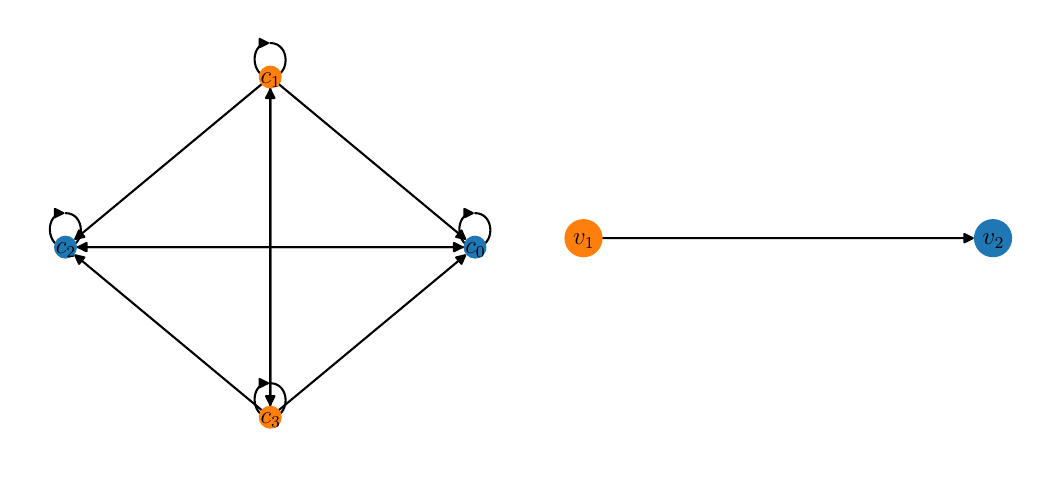}
        \caption{\textit{Left panel}: Example of a graph \(G_4\) that corresponds to a nonergodic Markov process with \(\hat{\mathcal{M}}_4\) and thus to a nonergodic continuous process. Nodes belonging to different SCCs are drawn in different colors. Transitions from orange to blue are possible, but not vice-versa. \textit{Right panel}: The condensation of the graph shown on the left. It contains two supernodes \(v_1\) and \(v_2\), corresponding to the two SCCs of \(G_4\). There is an edge from \(v_1\) to \(v_2\) (but not vice-versa), because there are edges in \(G_4\) leading from the orange set of nodes to the blue one (but not vice-versa). The condensation is indeed acyclic as there are no circular paths. \(v_2\) is a leaf node of the condensation, because no transitions out of \(v_2\) are possible. Thus, it defines an invariant subset for the continuous process.}
        \label{fig:nonergodic_graph_example_small}
    \end{figure}

     \begin{figure}[tbhp!]
        \centering
        \includegraphics[width=\linewidth]{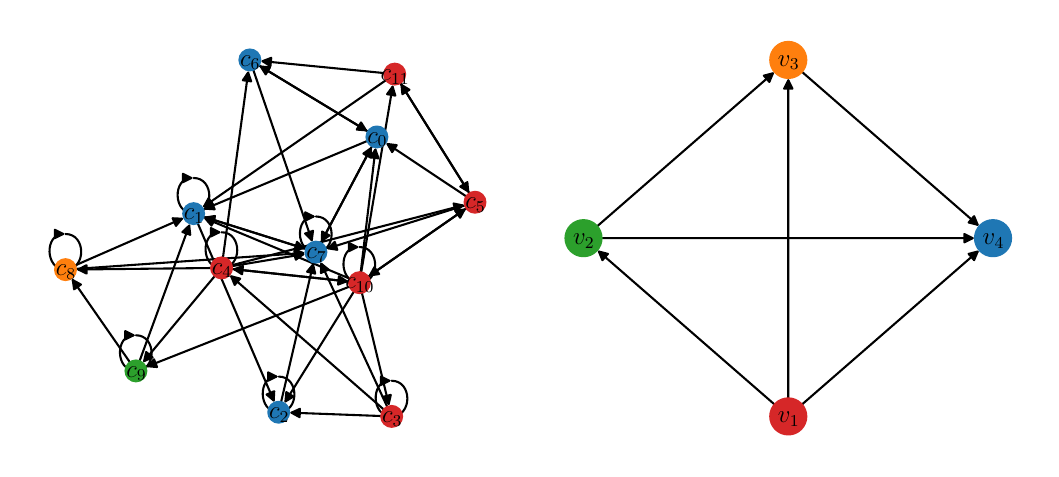}
        \caption{\textit{Left panel}: More complicated example for a graph \(G_{12}\) that corresponds to a nonergodic Markov process with \(\hat{\mathcal{M}}_{12}\) and thus to a nonergodic continuous process. Nodes belonging to different SCCs are drawn in different colors. \textit{Right panel}: The condensation of the graph shown on the left. It contains four supernodes, corresponding to the four SCCs of \(G_{12}\). Edges of the condensation correspond to possible transitions between SCCs. The condensation is acyclic, as there are no circular paths. The condensation has a single leaf node (blue, \(v_4\)) which does not have any outgoing edges. The leaf node again defines an invariant subset of the continuous process.}
        \label{fig:nonergodic_graph_example_large}
    \end{figure}
   
This is because a matrix element \([\hat{\mathcal{M}}_{N}]_{i, k}\) is non-zero (\((k, i) \in E_N\)) if and only if one of the maps permits a transition \(k \rightarrow i\) with nonzero probability,
    \begin{align}
        \sum_{\mu \in \{-, +\}}P_{\mu}(\theta_k) |f_\mu(c_i) \cap c_k| > 0.
    \end{align}
Calculating the overlap \(|f_\mu(c_i) \cap c_k|\) does not involve any discretization-induced approximation for the angles, since the intervals \(c_i\) and \(f_\mu(c_i)\) are continuous. In other words, this continuous mapping between the intervals is not equivalent to a discrete mapping between the labels $i$ of the cells. The discretization of the GC into a finite number of intervals establishes the resolution of an ``instrument'' exploring the mapping between the continuous angles. 
Therefore, our nonergodicity construction leads to a statement about the continuous process and is independent of the fact that the subset \(v_2\) was found by discretizing the GC.
Examples of the finite-$N$ representation of an ergodic process and a non-ergodic one described by situation (ii) are shown in
Figs.~\ref{fig:ergodic_graph_example} and \ref{fig:nonergodic_graph_example_small}, respectively.

The argument is readily generalized to situations where \(G_N\) splits into more than two SCCs. 
Formally, by using the language of the graph theory, one can define one ``supernode'' for each SCC of \(G_N\) and introduce a directed edge between two supernodes, if there exists a corresponding transition between the two represented sets of nodes~\cite{scc_condensation}. This defines the acyclic \textit{condensation} graph of \(G_N\)~\cite{partial_order_dags}. 
The construction of the condensation is illustrated in Fig.~\ref{fig:nonergodic_graph_example_small} and also for a more complex example with more than two SCCs in Fig.~\ref{fig:nonergodic_graph_example_large}.    
The leaf nodes (nodes without outgoing edges) in this graph correspond to invariant subsets of the GC. In the steady state, all of the weight of the distribution is accumulated in the invariant subset corresponding to the leaf nodes, with zeros everywhere else.

\bibliography{bibliography.bib}

\begin{thebibliography}{158}%
\makeatletter
\providecommand \@ifxundefined [1]{%
 \@ifx{#1\undefined}
}%
\providecommand \@ifnum [1]{%
 \ifnum #1\expandafter \@firstoftwo
 \else \expandafter \@secondoftwo
 \fi
}%
\providecommand \@ifx [1]{%
 \ifx #1\expandafter \@firstoftwo
 \else \expandafter \@secondoftwo
 \fi
}%
\providecommand \natexlab [1]{#1}%
\providecommand \enquote  [1]{``#1''}%
\providecommand \bibnamefont  [1]{#1}%
\providecommand \bibfnamefont [1]{#1}%
\providecommand \citenamefont [1]{#1}%
\providecommand \href@noop [0]{\@secondoftwo}%
\providecommand \href [0]{\begingroup \@sanitize@url \@href}%
\providecommand \@href[1]{\@@startlink{#1}\@@href}%
\providecommand \@@href[1]{\endgroup#1\@@endlink}%
\providecommand \@sanitize@url [0]{\catcode `\\12\catcode `\$12\catcode
  `\&12\catcode `\#12\catcode `\^12\catcode `\_12\catcode `\%12\relax}%
\providecommand \@@startlink[1]{}%
\providecommand \@@endlink[0]{}%
\providecommand \url  [0]{\begingroup\@sanitize@url \@url }%
\providecommand \@url [1]{\endgroup\@href {#1}{\urlprefix }}%
\providecommand \urlprefix  [0]{URL }%
\providecommand \Eprint [0]{\href }%
\providecommand \doibase [0]{https://doi.org/}%
\providecommand \selectlanguage [0]{\@gobble}%
\providecommand \bibinfo  [0]{\@secondoftwo}%
\providecommand \bibfield  [0]{\@secondoftwo}%
\providecommand \translation [1]{[#1]}%
\providecommand \BibitemOpen [0]{}%
\providecommand \bibitemStop [0]{}%
\providecommand \bibitemNoStop [0]{.\EOS\space}%
\providecommand \EOS [0]{\spacefactor3000\relax}%
\providecommand \BibitemShut  [1]{\csname bibitem#1\endcsname}%
\let\auto@bib@innerbib\@empty
\bibitem [{\citenamefont {Anderson}(1958)}]{Anderson58}%
  \BibitemOpen
  \bibfield  {author} {\bibinfo {author} {\bibfnamefont {P.~W.}\ \bibnamefont
  {Anderson}},\ }\bibfield  {title} {\bibinfo {title} {Absence of diffusion in
  certain random lattices},\ }\href {https://doi.org/10.1103/PhysRev.109.1492}
  {\bibfield  {journal} {\bibinfo  {journal} {Phys. Rev.}\ }\textbf {\bibinfo
  {volume} {109}},\ \bibinfo {pages} {1492} (\bibinfo {year}
  {1958})}\BibitemShut {NoStop}%
\bibitem [{\citenamefont {Evers}\ and\ \citenamefont
  {Mirlin}(2008)}]{evers_2008}%
  \BibitemOpen
  \bibfield  {author} {\bibinfo {author} {\bibfnamefont {F.}~\bibnamefont
  {Evers}}\ and\ \bibinfo {author} {\bibfnamefont {A.~D.}\ \bibnamefont
  {Mirlin}},\ }\bibfield  {title} {\bibinfo {title} {Anderson transitions},\
  }\href {https://doi.org/10.1103/RevModPhys.80.1355} {\bibfield  {journal}
  {\bibinfo  {journal} {Reviews of Modern Physics}\ }\textbf {\bibinfo {volume}
  {80}},\ \bibinfo {pages} {1355} (\bibinfo {year} {2008})}\BibitemShut
  {NoStop}%
\bibitem [{\citenamefont {Li}\ \emph {et~al.}(2018)\citenamefont {Li},
  \citenamefont {Chen},\ and\ \citenamefont {Fisher}}]{Li2018a}%
  \BibitemOpen
  \bibfield  {author} {\bibinfo {author} {\bibfnamefont {Y.}~\bibnamefont
  {Li}}, \bibinfo {author} {\bibfnamefont {X.}~\bibnamefont {Chen}},\ and\
  \bibinfo {author} {\bibfnamefont {M.~P.~A.}\ \bibnamefont {Fisher}},\
  }\bibfield  {title} {\bibinfo {title} {Quantum {Z}eno effect and the
  many-body entanglement transition},\ }\href
  {https://doi.org/10.1103/PhysRevB.98.205136} {\bibfield  {journal} {\bibinfo
  {journal} {Phys. Rev. B}\ }\textbf {\bibinfo {volume} {98}},\ \bibinfo
  {pages} {205136} (\bibinfo {year} {2018})}\BibitemShut {NoStop}%
\bibitem [{\citenamefont {Skinner}\ \emph {et~al.}(2019)\citenamefont
  {Skinner}, \citenamefont {Ruhman},\ and\ \citenamefont
  {Nahum}}]{Skinner2019a}%
  \BibitemOpen
  \bibfield  {author} {\bibinfo {author} {\bibfnamefont {B.}~\bibnamefont
  {Skinner}}, \bibinfo {author} {\bibfnamefont {J.}~\bibnamefont {Ruhman}},\
  and\ \bibinfo {author} {\bibfnamefont {A.}~\bibnamefont {Nahum}},\ }\bibfield
   {title} {\bibinfo {title} {Measurement-induced phase transitions in the
  dynamics of entanglement},\ }\href
  {https://doi.org/10.1103/PhysRevX.9.031009} {\bibfield  {journal} {\bibinfo
  {journal} {Phys. Rev. X}\ }\textbf {\bibinfo {volume} {9}},\ \bibinfo {pages}
  {031009} (\bibinfo {year} {2019})}\BibitemShut {NoStop}%
\bibitem [{\citenamefont {Chan}\ \emph {et~al.}(2019)\citenamefont {Chan},
  \citenamefont {Nandkishore}, \citenamefont {Pretko},\ and\ \citenamefont
  {Smith}}]{Chan2019a}%
  \BibitemOpen
  \bibfield  {author} {\bibinfo {author} {\bibfnamefont {A.}~\bibnamefont
  {Chan}}, \bibinfo {author} {\bibfnamefont {R.~M.}\ \bibnamefont
  {Nandkishore}}, \bibinfo {author} {\bibfnamefont {M.}~\bibnamefont
  {Pretko}},\ and\ \bibinfo {author} {\bibfnamefont {G.}~\bibnamefont
  {Smith}},\ }\bibfield  {title} {\bibinfo {title} {Unitary-projective
  entanglement dynamics},\ }\href {https://doi.org/10.1103/PhysRevB.99.224307}
  {\bibfield  {journal} {\bibinfo  {journal} {Phys. Rev. B}\ }\textbf {\bibinfo
  {volume} {99}},\ \bibinfo {pages} {224307} (\bibinfo {year}
  {2019})}\BibitemShut {NoStop}%
\bibitem [{\citenamefont {Szyniszewski}\ \emph {et~al.}(2019)\citenamefont
  {Szyniszewski}, \citenamefont {Romito},\ and\ \citenamefont
  {Schomerus}}]{Szyniszewski2019a}%
  \BibitemOpen
  \bibfield  {author} {\bibinfo {author} {\bibfnamefont {M.}~\bibnamefont
  {Szyniszewski}}, \bibinfo {author} {\bibfnamefont {A.}~\bibnamefont
  {Romito}},\ and\ \bibinfo {author} {\bibfnamefont {H.}~\bibnamefont
  {Schomerus}},\ }\bibfield  {title} {\bibinfo {title} {Entanglement transition
  from variable-strength weak measurements},\ }\href
  {https://doi.org/10.1103/PhysRevB.100.064204} {\bibfield  {journal} {\bibinfo
   {journal} {Phys. Rev. B}\ }\textbf {\bibinfo {volume} {100}},\ \bibinfo
  {pages} {064204} (\bibinfo {year} {2019})}\BibitemShut {NoStop}%
\bibitem [{\citenamefont {Fisher}\ \emph {et~al.}(2023)\citenamefont {Fisher},
  \citenamefont {Khemani}, \citenamefont {Nahum},\ and\ \citenamefont
  {Vijay}}]{Fisher2022}%
  \BibitemOpen
  \bibfield  {author} {\bibinfo {author} {\bibfnamefont {M.~P.}\ \bibnamefont
  {Fisher}}, \bibinfo {author} {\bibfnamefont {V.}~\bibnamefont {Khemani}},
  \bibinfo {author} {\bibfnamefont {A.}~\bibnamefont {Nahum}},\ and\ \bibinfo
  {author} {\bibfnamefont {S.}~\bibnamefont {Vijay}},\ }\bibfield  {title}
  {\bibinfo {title} {Random quantum circuits},\ }\href
  {https://doi.org/10.1146/annurev-conmatphys-031720-030658} {\bibfield
  {journal} {\bibinfo  {journal} {Annual Review of Condensed Matter Physics}\
  }\textbf {\bibinfo {volume} {14}},\ \bibinfo {pages} {335} (\bibinfo {year}
  {2023})},\ \Eprint
  {https://arxiv.org/abs/https://doi.org/10.1146/annurev-conmatphys-031720-030658}
  {https://doi.org/10.1146/annurev-conmatphys-031720-030658} \BibitemShut
  {NoStop}%
\bibitem [{\citenamefont {Aharonov}(2000)}]{Aharonov2000a}%
  \BibitemOpen
  \bibfield  {author} {\bibinfo {author} {\bibfnamefont {D.}~\bibnamefont
  {Aharonov}},\ }\bibfield  {title} {\bibinfo {title} {Quantum to classical
  phase transition in noisy quantum computers},\ }\href
  {https://doi.org/10.1103/PhysRevA.62.062311} {\bibfield  {journal} {\bibinfo
  {journal} {Phys. Rev. A}\ }\textbf {\bibinfo {volume} {62}},\ \bibinfo
  {pages} {062311} (\bibinfo {year} {2000})}\BibitemShut {NoStop}%
\bibitem [{\citenamefont {Preskill}(2018)}]{Preskill2018a}%
  \BibitemOpen
  \bibfield  {author} {\bibinfo {author} {\bibfnamefont {J.}~\bibnamefont
  {Preskill}},\ }\bibfield  {title} {\bibinfo {title} {Quantum computing in the
  {NISQ} era and beyond},\ }\href {https://doi.org/10.22331/q-2018-08-06-79}
  {\bibfield  {journal} {\bibinfo  {journal} {{Quantum}}\ }\textbf {\bibinfo
  {volume} {2}},\ \bibinfo {pages} {79} (\bibinfo {year} {2018})}\BibitemShut
  {NoStop}%
\bibitem [{\citenamefont {Bharti~\textit{et~al.}}(2022)}]{Bharti2022}%
  \BibitemOpen
  \bibfield  {author} {\bibinfo {author} {\bibfnamefont {K.}~\bibnamefont
  {Bharti~\textit{et~al.}}},\ }\bibfield  {title} {\bibinfo {title} {Noisy
  intermediate-scale quantum algorithms},\ }\href
  {https://doi.org/10.1103/RevModPhys.94.015004} {\bibfield  {journal}
  {\bibinfo  {journal} {Reviews of Modern Physics}\ }\textbf {\bibinfo {volume}
  {94}},\ \bibinfo {pages} {015004} (\bibinfo {year} {2022})}\BibitemShut
  {NoStop}%
\bibitem [{\citenamefont {Li}\ \emph {et~al.}(2019)\citenamefont {Li},
  \citenamefont {Chen},\ and\ \citenamefont {Fisher}}]{Li2019a}%
  \BibitemOpen
  \bibfield  {author} {\bibinfo {author} {\bibfnamefont {Y.}~\bibnamefont
  {Li}}, \bibinfo {author} {\bibfnamefont {X.}~\bibnamefont {Chen}},\ and\
  \bibinfo {author} {\bibfnamefont {M.~P.~A.}\ \bibnamefont {Fisher}},\
  }\bibfield  {title} {\bibinfo {title} {Measurement-driven entanglement
  transition in hybrid quantum circuits},\ }\href
  {https://doi.org/10.1103/PhysRevB.100.134306} {\bibfield  {journal} {\bibinfo
   {journal} {Phys. Rev. B}\ }\textbf {\bibinfo {volume} {100}},\ \bibinfo
  {pages} {134306} (\bibinfo {year} {2019})}\BibitemShut {NoStop}%
\bibitem [{\citenamefont {Bao}\ \emph {et~al.}(2020)\citenamefont {Bao},
  \citenamefont {Choi},\ and\ \citenamefont {Altman}}]{Bao2020a}%
  \BibitemOpen
  \bibfield  {author} {\bibinfo {author} {\bibfnamefont {Y.}~\bibnamefont
  {Bao}}, \bibinfo {author} {\bibfnamefont {S.}~\bibnamefont {Choi}},\ and\
  \bibinfo {author} {\bibfnamefont {E.}~\bibnamefont {Altman}},\ }\bibfield
  {title} {\bibinfo {title} {Theory of the phase transition in random unitary
  circuits with measurements},\ }\href
  {https://doi.org/10.1103/PhysRevB.101.104301} {\bibfield  {journal} {\bibinfo
   {journal} {Phys. Rev. B}\ }\textbf {\bibinfo {volume} {101}},\ \bibinfo
  {pages} {104301} (\bibinfo {year} {2020})}\BibitemShut {NoStop}%
\bibitem [{\citenamefont {Choi}\ \emph {et~al.}(2020)\citenamefont {Choi},
  \citenamefont {Bao}, \citenamefont {Qi},\ and\ \citenamefont
  {Altman}}]{Choi2020a}%
  \BibitemOpen
  \bibfield  {author} {\bibinfo {author} {\bibfnamefont {S.}~\bibnamefont
  {Choi}}, \bibinfo {author} {\bibfnamefont {Y.}~\bibnamefont {Bao}}, \bibinfo
  {author} {\bibfnamefont {X.-L.}\ \bibnamefont {Qi}},\ and\ \bibinfo {author}
  {\bibfnamefont {E.}~\bibnamefont {Altman}},\ }\bibfield  {title} {\bibinfo
  {title} {Quantum error correction in scrambling dynamics and
  measurement-induced phase transition},\ }\href
  {https://doi.org/10.1103/PhysRevLett.125.030505} {\bibfield  {journal}
  {\bibinfo  {journal} {Phys. Rev. Lett.}\ }\textbf {\bibinfo {volume} {125}},\
  \bibinfo {pages} {030505} (\bibinfo {year} {2020})}\BibitemShut {NoStop}%
\bibitem [{\citenamefont {Gullans}\ and\ \citenamefont
  {Huse}(2020{\natexlab{a}})}]{Gullans2020a}%
  \BibitemOpen
  \bibfield  {author} {\bibinfo {author} {\bibfnamefont {M.~J.}\ \bibnamefont
  {Gullans}}\ and\ \bibinfo {author} {\bibfnamefont {D.~A.}\ \bibnamefont
  {Huse}},\ }\bibfield  {title} {\bibinfo {title} {Dynamical purification phase
  transition induced by quantum measurements},\ }\href
  {https://doi.org/10.1103/PhysRevX.10.041020} {\bibfield  {journal} {\bibinfo
  {journal} {Phys. Rev. X}\ }\textbf {\bibinfo {volume} {10}},\ \bibinfo
  {pages} {041020} (\bibinfo {year} {2020}{\natexlab{a}})}\BibitemShut
  {NoStop}%
\bibitem [{\citenamefont {Gullans}\ and\ \citenamefont
  {Huse}(2020{\natexlab{b}})}]{Gullans2020b}%
  \BibitemOpen
  \bibfield  {author} {\bibinfo {author} {\bibfnamefont {M.~J.}\ \bibnamefont
  {Gullans}}\ and\ \bibinfo {author} {\bibfnamefont {D.~A.}\ \bibnamefont
  {Huse}},\ }\bibfield  {title} {\bibinfo {title} {Scalable probes of
  measurement-induced criticality},\ }\href
  {https://doi.org/10.1103/PhysRevLett.125.070606} {\bibfield  {journal}
  {\bibinfo  {journal} {Phys. Rev. Lett.}\ }\textbf {\bibinfo {volume} {125}},\
  \bibinfo {pages} {070606} (\bibinfo {year} {2020}{\natexlab{b}})}\BibitemShut
  {NoStop}%
\bibitem [{\citenamefont {Jian}\ \emph {et~al.}(2020)\citenamefont {Jian},
  \citenamefont {You}, \citenamefont {Vasseur},\ and\ \citenamefont
  {Ludwig}}]{Jian2020a}%
  \BibitemOpen
  \bibfield  {author} {\bibinfo {author} {\bibfnamefont {C.-M.}\ \bibnamefont
  {Jian}}, \bibinfo {author} {\bibfnamefont {Y.-Z.}\ \bibnamefont {You}},
  \bibinfo {author} {\bibfnamefont {R.}~\bibnamefont {Vasseur}},\ and\ \bibinfo
  {author} {\bibfnamefont {A.~W.~W.}\ \bibnamefont {Ludwig}},\ }\bibfield
  {title} {\bibinfo {title} {Measurement-induced criticality in random quantum
  circuits},\ }\href {https://doi.org/10.1103/PhysRevB.101.104302} {\bibfield
  {journal} {\bibinfo  {journal} {Phys. Rev. B}\ }\textbf {\bibinfo {volume}
  {101}},\ \bibinfo {pages} {104302} (\bibinfo {year} {2020})}\BibitemShut
  {NoStop}%
\bibitem [{\citenamefont {Zabalo}\ \emph {et~al.}(2020)\citenamefont {Zabalo},
  \citenamefont {Gullans}, \citenamefont {Wilson}, \citenamefont
  {Gopalakrishnan}, \citenamefont {Huse},\ and\ \citenamefont
  {Pixley}}]{Zabalo2020a}%
  \BibitemOpen
  \bibfield  {author} {\bibinfo {author} {\bibfnamefont {A.}~\bibnamefont
  {Zabalo}}, \bibinfo {author} {\bibfnamefont {M.~J.}\ \bibnamefont {Gullans}},
  \bibinfo {author} {\bibfnamefont {J.~H.}\ \bibnamefont {Wilson}}, \bibinfo
  {author} {\bibfnamefont {S.}~\bibnamefont {Gopalakrishnan}}, \bibinfo
  {author} {\bibfnamefont {D.~A.}\ \bibnamefont {Huse}},\ and\ \bibinfo
  {author} {\bibfnamefont {J.~H.}\ \bibnamefont {Pixley}},\ }\bibfield  {title}
  {\bibinfo {title} {Critical properties of the measurement-induced transition
  in random quantum circuits},\ }\href
  {https://doi.org/10.1103/PhysRevB.101.060301} {\bibfield  {journal} {\bibinfo
   {journal} {Phys. Rev. B}\ }\textbf {\bibinfo {volume} {101}},\ \bibinfo
  {pages} {060301(R)} (\bibinfo {year} {2020})}\BibitemShut {NoStop}%
\bibitem [{\citenamefont {Iaconis}\ \emph {et~al.}(2020)\citenamefont
  {Iaconis}, \citenamefont {Lucas},\ and\ \citenamefont {Chen}}]{Iaconis2020a}%
  \BibitemOpen
  \bibfield  {author} {\bibinfo {author} {\bibfnamefont {J.}~\bibnamefont
  {Iaconis}}, \bibinfo {author} {\bibfnamefont {A.}~\bibnamefont {Lucas}},\
  and\ \bibinfo {author} {\bibfnamefont {X.}~\bibnamefont {Chen}},\ }\bibfield
  {title} {\bibinfo {title} {Measurement-induced phase transitions in quantum
  automaton circuits},\ }\href {https://doi.org/10.1103/PhysRevB.102.224311}
  {\bibfield  {journal} {\bibinfo  {journal} {Phys. Rev. B}\ }\textbf {\bibinfo
  {volume} {102}},\ \bibinfo {pages} {224311} (\bibinfo {year}
  {2020})}\BibitemShut {NoStop}%
\bibitem [{\citenamefont {Turkeshi}\ \emph {et~al.}(2020)\citenamefont
  {Turkeshi}, \citenamefont {Fazio},\ and\ \citenamefont
  {Dalmonte}}]{Turkeshi2020a}%
  \BibitemOpen
  \bibfield  {author} {\bibinfo {author} {\bibfnamefont {X.}~\bibnamefont
  {Turkeshi}}, \bibinfo {author} {\bibfnamefont {R.}~\bibnamefont {Fazio}},\
  and\ \bibinfo {author} {\bibfnamefont {M.}~\bibnamefont {Dalmonte}},\
  }\bibfield  {title} {\bibinfo {title} {Measurement-induced criticality in
  $(2+1)$-dimensional hybrid quantum circuits},\ }\href
  {https://doi.org/10.1103/PhysRevB.102.014315} {\bibfield  {journal} {\bibinfo
   {journal} {Phys. Rev. B}\ }\textbf {\bibinfo {volume} {102}},\ \bibinfo
  {pages} {014315} (\bibinfo {year} {2020})}\BibitemShut {NoStop}%
\bibitem [{\citenamefont {Zhang}\ \emph {et~al.}(2020)\citenamefont {Zhang},
  \citenamefont {Reyes}, \citenamefont {Kourtis}, \citenamefont {Chamon},
  \citenamefont {Mucciolo},\ and\ \citenamefont {Ruckenstein}}]{Zhang2020c}%
  \BibitemOpen
  \bibfield  {author} {\bibinfo {author} {\bibfnamefont {L.}~\bibnamefont
  {Zhang}}, \bibinfo {author} {\bibfnamefont {J.~A.}\ \bibnamefont {Reyes}},
  \bibinfo {author} {\bibfnamefont {S.}~\bibnamefont {Kourtis}}, \bibinfo
  {author} {\bibfnamefont {C.}~\bibnamefont {Chamon}}, \bibinfo {author}
  {\bibfnamefont {E.~R.}\ \bibnamefont {Mucciolo}},\ and\ \bibinfo {author}
  {\bibfnamefont {A.~E.}\ \bibnamefont {Ruckenstein}},\ }\bibfield  {title}
  {\bibinfo {title} {Nonuniversal entanglement level statistics in
  projection-driven quantum circuits},\ }\href
  {https://doi.org/10.1103/PhysRevB.101.235104} {\bibfield  {journal} {\bibinfo
   {journal} {Phys. Rev. B}\ }\textbf {\bibinfo {volume} {101}},\ \bibinfo
  {pages} {235104} (\bibinfo {year} {2020})}\BibitemShut {NoStop}%
\bibitem [{\citenamefont {Nahum}\ \emph {et~al.}(2021)\citenamefont {Nahum},
  \citenamefont {Roy}, \citenamefont {Skinner},\ and\ \citenamefont
  {Ruhman}}]{Nahum2021a}%
  \BibitemOpen
  \bibfield  {author} {\bibinfo {author} {\bibfnamefont {A.}~\bibnamefont
  {Nahum}}, \bibinfo {author} {\bibfnamefont {S.}~\bibnamefont {Roy}}, \bibinfo
  {author} {\bibfnamefont {B.}~\bibnamefont {Skinner}},\ and\ \bibinfo {author}
  {\bibfnamefont {J.}~\bibnamefont {Ruhman}},\ }\bibfield  {title} {\bibinfo
  {title} {Measurement and entanglement phase transitions in all-to-all quantum
  circuits, on quantum trees, and in {Landau-Ginsburg} theory},\ }\href
  {https://doi.org/10.1103/PRXQuantum.2.010352} {\bibfield  {journal} {\bibinfo
   {journal} {PRX Quantum}\ }\textbf {\bibinfo {volume} {2}},\ \bibinfo {pages}
  {010352} (\bibinfo {year} {2021})}\BibitemShut {NoStop}%
\bibitem [{\citenamefont {Ippoliti}\ \emph {et~al.}(2021)\citenamefont
  {Ippoliti}, \citenamefont {Gullans}, \citenamefont {Gopalakrishnan},
  \citenamefont {Huse},\ and\ \citenamefont {Khemani}}]{Ippoliti2021a}%
  \BibitemOpen
  \bibfield  {author} {\bibinfo {author} {\bibfnamefont {M.}~\bibnamefont
  {Ippoliti}}, \bibinfo {author} {\bibfnamefont {M.~J.}\ \bibnamefont
  {Gullans}}, \bibinfo {author} {\bibfnamefont {S.}~\bibnamefont
  {Gopalakrishnan}}, \bibinfo {author} {\bibfnamefont {D.~A.}\ \bibnamefont
  {Huse}},\ and\ \bibinfo {author} {\bibfnamefont {V.}~\bibnamefont
  {Khemani}},\ }\bibfield  {title} {\bibinfo {title} {Entanglement phase
  transitions in measurement-only dynamics},\ }\href
  {https://doi.org/10.1103/PhysRevX.11.011030} {\bibfield  {journal} {\bibinfo
  {journal} {Phys. Rev. X}\ }\textbf {\bibinfo {volume} {11}},\ \bibinfo
  {pages} {011030} (\bibinfo {year} {2021})}\BibitemShut {NoStop}%
\bibitem [{\citenamefont {Ippoliti}\ and\ \citenamefont
  {Khemani}(2021)}]{Ippoliti2021b}%
  \BibitemOpen
  \bibfield  {author} {\bibinfo {author} {\bibfnamefont {M.}~\bibnamefont
  {Ippoliti}}\ and\ \bibinfo {author} {\bibfnamefont {V.}~\bibnamefont
  {Khemani}},\ }\bibfield  {title} {\bibinfo {title} {Postselection-free
  entanglement dynamics via spacetime duality},\ }\href
  {https://doi.org/10.1103/PhysRevLett.126.060501} {\bibfield  {journal}
  {\bibinfo  {journal} {Phys. Rev. Lett.}\ }\textbf {\bibinfo {volume} {126}},\
  \bibinfo {pages} {060501} (\bibinfo {year} {2021})}\BibitemShut {NoStop}%
\bibitem [{\citenamefont {Lavasani}\ \emph
  {et~al.}(2021{\natexlab{a}})\citenamefont {Lavasani}, \citenamefont
  {Alavirad},\ and\ \citenamefont {Barkeshli}}]{Lavasani2021a}%
  \BibitemOpen
  \bibfield  {author} {\bibinfo {author} {\bibfnamefont {A.}~\bibnamefont
  {Lavasani}}, \bibinfo {author} {\bibfnamefont {Y.}~\bibnamefont {Alavirad}},\
  and\ \bibinfo {author} {\bibfnamefont {M.}~\bibnamefont {Barkeshli}},\
  }\bibfield  {title} {\bibinfo {title} {Measurement-induced topological
  entanglement transitions in symmetric random quantum circuits},\ }\href
  {https://doi.org/10.1038/s41567-020-01112-z} {\bibfield  {journal} {\bibinfo
  {journal} {Nat. Phys.}\ }\textbf {\bibinfo {volume} {17}},\ \bibinfo {pages}
  {342} (\bibinfo {year} {2021}{\natexlab{a}})}\BibitemShut {NoStop}%
\bibitem [{\citenamefont {Lavasani}\ \emph
  {et~al.}(2021{\natexlab{b}})\citenamefont {Lavasani}, \citenamefont
  {Alavirad},\ and\ \citenamefont {Barkeshli}}]{Lavasani2021b}%
  \BibitemOpen
  \bibfield  {author} {\bibinfo {author} {\bibfnamefont {A.}~\bibnamefont
  {Lavasani}}, \bibinfo {author} {\bibfnamefont {Y.}~\bibnamefont {Alavirad}},\
  and\ \bibinfo {author} {\bibfnamefont {M.}~\bibnamefont {Barkeshli}},\
  }\bibfield  {title} {\bibinfo {title} {Topological order and criticality in
  $(2+1)\mathrm{D}$ monitored random quantum circuits},\ }\href
  {https://doi.org/10.1103/PhysRevLett.127.235701} {\bibfield  {journal}
  {\bibinfo  {journal} {Phys. Rev. Lett.}\ }\textbf {\bibinfo {volume} {127}},\
  \bibinfo {pages} {235701} (\bibinfo {year} {2021}{\natexlab{b}})}\BibitemShut
  {NoStop}%
\bibitem [{\citenamefont {Sang}\ and\ \citenamefont {Hsieh}(2021)}]{Sang2021a}%
  \BibitemOpen
  \bibfield  {author} {\bibinfo {author} {\bibfnamefont {S.}~\bibnamefont
  {Sang}}\ and\ \bibinfo {author} {\bibfnamefont {T.~H.}\ \bibnamefont
  {Hsieh}},\ }\bibfield  {title} {\bibinfo {title} {Measurement-protected
  quantum phases},\ }\href {https://doi.org/10.1103/PhysRevResearch.3.023200}
  {\bibfield  {journal} {\bibinfo  {journal} {Phys. Rev. Research}\ }\textbf
  {\bibinfo {volume} {3}},\ \bibinfo {pages} {023200} (\bibinfo {year}
  {2021})}\BibitemShut {NoStop}%
\bibitem [{\citenamefont {Block}\ \emph {et~al.}(2022)\citenamefont {Block},
  \citenamefont {Bao}, \citenamefont {Choi}, \citenamefont {Altman},\ and\
  \citenamefont {Yao}}]{Block2022a}%
  \BibitemOpen
  \bibfield  {author} {\bibinfo {author} {\bibfnamefont {M.}~\bibnamefont
  {Block}}, \bibinfo {author} {\bibfnamefont {Y.}~\bibnamefont {Bao}}, \bibinfo
  {author} {\bibfnamefont {S.}~\bibnamefont {Choi}}, \bibinfo {author}
  {\bibfnamefont {E.}~\bibnamefont {Altman}},\ and\ \bibinfo {author}
  {\bibfnamefont {N.~Y.}\ \bibnamefont {Yao}},\ }\bibfield  {title} {\bibinfo
  {title} {Measurement-induced transition in long-range interacting quantum
  circuits},\ }\href {https://doi.org/10.1103/PhysRevLett.128.010604}
  {\bibfield  {journal} {\bibinfo  {journal} {Phys. Rev. Lett.}\ }\textbf
  {\bibinfo {volume} {128}},\ \bibinfo {pages} {010604} (\bibinfo {year}
  {2022})}\BibitemShut {NoStop}%
\bibitem [{\citenamefont {Sharma}\ \emph {et~al.}(2022)\citenamefont {Sharma},
  \citenamefont {Turkeshi}, \citenamefont {Fazio},\ and\ \citenamefont
  {Dalmonte}}]{Sharma2022}%
  \BibitemOpen
  \bibfield  {author} {\bibinfo {author} {\bibfnamefont {S.}~\bibnamefont
  {Sharma}}, \bibinfo {author} {\bibfnamefont {X.}~\bibnamefont {Turkeshi}},
  \bibinfo {author} {\bibfnamefont {R.}~\bibnamefont {Fazio}},\ and\ \bibinfo
  {author} {\bibfnamefont {M.}~\bibnamefont {Dalmonte}},\ }\bibfield  {title}
  {\bibinfo {title} {{Measurement-induced criticality in extended and
  long-range unitary circuits}},\ }\href
  {https://doi.org/10.21468/SciPostPhysCore.5.2.023} {\bibfield  {journal}
  {\bibinfo  {journal} {SciPost Phys. Core}\ }\textbf {\bibinfo {volume} {5}},\
  \bibinfo {pages} {023} (\bibinfo {year} {2022})}\BibitemShut {NoStop}%
\bibitem [{\citenamefont {Jian}\ \emph {et~al.}(2023)\citenamefont {Jian},
  \citenamefont {Shapourian}, \citenamefont {Bauer},\ and\ \citenamefont
  {Ludwig}}]{Jian2023}%
  \BibitemOpen
  \bibfield  {author} {\bibinfo {author} {\bibfnamefont {C.-M.}\ \bibnamefont
  {Jian}}, \bibinfo {author} {\bibfnamefont {H.}~\bibnamefont {Shapourian}},
  \bibinfo {author} {\bibfnamefont {B.}~\bibnamefont {Bauer}},\ and\ \bibinfo
  {author} {\bibfnamefont {A.~W.~W.}\ \bibnamefont {Ludwig}},\ }\href
  {https://doi.org/10.48550/arxiv.2302.09094} {\bibinfo {title}
  {Measurement-induced entanglement transitions in quantum circuits of
  non-interacting fermions: {Born}-rule versus forced measurements}} (\bibinfo
  {year} {2023}),\ \Eprint {https://arxiv.org/abs/2302.09094}
  {arXiv:2302.09094} \BibitemShut {NoStop}%
\bibitem [{\citenamefont {Kelly}\ \emph {et~al.}(2023)\citenamefont {Kelly},
  \citenamefont {Poschinger}, \citenamefont {Schmidt-Kaler}, \citenamefont
  {Fisher},\ and\ \citenamefont {Marino}}]{Kelly2023}%
  \BibitemOpen
  \bibfield  {author} {\bibinfo {author} {\bibfnamefont {S.~P.}\ \bibnamefont
  {Kelly}}, \bibinfo {author} {\bibfnamefont {U.}~\bibnamefont {Poschinger}},
  \bibinfo {author} {\bibfnamefont {F.}~\bibnamefont {Schmidt-Kaler}}, \bibinfo
  {author} {\bibfnamefont {M.~P.~A.}\ \bibnamefont {Fisher}},\ and\ \bibinfo
  {author} {\bibfnamefont {J.}~\bibnamefont {Marino}},\ }\bibfield  {title}
  {\bibinfo {title} {{Coherence requirements for quantum communication from
  hybrid circuit dynamics}},\ }\href
  {https://doi.org/10.21468/SciPostPhys.15.6.250} {\bibfield  {journal}
  {\bibinfo  {journal} {SciPost Phys.}\ }\textbf {\bibinfo {volume} {15}},\
  \bibinfo {pages} {250} (\bibinfo {year} {2023})}\BibitemShut {NoStop}%
\bibitem [{\citenamefont {Cao}\ \emph {et~al.}(2019)\citenamefont {Cao},
  \citenamefont {Tilloy},\ and\ \citenamefont {{De~Luca}}}]{Cao2019a}%
  \BibitemOpen
  \bibfield  {author} {\bibinfo {author} {\bibfnamefont {X.}~\bibnamefont
  {Cao}}, \bibinfo {author} {\bibfnamefont {A.}~\bibnamefont {Tilloy}},\ and\
  \bibinfo {author} {\bibfnamefont {A.}~\bibnamefont {{De~Luca}}},\ }\bibfield
  {title} {\bibinfo {title} {Entanglement in a fermion chain under continuous
  monitoring},\ }\href {https://doi.org/10.21468/SciPostPhys.7.2.024}
  {\bibfield  {journal} {\bibinfo  {journal} {SciPost Phys.}\ }\textbf
  {\bibinfo {volume} {7}},\ \bibinfo {pages} {024} (\bibinfo {year}
  {2019})}\BibitemShut {NoStop}%
\bibitem [{\citenamefont {Alberton}\ \emph {et~al.}(2021)\citenamefont
  {Alberton}, \citenamefont {Buchhold},\ and\ \citenamefont
  {Diehl}}]{Alberton2021a}%
  \BibitemOpen
  \bibfield  {author} {\bibinfo {author} {\bibfnamefont {O.}~\bibnamefont
  {Alberton}}, \bibinfo {author} {\bibfnamefont {M.}~\bibnamefont {Buchhold}},\
  and\ \bibinfo {author} {\bibfnamefont {S.}~\bibnamefont {Diehl}},\ }\bibfield
   {title} {\bibinfo {title} {Entanglement transition in a monitored
  free-fermion chain: {F}rom extended criticality to area law},\ }\href
  {https://doi.org/10.1103/PhysRevLett.126.170602} {\bibfield  {journal}
  {\bibinfo  {journal} {Phys. Rev. Lett.}\ }\textbf {\bibinfo {volume} {126}},\
  \bibinfo {pages} {170602} (\bibinfo {year} {2021})}\BibitemShut {NoStop}%
\bibitem [{\citenamefont {Chen}\ \emph {et~al.}(2020)\citenamefont {Chen},
  \citenamefont {Li}, \citenamefont {Fisher},\ and\ \citenamefont
  {Lucas}}]{Chen2020a}%
  \BibitemOpen
  \bibfield  {author} {\bibinfo {author} {\bibfnamefont {X.}~\bibnamefont
  {Chen}}, \bibinfo {author} {\bibfnamefont {Y.}~\bibnamefont {Li}}, \bibinfo
  {author} {\bibfnamefont {M.~P.~A.}\ \bibnamefont {Fisher}},\ and\ \bibinfo
  {author} {\bibfnamefont {A.}~\bibnamefont {Lucas}},\ }\bibfield  {title}
  {\bibinfo {title} {Emergent conformal symmetry in nonunitary random dynamics
  of free fermions},\ }\href {https://doi.org/10.1103/PhysRevResearch.2.033017}
  {\bibfield  {journal} {\bibinfo  {journal} {Phys. Rev. Research}\ }\textbf
  {\bibinfo {volume} {2}},\ \bibinfo {pages} {033017} (\bibinfo {year}
  {2020})}\BibitemShut {NoStop}%
\bibitem [{\citenamefont {Tang}\ \emph {et~al.}(2021)\citenamefont {Tang},
  \citenamefont {Chen},\ and\ \citenamefont {Zhu}}]{Tang2021a}%
  \BibitemOpen
  \bibfield  {author} {\bibinfo {author} {\bibfnamefont {Q.}~\bibnamefont
  {Tang}}, \bibinfo {author} {\bibfnamefont {X.}~\bibnamefont {Chen}},\ and\
  \bibinfo {author} {\bibfnamefont {W.}~\bibnamefont {Zhu}},\ }\bibfield
  {title} {\bibinfo {title} {Quantum criticality in the nonunitary dynamics of
  $(2+1)$-dimensional free fermions},\ }\href
  {https://doi.org/10.1103/PhysRevB.103.174303} {\bibfield  {journal} {\bibinfo
   {journal} {Phys. Rev. B}\ }\textbf {\bibinfo {volume} {103}},\ \bibinfo
  {pages} {174303} (\bibinfo {year} {2021})}\BibitemShut {NoStop}%
\bibitem [{\citenamefont {Coppola}\ \emph {et~al.}(2022)\citenamefont
  {Coppola}, \citenamefont {Tirrito}, \citenamefont {Karevski},\ and\
  \citenamefont {Collura}}]{Coppola2022}%
  \BibitemOpen
  \bibfield  {author} {\bibinfo {author} {\bibfnamefont {M.}~\bibnamefont
  {Coppola}}, \bibinfo {author} {\bibfnamefont {E.}~\bibnamefont {Tirrito}},
  \bibinfo {author} {\bibfnamefont {D.}~\bibnamefont {Karevski}},\ and\
  \bibinfo {author} {\bibfnamefont {M.}~\bibnamefont {Collura}},\ }\bibfield
  {title} {\bibinfo {title} {Growth of entanglement entropy under local
  projective measurements},\ }\href
  {https://doi.org/10.1103/PhysRevB.105.094303} {\bibfield  {journal} {\bibinfo
   {journal} {Phys. Rev. B}\ }\textbf {\bibinfo {volume} {105}},\ \bibinfo
  {pages} {094303} (\bibinfo {year} {2022})}\BibitemShut {NoStop}%
\bibitem [{\citenamefont {Ladewig}\ \emph {et~al.}(2022)\citenamefont
  {Ladewig}, \citenamefont {Diehl},\ and\ \citenamefont
  {Buchhold}}]{Ladewig2022}%
  \BibitemOpen
  \bibfield  {author} {\bibinfo {author} {\bibfnamefont {B.}~\bibnamefont
  {Ladewig}}, \bibinfo {author} {\bibfnamefont {S.}~\bibnamefont {Diehl}},\
  and\ \bibinfo {author} {\bibfnamefont {M.}~\bibnamefont {Buchhold}},\
  }\bibfield  {title} {\bibinfo {title} {Monitored open fermion dynamics:
  {E}xploring the interplay of measurement, decoherence, and free {Hamiltonian}
  evolution},\ }\href {https://doi.org/10.1103/PhysRevResearch.4.033001}
  {\bibfield  {journal} {\bibinfo  {journal} {Phys. Rev. Research}\ }\textbf
  {\bibinfo {volume} {4}},\ \bibinfo {pages} {033001} (\bibinfo {year}
  {2022})}\BibitemShut {NoStop}%
\bibitem [{\citenamefont {Carollo}\ and\ \citenamefont
  {Alba}(2022)}]{Carollo2022}%
  \BibitemOpen
  \bibfield  {author} {\bibinfo {author} {\bibfnamefont {F.}~\bibnamefont
  {Carollo}}\ and\ \bibinfo {author} {\bibfnamefont {V.}~\bibnamefont {Alba}},\
  }\bibfield  {title} {\bibinfo {title} {Entangled multiplets and spreading of
  quantum correlations in a continuously monitored tight-binding chain},\
  }\href {https://doi.org/10.1103/PhysRevB.106.L220304} {\bibfield  {journal}
  {\bibinfo  {journal} {Phys. Rev. B}\ }\textbf {\bibinfo {volume} {106}},\
  \bibinfo {pages} {L220304} (\bibinfo {year} {2022})}\BibitemShut {NoStop}%
\bibitem [{\citenamefont {Buchhold}\ \emph {et~al.}(2022)\citenamefont
  {Buchhold}, \citenamefont {M{\"u}ller},\ and\ \citenamefont
  {Diehl}}]{Buchhold2022}%
  \BibitemOpen
  \bibfield  {author} {\bibinfo {author} {\bibfnamefont {M.}~\bibnamefont
  {Buchhold}}, \bibinfo {author} {\bibfnamefont {T.}~\bibnamefont
  {M{\"u}ller}},\ and\ \bibinfo {author} {\bibfnamefont {S.}~\bibnamefont
  {Diehl}},\ }\href {https://doi.org/10.48550/arxiv.2208.10506} {\bibinfo
  {title} {Revealing measurement-induced phase transitions by pre-selection}}
  (\bibinfo {year} {2022}),\ \Eprint {https://arxiv.org/abs/2208.10506}
  {arXiv:2208.10506} \BibitemShut {NoStop}%
\bibitem [{\citenamefont {Yang}\ \emph
  {et~al.}(2023{\natexlab{a}})\citenamefont {Yang}, \citenamefont {Zuo},\ and\
  \citenamefont {Liu}}]{Yang2022}%
  \BibitemOpen
  \bibfield  {author} {\bibinfo {author} {\bibfnamefont {Q.}~\bibnamefont
  {Yang}}, \bibinfo {author} {\bibfnamefont {Y.}~\bibnamefont {Zuo}},\ and\
  \bibinfo {author} {\bibfnamefont {D.~E.}\ \bibnamefont {Liu}},\ }\bibfield
  {title} {\bibinfo {title} {Keldysh nonlinear sigma model for a free-fermion
  gas under continuous measurements},\ }\href
  {https://doi.org/10.1103/PhysRevResearch.5.033174} {\bibfield  {journal}
  {\bibinfo  {journal} {Phys. Rev. Res.}\ }\textbf {\bibinfo {volume} {5}},\
  \bibinfo {pages} {033174} (\bibinfo {year} {2023}{\natexlab{a}})}\BibitemShut
  {NoStop}%
\bibitem [{\citenamefont {Szyniszewski}\ \emph {et~al.}(2023)\citenamefont
  {Szyniszewski}, \citenamefont {Lunt},\ and\ \citenamefont
  {Pal}}]{Szyniszewski2022}%
  \BibitemOpen
  \bibfield  {author} {\bibinfo {author} {\bibfnamefont {M.}~\bibnamefont
  {Szyniszewski}}, \bibinfo {author} {\bibfnamefont {O.}~\bibnamefont {Lunt}},\
  and\ \bibinfo {author} {\bibfnamefont {A.}~\bibnamefont {Pal}},\ }\bibfield
  {title} {\bibinfo {title} {Disordered monitored free fermions},\ }\href
  {https://doi.org/10.1103/PhysRevB.108.165126} {\bibfield  {journal} {\bibinfo
   {journal} {Phys. Rev. B}\ }\textbf {\bibinfo {volume} {108}},\ \bibinfo
  {pages} {165126} (\bibinfo {year} {2023})}\BibitemShut {NoStop}%
\bibitem [{\citenamefont {Buchhold}\ \emph {et~al.}(2021)\citenamefont
  {Buchhold}, \citenamefont {Minoguchi}, \citenamefont {Altland},\ and\
  \citenamefont {Diehl}}]{Buchhold2021a}%
  \BibitemOpen
  \bibfield  {author} {\bibinfo {author} {\bibfnamefont {M.}~\bibnamefont
  {Buchhold}}, \bibinfo {author} {\bibfnamefont {Y.}~\bibnamefont {Minoguchi}},
  \bibinfo {author} {\bibfnamefont {A.}~\bibnamefont {Altland}},\ and\ \bibinfo
  {author} {\bibfnamefont {S.}~\bibnamefont {Diehl}},\ }\bibfield  {title}
  {\bibinfo {title} {Effective theory for the measurement-induced phase
  transition of {Dirac} fermions},\ }\href
  {https://doi.org/10.1103/PhysRevX.11.041004} {\bibfield  {journal} {\bibinfo
  {journal} {Phys. Rev. X}\ }\textbf {\bibinfo {volume} {11}},\ \bibinfo
  {pages} {041004} (\bibinfo {year} {2021})}\BibitemShut {NoStop}%
\bibitem [{\citenamefont {Van~Regemortel}\ \emph {et~al.}(2021)\citenamefont
  {Van~Regemortel}, \citenamefont {Cian}, \citenamefont {Seif}, \citenamefont
  {Dehghani},\ and\ \citenamefont {Hafezi}}]{VanRegemortel2021a}%
  \BibitemOpen
  \bibfield  {author} {\bibinfo {author} {\bibfnamefont {M.}~\bibnamefont
  {Van~Regemortel}}, \bibinfo {author} {\bibfnamefont {Z.-P.}\ \bibnamefont
  {Cian}}, \bibinfo {author} {\bibfnamefont {A.}~\bibnamefont {Seif}}, \bibinfo
  {author} {\bibfnamefont {H.}~\bibnamefont {Dehghani}},\ and\ \bibinfo
  {author} {\bibfnamefont {M.}~\bibnamefont {Hafezi}},\ }\bibfield  {title}
  {\bibinfo {title} {Entanglement entropy scaling transition under competing
  monitoring protocols},\ }\href
  {https://doi.org/10.1103/PhysRevLett.126.123604} {\bibfield  {journal}
  {\bibinfo  {journal} {Phys. Rev. Lett.}\ }\textbf {\bibinfo {volume} {126}},\
  \bibinfo {pages} {123604} (\bibinfo {year} {2021})}\BibitemShut {NoStop}%
\bibitem [{\citenamefont {Gal}\ \emph {et~al.}(2023)\citenamefont {Gal},
  \citenamefont {Turkeshi},\ and\ \citenamefont {Schirò}}]{Youenn2023}%
  \BibitemOpen
  \bibfield  {author} {\bibinfo {author} {\bibfnamefont {Y.~L.}\ \bibnamefont
  {Gal}}, \bibinfo {author} {\bibfnamefont {X.}~\bibnamefont {Turkeshi}},\ and\
  \bibinfo {author} {\bibfnamefont {M.}~\bibnamefont {Schirò}},\ }\bibfield
  {title} {\bibinfo {title} {{Volume-to-area law entanglement transition in a
  non-Hermitian free fermionic chain}},\ }\href
  {https://doi.org/10.21468/SciPostPhys.14.5.138} {\bibfield  {journal}
  {\bibinfo  {journal} {SciPost Phys.}\ }\textbf {\bibinfo {volume} {14}},\
  \bibinfo {pages} {138} (\bibinfo {year} {2023})}\BibitemShut {NoStop}%
\bibitem [{\citenamefont {L\'oio}\ \emph {et~al.}(2023)\citenamefont {L\'oio},
  \citenamefont {De~Luca}, \citenamefont {De~Nardis},\ and\ \citenamefont
  {Turkeshi}}]{Loio2023}%
  \BibitemOpen
  \bibfield  {author} {\bibinfo {author} {\bibfnamefont {H.}~\bibnamefont
  {L\'oio}}, \bibinfo {author} {\bibfnamefont {A.}~\bibnamefont {De~Luca}},
  \bibinfo {author} {\bibfnamefont {J.}~\bibnamefont {De~Nardis}},\ and\
  \bibinfo {author} {\bibfnamefont {X.}~\bibnamefont {Turkeshi}},\ }\bibfield
  {title} {\bibinfo {title} {Purification timescales in monitored fermions},\
  }\href {https://doi.org/10.1103/PhysRevB.108.L020306} {\bibfield  {journal}
  {\bibinfo  {journal} {Phys. Rev. B}\ }\textbf {\bibinfo {volume} {108}},\
  \bibinfo {pages} {L020306} (\bibinfo {year} {2023})}\BibitemShut {NoStop}%
\bibitem [{\citenamefont {Turkeshi}\ \emph
  {et~al.}(2022{\natexlab{a}})\citenamefont {Turkeshi}, \citenamefont
  {Piroli},\ and\ \citenamefont {Schir\'o}}]{Turkeshi2022b}%
  \BibitemOpen
  \bibfield  {author} {\bibinfo {author} {\bibfnamefont {X.}~\bibnamefont
  {Turkeshi}}, \bibinfo {author} {\bibfnamefont {L.}~\bibnamefont {Piroli}},\
  and\ \bibinfo {author} {\bibfnamefont {M.}~\bibnamefont {Schir\'o}},\
  }\bibfield  {title} {\bibinfo {title} {Enhanced entanglement negativity in
  boundary-driven monitored fermionic chains},\ }\href
  {https://doi.org/10.1103/PhysRevB.106.024304} {\bibfield  {journal} {\bibinfo
   {journal} {Phys. Rev. B}\ }\textbf {\bibinfo {volume} {106}},\ \bibinfo
  {pages} {024304} (\bibinfo {year} {2022}{\natexlab{a}})}\BibitemShut
  {NoStop}%
\bibitem [{\citenamefont {Kells}\ \emph {et~al.}(2023)\citenamefont {Kells},
  \citenamefont {Meidan},\ and\ \citenamefont {Romito}}]{Kells2023}%
  \BibitemOpen
  \bibfield  {author} {\bibinfo {author} {\bibfnamefont {G.}~\bibnamefont
  {Kells}}, \bibinfo {author} {\bibfnamefont {D.}~\bibnamefont {Meidan}},\ and\
  \bibinfo {author} {\bibfnamefont {A.}~\bibnamefont {Romito}},\ }\bibfield
  {title} {\bibinfo {title} {{Topological transitions in weakly monitored free
  fermions}},\ }\href {https://doi.org/10.21468/SciPostPhys.14.3.031}
  {\bibfield  {journal} {\bibinfo  {journal} {SciPost Phys.}\ }\textbf
  {\bibinfo {volume} {14}},\ \bibinfo {pages} {031} (\bibinfo {year}
  {2023})}\BibitemShut {NoStop}%
\bibitem [{\citenamefont {Fava}\ \emph {et~al.}(2023)\citenamefont {Fava},
  \citenamefont {Piroli}, \citenamefont {Swann}, \citenamefont {Bernard},\ and\
  \citenamefont {Nahum}}]{Fava2023}%
  \BibitemOpen
  \bibfield  {author} {\bibinfo {author} {\bibfnamefont {M.}~\bibnamefont
  {Fava}}, \bibinfo {author} {\bibfnamefont {L.}~\bibnamefont {Piroli}},
  \bibinfo {author} {\bibfnamefont {T.}~\bibnamefont {Swann}}, \bibinfo
  {author} {\bibfnamefont {D.}~\bibnamefont {Bernard}},\ and\ \bibinfo {author}
  {\bibfnamefont {A.}~\bibnamefont {Nahum}},\ }\bibfield  {title} {\bibinfo
  {title} {Nonlinear sigma models for monitored dynamics of free fermions},\
  }\href {https://doi.org/10.1103/PhysRevX.13.041045} {\bibfield  {journal}
  {\bibinfo  {journal} {Phys. Rev. X}\ }\textbf {\bibinfo {volume} {13}},\
  \bibinfo {pages} {041045} (\bibinfo {year} {2023})}\BibitemShut {NoStop}%
\bibitem [{\citenamefont {Swann}\ \emph {et~al.}(2023)\citenamefont {Swann},
  \citenamefont {Bernard},\ and\ \citenamefont {Nahum}}]{Swann2023}%
  \BibitemOpen
  \bibfield  {author} {\bibinfo {author} {\bibfnamefont {T.}~\bibnamefont
  {Swann}}, \bibinfo {author} {\bibfnamefont {D.}~\bibnamefont {Bernard}},\
  and\ \bibinfo {author} {\bibfnamefont {A.}~\bibnamefont {Nahum}},\ }\href
  {https://doi.org/10.48550/arxiv.2302.12212} {\bibinfo {title} {Spacetime
  picture for entanglement generation in noisy fermion chains}} (\bibinfo
  {year} {2023}),\ \Eprint {https://arxiv.org/abs/2302.12212}
  {arXiv:2302.12212} \BibitemShut {NoStop}%
\bibitem [{\citenamefont {Merritt}\ and\ \citenamefont
  {Fidkowski}(2023)}]{Merritt2023}%
  \BibitemOpen
  \bibfield  {author} {\bibinfo {author} {\bibfnamefont {J.}~\bibnamefont
  {Merritt}}\ and\ \bibinfo {author} {\bibfnamefont {L.}~\bibnamefont
  {Fidkowski}},\ }\bibfield  {title} {\bibinfo {title} {Entanglement
  transitions with free fermions},\ }\href
  {https://doi.org/10.1103/PhysRevB.107.064303} {\bibfield  {journal} {\bibinfo
   {journal} {Phys. Rev. B}\ }\textbf {\bibinfo {volume} {107}},\ \bibinfo
  {pages} {064303} (\bibinfo {year} {2023})}\BibitemShut {NoStop}%
\bibitem [{\citenamefont {Poboiko}\ \emph {et~al.}(2023)\citenamefont
  {Poboiko}, \citenamefont {P\"opperl}, \citenamefont {Gornyi},\ and\
  \citenamefont {Mirlin}}]{Poboiko2023}%
  \BibitemOpen
  \bibfield  {author} {\bibinfo {author} {\bibfnamefont {I.}~\bibnamefont
  {Poboiko}}, \bibinfo {author} {\bibfnamefont {P.}~\bibnamefont {P\"opperl}},
  \bibinfo {author} {\bibfnamefont {I.~V.}\ \bibnamefont {Gornyi}},\ and\
  \bibinfo {author} {\bibfnamefont {A.~D.}\ \bibnamefont {Mirlin}},\ }\bibfield
   {title} {\bibinfo {title} {Theory of free fermions under random projective
  measurements},\ }\href {https://doi.org/10.1103/PhysRevX.13.041046}
  {\bibfield  {journal} {\bibinfo  {journal} {Phys. Rev. X}\ }\textbf {\bibinfo
  {volume} {13}},\ \bibinfo {pages} {041046} (\bibinfo {year}
  {2023})}\BibitemShut {NoStop}%
\bibitem [{\citenamefont {Poboiko}\ \emph {et~al.}(2024)\citenamefont
  {Poboiko}, \citenamefont {Gornyi},\ and\ \citenamefont
  {Mirlin}}]{poboiko2023a}%
  \BibitemOpen
  \bibfield  {author} {\bibinfo {author} {\bibfnamefont {I.}~\bibnamefont
  {Poboiko}}, \bibinfo {author} {\bibfnamefont {I.~V.}\ \bibnamefont
  {Gornyi}},\ and\ \bibinfo {author} {\bibfnamefont {A.~D.}\ \bibnamefont
  {Mirlin}},\ }\bibfield  {title} {\bibinfo {title} {Measurement-induced phase
  transition for free fermions above one dimension},\ }\href
  {https://doi.org/10.1103/PhysRevLett.132.110403} {\bibfield  {journal}
  {\bibinfo  {journal} {Phys. Rev. Lett.}\ }\textbf {\bibinfo {volume} {132}},\
  \bibinfo {pages} {110403} (\bibinfo {year} {2024})}\BibitemShut {NoStop}%
\bibitem [{\citenamefont {Chahine}\ and\ \citenamefont
  {Buchhold}(2023)}]{chahine2023}%
  \BibitemOpen
  \bibfield  {author} {\bibinfo {author} {\bibfnamefont {K.}~\bibnamefont
  {Chahine}}\ and\ \bibinfo {author} {\bibfnamefont {M.}~\bibnamefont
  {Buchhold}},\ }\href@noop {} {\bibinfo {title} {Entanglement phases,
  localization and multifractality of monitored free fermions in two
  dimensions}} (\bibinfo {year} {2023}),\ \Eprint
  {https://arxiv.org/abs/2309.12391} {arXiv:2309.12391 [cond-mat.str-el]}
  \BibitemShut {NoStop}%
\bibitem [{\citenamefont {Jin}\ and\ \citenamefont {Martin}(2023)}]{jin2023}%
  \BibitemOpen
  \bibfield  {author} {\bibinfo {author} {\bibfnamefont {T.}~\bibnamefont
  {Jin}}\ and\ \bibinfo {author} {\bibfnamefont {D.~G.}\ \bibnamefont
  {Martin}},\ }\href@noop {} {\bibinfo {title} {Measurement-induced phase
  transition in a single-body tight-binding model}} (\bibinfo {year} {2023}),\
  \Eprint {https://arxiv.org/abs/2309.15034} {arXiv:2309.15034 [quant-ph]}
  \BibitemShut {NoStop}%
\bibitem [{\citenamefont {Lang}\ and\ \citenamefont
  {B\"uchler}(2020)}]{Lang2020a}%
  \BibitemOpen
  \bibfield  {author} {\bibinfo {author} {\bibfnamefont {N.}~\bibnamefont
  {Lang}}\ and\ \bibinfo {author} {\bibfnamefont {H.~P.}\ \bibnamefont
  {B\"uchler}},\ }\bibfield  {title} {\bibinfo {title} {Entanglement transition
  in the projective transverse field {Ising} model},\ }\href
  {https://doi.org/10.1103/PhysRevB.102.094204} {\bibfield  {journal} {\bibinfo
   {journal} {Phys. Rev. B}\ }\textbf {\bibinfo {volume} {102}},\ \bibinfo
  {pages} {094204} (\bibinfo {year} {2020})}\BibitemShut {NoStop}%
\bibitem [{\citenamefont {Rossini}\ and\ \citenamefont
  {Vicari}(2020)}]{Rossini2020a}%
  \BibitemOpen
  \bibfield  {author} {\bibinfo {author} {\bibfnamefont {D.}~\bibnamefont
  {Rossini}}\ and\ \bibinfo {author} {\bibfnamefont {E.}~\bibnamefont
  {Vicari}},\ }\bibfield  {title} {\bibinfo {title} {Measurement-induced
  dynamics of many-body systems at quantum criticality},\ }\href
  {https://doi.org/10.1103/PhysRevB.102.035119} {\bibfield  {journal} {\bibinfo
   {journal} {Phys. Rev. B}\ }\textbf {\bibinfo {volume} {102}},\ \bibinfo
  {pages} {035119} (\bibinfo {year} {2020})}\BibitemShut {NoStop}%
\bibitem [{\citenamefont {Biella}\ and\ \citenamefont
  {Schir{\`o}}(2021)}]{Biella2021a}%
  \BibitemOpen
  \bibfield  {author} {\bibinfo {author} {\bibfnamefont {A.}~\bibnamefont
  {Biella}}\ and\ \bibinfo {author} {\bibfnamefont {M.}~\bibnamefont
  {Schir{\`o}}},\ }\bibfield  {title} {\bibinfo {title} {Many-body quantum
  {Z}eno effect and measurement-induced subradiance transition},\ }\href
  {https://doi.org/10.22331/q-2021-08-19-528} {\bibfield  {journal} {\bibinfo
  {journal} {{Quantum}}\ }\textbf {\bibinfo {volume} {5}},\ \bibinfo {pages}
  {528} (\bibinfo {year} {2021})}\BibitemShut {NoStop}%
\bibitem [{\citenamefont {Turkeshi}\ \emph {et~al.}(2021)\citenamefont
  {Turkeshi}, \citenamefont {Biella}, \citenamefont {Fazio}, \citenamefont
  {Dalmonte},\ and\ \citenamefont {Schir{\`o}}}]{Turkeshi2021}%
  \BibitemOpen
  \bibfield  {author} {\bibinfo {author} {\bibfnamefont {X.}~\bibnamefont
  {Turkeshi}}, \bibinfo {author} {\bibfnamefont {A.}~\bibnamefont {Biella}},
  \bibinfo {author} {\bibfnamefont {R.}~\bibnamefont {Fazio}}, \bibinfo
  {author} {\bibfnamefont {M.}~\bibnamefont {Dalmonte}},\ and\ \bibinfo
  {author} {\bibfnamefont {M.}~\bibnamefont {Schir{\`o}}},\ }\bibfield  {title}
  {\bibinfo {title} {Measurement-induced entanglement transitions in the
  quantum {Ising} chain: From infinite to zero clicks},\ }\href
  {https://doi.org/10.1103/PhysRevB.103.224210} {\bibfield  {journal} {\bibinfo
   {journal} {Phys. Rev. B}\ }\textbf {\bibinfo {volume} {103}},\ \bibinfo
  {pages} {224210} (\bibinfo {year} {2021})}\BibitemShut {NoStop}%
\bibitem [{\citenamefont {Tirrito}\ \emph {et~al.}(2023)\citenamefont
  {Tirrito}, \citenamefont {Santini}, \citenamefont {Fazio},\ and\
  \citenamefont {Collura}}]{Tirrito2022}%
  \BibitemOpen
  \bibfield  {author} {\bibinfo {author} {\bibfnamefont {E.}~\bibnamefont
  {Tirrito}}, \bibinfo {author} {\bibfnamefont {A.}~\bibnamefont {Santini}},
  \bibinfo {author} {\bibfnamefont {R.}~\bibnamefont {Fazio}},\ and\ \bibinfo
  {author} {\bibfnamefont {M.}~\bibnamefont {Collura}},\ }\bibfield  {title}
  {\bibinfo {title} {{Full counting statistics as probe of measurement-induced
  transitions in the quantum Ising chain}},\ }\href
  {https://doi.org/10.21468/SciPostPhys.15.3.096} {\bibfield  {journal}
  {\bibinfo  {journal} {SciPost Phys.}\ }\textbf {\bibinfo {volume} {15}},\
  \bibinfo {pages} {096} (\bibinfo {year} {2023})}\BibitemShut {NoStop}%
\bibitem [{\citenamefont {Yang}\ \emph
  {et~al.}(2023{\natexlab{b}})\citenamefont {Yang}, \citenamefont {Mao},\ and\
  \citenamefont {Jian}}]{Yang2023}%
  \BibitemOpen
  \bibfield  {author} {\bibinfo {author} {\bibfnamefont {Z.}~\bibnamefont
  {Yang}}, \bibinfo {author} {\bibfnamefont {D.}~\bibnamefont {Mao}},\ and\
  \bibinfo {author} {\bibfnamefont {C.-M.}\ \bibnamefont {Jian}},\ }\bibfield
  {title} {\bibinfo {title} {Entanglement in a one-dimensional critical state
  after measurements},\ }\href {https://doi.org/10.1103/PhysRevB.108.165120}
  {\bibfield  {journal} {\bibinfo  {journal} {Phys. Rev. B}\ }\textbf {\bibinfo
  {volume} {108}},\ \bibinfo {pages} {165120} (\bibinfo {year}
  {2023}{\natexlab{b}})}\BibitemShut {NoStop}%
\bibitem [{\citenamefont {Weinstein}\ \emph {et~al.}(2023)\citenamefont
  {Weinstein}, \citenamefont {Sajith}, \citenamefont {Altman},\ and\
  \citenamefont {Garratt}}]{Weinstein2023}%
  \BibitemOpen
  \bibfield  {author} {\bibinfo {author} {\bibfnamefont {Z.}~\bibnamefont
  {Weinstein}}, \bibinfo {author} {\bibfnamefont {R.}~\bibnamefont {Sajith}},
  \bibinfo {author} {\bibfnamefont {E.}~\bibnamefont {Altman}},\ and\ \bibinfo
  {author} {\bibfnamefont {S.~J.}\ \bibnamefont {Garratt}},\ }\bibfield
  {title} {\bibinfo {title} {Nonlocality and entanglement in measured critical
  quantum ising chains},\ }\href {https://doi.org/10.1103/PhysRevB.107.245132}
  {\bibfield  {journal} {\bibinfo  {journal} {Phys. Rev. B}\ }\textbf {\bibinfo
  {volume} {107}},\ \bibinfo {pages} {245132} (\bibinfo {year}
  {2023})}\BibitemShut {NoStop}%
\bibitem [{\citenamefont {Murciano}\ \emph {et~al.}(2023)\citenamefont
  {Murciano}, \citenamefont {Sala}, \citenamefont {Liu}, \citenamefont {Mong},\
  and\ \citenamefont {Alicea}}]{Murciano2023}%
  \BibitemOpen
  \bibfield  {author} {\bibinfo {author} {\bibfnamefont {S.}~\bibnamefont
  {Murciano}}, \bibinfo {author} {\bibfnamefont {P.}~\bibnamefont {Sala}},
  \bibinfo {author} {\bibfnamefont {Y.}~\bibnamefont {Liu}}, \bibinfo {author}
  {\bibfnamefont {R.~S.~K.}\ \bibnamefont {Mong}},\ and\ \bibinfo {author}
  {\bibfnamefont {J.}~\bibnamefont {Alicea}},\ }\bibfield  {title} {\bibinfo
  {title} {Measurement-altered ising quantum criticality},\ }\href
  {https://doi.org/10.1103/PhysRevX.13.041042} {\bibfield  {journal} {\bibinfo
  {journal} {Phys. Rev. X}\ }\textbf {\bibinfo {volume} {13}},\ \bibinfo
  {pages} {041042} (\bibinfo {year} {2023})}\BibitemShut {NoStop}%
\bibitem [{\citenamefont {Sierant}\ \emph {et~al.}(2022)\citenamefont
  {Sierant}, \citenamefont {Chiriac{\`o}}, \citenamefont {Surace},
  \citenamefont {Sharma}, \citenamefont {Turkeshi}, \citenamefont {Dalmonte},
  \citenamefont {Fazio},\ and\ \citenamefont {Pagano}}]{Sierant2022a}%
  \BibitemOpen
  \bibfield  {author} {\bibinfo {author} {\bibfnamefont {P.}~\bibnamefont
  {Sierant}}, \bibinfo {author} {\bibfnamefont {G.}~\bibnamefont
  {Chiriac{\`o}}}, \bibinfo {author} {\bibfnamefont {F.~M.}\ \bibnamefont
  {Surace}}, \bibinfo {author} {\bibfnamefont {S.}~\bibnamefont {Sharma}},
  \bibinfo {author} {\bibfnamefont {X.}~\bibnamefont {Turkeshi}}, \bibinfo
  {author} {\bibfnamefont {M.}~\bibnamefont {Dalmonte}}, \bibinfo {author}
  {\bibfnamefont {R.}~\bibnamefont {Fazio}},\ and\ \bibinfo {author}
  {\bibfnamefont {G.}~\bibnamefont {Pagano}},\ }\bibfield  {title} {\bibinfo
  {title} {Dissipative {Floquet} dynamics: from steady state to measurement
  induced criticality in trapped-ion chains},\ }\href
  {https://doi.org/10.22331/q-2022-02-02-638} {\bibfield  {journal} {\bibinfo
  {journal} {{Quantum}}\ }\textbf {\bibinfo {volume} {6}},\ \bibinfo {pages}
  {638} (\bibinfo {year} {2022})}\BibitemShut {NoStop}%
\bibitem [{\citenamefont {Turkeshi}\ \emph
  {et~al.}(2022{\natexlab{b}})\citenamefont {Turkeshi}, \citenamefont
  {Dalmonte}, \citenamefont {Fazio},\ and\ \citenamefont
  {Schir\`o}}]{Turkeshi2022a}%
  \BibitemOpen
  \bibfield  {author} {\bibinfo {author} {\bibfnamefont {X.}~\bibnamefont
  {Turkeshi}}, \bibinfo {author} {\bibfnamefont {M.}~\bibnamefont {Dalmonte}},
  \bibinfo {author} {\bibfnamefont {R.}~\bibnamefont {Fazio}},\ and\ \bibinfo
  {author} {\bibfnamefont {M.}~\bibnamefont {Schir\`o}},\ }\bibfield  {title}
  {\bibinfo {title} {Entanglement transitions from stochastic resetting of
  non-{H}ermitian quasiparticles},\ }\href
  {https://doi.org/10.1103/PhysRevB.105.L241114} {\bibfield  {journal}
  {\bibinfo  {journal} {Phys. Rev. B}\ }\textbf {\bibinfo {volume} {105}},\
  \bibinfo {pages} {L241114} (\bibinfo {year}
  {2022}{\natexlab{b}})}\BibitemShut {NoStop}%
\bibitem [{\citenamefont {Tang}\ and\ \citenamefont {Zhu}(2020)}]{Tang2020a}%
  \BibitemOpen
  \bibfield  {author} {\bibinfo {author} {\bibfnamefont {Q.}~\bibnamefont
  {Tang}}\ and\ \bibinfo {author} {\bibfnamefont {W.}~\bibnamefont {Zhu}},\
  }\bibfield  {title} {\bibinfo {title} {Measurement-induced phase transition:
  A case study in the nonintegrable model by density-matrix renormalization
  group calculations},\ }\href
  {https://doi.org/10.1103/PhysRevResearch.2.013022} {\bibfield  {journal}
  {\bibinfo  {journal} {Phys. Rev. Research}\ }\textbf {\bibinfo {volume}
  {2}},\ \bibinfo {pages} {013022} (\bibinfo {year} {2020})}\BibitemShut
  {NoStop}%
\bibitem [{\citenamefont {Goto}\ and\ \citenamefont
  {Danshita}(2020)}]{Goto2020a}%
  \BibitemOpen
  \bibfield  {author} {\bibinfo {author} {\bibfnamefont {S.}~\bibnamefont
  {Goto}}\ and\ \bibinfo {author} {\bibfnamefont {I.}~\bibnamefont
  {Danshita}},\ }\bibfield  {title} {\bibinfo {title} {Measurement-induced
  transitions of the entanglement scaling law in ultracold gases with
  controllable dissipation},\ }\href
  {https://doi.org/10.1103/PhysRevA.102.033316} {\bibfield  {journal} {\bibinfo
   {journal} {Phys. Rev. A}\ }\textbf {\bibinfo {volume} {102}},\ \bibinfo
  {pages} {033316} (\bibinfo {year} {2020})}\BibitemShut {NoStop}%
\bibitem [{\citenamefont {Fuji}\ and\ \citenamefont
  {Ashida}(2020)}]{Fuji2020a}%
  \BibitemOpen
  \bibfield  {author} {\bibinfo {author} {\bibfnamefont {Y.}~\bibnamefont
  {Fuji}}\ and\ \bibinfo {author} {\bibfnamefont {Y.}~\bibnamefont {Ashida}},\
  }\bibfield  {title} {\bibinfo {title} {Measurement-induced quantum
  criticality under continuous monitoring},\ }\href
  {https://doi.org/10.1103/PhysRevB.102.054302} {\bibfield  {journal} {\bibinfo
   {journal} {Phys. Rev. B}\ }\textbf {\bibinfo {volume} {102}},\ \bibinfo
  {pages} {054302} (\bibinfo {year} {2020})}\BibitemShut {NoStop}%
\bibitem [{\citenamefont {Jian}\ \emph {et~al.}(2021)\citenamefont {Jian},
  \citenamefont {Liu}, \citenamefont {Chen}, \citenamefont {Swingle},\ and\
  \citenamefont {Zhang}}]{Jian2021a}%
  \BibitemOpen
  \bibfield  {author} {\bibinfo {author} {\bibfnamefont {S.-K.}\ \bibnamefont
  {Jian}}, \bibinfo {author} {\bibfnamefont {C.}~\bibnamefont {Liu}}, \bibinfo
  {author} {\bibfnamefont {X.}~\bibnamefont {Chen}}, \bibinfo {author}
  {\bibfnamefont {B.}~\bibnamefont {Swingle}},\ and\ \bibinfo {author}
  {\bibfnamefont {P.}~\bibnamefont {Zhang}},\ }\bibfield  {title} {\bibinfo
  {title} {Measurement-induced phase transition in the monitored
  {Sachdev-Ye-Kitaev} model},\ }\href
  {https://doi.org/10.1103/PhysRevLett.127.140601} {\bibfield  {journal}
  {\bibinfo  {journal} {Phys. Rev. Lett.}\ }\textbf {\bibinfo {volume} {127}},\
  \bibinfo {pages} {140601} (\bibinfo {year} {2021})}\BibitemShut {NoStop}%
\bibitem [{\citenamefont {Altland}\ \emph {et~al.}(2022)\citenamefont
  {Altland}, \citenamefont {Buchhold}, \citenamefont {Diehl},\ and\
  \citenamefont {Micklitz}}]{Altland2022}%
  \BibitemOpen
  \bibfield  {author} {\bibinfo {author} {\bibfnamefont {A.}~\bibnamefont
  {Altland}}, \bibinfo {author} {\bibfnamefont {M.}~\bibnamefont {Buchhold}},
  \bibinfo {author} {\bibfnamefont {S.}~\bibnamefont {Diehl}},\ and\ \bibinfo
  {author} {\bibfnamefont {T.}~\bibnamefont {Micklitz}},\ }\bibfield  {title}
  {\bibinfo {title} {Dynamics of measured many-body quantum chaotic systems},\
  }\href {https://doi.org/10.1103/PhysRevResearch.4.L022066} {\bibfield
  {journal} {\bibinfo  {journal} {Phys. Rev. Research}\ }\textbf {\bibinfo
  {volume} {4}},\ \bibinfo {pages} {L022066} (\bibinfo {year}
  {2022})}\BibitemShut {NoStop}%
\bibitem [{\citenamefont {Doggen}\ \emph {et~al.}(2022)\citenamefont {Doggen},
  \citenamefont {Gefen}, \citenamefont {Gornyi}, \citenamefont {Mirlin},\ and\
  \citenamefont {Polyakov}}]{Doggen2022a}%
  \BibitemOpen
  \bibfield  {author} {\bibinfo {author} {\bibfnamefont {E.~V.~H.}\
  \bibnamefont {Doggen}}, \bibinfo {author} {\bibfnamefont {Y.}~\bibnamefont
  {Gefen}}, \bibinfo {author} {\bibfnamefont {I.~V.}\ \bibnamefont {Gornyi}},
  \bibinfo {author} {\bibfnamefont {A.~D.}\ \bibnamefont {Mirlin}},\ and\
  \bibinfo {author} {\bibfnamefont {D.~G.}\ \bibnamefont {Polyakov}},\
  }\bibfield  {title} {\bibinfo {title} {Generalized quantum measurements with
  matrix product states: Entanglement phase transition and clusterization},\
  }\href {https://doi.org/10.1103/PhysRevResearch.4.023146} {\bibfield
  {journal} {\bibinfo  {journal} {Phys. Rev. Res.}\ }\textbf {\bibinfo {volume}
  {4}},\ \bibinfo {pages} {023146} (\bibinfo {year} {2022})}\BibitemShut
  {NoStop}%
\bibitem [{\citenamefont {Doggen}\ \emph {et~al.}(2023)\citenamefont {Doggen},
  \citenamefont {Gefen}, \citenamefont {Gornyi}, \citenamefont {Mirlin},\ and\
  \citenamefont {Polyakov}}]{Doggen2023}%
  \BibitemOpen
  \bibfield  {author} {\bibinfo {author} {\bibfnamefont {E.~V.~H.}\
  \bibnamefont {Doggen}}, \bibinfo {author} {\bibfnamefont {Y.}~\bibnamefont
  {Gefen}}, \bibinfo {author} {\bibfnamefont {I.~V.}\ \bibnamefont {Gornyi}},
  \bibinfo {author} {\bibfnamefont {A.~D.}\ \bibnamefont {Mirlin}},\ and\
  \bibinfo {author} {\bibfnamefont {D.~G.}\ \bibnamefont {Polyakov}},\
  }\bibfield  {title} {\bibinfo {title} {Evolution of many-body systems under
  ancilla quantum measurements},\ }\href
  {https://doi.org/10.1103/PhysRevB.107.214203} {\bibfield  {journal} {\bibinfo
   {journal} {Phys. Rev. B}\ }\textbf {\bibinfo {volume} {107}},\ \bibinfo
  {pages} {214203} (\bibinfo {year} {2023})}\BibitemShut {NoStop}%
\bibitem [{\citenamefont {Lunt}\ and\ \citenamefont {Pal}(2020)}]{Lunt2020a}%
  \BibitemOpen
  \bibfield  {author} {\bibinfo {author} {\bibfnamefont {O.}~\bibnamefont
  {Lunt}}\ and\ \bibinfo {author} {\bibfnamefont {A.}~\bibnamefont {Pal}},\
  }\bibfield  {title} {\bibinfo {title} {Measurement-induced entanglement
  transitions in many-body localized systems},\ }\href
  {https://doi.org/10.1103/PhysRevResearch.2.043072} {\bibfield  {journal}
  {\bibinfo  {journal} {Phys. Rev. Research}\ }\textbf {\bibinfo {volume}
  {2}},\ \bibinfo {pages} {043072} (\bibinfo {year} {2020})}\BibitemShut
  {NoStop}%
\bibitem [{\citenamefont {P\"opperl}\ \emph {et~al.}(2023)\citenamefont
  {P\"opperl}, \citenamefont {Gornyi},\ and\ \citenamefont {Gefen}}]{Paul2023}%
  \BibitemOpen
  \bibfield  {author} {\bibinfo {author} {\bibfnamefont {P.}~\bibnamefont
  {P\"opperl}}, \bibinfo {author} {\bibfnamefont {I.~V.}\ \bibnamefont
  {Gornyi}},\ and\ \bibinfo {author} {\bibfnamefont {Y.}~\bibnamefont
  {Gefen}},\ }\bibfield  {title} {\bibinfo {title} {Measurements on an
  {A}nderson chain},\ }\href {https://doi.org/10.1103/PhysRevB.107.174203}
  {\bibfield  {journal} {\bibinfo  {journal} {Phys. Rev. B}\ }\textbf {\bibinfo
  {volume} {107}},\ \bibinfo {pages} {174203} (\bibinfo {year}
  {2023})}\BibitemShut {NoStop}%
\bibitem [{\citenamefont {Yamamoto}\ and\ \citenamefont
  {Hamazaki}(2023)}]{Yamamoto2023}%
  \BibitemOpen
  \bibfield  {author} {\bibinfo {author} {\bibfnamefont {K.}~\bibnamefont
  {Yamamoto}}\ and\ \bibinfo {author} {\bibfnamefont {R.}~\bibnamefont
  {Hamazaki}},\ }\bibfield  {title} {\bibinfo {title} {Localization properties
  in disordered quantum many-body dynamics under continuous measurement},\
  }\href {https://doi.org/10.1103/PhysRevB.107.L220201} {\bibfield  {journal}
  {\bibinfo  {journal} {Phys. Rev. B}\ }\textbf {\bibinfo {volume} {107}},\
  \bibinfo {pages} {L220201} (\bibinfo {year} {2023})}\BibitemShut {NoStop}%
\bibitem [{\citenamefont {Noel}\ \emph {et~al.}(2022)\citenamefont {Noel},
  \citenamefont {Niroula}, \citenamefont {Zhu}, \citenamefont {Risinger},
  \citenamefont {Egan}, \citenamefont {Biswas}, \citenamefont {Cetina},
  \citenamefont {Gorshkov}, \citenamefont {Gullans}, \citenamefont {Huse},\
  and\ \citenamefont {Monroe}}]{Noel2022a}%
  \BibitemOpen
  \bibfield  {author} {\bibinfo {author} {\bibfnamefont {C.}~\bibnamefont
  {Noel}}, \bibinfo {author} {\bibfnamefont {P.}~\bibnamefont {Niroula}},
  \bibinfo {author} {\bibfnamefont {D.}~\bibnamefont {Zhu}}, \bibinfo {author}
  {\bibfnamefont {A.}~\bibnamefont {Risinger}}, \bibinfo {author}
  {\bibfnamefont {L.}~\bibnamefont {Egan}}, \bibinfo {author} {\bibfnamefont
  {D.}~\bibnamefont {Biswas}}, \bibinfo {author} {\bibfnamefont
  {M.}~\bibnamefont {Cetina}}, \bibinfo {author} {\bibfnamefont {A.~V.}\
  \bibnamefont {Gorshkov}}, \bibinfo {author} {\bibfnamefont {M.~J.}\
  \bibnamefont {Gullans}}, \bibinfo {author} {\bibfnamefont {D.~A.}\
  \bibnamefont {Huse}},\ and\ \bibinfo {author} {\bibfnamefont
  {C.}~\bibnamefont {Monroe}},\ }\bibfield  {title} {\bibinfo {title}
  {Measurement-induced quantum phases realized in a trapped-ion quantum
  computer},\ }\href {https://doi.org/10.1038/s41567-022-01619-7} {\bibfield
  {journal} {\bibinfo  {journal} {Nat. Phys.}\ }\textbf {\bibinfo {volume}
  {18}},\ \bibinfo {pages} {760} (\bibinfo {year} {2022})}\BibitemShut
  {NoStop}%
\bibitem [{\citenamefont {Koh}\ \emph {et~al.}(2023)\citenamefont {Koh},
  \citenamefont {Sun}, \citenamefont {Motta},\ and\ \citenamefont
  {Minnich}}]{Koh2022}%
  \BibitemOpen
  \bibfield  {author} {\bibinfo {author} {\bibfnamefont {J.~M.}\ \bibnamefont
  {Koh}}, \bibinfo {author} {\bibfnamefont {S.-N.}\ \bibnamefont {Sun}},
  \bibinfo {author} {\bibfnamefont {M.}~\bibnamefont {Motta}},\ and\ \bibinfo
  {author} {\bibfnamefont {A.~J.}\ \bibnamefont {Minnich}},\ }\bibfield
  {title} {\bibinfo {title} {Measurement-induced entanglement phase transition
  on a superconducting quantum processor with mid-circuit readout},\ }\bibfield
   {journal} {\bibinfo  {journal} {Nat. Phys.}\ }\href
  {https://doi.org/10.1038/s41567-023-02076-6} {10.1038/s41567-023-02076-6}
  (\bibinfo {year} {2023})\BibitemShut {NoStop}%
\bibitem [{\citenamefont {Hoke~\textit{et al.,} {Google Quantum AI and
  Collaborators}}(2023)}]{Hoke2023}%
  \BibitemOpen
  \bibfield  {author} {\bibinfo {author} {\bibfnamefont {J.~C.}\ \bibnamefont
  {Hoke~\textit{et al.,} {Google Quantum AI and Collaborators}}},\ }\bibfield
  {title} {\bibinfo {title} {Measurement-induced entanglement and teleportation
  on a noisy quantum processor},\ }\href
  {https://doi.org/10.1038/s41586-023-06505-7} {\bibfield  {journal} {\bibinfo
  {journal} {Nature}\ }\textbf {\bibinfo {volume} {622}},\ \bibinfo {pages}
  {481} (\bibinfo {year} {2023})}\BibitemShut {NoStop}%
\bibitem [{\citenamefont {Snizhko}\ \emph {et~al.}(2020)\citenamefont
  {Snizhko}, \citenamefont {Kumar},\ and\ \citenamefont
  {Romito}}]{snizhko_2020}%
  \BibitemOpen
  \bibfield  {author} {\bibinfo {author} {\bibfnamefont {K.}~\bibnamefont
  {Snizhko}}, \bibinfo {author} {\bibfnamefont {P.}~\bibnamefont {Kumar}},\
  and\ \bibinfo {author} {\bibfnamefont {A.}~\bibnamefont {Romito}},\
  }\bibfield  {title} {\bibinfo {title} {Quantum {Z}eno effect appears in
  stages},\ }\href {https://doi.org/10.1103/PhysRevResearch.2.033512}
  {\bibfield  {journal} {\bibinfo  {journal} {Physical Review Research}\
  }\textbf {\bibinfo {volume} {2}},\ \bibinfo {pages} {033512} (\bibinfo {year}
  {2020})}\BibitemShut {NoStop}%
\bibitem [{Note1()}]{Note1}%
  \BibitemOpen
  \bibinfo {note} {In parallel to this work, a monograph containing a chapter
  about the project was written and submitted as a doctoral thesis by
  P.P.~\cite {thesis_paul}}\BibitemShut {NoStop}%
\bibitem [{\citenamefont {Loss}\ and\ \citenamefont
  {DiVincenzo}(1998)}]{Loss1998}%
  \BibitemOpen
  \bibfield  {author} {\bibinfo {author} {\bibfnamefont {D.}~\bibnamefont
  {Loss}}\ and\ \bibinfo {author} {\bibfnamefont {D.~P.}\ \bibnamefont
  {DiVincenzo}},\ }\bibfield  {title} {\bibinfo {title} {Quantum computation
  with quantum dots},\ }\href {https://doi.org/10.1103/PhysRevA.57.120}
  {\bibfield  {journal} {\bibinfo  {journal} {Phys. Rev. A}\ }\textbf {\bibinfo
  {volume} {57}},\ \bibinfo {pages} {120} (\bibinfo {year} {1998})}\BibitemShut
  {NoStop}%
\bibitem [{\citenamefont {Murch}\ \emph {et~al.}(2013)\citenamefont {Murch},
  \citenamefont {Weber}, \citenamefont {Macklin},\ and\ \citenamefont
  {Siddiqi}}]{qubit_quantum_traj_exp}%
  \BibitemOpen
  \bibfield  {author} {\bibinfo {author} {\bibfnamefont {K.~W.}\ \bibnamefont
  {Murch}}, \bibinfo {author} {\bibfnamefont {S.~J.}\ \bibnamefont {Weber}},
  \bibinfo {author} {\bibfnamefont {C.}~\bibnamefont {Macklin}},\ and\ \bibinfo
  {author} {\bibfnamefont {I.}~\bibnamefont {Siddiqi}},\ }\bibfield  {title}
  {\bibinfo {title} {Observing single quantum trajectories of a superconducting
  quantum bit},\ }\href {https://doi.org/10.1038/nature12539} {\bibfield
  {journal} {\bibinfo  {journal} {Nature}\ }\textbf {\bibinfo {volume} {502}},\
  \bibinfo {pages} {211} (\bibinfo {year} {2013})}\BibitemShut {NoStop}%
\bibitem [{\citenamefont {Minev}\ \emph {et~al.}(2019)\citenamefont {Minev},
  \citenamefont {Mundhada}, \citenamefont {Shankar}, \citenamefont {Reinhold},
  \citenamefont {Guti{\'{e}}rrez-J{\'{a}}uregui}, \citenamefont {Schoelkopf},
  \citenamefont {Mirrahimi}, \citenamefont {Carmichael},\ and\ \citenamefont
  {Devoret}}]{quantum_jumps}%
  \BibitemOpen
  \bibfield  {author} {\bibinfo {author} {\bibfnamefont {Z.~K.}\ \bibnamefont
  {Minev}}, \bibinfo {author} {\bibfnamefont {S.~O.}\ \bibnamefont {Mundhada}},
  \bibinfo {author} {\bibfnamefont {S.}~\bibnamefont {Shankar}}, \bibinfo
  {author} {\bibfnamefont {P.}~\bibnamefont {Reinhold}}, \bibinfo {author}
  {\bibfnamefont {R.}~\bibnamefont {Guti{\'{e}}rrez-J{\'{a}}uregui}}, \bibinfo
  {author} {\bibfnamefont {R.~J.}\ \bibnamefont {Schoelkopf}}, \bibinfo
  {author} {\bibfnamefont {M.}~\bibnamefont {Mirrahimi}}, \bibinfo {author}
  {\bibfnamefont {H.~J.}\ \bibnamefont {Carmichael}},\ and\ \bibinfo {author}
  {\bibfnamefont {M.~H.}\ \bibnamefont {Devoret}},\ }\bibfield  {title}
  {\bibinfo {title} {To catch and reverse a quantum jump mid-flight},\ }\href
  {https://doi.org/10.1038/s41586-019-1287-z} {\bibfield  {journal} {\bibinfo
  {journal} {Nature}\ }\textbf {\bibinfo {volume} {570}},\ \bibinfo {pages}
  {200} (\bibinfo {year} {2019})}\BibitemShut {NoStop}%
\bibitem [{\citenamefont {Koolstra}\ \emph {et~al.}(2022)\citenamefont
  {Koolstra}, \citenamefont {Stevenson}, \citenamefont {Barzili}, \citenamefont
  {Burns}, \citenamefont {Siva}, \citenamefont {Greenfield}, \citenamefont
  {Livingston}, \citenamefont {Hashim}, \citenamefont {Naik}, \citenamefont
  {Kreikebaum}, \citenamefont {O'Brien}, \citenamefont {Santiago},
  \citenamefont {Dressel},\ and\ \citenamefont
  {Siddiqi}}]{quantum_trajectory_tracking_nn}%
  \BibitemOpen
  \bibfield  {author} {\bibinfo {author} {\bibfnamefont {G.}~\bibnamefont
  {Koolstra}}, \bibinfo {author} {\bibfnamefont {N.}~\bibnamefont {Stevenson}},
  \bibinfo {author} {\bibfnamefont {S.}~\bibnamefont {Barzili}}, \bibinfo
  {author} {\bibfnamefont {L.}~\bibnamefont {Burns}}, \bibinfo {author}
  {\bibfnamefont {K.}~\bibnamefont {Siva}}, \bibinfo {author} {\bibfnamefont
  {S.}~\bibnamefont {Greenfield}}, \bibinfo {author} {\bibfnamefont
  {W.}~\bibnamefont {Livingston}}, \bibinfo {author} {\bibfnamefont
  {A.}~\bibnamefont {Hashim}}, \bibinfo {author} {\bibfnamefont {R.~K.}\
  \bibnamefont {Naik}}, \bibinfo {author} {\bibfnamefont {J.~M.}\ \bibnamefont
  {Kreikebaum}}, \bibinfo {author} {\bibfnamefont {K.~P.}\ \bibnamefont
  {O'Brien}}, \bibinfo {author} {\bibfnamefont {D.~I.}\ \bibnamefont
  {Santiago}}, \bibinfo {author} {\bibfnamefont {J.}~\bibnamefont {Dressel}},\
  and\ \bibinfo {author} {\bibfnamefont {I.}~\bibnamefont {Siddiqi}},\
  }\bibfield  {title} {\bibinfo {title} {Monitoring fast superconducting qubit
  dynamics using a neural network},\ }\href
  {https://doi.org/10.1103/PhysRevX.12.031017} {\bibfield  {journal} {\bibinfo
  {journal} {Phys. Rev. X}\ }\textbf {\bibinfo {volume} {12}},\ \bibinfo
  {pages} {031017} (\bibinfo {year} {2022})}\BibitemShut {NoStop}%
\bibitem [{\citenamefont {Wang}\ \emph {et~al.}(2022)\citenamefont {Wang},
  \citenamefont {Snizhko}, \citenamefont {Romito}, \citenamefont {Gefen},\ and\
  \citenamefont {Murch}}]{topological_transition_weak_measurements_experiment}%
  \BibitemOpen
  \bibfield  {author} {\bibinfo {author} {\bibfnamefont {Y.}~\bibnamefont
  {Wang}}, \bibinfo {author} {\bibfnamefont {K.}~\bibnamefont {Snizhko}},
  \bibinfo {author} {\bibfnamefont {A.}~\bibnamefont {Romito}}, \bibinfo
  {author} {\bibfnamefont {Y.}~\bibnamefont {Gefen}},\ and\ \bibinfo {author}
  {\bibfnamefont {K.}~\bibnamefont {Murch}},\ }\bibfield  {title} {\bibinfo
  {title} {Observing a topological transition in weak-measurement-induced
  geometric phases},\ }\href {https://doi.org/10.1103/PhysRevResearch.4.023179}
  {\bibfield  {journal} {\bibinfo  {journal} {Phys. Rev. Res.}\ }\textbf
  {\bibinfo {volume} {4}},\ \bibinfo {pages} {023179} (\bibinfo {year}
  {2022})}\BibitemShut {NoStop}%
\bibitem [{\citenamefont {Ferrer-Garcia}\ \emph {et~al.}(2023)\citenamefont
  {Ferrer-Garcia}, \citenamefont {Snizhko}, \citenamefont {D’Errico},
  \citenamefont {Romito}, \citenamefont {Gefen},\ and\ \citenamefont
  {Karimi}}]{topological_transitions_optical_experiment}%
  \BibitemOpen
  \bibfield  {author} {\bibinfo {author} {\bibfnamefont {M.~F.}\ \bibnamefont
  {Ferrer-Garcia}}, \bibinfo {author} {\bibfnamefont {K.}~\bibnamefont
  {Snizhko}}, \bibinfo {author} {\bibfnamefont {A.}~\bibnamefont {D’Errico}},
  \bibinfo {author} {\bibfnamefont {A.}~\bibnamefont {Romito}}, \bibinfo
  {author} {\bibfnamefont {Y.}~\bibnamefont {Gefen}},\ and\ \bibinfo {author}
  {\bibfnamefont {E.}~\bibnamefont {Karimi}},\ }\bibfield  {title} {\bibinfo
  {title} {Topological transitions of the generalized pancharatnam-berry
  phase},\ }\href {https://doi.org/10.1126/sciadv.adg6810} {\bibfield
  {journal} {\bibinfo  {journal} {Science Advances}\ }\textbf {\bibinfo
  {volume} {9}},\ \bibinfo {pages} {eadg6810} (\bibinfo {year} {2023})},\
  \Eprint
  {https://arxiv.org/abs/https://www.science.org/doi/pdf/10.1126/sciadv.adg6810}
  {https://www.science.org/doi/pdf/10.1126/sciadv.adg6810} \BibitemShut
  {NoStop}%
\bibitem [{\citenamefont {Gurvitz}(1997)}]{Gurvitz1997}%
  \BibitemOpen
  \bibfield  {author} {\bibinfo {author} {\bibfnamefont {S.~A.}\ \bibnamefont
  {Gurvitz}},\ }\bibfield  {title} {\bibinfo {title} {Measurements with a
  noninvasive detector and dephasing mechanism},\ }\href
  {https://doi.org/10.1103/PhysRevB.56.15215} {\bibfield  {journal} {\bibinfo
  {journal} {Phys. Rev. B}\ }\textbf {\bibinfo {volume} {56}},\ \bibinfo
  {pages} {15215} (\bibinfo {year} {1997})}\BibitemShut {NoStop}%
\bibitem [{\citenamefont {Korotkov}(1999)}]{Korotkov1999}%
  \BibitemOpen
  \bibfield  {author} {\bibinfo {author} {\bibfnamefont {A.~N.}\ \bibnamefont
  {Korotkov}},\ }\bibfield  {title} {\bibinfo {title} {Continuous quantum
  measurement of a double dot},\ }\href
  {https://doi.org/10.1103/PhysRevB.60.5737} {\bibfield  {journal} {\bibinfo
  {journal} {Phys. Rev. B}\ }\textbf {\bibinfo {volume} {60}},\ \bibinfo
  {pages} {5737} (\bibinfo {year} {1999})}\BibitemShut {NoStop}%
\bibitem [{\citenamefont {Korotkov}(2001)}]{Korotkov2001}%
  \BibitemOpen
  \bibfield  {author} {\bibinfo {author} {\bibfnamefont {A.~N.}\ \bibnamefont
  {Korotkov}},\ }\bibfield  {title} {\bibinfo {title} {Selective quantum
  evolution of a qubit state due to continuous measurement},\ }\href
  {https://doi.org/10.1103/PhysRevB.63.115403} {\bibfield  {journal} {\bibinfo
  {journal} {Phys. Rev. B}\ }\textbf {\bibinfo {volume} {63}},\ \bibinfo
  {pages} {115403} (\bibinfo {year} {2001})}\BibitemShut {NoStop}%
\bibitem [{\citenamefont {Goan}\ \emph {et~al.}(2001)\citenamefont {Goan},
  \citenamefont {Milburn}, \citenamefont {Wiseman},\ and\ \citenamefont
  {Sun}}]{Goan2001}%
  \BibitemOpen
  \bibfield  {author} {\bibinfo {author} {\bibfnamefont {H.-S.}\ \bibnamefont
  {Goan}}, \bibinfo {author} {\bibfnamefont {G.~J.}\ \bibnamefont {Milburn}},
  \bibinfo {author} {\bibfnamefont {H.~M.}\ \bibnamefont {Wiseman}},\ and\
  \bibinfo {author} {\bibfnamefont {H.~B.}\ \bibnamefont {Sun}},\ }\bibfield
  {title} {\bibinfo {title} {Continuous quantum measurement of two coupled
  quantum dots using a point contact: A quantum trajectory approach},\ }\href
  {https://doi.org/10.1103/PhysRevB.63.125326} {\bibfield  {journal} {\bibinfo
  {journal} {Phys. Rev. B}\ }\textbf {\bibinfo {volume} {63}},\ \bibinfo
  {pages} {125326} (\bibinfo {year} {2001})}\BibitemShut {NoStop}%
\bibitem [{\citenamefont {Shpitalnik}\ \emph {et~al.}(2008)\citenamefont
  {Shpitalnik}, \citenamefont {Gefen},\ and\ \citenamefont
  {Romito}}]{Shpitalnik2008}%
  \BibitemOpen
  \bibfield  {author} {\bibinfo {author} {\bibfnamefont {V.}~\bibnamefont
  {Shpitalnik}}, \bibinfo {author} {\bibfnamefont {Y.}~\bibnamefont {Gefen}},\
  and\ \bibinfo {author} {\bibfnamefont {A.}~\bibnamefont {Romito}},\
  }\bibfield  {title} {\bibinfo {title} {Tomography of many-body weak values:
  {M}ach-{Z}ehnder interferometry},\ }\href
  {https://doi.org/10.1103/PhysRevLett.101.226802} {\bibfield  {journal}
  {\bibinfo  {journal} {Phys. Rev. Lett.}\ }\textbf {\bibinfo {volume} {101}},\
  \bibinfo {pages} {226802} (\bibinfo {year} {2008})}\BibitemShut {NoStop}%
\bibitem [{\citenamefont {Romito}\ \emph {et~al.}(2008)\citenamefont {Romito},
  \citenamefont {Gefen},\ and\ \citenamefont {Blanter}}]{Romito2008}%
  \BibitemOpen
  \bibfield  {author} {\bibinfo {author} {\bibfnamefont {A.}~\bibnamefont
  {Romito}}, \bibinfo {author} {\bibfnamefont {Y.}~\bibnamefont {Gefen}},\ and\
  \bibinfo {author} {\bibfnamefont {Y.~M.}\ \bibnamefont {Blanter}},\
  }\bibfield  {title} {\bibinfo {title} {Weak values of electron spin in a
  double quantum dot},\ }\href {https://doi.org/10.1103/PhysRevLett.100.056801}
  {\bibfield  {journal} {\bibinfo  {journal} {Phys. Rev. Lett.}\ }\textbf
  {\bibinfo {volume} {100}},\ \bibinfo {pages} {056801} (\bibinfo {year}
  {2008})}\BibitemShut {NoStop}%
\bibitem [{\citenamefont {Kwapi\ifmmode~\acute{n}\else \'{n}\fi{}ski}\ and\
  \citenamefont {Taranko}(2012)}]{Taranko2012}%
  \BibitemOpen
  \bibfield  {author} {\bibinfo {author} {\bibfnamefont {T.}~\bibnamefont
  {Kwapi\ifmmode~\acute{n}\else \'{n}\fi{}ski}}\ and\ \bibinfo {author}
  {\bibfnamefont {R.}~\bibnamefont {Taranko}},\ }\bibfield  {title} {\bibinfo
  {title} {Quantum wire as a charge-qubit detector},\ }\href
  {https://doi.org/10.1103/PhysRevA.86.052338} {\bibfield  {journal} {\bibinfo
  {journal} {Phys. Rev. A}\ }\textbf {\bibinfo {volume} {86}},\ \bibinfo
  {pages} {052338} (\bibinfo {year} {2012})}\BibitemShut {NoStop}%
\bibitem [{\citenamefont {Li}\ \emph {et~al.}(2014)\citenamefont {Li},
  \citenamefont {Ren},\ and\ \citenamefont {Sinitsyn}}]{li_2014}%
  \BibitemOpen
  \bibfield  {author} {\bibinfo {author} {\bibfnamefont {F.}~\bibnamefont
  {Li}}, \bibinfo {author} {\bibfnamefont {J.}~\bibnamefont {Ren}},\ and\
  \bibinfo {author} {\bibfnamefont {N.~A.}\ \bibnamefont {Sinitsyn}},\
  }\bibfield  {title} {\bibinfo {title} {Quantum {Z}eno effect as a topological
  phase transition in full counting statistics and spin noise spectroscopy},\
  }\href {https://doi.org/10.1209/0295-5075/105/27001} {\bibfield  {journal}
  {\bibinfo  {journal} {Europhysics Letters}\ }\textbf {\bibinfo {volume}
  {105}},\ \bibinfo {pages} {27001} (\bibinfo {year} {2014})}\BibitemShut
  {NoStop}%
\bibitem [{\citenamefont {Barbarino}\ \emph {et~al.}(2019)\citenamefont
  {Barbarino}, \citenamefont {Fazio}, \citenamefont {Vedral},\ and\
  \citenamefont {Gefen}}]{Barbarino2019}%
  \BibitemOpen
  \bibfield  {author} {\bibinfo {author} {\bibfnamefont {S.}~\bibnamefont
  {Barbarino}}, \bibinfo {author} {\bibfnamefont {R.}~\bibnamefont {Fazio}},
  \bibinfo {author} {\bibfnamefont {V.}~\bibnamefont {Vedral}},\ and\ \bibinfo
  {author} {\bibfnamefont {Y.}~\bibnamefont {Gefen}},\ }\bibfield  {title}
  {\bibinfo {title} {Engineering statistical transmutation of identical quantum
  particles},\ }\href {https://doi.org/10.1103/PhysRevB.99.045430} {\bibfield
  {journal} {\bibinfo  {journal} {Phys. Rev. B}\ }\textbf {\bibinfo {volume}
  {99}},\ \bibinfo {pages} {045430} (\bibinfo {year} {2019})}\BibitemShut
  {NoStop}%
\bibitem [{\citenamefont {Esin}\ \emph {et~al.}(2020)\citenamefont {Esin},
  \citenamefont {Romito},\ and\ \citenamefont {Gefen}}]{Esin2020}%
  \BibitemOpen
  \bibfield  {author} {\bibinfo {author} {\bibfnamefont {I.}~\bibnamefont
  {Esin}}, \bibinfo {author} {\bibfnamefont {A.}~\bibnamefont {Romito}},\ and\
  \bibinfo {author} {\bibfnamefont {Y.}~\bibnamefont {Gefen}},\ }\bibfield
  {title} {\bibinfo {title} {Detection of quantum interference without an
  interference pattern},\ }\href
  {https://doi.org/10.1103/PhysRevLett.125.020405} {\bibfield  {journal}
  {\bibinfo  {journal} {Phys. Rev. Lett.}\ }\textbf {\bibinfo {volume} {125}},\
  \bibinfo {pages} {020405} (\bibinfo {year} {2020})}\BibitemShut {NoStop}%
\bibitem [{\citenamefont {Kumar}\ \emph {et~al.}(2020)\citenamefont {Kumar},
  \citenamefont {Romito},\ and\ \citenamefont {Snizhko}}]{Kumar2020}%
  \BibitemOpen
  \bibfield  {author} {\bibinfo {author} {\bibfnamefont {P.}~\bibnamefont
  {Kumar}}, \bibinfo {author} {\bibfnamefont {A.}~\bibnamefont {Romito}},\ and\
  \bibinfo {author} {\bibfnamefont {K.}~\bibnamefont {Snizhko}},\ }\bibfield
  {title} {\bibinfo {title} {Quantum {Z}eno effect with partial measurement and
  noisy dynamics},\ }\href {https://doi.org/10.1103/PhysRevResearch.2.043420}
  {\bibfield  {journal} {\bibinfo  {journal} {Phys. Rev. Res.}\ }\textbf
  {\bibinfo {volume} {2}},\ \bibinfo {pages} {043420} (\bibinfo {year}
  {2020})}\BibitemShut {NoStop}%
\bibitem [{\citenamefont {Roy}\ \emph {et~al.}(2020)\citenamefont {Roy},
  \citenamefont {Chalker}, \citenamefont {Gornyi},\ and\ \citenamefont
  {Gefen}}]{Roy2020}%
  \BibitemOpen
  \bibfield  {author} {\bibinfo {author} {\bibfnamefont {S.}~\bibnamefont
  {Roy}}, \bibinfo {author} {\bibfnamefont {J.~T.}\ \bibnamefont {Chalker}},
  \bibinfo {author} {\bibfnamefont {I.~V.}\ \bibnamefont {Gornyi}},\ and\
  \bibinfo {author} {\bibfnamefont {Y.}~\bibnamefont {Gefen}},\ }\bibfield
  {title} {\bibinfo {title} {Measurement-induced steering of quantum systems},\
  }\href {https://doi.org/10.1103/PhysRevResearch.2.033347} {\bibfield
  {journal} {\bibinfo  {journal} {Phys. Rev. Research}\ }\textbf {\bibinfo
  {volume} {2}},\ \bibinfo {pages} {033347} (\bibinfo {year}
  {2020})}\BibitemShut {NoStop}%
\bibitem [{\citenamefont {Herasymenko}\ \emph {et~al.}(2023)\citenamefont
  {Herasymenko}, \citenamefont {Gornyi},\ and\ \citenamefont
  {Gefen}}]{Herasymenko2021}%
  \BibitemOpen
  \bibfield  {author} {\bibinfo {author} {\bibfnamefont {Y.}~\bibnamefont
  {Herasymenko}}, \bibinfo {author} {\bibfnamefont {I.}~\bibnamefont
  {Gornyi}},\ and\ \bibinfo {author} {\bibfnamefont {Y.}~\bibnamefont
  {Gefen}},\ }\bibfield  {title} {\bibinfo {title} {Measurement-driven
  navigation in many-body {H}ilbert space: Active-decision steering},\ }\href
  {https://doi.org/10.1103/PRXQuantum.4.020347} {\bibfield  {journal} {\bibinfo
   {journal} {PRX Quantum}\ }\textbf {\bibinfo {volume} {4}},\ \bibinfo {pages}
  {020347} (\bibinfo {year} {2023})}\BibitemShut {NoStop}%
\bibitem [{\citenamefont {Dubey}\ \emph {et~al.}(2023)\citenamefont {Dubey},
  \citenamefont {Chetrite},\ and\ \citenamefont {Dhar}}]{dubey_2023}%
  \BibitemOpen
  \bibfield  {author} {\bibinfo {author} {\bibfnamefont {V.}~\bibnamefont
  {Dubey}}, \bibinfo {author} {\bibfnamefont {R.}~\bibnamefont {Chetrite}},\
  and\ \bibinfo {author} {\bibfnamefont {A.}~\bibnamefont {Dhar}},\ }\bibfield
  {title} {\bibinfo {title} {Quantum resetting in continuous measurement
  induced dynamics of a qubit},\ }\href
  {https://doi.org/10.1088/1751-8121/acc290} {\bibfield  {journal} {\bibinfo
  {journal} {Journal of Physics A: Mathematical and Theoretical}\ }\textbf
  {\bibinfo {volume} {56}},\ \bibinfo {pages} {154001} (\bibinfo {year}
  {2023})}\BibitemShut {NoStop}%
\bibitem [{\citenamefont {Friedman}\ \emph {et~al.}(2023)\citenamefont
  {Friedman}, \citenamefont {Hart},\ and\ \citenamefont
  {Nandkishore}}]{Friedman2022}%
  \BibitemOpen
  \bibfield  {author} {\bibinfo {author} {\bibfnamefont {A.~J.}\ \bibnamefont
  {Friedman}}, \bibinfo {author} {\bibfnamefont {O.}~\bibnamefont {Hart}},\
  and\ \bibinfo {author} {\bibfnamefont {R.}~\bibnamefont {Nandkishore}},\
  }\bibfield  {title} {\bibinfo {title} {Measurement-induced phases of matter
  require feedback},\ }\href {https://doi.org/10.1103/PRXQuantum.4.040309}
  {\bibfield  {journal} {\bibinfo  {journal} {PRX Quantum}\ }\textbf {\bibinfo
  {volume} {4}},\ \bibinfo {pages} {040309} (\bibinfo {year}
  {2023})}\BibitemShut {NoStop}%
\bibitem [{\citenamefont {Kumar}\ \emph {et~al.}(2022)\citenamefont {Kumar},
  \citenamefont {Snizhko}, \citenamefont {Gefen},\ and\ \citenamefont
  {Rosenow}}]{Rosenow2022}%
  \BibitemOpen
  \bibfield  {author} {\bibinfo {author} {\bibfnamefont {P.}~\bibnamefont
  {Kumar}}, \bibinfo {author} {\bibfnamefont {K.}~\bibnamefont {Snizhko}},
  \bibinfo {author} {\bibfnamefont {Y.}~\bibnamefont {Gefen}},\ and\ \bibinfo
  {author} {\bibfnamefont {B.}~\bibnamefont {Rosenow}},\ }\bibfield  {title}
  {\bibinfo {title} {Optimized steering: Quantum state engineering and
  exceptional points},\ }\href {https://doi.org/10.1103/PhysRevA.105.L010203}
  {\bibfield  {journal} {\bibinfo  {journal} {Phys. Rev. A}\ }\textbf {\bibinfo
  {volume} {105}},\ \bibinfo {pages} {L010203} (\bibinfo {year}
  {2022})}\BibitemShut {NoStop}%
\bibitem [{\citenamefont {Martín-Vázquez}\ \emph {et~al.}(2023)\citenamefont
  {Martín-Vázquez}, \citenamefont {Tolppanen},\ and\ \citenamefont
  {Silveri}}]{Silveri2023}%
  \BibitemOpen
  \bibfield  {author} {\bibinfo {author} {\bibfnamefont {G.}~\bibnamefont
  {Martín-Vázquez}}, \bibinfo {author} {\bibfnamefont {T.}~\bibnamefont
  {Tolppanen}},\ and\ \bibinfo {author} {\bibfnamefont {M.}~\bibnamefont
  {Silveri}},\ }\href {https://doi.org/10.48550/arxiv.2302.02934} {\bibinfo
  {title} {Phase transitions induced by standard and predetermined measurements
  in transmon arrays}} (\bibinfo {year} {2023}),\ \Eprint
  {https://arxiv.org/abs/2302.02934} {arXiv:2302.02934 [quant-ph]} \BibitemShut
  {NoStop}%
\bibitem [{\citenamefont {Medina-Guerra}\ \emph {et~al.}(2023)\citenamefont
  {Medina-Guerra}, \citenamefont {Kumar}, \citenamefont {Gornyi},\ and\
  \citenamefont {Gefen}}]{MedinaGuerra2023}%
  \BibitemOpen
  \bibfield  {author} {\bibinfo {author} {\bibfnamefont {E.}~\bibnamefont
  {Medina-Guerra}}, \bibinfo {author} {\bibfnamefont {P.}~\bibnamefont
  {Kumar}}, \bibinfo {author} {\bibfnamefont {I.~V.}\ \bibnamefont {Gornyi}},\
  and\ \bibinfo {author} {\bibfnamefont {Y.}~\bibnamefont {Gefen}},\
  }\href@noop {} {\bibinfo {title} {Quantum state engineering by steering in
  the presence of errors}} (\bibinfo {year} {2023}),\ \Eprint
  {https://arxiv.org/abs/2303.16329} {arXiv:2303.16329 [quant-ph]} \BibitemShut
  {NoStop}%
\bibitem [{\citenamefont {Morales}\ \emph {et~al.}(2024)\citenamefont
  {Morales}, \citenamefont {Gefen}, \citenamefont {Gornyi}, \citenamefont
  {Zazunov},\ and\ \citenamefont {Egger}}]{Morales2023}%
  \BibitemOpen
  \bibfield  {author} {\bibinfo {author} {\bibfnamefont {S.}~\bibnamefont
  {Morales}}, \bibinfo {author} {\bibfnamefont {Y.}~\bibnamefont {Gefen}},
  \bibinfo {author} {\bibfnamefont {I.}~\bibnamefont {Gornyi}}, \bibinfo
  {author} {\bibfnamefont {A.}~\bibnamefont {Zazunov}},\ and\ \bibinfo {author}
  {\bibfnamefont {R.}~\bibnamefont {Egger}},\ }\bibfield  {title} {\bibinfo
  {title} {Engineering unsteerable quantum states with active feedback},\
  }\href {https://doi.org/10.1103/PhysRevResearch.6.013244} {\bibfield
  {journal} {\bibinfo  {journal} {Phys. Rev. Res.}\ }\textbf {\bibinfo {volume}
  {6}},\ \bibinfo {pages} {013244} (\bibinfo {year} {2024})}\BibitemShut
  {NoStop}%
\bibitem [{\citenamefont {Lichtenberg}\ and\ \citenamefont
  {Lieberman}(1992)}]{lichtenberg_1992}%
  \BibitemOpen
  \bibfield  {author} {\bibinfo {author} {\bibfnamefont {A.~J.}\ \bibnamefont
  {Lichtenberg}}\ and\ \bibinfo {author} {\bibfnamefont {M.~A.}\ \bibnamefont
  {Lieberman}},\ }\href {https://doi.org/10.1007/978-1-4757-2184-3} {\emph
  {\bibinfo {title} {Regular and {Chaotic} {Dynamics}}}},\ edited by\ \bibinfo
  {editor} {\bibfnamefont {F.}~\bibnamefont {John}}, \bibinfo {editor}
  {\bibfnamefont {J.~E.}\ \bibnamefont {Marsden}},\ and\ \bibinfo {editor}
  {\bibfnamefont {L.}~\bibnamefont {Sirovich}},\ \bibinfo {series} {Applied
  {Mathematical} {Sciences}}, Vol.~\bibinfo {volume} {38}\ (\bibinfo
  {publisher} {Springer},\ \bibinfo {address} {New York, NY},\ \bibinfo {year}
  {1992})\BibitemShut {NoStop}%
\bibitem [{Note2()}]{Note2}%
  \BibitemOpen
  \bibinfo {note} {In principle, there might be some fine-tuned cases where
  some endpoints for different branches correspond to the same
  state}\BibitemShut {NoStop}%
\bibitem [{\citenamefont {Furstenberg}(1963)}]{Furstenberg1963}%
  \BibitemOpen
  \bibfield  {author} {\bibinfo {author} {\bibfnamefont {H.}~\bibnamefont
  {Furstenberg}},\ }\bibfield  {title} {\bibinfo {title} {Noncommuting random
  products},\ }\href
  {https://doi.org/https://doi.org/10.1090/S0002-9947-1963-0163345-0}
  {\bibfield  {journal} {\bibinfo  {journal} {Trans. Amer. Math. Soc.}\
  }\textbf {\bibinfo {volume} {108}},\ \bibinfo {pages} {377} (\bibinfo {year}
  {1963})}\BibitemShut {NoStop}%
\bibitem [{\citenamefont {Berezinskii}\ and\ \citenamefont
  {Gor'kov}(1979)}]{Berezinskii1979}%
  \BibitemOpen
  \bibfield  {author} {\bibinfo {author} {\bibfnamefont {V.}~\bibnamefont
  {Berezinskii}}\ and\ \bibinfo {author} {\bibfnamefont {L.}~\bibnamefont
  {Gor'kov}},\ }\bibfield  {title} {\bibinfo {title} {On the theory of
  electrons localized in the field of defects},\ }\href
  {http://www.jetp.ras.ru/cgi-bin/e/index/e/50/6/p1209?a=list} {\bibfield
  {journal} {\bibinfo  {journal} {Sov. Phys. JETP}\ }\textbf {\bibinfo {volume}
  {50}},\ \bibinfo {pages} {1209} (\bibinfo {year} {1979})}\BibitemShut
  {NoStop}%
\bibitem [{\citenamefont {Perel'}\ and\ \citenamefont
  {Polyakov}(1984)}]{Perel1984}%
  \BibitemOpen
  \bibfield  {author} {\bibinfo {author} {\bibfnamefont {V.}~\bibnamefont
  {Perel'}}\ and\ \bibinfo {author} {\bibfnamefont {D.}~\bibnamefont
  {Polyakov}},\ }\bibfield  {title} {\bibinfo {title} {Probability distribution
  for the transmission of an electron through a chain of randomly placed
  centers},\ }\href {http://jetp.ras.ru/cgi-bin/e/index/e/59/1/p204?a=list}
  {\bibfield  {journal} {\bibinfo  {journal} {Sov. Phys. JETP}\ }\textbf
  {\bibinfo {volume} {59}},\ \bibinfo {pages} {204} (\bibinfo {year}
  {1984})}\BibitemShut {NoStop}%
\bibitem [{\citenamefont {Dmitriev}(1989)}]{Dmitriev1989}%
  \BibitemOpen
  \bibfield  {author} {\bibinfo {author} {\bibfnamefont {A.~P.}\ \bibnamefont
  {Dmitriev}},\ }\bibfield  {title} {\bibinfo {title} {Transmission of an
  electron through a finite chain of periodic disordered random scatterers},\
  }\href {http://jetp.ras.ru/cgi-bin/e/index/e/68/1/p132?a=list} {\bibfield
  {journal} {\bibinfo  {journal} {Sov. Phys. JETP}\ }\textbf {\bibinfo {volume}
  {68}},\ \bibinfo {pages} {132} (\bibinfo {year} {1989})}\BibitemShut
  {NoStop}%
\bibitem [{\citenamefont {Lifshits}\ \emph {et~al.}(1988)\citenamefont
  {Lifshits}, \citenamefont {Gredeskul},\ and\ \citenamefont
  {Pastur}}]{Pastur-Book}%
  \BibitemOpen
  \bibfield  {author} {\bibinfo {author} {\bibfnamefont {I.~M.}\ \bibnamefont
  {Lifshits}}, \bibinfo {author} {\bibfnamefont {S.~A.}\ \bibnamefont
  {Gredeskul}},\ and\ \bibinfo {author} {\bibfnamefont {L.~A.}\ \bibnamefont
  {Pastur}},\ }\href
  {https://www.worldcat.org/title/introduction-to-the-theory-of-disordered-systems/oclc/423264626}
  {\emph {\bibinfo {title} {Introduction to the theory of disordered
  systems}}}\ (\bibinfo  {publisher} {Wiley, New Yorkg},\ \bibinfo {year}
  {1988})\BibitemShut {NoStop}%
\bibitem [{\citenamefont {Comtet}\ \emph {et~al.}(2013)\citenamefont {Comtet},
  \citenamefont {Texier},\ and\ \citenamefont {Tourigny}}]{Comtet2013}%
  \BibitemOpen
  \bibfield  {author} {\bibinfo {author} {\bibfnamefont {A.}~\bibnamefont
  {Comtet}}, \bibinfo {author} {\bibfnamefont {C.}~\bibnamefont {Texier}},\
  and\ \bibinfo {author} {\bibfnamefont {Y.}~\bibnamefont {Tourigny}},\
  }\bibfield  {title} {\bibinfo {title} {Lyapunov exponents, one-dimensional
  {A}nderson localization and products of random matrices},\ }\href
  {https://doi.org/10.1088/1751-8113/46/25/254003} {\bibfield  {journal}
  {\bibinfo  {journal} {J. Phys. A: Math. Theor.}\ }\textbf {\bibinfo {volume}
  {46}},\ \bibinfo {pages} {254003} (\bibinfo {year} {2013})}\BibitemShut
  {NoStop}%
\bibitem [{\citenamefont {Hufton}\ \emph {et~al.}(2016)\citenamefont {Hufton},
  \citenamefont {Lin}, \citenamefont {Galla},\ and\ \citenamefont
  {McKane}}]{Hufton2016}%
  \BibitemOpen
  \bibfield  {author} {\bibinfo {author} {\bibfnamefont {P.~G.}\ \bibnamefont
  {Hufton}}, \bibinfo {author} {\bibfnamefont {Y.~T.}\ \bibnamefont {Lin}},
  \bibinfo {author} {\bibfnamefont {T.}~\bibnamefont {Galla}},\ and\ \bibinfo
  {author} {\bibfnamefont {A.~J.}\ \bibnamefont {McKane}},\ }\bibfield  {title}
  {\bibinfo {title} {Intrinsic noise in systems with switching environments},\
  }\href {https://doi.org/10.1103/PhysRevE.93.052119} {\bibfield  {journal}
  {\bibinfo  {journal} {Phys. Rev. E}\ }\textbf {\bibinfo {volume} {93}},\
  \bibinfo {pages} {052119} (\bibinfo {year} {2016})}\BibitemShut {NoStop}%
\bibitem [{\citenamefont {Saakian}(2017)}]{saakian_2017}%
  \BibitemOpen
  \bibfield  {author} {\bibinfo {author} {\bibfnamefont {D.~B.}\ \bibnamefont
  {Saakian}},\ }\bibfield  {title} {\bibinfo {title} {Exact solution of the
  hidden {Markov} processes},\ }\href
  {https://doi.org/10.1103/PhysRevE.96.052112} {\bibfield  {journal} {\bibinfo
  {journal} {Physical Review E}\ }\textbf {\bibinfo {volume} {96}},\ \bibinfo
  {pages} {052112} (\bibinfo {year} {2017})}\BibitemShut {NoStop}%
\bibitem [{\citenamefont {Saakian}(2018)}]{saakian_2018}%
  \BibitemOpen
  \bibfield  {author} {\bibinfo {author} {\bibfnamefont {D.~B.}\ \bibnamefont
  {Saakian}},\ }\bibfield  {title} {\bibinfo {title} {Semianalytical solution
  of the random-product problem of matrices and discrete-time random
  evolution},\ }\href {https://doi.org/10.1103/PhysRevE.98.062115} {\bibfield
  {journal} {\bibinfo  {journal} {Physical Review E}\ }\textbf {\bibinfo
  {volume} {98}},\ \bibinfo {pages} {062115} (\bibinfo {year}
  {2018})}\BibitemShut {NoStop}%
\bibitem [{\citenamefont {Poghosyan}\ and\ \citenamefont
  {Saakian}(2021)}]{infinite_series_of_singularities}%
  \BibitemOpen
  \bibfield  {author} {\bibinfo {author} {\bibfnamefont {R.}~\bibnamefont
  {Poghosyan}}\ and\ \bibinfo {author} {\bibfnamefont {D.~B.}\ \bibnamefont
  {Saakian}},\ }\bibfield  {title} {\bibinfo {title} {Infinite series of
  singularities in the correlated random matrices product},\ }\bibfield
  {journal} {\bibinfo  {journal} {Frontiers in Physics}\ }\textbf {\bibinfo
  {volume} {9}},\ \href {https://doi.org/10.3389/fphy.2021.678805}
  {10.3389/fphy.2021.678805} (\bibinfo {year} {2021})\BibitemShut {NoStop}%
\bibitem [{\citenamefont {Mineo}\ \emph {et~al.}(2023)\citenamefont {Mineo},
  \citenamefont {Suvorov},\ and\ \citenamefont
  {Saakian}}]{david_random_matrices_2}%
  \BibitemOpen
  \bibfield  {author} {\bibinfo {author} {\bibfnamefont {H.}~\bibnamefont
  {Mineo}}, \bibinfo {author} {\bibfnamefont {V.}~\bibnamefont {Suvorov}},\
  and\ \bibinfo {author} {\bibfnamefont {D.~B.}\ \bibnamefont {Saakian}},\
  }\bibfield  {title} {\bibinfo {title} {Investigation of the product of random
  matrices and related evolution models},\ }\href
  {https://www.mdpi.com/2227-7390/11/15/3430} {\bibfield  {journal} {\bibinfo
  {journal} {Mathematics}\ }\textbf {\bibinfo {volume} {11}} (\bibinfo {year}
  {2023})}\BibitemShut {NoStop}%
\bibitem [{\citenamefont {Beenakker}(1997)}]{BeenakkerRMP97}%
  \BibitemOpen
  \bibfield  {author} {\bibinfo {author} {\bibfnamefont {C.~W.~J.}\
  \bibnamefont {Beenakker}},\ }\bibfield  {title} {\bibinfo {title}
  {Random-matrix theory of quantum transport},\ }\href
  {https://doi.org/10.1103/RevModPhys.69.731} {\bibfield  {journal} {\bibinfo
  {journal} {Rev. Mod. Phys.}\ }\textbf {\bibinfo {volume} {69}},\ \bibinfo
  {pages} {731} (\bibinfo {year} {1997})}\BibitemShut {NoStop}%
\bibitem [{\citenamefont {Morlet}\ and\ \citenamefont
  {Lorenz}(1992)}]{numerical_solution_of_circle_equation}%
  \BibitemOpen
  \bibfield  {author} {\bibinfo {author} {\bibfnamefont {A.~C.}\ \bibnamefont
  {Morlet}}\ and\ \bibinfo {author} {\bibfnamefont {J.}~\bibnamefont
  {Lorenz}},\ }\bibfield  {title} {\bibinfo {title} {Numerical solution of a
  functional equation on a circle},\ }\href {https://doi.org/10.1137/0729098}
  {\bibfield  {journal} {\bibinfo  {journal} {SIAM Journal on Numerical
  Analysis}\ }\textbf {\bibinfo {volume} {29}},\ \bibinfo {pages} {1741}
  (\bibinfo {year} {1992})}\BibitemShut {NoStop}%
\bibitem [{\citenamefont {Hofstadter}(1976)}]{hofstadter_1976}%
  \BibitemOpen
  \bibfield  {author} {\bibinfo {author} {\bibfnamefont {D.~R.}\ \bibnamefont
  {Hofstadter}},\ }\bibfield  {title} {\bibinfo {title} {Energy levels and wave
  functions of {B}loch electrons in rational and irrational magnetic fields},\
  }\href {https://doi.org/10.1103/PhysRevB.14.2239} {\bibfield  {journal}
  {\bibinfo  {journal} {Physical Review B}\ }\textbf {\bibinfo {volume} {14}},\
  \bibinfo {pages} {2239} (\bibinfo {year} {1976})}\BibitemShut {NoStop}%
\bibitem [{\citenamefont {Harper}(1955)}]{harper_1955}%
  \BibitemOpen
  \bibfield  {author} {\bibinfo {author} {\bibfnamefont {P.~G.}\ \bibnamefont
  {Harper}},\ }\bibfield  {title} {\bibinfo {title} {Single {Band} {Motion} of
  {Conduction} {Electrons} in a {Uniform} {Magnetic} {Field}},\ }\href
  {https://doi.org/10.1088/0370-1298/68/10/304} {\bibfield  {journal} {\bibinfo
   {journal} {Proceedings of the Physical Society. Section A}\ }\textbf
  {\bibinfo {volume} {68}},\ \bibinfo {pages} {874} (\bibinfo {year}
  {1955})}\BibitemShut {NoStop}%
\bibitem [{\citenamefont {Mirlin}\ and\ \citenamefont
  {Fyodorov}(1994)}]{ldos_anderson_transition}%
  \BibitemOpen
  \bibfield  {author} {\bibinfo {author} {\bibfnamefont {A.~D.}\ \bibnamefont
  {Mirlin}}\ and\ \bibinfo {author} {\bibfnamefont {Y.~V.}\ \bibnamefont
  {Fyodorov}},\ }\bibfield  {title} {\bibinfo {title} {Distribution of local
  densities of states, order parameter function, and critical behavior near the
  {A}nderson transition},\ }\href {https://doi.org/10.1103/PhysRevLett.72.526}
  {\bibfield  {journal} {\bibinfo  {journal} {Phys. Rev. Lett.}\ }\textbf
  {\bibinfo {volume} {72}},\ \bibinfo {pages} {526} (\bibinfo {year}
  {1994})}\BibitemShut {NoStop}%
\bibitem [{\citenamefont {Yakubo}\ and\ \citenamefont
  {Mizutaka}(2012)}]{anderson_transition_numerics}%
  \BibitemOpen
  \bibfield  {author} {\bibinfo {author} {\bibfnamefont {K.}~\bibnamefont
  {Yakubo}}\ and\ \bibinfo {author} {\bibfnamefont {S.}~\bibnamefont
  {Mizutaka}},\ }\bibfield  {title} {\bibinfo {title} {Testing the order
  parameter of the {A}nderson transition},\ }\href
  {https://doi.org/10.1143/JPSJ.81.104707} {\bibfield  {journal} {\bibinfo
  {journal} {Journal of the Physical Society of Japan}\ }\textbf {\bibinfo
  {volume} {81}},\ \bibinfo {pages} {104707} (\bibinfo {year} {2012})},\
  \Eprint {https://arxiv.org/abs/https://doi.org/10.1143/JPSJ.81.104707}
  {https://doi.org/10.1143/JPSJ.81.104707} \BibitemShut {NoStop}%
\bibitem [{\citenamefont {Dobrosavljević}\ \emph {et~al.}(2003)\citenamefont
  {Dobrosavljević}, \citenamefont {Pastor},\ and\ \citenamefont
  {Nikolić}}]{anderson_transition_introduction}%
  \BibitemOpen
  \bibfield  {author} {\bibinfo {author} {\bibfnamefont {V.}~\bibnamefont
  {Dobrosavljević}}, \bibinfo {author} {\bibfnamefont {A.~A.}\ \bibnamefont
  {Pastor}},\ and\ \bibinfo {author} {\bibfnamefont {B.~K.}\ \bibnamefont
  {Nikolić}},\ }\bibfield  {title} {\bibinfo {title} {Typical medium theory of
  {A}nderson localization: {A} local order parameter approach to
  strong-disorder effects},\ }\href {https://doi.org/10.1209/epl/i2003-00364-5}
  {\bibfield  {journal} {\bibinfo  {journal} {Europhysics Letters}\ }\textbf
  {\bibinfo {volume} {62}},\ \bibinfo {pages} {76} (\bibinfo {year}
  {2003})}\BibitemShut {NoStop}%
\bibitem [{Note3()}]{Note3}%
  \BibitemOpen
  \bibinfo {note} {We cover the curve resulting from connecting the data
  points, not the points themselves. A single spike of height \(h\) thus
  contributes \(h / m\) boxes, not one box.}\BibitemShut {Stop}%
\bibitem [{\citenamefont {Falconer}(2003)}]{box_counting}%
  \BibitemOpen
  \bibfield  {author} {\bibinfo {author} {\bibfnamefont {K.}~\bibnamefont
  {Falconer}},\ }\href {https://doi.org/10.1002/0470013850} {\emph {\bibinfo
  {title} {Fractal Geometry}}}\ (\bibinfo  {publisher} {Wiley},\ \bibinfo
  {year} {2003})\BibitemShut {NoStop}%
\bibitem [{\citenamefont {Chhabra}\ and\ \citenamefont
  {Jensen}(1989)}]{singularity_spectrum_fractals}%
  \BibitemOpen
  \bibfield  {author} {\bibinfo {author} {\bibfnamefont {A.}~\bibnamefont
  {Chhabra}}\ and\ \bibinfo {author} {\bibfnamefont {R.~V.}\ \bibnamefont
  {Jensen}},\ }\bibfield  {title} {\bibinfo {title} {Direct determination of
  the f(\ensuremath{\alpha}) singularity spectrum},\ }\href
  {https://doi.org/10.1103/PhysRevLett.62.1327} {\bibfield  {journal} {\bibinfo
   {journal} {Phys. Rev. Lett.}\ }\textbf {\bibinfo {volume} {62}},\ \bibinfo
  {pages} {1327} (\bibinfo {year} {1989})}\BibitemShut {NoStop}%
\bibitem [{\citenamefont {Halsey}\ \emph {et~al.}(1986)\citenamefont {Halsey},
  \citenamefont {Jensen}, \citenamefont {Kadanoff}, \citenamefont {Procaccia},\
  and\ \citenamefont {Shraiman}}]{fractal_dimension_singularity_powerlaw}%
  \BibitemOpen
  \bibfield  {author} {\bibinfo {author} {\bibfnamefont {T.~C.}\ \bibnamefont
  {Halsey}}, \bibinfo {author} {\bibfnamefont {M.~H.}\ \bibnamefont {Jensen}},
  \bibinfo {author} {\bibfnamefont {L.~P.}\ \bibnamefont {Kadanoff}}, \bibinfo
  {author} {\bibfnamefont {I.}~\bibnamefont {Procaccia}},\ and\ \bibinfo
  {author} {\bibfnamefont {B.~I.}\ \bibnamefont {Shraiman}},\ }\bibfield
  {title} {\bibinfo {title} {Fractal measures and their singularities: The
  characterization of strange sets},\ }\href
  {https://doi.org/10.1103/PhysRevA.33.1141} {\bibfield  {journal} {\bibinfo
  {journal} {Phys. Rev. A}\ }\textbf {\bibinfo {volume} {33}},\ \bibinfo
  {pages} {1141} (\bibinfo {year} {1986})}\BibitemShut {NoStop}%
\bibitem [{\citenamefont {Hentschel}\ and\ \citenamefont
  {Procaccia}(1983)}]{geometric_multifractality}%
  \BibitemOpen
  \bibfield  {author} {\bibinfo {author} {\bibfnamefont {H.}~\bibnamefont
  {Hentschel}}\ and\ \bibinfo {author} {\bibfnamefont {I.}~\bibnamefont
  {Procaccia}},\ }\bibfield  {title} {\bibinfo {title} {The infinite number of
  generalized dimensions of fractals and strange attractors},\ }\href
  {https://doi.org/https://doi.org/10.1016/0167-2789(83)90235-X} {\bibfield
  {journal} {\bibinfo  {journal} {Physica D: Nonlinear Phenomena}\ }\textbf
  {\bibinfo {volume} {8}},\ \bibinfo {pages} {435} (\bibinfo {year}
  {1983})}\BibitemShut {NoStop}%
\bibitem [{\citenamefont
  {Parzen}(1999)}]{different_notions_of_markov_ergodicity}%
  \BibitemOpen
  \bibfield  {author} {\bibinfo {author} {\bibfnamefont {E.}~\bibnamefont
  {Parzen}},\ }\bibinfo {title} {Stochastic processes, 6. {M}arkov chains:
  Discrete parameter},\ in\ \href {https://doi.org/10.1137/1.9781611971125.ch6}
  {\emph {\bibinfo {booktitle} {Classics in Applied Mathematics}}}\ (\bibinfo
  {publisher} {Society for Industrial and Applied Mathematics},\ \bibinfo
  {year} {1999})\ pp.\ \bibinfo {pages} {187--275}\BibitemShut {NoStop}%
\bibitem [{\citenamefont {Isaacson}(1979)}]{geometric_ergodicity}%
  \BibitemOpen
  \bibfield  {author} {\bibinfo {author} {\bibfnamefont {D.}~\bibnamefont
  {Isaacson}},\ }\bibfield  {title} {\bibinfo {title} {A characterization of
  geometric ergodicity},\ }\href {https://doi.org/10.1007/bf00535499}
  {\bibfield  {journal} {\bibinfo  {journal} {Zeitschrift f\"ur
  Wahrscheinlichkeitstheorie und Verwandte Gebiete}\ }\textbf {\bibinfo
  {volume} {49}},\ \bibinfo {pages} {267} (\bibinfo {year} {1979})}\BibitemShut
  {NoStop}%
\bibitem [{Note4()}]{Note4}%
  \BibitemOpen
  \bibinfo {note} {The apparently most common notion of ergodicity of a Markov
  chain requires aperiodicity as well~\cite
  {different_notions_of_markov_ergodicity}. This would exclude for example our
  period-2-trajectory case. Irreducibility and aperiodicity together imply a
  unique stationary distribution of the Markov process. However, by including a
  time average in the definition of the GC distribution, we can define a
  universal distribution for a set of parameters without this
  requirement.}\BibitemShut {Stop}%
\bibitem [{\citenamefont {Walters}(1982)}]{ergodic_theory_dyn_sys}%
  \BibitemOpen
  \bibfield  {author} {\bibinfo {author} {\bibfnamefont {P.}~\bibnamefont
  {Walters}},\ }\href@noop {} {\emph {\bibinfo {title} {An introduction to
  ergodic theory}}},\ Graduate texts in mathematics ; 79\ (\bibinfo
  {publisher} {Springer},\ \bibinfo {address} {New York},\ \bibinfo {year}
  {1982})\BibitemShut {NoStop}%
\bibitem [{Note5()}]{Note5}%
  \BibitemOpen
  \bibinfo {note} {Note that power-iteration does not necessarily
  work.}\BibitemShut {Stop}%
\bibitem [{\citenamefont {Tarjan}(1972)}]{scc_alg_1}%
  \BibitemOpen
  \bibfield  {author} {\bibinfo {author} {\bibfnamefont {R.}~\bibnamefont
  {Tarjan}},\ }\bibfield  {title} {\bibinfo {title} {Depth-first search and
  linear graph algorithms},\ }\href {https://doi.org/10.1137/0201010}
  {\bibfield  {journal} {\bibinfo  {journal} {SIAM Journal on Computing}\
  }\textbf {\bibinfo {volume} {1}},\ \bibinfo {pages} {146} (\bibinfo {year}
  {1972})},\ \Eprint {https://arxiv.org/abs/https://doi.org/10.1137/0201010}
  {https://doi.org/10.1137/0201010} \BibitemShut {NoStop}%
\bibitem [{\citenamefont {Nuutila}\ and\ \citenamefont
  {Soisalon-Soininen}(1994)}]{scc_alg_2}%
  \BibitemOpen
  \bibfield  {author} {\bibinfo {author} {\bibfnamefont {E.}~\bibnamefont
  {Nuutila}}\ and\ \bibinfo {author} {\bibfnamefont {E.}~\bibnamefont
  {Soisalon-Soininen}},\ }\bibfield  {title} {\bibinfo {title} {On finding the
  strongly connected components in a directed graph},\ }\href
  {https://doi.org/https://doi.org/10.1016/0020-0190(94)90047-7} {\bibfield
  {journal} {\bibinfo  {journal} {Information Processing Letters}\ }\textbf
  {\bibinfo {volume} {49}},\ \bibinfo {pages} {9} (\bibinfo {year}
  {1994})}\BibitemShut {NoStop}%
\bibitem [{Note6()}]{Note6}%
  \BibitemOpen
  \bibinfo {note} {We use the Python library \protect \textit {NetworkX}~\cite
  {networkx} to find SCCs}\BibitemShut {NoStop}%
\bibitem [{Note7()}]{Note7}%
  \BibitemOpen
  \bibinfo {note} {The existence of such a region is constructively proven with
  analogous reasoning, if we find more than one SCC in \(G_{N}\) at any
  \(N\).}\BibitemShut {Stop}%
\bibitem [{Note8()}]{Note8}%
  \BibitemOpen
  \bibinfo {note} {There are two discretization steps, one to find the
  stationary solution of the discretized Master-equation and another one to
  find the histogram of heights from which the maximum is
  extracted.}\BibitemShut {Stop}%
\bibitem [{Note9()}]{Note9}%
  \BibitemOpen
  \bibinfo {note} {Technically, none of the discussed properties of the maps
  prevents the second derivative from changing sign between those points, such
  that attraction at the ``wrong'' eigenangle does not imply attraction
  everywhere between the eigenangles. We ignore this possibility in our
  estimate.}\BibitemShut {Stop}%
\bibitem [{Note10()}]{Note10}%
  \BibitemOpen
  \bibinfo {note} {See the condensation graph in Fig.~\ref
  {fig:nonergodic_graph_example_large} of Appendix~\ref
  {sec:non_erg_elaboration} as an example of such leakage to the blue
  supernode. Without leakage those complementary regions would themselves form
  an invariant subset.}\BibitemShut {Stop}%
\bibitem [{Note11()}]{Note11}%
  \BibitemOpen
  \bibinfo {note} {We note that, without further conditions, it is possible to
  miss invariant subspaces with a given finite resolution \(N\). For these
  missed intervals, the localization condition may not be fulfilled, and the
  ``granular'' structure may appear in the limit \(N \rightarrow \infty \). At
  the same time, since Eq.~\protect \eqref {eq:nonergodicity_localization}
  gives only a sufficient condition for localization, it is possible that the
  localized phase extends beyond the dark regions in the right panel of
  Fig.~\ref {fig:ergodicity_calc} .}\BibitemShut {Stop}%
\bibitem [{Note12()}]{Note12}%
  \BibitemOpen
  \bibinfo {note} {With our semi-analytical explanation of different regimes in
  the parameter space, we do not describe every detail of the phase diagrams.
  One particularly prominent feature is visible in the lower-left corner of the
  \(h_{\protect \rm max}\), PR, and fractal dimension diagrams: Regions of
  delocalization are broken by almost horizontal lines of localization
  (category two for \(h_{\protect \rm max}\), \(d \approx 1\) for the fractal
  dimension). This frequent change in the behavior of the distributions may be
  attributed to crossings of other types of commensurability, which we did not
  address here.}\BibitemShut {Stop}%
\bibitem [{\citenamefont {Livingston}\ \emph {et~al.}(2022)\citenamefont
  {Livingston}, \citenamefont {Blok}, \citenamefont {Flurin}, \citenamefont
  {Dressel}, \citenamefont {Jordan},\ and\ \citenamefont
  {Siddiqi}}]{continuous_error_correction_experiment}%
  \BibitemOpen
  \bibfield  {author} {\bibinfo {author} {\bibfnamefont {W.~P.}\ \bibnamefont
  {Livingston}}, \bibinfo {author} {\bibfnamefont {M.~S.}\ \bibnamefont
  {Blok}}, \bibinfo {author} {\bibfnamefont {E.}~\bibnamefont {Flurin}},
  \bibinfo {author} {\bibfnamefont {J.}~\bibnamefont {Dressel}}, \bibinfo
  {author} {\bibfnamefont {A.~N.}\ \bibnamefont {Jordan}},\ and\ \bibinfo
  {author} {\bibfnamefont {I.}~\bibnamefont {Siddiqi}},\ }\bibfield  {title}
  {\bibinfo {title} {Experimental demonstration of continuous quantum error
  correction},\ }\bibfield  {journal} {\bibinfo  {journal} {Nature
  Communications}\ }\textbf {\bibinfo {volume} {13}},\ \href
  {https://doi.org/10.1038/s41467-022-29906-0} {10.1038/s41467-022-29906-0}
  (\bibinfo {year} {2022})\BibitemShut {NoStop}%
\bibitem [{\citenamefont {Blumenthal}\ \emph {et~al.}(2022)\citenamefont
  {Blumenthal}, \citenamefont {Mor}, \citenamefont {Diringer}, \citenamefont
  {Martin}, \citenamefont {Lewalle}, \citenamefont {Burgarth}, \citenamefont
  {Whaley},\ and\ \citenamefont
  {Hacohen-Gourgy}}]{continuous_control_experiment}%
  \BibitemOpen
  \bibfield  {author} {\bibinfo {author} {\bibfnamefont {E.}~\bibnamefont
  {Blumenthal}}, \bibinfo {author} {\bibfnamefont {C.}~\bibnamefont {Mor}},
  \bibinfo {author} {\bibfnamefont {A.~A.}\ \bibnamefont {Diringer}}, \bibinfo
  {author} {\bibfnamefont {L.~S.}\ \bibnamefont {Martin}}, \bibinfo {author}
  {\bibfnamefont {P.}~\bibnamefont {Lewalle}}, \bibinfo {author} {\bibfnamefont
  {D.}~\bibnamefont {Burgarth}}, \bibinfo {author} {\bibfnamefont {K.~B.}\
  \bibnamefont {Whaley}},\ and\ \bibinfo {author} {\bibfnamefont
  {S.}~\bibnamefont {Hacohen-Gourgy}},\ }\bibfield  {title} {\bibinfo {title}
  {Demonstration of universal control between non-interacting qubits using the
  {Q}uantum {Z}eno effect},\ }\bibfield  {journal} {\bibinfo  {journal} {npj
  Quantum Information}\ }\textbf {\bibinfo {volume} {8}},\ \href
  {https://doi.org/10.1038/s41534-022-00594-4} {10.1038/s41534-022-00594-4}
  (\bibinfo {year} {2022})\BibitemShut {NoStop}%
\bibitem [{\citenamefont {Yevtushenko}\ and\ \citenamefont
  {Kravtsov}(2003)}]{yevtushenko_virial_2003}%
  \BibitemOpen
  \bibfield  {author} {\bibinfo {author} {\bibfnamefont {O.}~\bibnamefont
  {Yevtushenko}}\ and\ \bibinfo {author} {\bibfnamefont {V.~E.}\ \bibnamefont
  {Kravtsov}},\ }\bibfield  {title} {\bibinfo {title} {Virial expansion for
  almost diagonal random matrices},\ }\href
  {https://doi.org/10.1088/0305-4470/36/30/305} {\bibfield  {journal} {\bibinfo
   {journal} {Journal of Physics A: Mathematical and General}\ }\textbf
  {\bibinfo {volume} {36}},\ \bibinfo {pages} {8265} (\bibinfo {year}
  {2003})}\BibitemShut {NoStop}%
\bibitem [{\citenamefont {Yevtushenko}\ and\ \citenamefont
  {Ossipov}(2007)}]{yevtushenko_2007}%
  \BibitemOpen
  \bibfield  {author} {\bibinfo {author} {\bibfnamefont {O.}~\bibnamefont
  {Yevtushenko}}\ and\ \bibinfo {author} {\bibfnamefont {A.}~\bibnamefont
  {Ossipov}},\ }\bibfield  {title} {\bibinfo {title} {A supersymmetry approach
  to almost diagonal random matrices},\ }\href
  {https://doi.org/10.1088/1751-8113/40/18/002} {\bibfield  {journal} {\bibinfo
   {journal} {Journal of Physics A: Mathematical and Theoretical}\ }\textbf
  {\bibinfo {volume} {40}},\ \bibinfo {pages} {4691} (\bibinfo {year}
  {2007})}\BibitemShut {NoStop}%
\bibitem [{\citenamefont {Kronmüller}\ \emph {et~al.}(2010)\citenamefont
  {Kronmüller}, \citenamefont {Yevtushenko},\ and\ \citenamefont
  {Cuevas}}]{kronmuller_2010}%
  \BibitemOpen
  \bibfield  {author} {\bibinfo {author} {\bibfnamefont {S.}~\bibnamefont
  {Kronmüller}}, \bibinfo {author} {\bibfnamefont {O.~M.}\ \bibnamefont
  {Yevtushenko}},\ and\ \bibinfo {author} {\bibfnamefont {E.}~\bibnamefont
  {Cuevas}},\ }\bibfield  {title} {\bibinfo {title} {Supersymmetric virial
  expansion for time-reversal invariant disordered systems},\ }\href
  {https://doi.org/10.1088/1751-8113/43/7/075001} {\bibfield  {journal}
  {\bibinfo  {journal} {Journal of Physics A: Mathematical and Theoretical}\
  }\textbf {\bibinfo {volume} {43}},\ \bibinfo {pages} {075001} (\bibinfo
  {year} {2010})}\BibitemShut {NoStop}%
\bibitem [{\citenamefont {Kravtsov}\ \emph {et~al.}(2010)\citenamefont
  {Kravtsov}, \citenamefont {Ossipov}, \citenamefont {Yevtushenko},\ and\
  \citenamefont {Cuevas}}]{kravtsov_2010}%
  \BibitemOpen
  \bibfield  {author} {\bibinfo {author} {\bibfnamefont {V.~E.}\ \bibnamefont
  {Kravtsov}}, \bibinfo {author} {\bibfnamefont {A.}~\bibnamefont {Ossipov}},
  \bibinfo {author} {\bibfnamefont {O.~M.}\ \bibnamefont {Yevtushenko}},\ and\
  \bibinfo {author} {\bibfnamefont {E.}~\bibnamefont {Cuevas}},\ }\bibfield
  {title} {\bibinfo {title} {Dynamical scaling for critical states: {Validity}
  of {Chalker}'s ansatz for strong fractality},\ }\href
  {https://doi.org/10.1103/PhysRevB.82.161102} {\bibfield  {journal} {\bibinfo
  {journal} {Physical Review B}\ }\textbf {\bibinfo {volume} {82}},\ \bibinfo
  {pages} {161102} (\bibinfo {year} {2010})}\BibitemShut {NoStop}%
\bibitem [{\citenamefont {Kravtsov}\ \emph {et~al.}(2011)\citenamefont
  {Kravtsov}, \citenamefont {Ossipov},\ and\ \citenamefont
  {Yevtushenko}}]{kravtsov_2011}%
  \BibitemOpen
  \bibfield  {author} {\bibinfo {author} {\bibfnamefont {V.~E.}\ \bibnamefont
  {Kravtsov}}, \bibinfo {author} {\bibfnamefont {A.}~\bibnamefont {Ossipov}},\
  and\ \bibinfo {author} {\bibfnamefont {O.~M.}\ \bibnamefont {Yevtushenko}},\
  }\bibfield  {title} {\bibinfo {title} {Return probability and scaling
  exponents in the critical random matrix ensemble},\ }\href
  {https://doi.org/10.1088/1751-8113/44/30/305003} {\bibfield  {journal}
  {\bibinfo  {journal} {Journal of Physics A: Mathematical and Theoretical}\
  }\textbf {\bibinfo {volume} {44}},\ \bibinfo {pages} {305003} (\bibinfo
  {year} {2011})}\BibitemShut {NoStop}%
\bibitem [{\citenamefont {Kravtsov}\ \emph {et~al.}(2012)\citenamefont
  {Kravtsov}, \citenamefont {Yevtushenko}, \citenamefont {Snajberk},\ and\
  \citenamefont {Cuevas}}]{kravtsov_2012}%
  \BibitemOpen
  \bibfield  {author} {\bibinfo {author} {\bibfnamefont {V.~E.}\ \bibnamefont
  {Kravtsov}}, \bibinfo {author} {\bibfnamefont {O.~M.}\ \bibnamefont
  {Yevtushenko}}, \bibinfo {author} {\bibfnamefont {P.}~\bibnamefont
  {Snajberk}},\ and\ \bibinfo {author} {\bibfnamefont {E.}~\bibnamefont
  {Cuevas}},\ }\bibfield  {title} {\bibinfo {title} {L{\'e}vy flights and
  multifractality in quantum critical diffusion and in classical random walks
  on fractals},\ }\href {https://doi.org/10.1103/PhysRevE.86.021136} {\bibfield
   {journal} {\bibinfo  {journal} {Physical Review E}\ }\textbf {\bibinfo
  {volume} {86}},\ \bibinfo {pages} {021136} (\bibinfo {year}
  {2012})}\BibitemShut {NoStop}%
\bibitem [{\citenamefont {Organick}(1966)}]{organick1966fortran}%
  \BibitemOpen
  \bibfield  {author} {\bibinfo {author} {\bibfnamefont {E.}~\bibnamefont
  {Organick}},\ }\href {https://books.google.de/books?id=89BWAAAAMAAJ} {\emph
  {\bibinfo {title} {A Fortran IV Primer}}},\ Addison-Wesley series in computer
  science and information processing\ (\bibinfo  {publisher} {Addison-Wesley},\
  \bibinfo {year} {1966})\BibitemShut {NoStop}%
\bibitem [{\citenamefont {Serfozo}(2009)}]{markov_matrices}%
  \BibitemOpen
  \bibfield  {author} {\bibinfo {author} {\bibfnamefont {R.}~\bibnamefont
  {Serfozo}},\ }\href {https://doi.org/10.1007/978-3-540-89332-5} {\emph
  {\bibinfo {title} {Basics of Applied Stochastic Processes}}}\ (\bibinfo
  {publisher} {Springer Berlin Heidelberg},\ \bibinfo {year}
  {2009})\BibitemShut {NoStop}%
\bibitem [{\citenamefont {Mises}\ and\ \citenamefont
  {Pollaczek-Geiringer}(1929)}]{power_iteration_1}%
  \BibitemOpen
  \bibfield  {author} {\bibinfo {author} {\bibfnamefont {R.~V.}\ \bibnamefont
  {Mises}}\ and\ \bibinfo {author} {\bibfnamefont {H.}~\bibnamefont
  {Pollaczek-Geiringer}},\ }\bibfield  {title} {\bibinfo {title} {Praktische
  {V}erfahren der {G}leichungsauflösung},\ }\href
  {https://doi.org/https://doi.org/10.1002/zamm.19290090206} {\bibfield
  {journal} {\bibinfo  {journal} {ZAMM - Journal of Applied Mathematics and
  Mechanics / Zeitschrift für Angewandte Mathematik und Mechanik}\ }\textbf
  {\bibinfo {volume} {9}},\ \bibinfo {pages} {152} (\bibinfo {year} {1929})},\
  \Eprint
  {https://arxiv.org/abs/https://onlinelibrary.wiley.com/doi/pdf/10.1002/zamm.19290090206}
  {https://onlinelibrary.wiley.com/doi/pdf/10.1002/zamm.19290090206}
  \BibitemShut {NoStop}%
\bibitem [{\citenamefont {Bronson}\ \emph {et~al.}(2014)\citenamefont
  {Bronson}, \citenamefont {Costa},\ and\ \citenamefont
  {Saccoman}}]{power_iteration_2}%
  \BibitemOpen
  \bibfield  {author} {\bibinfo {author} {\bibfnamefont {R.}~\bibnamefont
  {Bronson}}, \bibinfo {author} {\bibfnamefont {G.~B.}\ \bibnamefont {Costa}},\
  and\ \bibinfo {author} {\bibfnamefont {J.~T.}\ \bibnamefont {Saccoman}},\
  }\bibfield  {title} {\bibinfo {title} {Chapter 4 - eigenvalues, eigenvectors,
  and differential equations},\ }in\ \href
  {https://doi.org/https://doi.org/10.1016/B978-0-12-391420-0.00004-4} {\emph
  {\bibinfo {booktitle} {Linear Algebra (Third Edition)}}},\ \bibinfo {editor}
  {edited by\ \bibinfo {editor} {\bibfnamefont {R.}~\bibnamefont {Bronson}},
  \bibinfo {editor} {\bibfnamefont {G.~B.}\ \bibnamefont {Costa}},\ and\
  \bibinfo {editor} {\bibfnamefont {J.~T.}\ \bibnamefont {Saccoman}}}\
  (\bibinfo  {publisher} {Academic Press},\ \bibinfo {year} {2014})\ \bibinfo
  {edition} {third edition}\ ed.,\ pp.\ \bibinfo {pages} {237--288}\BibitemShut
  {NoStop}%
\bibitem [{\citenamefont {Farbey}(1966)}]{scc_condensation}%
  \BibitemOpen
  \bibfield  {author} {\bibinfo {author} {\bibfnamefont {B.~A.}\ \bibnamefont
  {Farbey}},\ }\bibfield  {title} {\bibinfo {title} {Structural models: An
  introduction to the theory of directed graphs},\ }\href
  {https://doi.org/10.1057/jors.1966.32} {\bibfield  {journal} {\bibinfo
  {journal} {Journal of the Operational Research Society}\ }\textbf {\bibinfo
  {volume} {17}},\ \bibinfo {pages} {202} (\bibinfo {year} {1966})},\ \Eprint
  {https://arxiv.org/abs/https://doi.org/10.1057/jors.1966.32}
  {https://doi.org/10.1057/jors.1966.32} \BibitemShut {NoStop}%
\bibitem [{\citenamefont {Kozen}(1992)}]{partial_order_dags}%
  \BibitemOpen
  \bibfield  {author} {\bibinfo {author} {\bibfnamefont {D.~C.}\ \bibnamefont
  {Kozen}},\ }\href {https://doi.org/10.1007/978-1-4612-4400-4} {\emph
  {\bibinfo {title} {The Design and Analysis of Algorithms}}}\ (\bibinfo
  {publisher} {Springer New York},\ \bibinfo {year} {1992})\BibitemShut
  {NoStop}%
\bibitem [{\citenamefont {P{\"o}pperl}(2023)}]{thesis_paul}%
  \BibitemOpen
  \bibfield  {author} {\bibinfo {author} {\bibfnamefont {P.}~\bibnamefont
  {P{\"o}pperl}},\ }\emph {\bibinfo {title} {Dynamics of disordered and
  measured systems}},\ \href {https://doi.org/10.5445/IR/1000164650} {Ph.D.
  thesis},\ \bibinfo  {school} {Karlsruher Institut für Technologie (KIT)}
  (\bibinfo {year} {2023}),\ \bibinfo {note} {47.12.01; LK 01}\BibitemShut
  {NoStop}%
\bibitem [{\citenamefont {Hagberg}\ \emph {et~al.}(2008)\citenamefont
  {Hagberg}, \citenamefont {Schult},\ and\ \citenamefont {Swart}}]{networkx}%
  \BibitemOpen
  \bibfield  {author} {\bibinfo {author} {\bibfnamefont {A.~A.}\ \bibnamefont
  {Hagberg}}, \bibinfo {author} {\bibfnamefont {D.~A.}\ \bibnamefont
  {Schult}},\ and\ \bibinfo {author} {\bibfnamefont {P.~J.}\ \bibnamefont
  {Swart}},\ }\bibfield  {title} {\bibinfo {title} {Exploring network
  structure, dynamics, and function using {N}etwork{X}},\ }in\ \href@noop {}
  {\emph {\bibinfo {booktitle} {Proceedings of the 7th Python in Science
  Conference}}},\ \bibinfo {editor} {edited by\ \bibinfo {editor}
  {\bibfnamefont {G.}~\bibnamefont {Varoquaux}}, \bibinfo {editor}
  {\bibfnamefont {T.}~\bibnamefont {Vaught}},\ and\ \bibinfo {editor}
  {\bibfnamefont {J.}~\bibnamefont {Millman}}}\ (\bibinfo {address} {Pasadena,
  CA USA},\ \bibinfo {year} {2008})\ pp.\ \bibinfo {pages} {11 --
  15}\BibitemShut {NoStop}%
\end{thebibliography}%

\end{document}